\documentclass[
     12pt,         
     a4paper,      
     DIV14,        
     liststotoc,   
     bibtotoc,     
     idxtotoc,     
     ]{scrreprt}

\usepackage[USenglish]{babel}
\usepackage{enumerate}
\usepackage[utf8]{inputenc}
\usepackage{changepage}
\usepackage[T1]{fontenc}
\usepackage{makeidx}
\usepackage{url}
\usepackage{doc}
\usepackage{amsmath}
\usepackage{graphicx}
\usepackage[framemethod=TikZ]{mdframed}
\usepackage{natbib}
\usepackage[bookmarks=true,colorlinks,pdfpagelabels,pdfstartview = FitH,bookmarksopen = true,bookmarksnumbered = true,linkcolor = black,plainpages = false,hypertexnames = false,citecolor = black,urlcolor=black]{hyperref} 
\usepackage{psfrag}
\usepackage{multirow}
\usepackage{gensymb}
\usepackage{float}
\usepackage{titlesec}
\titleformat{\chapter}[display]
  {\normalfont\huge\bfseries}
  {\chaptertitlename\ \thechapter}{20pt}{\Huge}
\titleformat{\section}
  {\normalfont\fontsize{14pt}{16pt}\selectfont\bfseries}{\thesection}{1em}{}
\titleformat{\subsection}
  {\normalfont\fontsize{12pt}{14pt}\selectfont\itshape}{\thesubsection}{1em}{}
\titlespacing*{\section} {0pt}{2.3ex}{2.3ex plus .2ex}
\titlespacing*{\subsection}{0pt}{1.5ex}{1.5ex plus .2ex}
\DeclareUnicodeCharacter{00A0}{ }

\makeatletter
\newcommand\footnoteref[1]{\protected@xdef\@thefnmark{\ref{#1}}\@footnotemark}
\makeatother

\def \apjs {{\rm {ApJS}}}

\def \aap {{\rm {A\&A}}}

\def \apjl{\rm {ApJL}}

\def \deg {$^\circ$}
\begin{document}

\pagenumbering{gobble}

\begin{titlepage}

  \vspace*{-1.7cm}
\begin{center}
\vspace*{1cm}
\textbf{ The papers about the main work in this dissertation have been published in peer-reviewed journals. 
  }
\end{center}
\clearpage

\vspace*{-1.7cm}
\begin{center}
\vspace*{1cm}
\textbf{
  {\large Dissertation\\[3ex]
  submitted to the\\[3ex]
  Combined Faculties of the Natural Sciences and Mathematics\\[3ex]
  of the Ruperto-Carola-University of Heidelberg, Germany\\[3ex]
  for the degree of \\[3ex]
  Doctor of Natural Sciences}\\
}

\end{center}

\vfill 

\begin{center}
  \textbf{
    {\large
      Put forward by\\[3ex]
      Fabo Feng\\[3ex]
\begin{tabular}[l]{rl}
Born in: & Hubei, China\\[3ex]
Oral examination: &21$^{st}$ January, 2015
\end{tabular}
}
}
\end{center}
\clearpage\mbox{}\clearpage

\vspace*{2.0cm}
\begin{center}
  \textbf{
    {\large
       Investigations into the impact of astronomical phenomena on \\[3ex]
       the terrestrial biosphere and climate
      }
  }
\end{center}

\vfill 

\begin{center}
  \textbf{
{\large
\begin{tabular}[l]{rl}
  Referees: \quad&Dr. Coryn Bailer-Jones\\[3ex]
            &Prof. Dr. Eva Grebel\\
\end{tabular}
}
}
\end{center}
\clearpage\mbox{}\clearpage

\begin{center}
  {\large\bf Abstract}\\
\end{center}
{\small
  This thesis assesses the influence of astronomical phenomena on the Earth's biosphere and climate. I examine in particular the relevance of both the path of the Sun through the Galaxy and the evolution of the Earth's orbital parameters in modulating non-terrestrial mechanisms. I build models to predict the extinction rate of species, the temporal variation of the impact cratering rate and ice sheet deglaciations, and then compare these models with other models within a Bayesian framework. I find that the temporal distribution of mass extinction events over the past 550\,Myr can be explained just as well by a uniform random distribution as by other models, such as variations in the stellar density local to the Sun arising from the Sun's orbit. Given the uncertainties in the Galaxy model and the Sun’s current phase space coordinates, as well as the errors in the geological data, it is not possible to draw a clear connection between terrestrial extinction and the solar motion. In a separate study, I find that the solar motion, which modulates the Galactic tidal forces imposed on Oort cloud comets, does not significantly influence this cratering rate. My dynamical models, together with the solar apex motion, can explain the anisotropic perihelia of long period comets without needing to invoke the existence of a Jupiter-mass solar companion. Finally, I find that variations in the Earth's obliquity play a dominant role in triggering terrestrial deglaciations over the past 2\,Myr. The precession of the equinoxes, in contrast, only becomes important in pacing large deglaciations after the transition from the 100-kyr dominant periodicity in the ice coverage to a 41-kyr dominant periodicity, which occurred 0.7\,Myr ago.

}

\vspace*{1.0cm}
\begin{center}
{\large\bf Zusammenfassung}\\
\end{center}
{\small
  Die vorliegende Doktorarbeit behandelt den Einfluss astronomischer Phänomene auf die Biosphäre und das Klima der Erde. Ich untersuche dabei im Besonderen die Sonnenbahn durch die Galaxie und die Evolution der Erdbahnparameter hinsichtlich ihrer Beeinflussung extraterrestrischer Mechanismen. Ich erstelle Modelle zur Vorhersage der Geschwindigkeit des Artensterbens, der zeitlichen Variation von Einschlagshäufigkeiten, sowie des Abschmelzens von Eisschilden, und vergleiche diese mittels Bayesscher Statistik mit alternativen Modellen. Ich schließe daraus, daß die zeitliche Verteilung der Massensterbeereignisse innerhalb der letzten 550 Millionen Jahre durch eine Gleichverteilung ebenso gut beschrieben wird wie durch andere Modelle, wie zum Beispiel der Veränderung der sonnennahen Sternendichte aufgrund der Sonnenbahn. In Anbetracht der Unsicherheiten, die der Modellierung unserer Galaxie und der gegenwärtigen Bahndaten der Sonne anhaften, aber auch durch Meßfehler in geologischen Daten, ist es nicht möglich eine klare Verbindung zwischen terrestrischer Extinktion und Sonnenbewegung herzustellen. Im Rahmen einer separaten Untersuchung stelle ich fest, daß die Bewegung der Sonne, welche Veränderungen der Gezeitenkräfte auf die Kometen der Oortschen Wolke bedingt, keinen bedeutenden Einfluss auf die Einschlagshäufigkeit hat. In Kombination mit der Bewegung des Sonnenapex können meine dynamischen Modelle die anisotropischen Perihelien langperiodischer Kometen erklären, ohne dabei auf die Existenz eines Sonnebegleiters mit Jupitermasse abzustellen.Abschließend ermittle ich, daß Veränderungen der Erdbahnneigung innerhalb der letzten 2 Millionen Jahre eine hervorragende Rolle beim Auslösen von Eisschildschmelzen spielen. Im Gegensatz dazu wird die Präzession der Äquinoktia für das Abschmelzen erst seit 0.7 Millionen jahren bedeutsam, nämlich nach dem Übergang von der 100- zur 41-tausendjährigen Schmelzperiode.

}
\clearpage\mbox{}\clearpage

\chapter*{}
\vspace*{2.0cm}
\begin{center}
{\LARGE\bf  Dedication}\\[5ex]
{\large\it to my wife, children and parents}
\end{center}
\clearpage\mbox{}\clearpage

\end{titlepage}

\pagenumbering{roman}
\tableofcontents
\cleardoublepage
\thispagestyle{empty}
\clearpage\mbox{}\clearpage

\pagenumbering{arabic} 

\chapter{Introduction}\label{cha:introduction}
The climate, geology, and life on the Earth have changed widely and constantly over the whole of Earth's history, which is closely related to the history of the cosmos. This integrated history of the whole universe is defined as ``Big History'' by an American historian, David Gilbert Christian. As Prof. Walter Alvarez, a well-known geologist, said, Big History is ``the attempt to understand, in a unified, interdisciplinary way, the history of cosmos, Earth, life, and humanity''. To understand this Big History better, I study the biological, geological and climatic history on the Earth in the context of the history of the cosmos . 

\section{The history of the Earth}\label{sec:history}

During the evolution of the Earth over the past 4.5 billion years, its biological, geological and climate systems have interacted with each other and been affected by extraterrestrial factors as well. These interactions were amplified or reduced through certain mechanisms in these systems and left imprints in the fossil record, impact craters, ice core, seafloor sediment, etc. For example, the asteroid and comet impacts on the Earth have produced an enormous amount of craters, and about 175 of them have been confirmed by researchers. These terrestrial craters and craters on the Moon can be used to reconstruct the bombardment history of the Earth. 

\subsection{History of life}\label{sec:life}

In the history of Earth's biological system, species originate and become extinct gradually, and sometimes abruptly, starting about 550\,Myr ago (i.e. the start of Phanerozoic eon), when diverse hard-shelled animals first appeared. Since then there has been an overall increase in the diversity of life, but with significant variation \citep{sepkoski02, rohde05, alroy08sci}. The rates of appearance and disappearance of species or genera are named speciation and extinction rate, respectively. A mass extinction event occurs when the extinction rate increases with respect to the speciation rate. In the same way, a radiation or explosion of species occurs when the speciation rate is larger than the origination rate. For example, nearly all present animal phyla appeared during the Cambrian explosion which start 542\,Myr ago. About 65\,Myr ago, the well known Cretaceous–Paleogene (or K-T) extinction event removed 75\% of all species \citep{jablonski94}, including dinosaurs.

Four groups of hypotheses have been proposed to explain these abrupt events and the long term variation in biodiversity. One theory is called the ``Red Queen hypothesis'' \citep{van-valen73,benton09}, which claims that the interactions between species, such as prey-predator dynamics, are responsible for the variation in biodiversity. Another theory, named the ``Court Jester hypothesis'' \citep{barnosky01,benton09}, proposes that the biodiversity variation is triggered by environmental changes, such as plate tectonics, atmospheric composition, and global climate and climate change \citep{sigurdsson88, crowley88, wignall01, marti05, feulner09, wignall09}. Third, extraterrestrial mechanisms could be involved, either through a direct impact on life or by changing the terrestrial climate. These mechanisms include variations in Earth's orbit \citep{hays76,muller00}, solar variability \citep{shaviv03,lockwood05,lockwood07}, asteroid or comet impacts \citep{shoemaker83,alvarez80,glen94}, cosmic rays \citep{shaviv05, sloan08}, supernovae (SNe) and gamma-ray burst \citep{ellis95,melott08} (for a review see \cite{bailer-jones09}). Finally, the apparent variation may be the result of uneven preservation and sampling bias \citep{raup72, alroy96,peters05}.  

These four types of cause of terrestrial extinction are not mutually exclusive. They are all likely to have played a role at some point, and furthermore may also have interacted with each other. Given the limited geological record and the corresponding preservation bias, untangling the relevance of these different causes for an individual event, such as a mass extinction, is difficult. More promising, however, might be an attempt to identify the overall, long-term significance of these potential causes of mass extinctions\footnote{This paragraph is adapted from \cite{feng13}, and from chapter \ref{cha:biodiversity} as well}.

Many potential extraterrestrial causes of mass extinctions are associated with the stellar density local to the Sun, and thus the occurrence rate of these phenomena are modulated by the Sun's motion in the Galaxy. For example, the Oort cloud comets located in the outer solar system are likely to be perturbed by stellar encounters, and thus be injected into Earth-crossing orbits when the Sun moves into the mid-plane of the Galaxy or enters a spiral arm. I use the local stellar density as a proxy to model the rate of extraterrestrial phenomena, which are related to the Sun's motion. Comparing this model with other models, I am able to assess the influence of the Sun's motion on the Earth's biodiversity, particularly on mass extinctions (see chapter \ref{cha:biodiversity} for details). 

\subsection{Geological history}\label{sec:geological}

Activities on the Earth, such as plate tectonics, and volcanic eruptions, have been recorded in different geological features. These activities are not independent of each other. For example, the drift of plates can cause earthquakes, intensify volcanic activity and build mountains. Geological activities can also be triggered by extraterrestrial phenomena, such as asteroid impacts and tidal forces from the Sun and the Moon. These extraterrestrial effects can be directly or indirectly recorded on the Earth. One direct geological record of extraterrestrial phenomena is impact craters, which are shaped by asteroid/comet impacts.

Comet or asteroid impacts on the Earth are potentially catastrophic events which could have a fundamental effect on terrestrial life.  While at least one extinction event and associated crater is well documented -- the K-T impact from 65\,Myr ago and the Chicxulub crater \citep{alvarez80,hildebrand91} -- a clear connection between other craters and extinction events is less well established. Nonetheless, we know of around 200 large impact craters on the Earth (ref. Earth Impact Database: \url{http://www.passc.net/EarthImpactDatabase/}), and doubtless the craters of many other impacts have either since eroded or are yet to be discovered \footnote{\label{note:comet}This paragraph is adapted from section 1.1 of \cite{feng13}}.

Many studies in the past have attempted to identify patterns in the temporal distribution of craters and/or mass extinction events. Some claim to have found a periodic component in the data (e.g.\ \citealp{alvarez84,raup84,rohde05,melott11}), although the reliability of these analyses is debated, and other studies have come to other conclusions (e.g.\ \citealp{grieve96,yabushita96,jetsu00,bailer-jones09,bailer-jones11,feng13})\footnote{See footnote \ref{note:comet}.}.

Of particular interest is whether these impacts are entirely random, or whether there are one or two dominant mechanisms which account for much of their temporal distribution. Such mechanisms need not be deterministic: stochastic models show characteristic distributions in their time series or frequency spectra (e.g.\ \citealp{bailer-jones12}). We are therefore interested in accounting not for the times of individual impacts, but for the impact rate as a function of time\footnote{See footnote \ref{note:comet}.}. 

The terrestrial impact rate, together with the preservation bias (older and smaller craters are unlikely to be preserved), determines the cratering rate. Few previous studies have modeled the cratering rate while taking into account the effect of the solar motion on the flux of Earth-crossing comets, and compared models using a robust inference method. I perform my research along this line of inquiry, and will further describe the method and results of this work in chapter \ref{cha:comet}. 

\subsection{Climate history}\label{sec:climate_history}

The Earth's paleoclimatological record contains a number of patterns which are generally thought to be caused by either the internal climate system or geological and astronomical factors outside the climate system. The marine $\delta^{18}$O (a measure of the ratio of $^{18}$O and $^{16}$O) record has shown that, while the Earth has been cooling for the last 40\,Myr, the ice ages marked by relatively cold (glacial) and warm (interglacial) periods began about 2.5\,Myr ago (known as the ``onset of Northern Hemisphere glaciation''; \citealt{raymo94}).

The most probable cause of the glacial cycles is the variation in the geometry of the Earth's orbit. The Earth's rotation and orbit can be gradually changed by the gravitational forces from the Sun, the Moon and other planets. These (semi-)periodic orbital variations can change the terrestrial climate by altering the amount and location of solar radiation reaching the Earth, a phenomenon known as solar forcing. Because the larger land masses in the Northern Hemisphere can respond to temperature change more quickly, the climate system is generally thought to be most sensitive to insolation near a latitude of 65 degrees North. This hypothesis is known as {\it Milankovitch's theory } \citep{milankovitch41}.  

With the exception of the last one million years, the glacial-interglacial cycles have been dominated by periodicities corresponding to the predicted cycles of orbital obliquity (41-kyr cycle) and axial precession (23-kyr cycle) \citep{raymo94}. The 100-kyr glacial-interglacial cycles dominate the glacial variations over the past million years \citep{broecker70}, and are claimed to be forced by eccentricity which has $\sim$100\,kyr periods, i.e. 95\,kyr and 125\,kyr \citep{imbrie80,paillard98,gildor00}. However, a linear climate response to eccentricity variations can not explain the transition from the 100-kyr cycles to the 41-kyr cycles at the mid-Pleistocene (about 0.8\,Myr ago), or generate 100-kyr sawtooth variations. This 100-kyr problem can be solved either by finding non-linear climate responses to climate forcings \citep{saltzman84,paillard98,huybers05,tziperman06} or invoking other possible climate drivers, such as solar activity \citep{sharma02}, cosmic rays \citep{kirkby04}, and variations in the inclination of the Earth's orbit \citep{muller97}.

The models that aims to reconstruct the glacial-interglacial cycles always consist of climate forcings (or extraterrestrial drivers of climate change) and responses. According to current studies, climate forcings usually determine the time of occurrence of a certain climate feature, such as ice-sheet deglaciations, when the climate system reaches a threshold such as a maximum ice volume (so-called {\it pacing model}). As is mentioned by \cite{huybers11}, dozens of pacing models are proposed but with a lack of means to choose among them. My current work aims to assess the roles of different forcings in triggering deglaciations over the past 2\,Myr by comparing pacing models, which are truncated by different forcings, in a Bayesian framework. I will explain this method thoroughly in chapter \ref{cha:climate}. 

\section{Extraterrestrial influence}\label{sec:extraterrestrial}

Because the Earth is not an isolated system, various astronomical phenomena can and do influence the biological, geological and climatic systems on the Earth. I will introduce some of them: cosmic ray and gamma-ray bursts, asteroid and cometary impacts, and solar variability. 

\subsection{Cosmic ray and gamma-ray burst}\label{sec:burst}

The most energetic particles (cosmic rays) and photons (gamma rays) on Earth are not made by the Large Hadron Collider (LHC), but by celestial bodies in the universe. The cosmic rays (CRs) are believed to have originated partly from the supernovae of massive stars \citep{ackermann13} and active galactic nuclei \citep{pierre10}. There are two types of gamma-ray bursts (GRB): short GRB and longer GRB. The former is believed to result from the collision of two neutron stars or a neutron star with a black hole \citep{nakar07}, while the latter are associated with supernovae \citep{woosley06} and star formation regions \citep{pontzen10}.

Cosmic rays might influence Earth’s climate if they play a significant role in cloud formation through the formation of cloud condensation nuclei \citep{carslaw02, kirkby07}. Secondary muons resulting from cosmic rays—as well as high-energy gamma rays from SNe—could kill organisms directly or damage their DNA \citep{thorsett95, scalo02, atri12}. Similarly, GRBs near the Solar system can also have a significant impact on the Earth's atmosphere, biosphere and climate \citep{thorsett95,scalo02,thomas05,dar98,melott11}.

Many studies have attempted to connect GRBs and CR bursts to climate change and mass extinctions, based on the assumption that the Earth are more likely to be bombarded by GRBs and CRs when the Sun crosses the spiral arms or mid-plane of the disc in the Galaxy \citep{bahcall85,leitch98,shaviv03,shaviv05}. However, without taking into account the uncertainties in the Sun's current location and velocity and the parameters of the Galactic potential, these studies cannot effectively confirm or exclude periodic or non-periodic models of climate change and mass extinctions (reviewed by \citealt{bailer-jones09}). By taking into account the uncertainties in the Sun's motion, \cite{domainko13} have statistically reconstructed a history of the influx variation of gamma-rays on the Earth, on the assumption that the GRBs are strongly associated with the globular star clusters. Nevertheless, this model did not properly model CRs and the gamma-ray flux from sources outside globular clusters. Considering the complexities in the extraterrestrial mechanisms of biological change, a better approach is to model the biological effect of both GRBs and CRs, as well as other astronomical mechanisms using the local stellar density of the Sun. The influence of the Sun's motion on these phenomena is analytically modeled in section \ref{sec:OBM_biodiversity}.

\subsection{Asteroid and comet impacts}\label{sec:impact}

There are two types of near-Earth objects (NEO) which can impact the Earth: near-Earth asteroids (NEAs) and near-Earth comets (NECs). The NEAs are delivered from the main-belt asteroids through orbital resonances between planets \citep{wetherill88,froeschle94,moons96}. For example, the asteroid impactor, which is responsible for the K-T extinction event may have been a member of the Baptistina asteroid family in the main belt \citep{bottke07, reddy09}.

The cometary impactors are mainly supplied by the Trojan population \citep{levison97}, the Kuiper belt \citep{levison94}, the scattered comet disk associated with the Kuiper belt \citep{duncan97}, and the Oort Cloud \citep{weissman96}. The first two populations and part of the third population are called ecliptic comets, since they tend to have inclinations close to the plane of the ecliptic, while part of the third population and the Oort Cloud comets are referred to as nearly isotropic comets (NICs; \citealt{weissman96,levison97,duncan97}), since they tend to have isotropic angular distributions \citep{bottke02}. 

Both NEAs and NECs could impact the Earth when they are delivered to Earth-crossing orbits by the gravitational forces from resonances of planets, stellar encounters with the Solar System, and the Galactic tide. About 10-30\% of the craters on the Earth are generated by NICs \citep{shoemaker83,weissman90,zahnle98}. Although NEAs dominate the population of terrestrial impactors, NICs have relatively larger velocities and thus cause more energetic and catastrophic impact events.

These catastrophic impact events can influence the climate and damage life on the Earth directly through blast, fires, earthquakes, and tsunamis, or indirectly by releasing stratospheric dust and sulfates. For example, the dust released by the impact event would remain in the atmosphere for a year or so and result in global cooling, a consequence similar to volcanic eruptions. The threshold energy for a global impact catastrophe (defined as an event that leads to the death of 25\% of the world's population) is about $2 \times 10^5$ Mt TNT (1\,Mt TNT = $4.2\times 10^{15}$J), equivalent to a stony object with a diameter of 2\,km striking at 20\,km s$^{-1}$ \citep{chapman94}. For comparison, the impactor that caused the Chicxulub crater (diameter 180\,km; associated with the K-T extinction event) had a kinetic energy of about $10^8$ Mt TNT \citep{toon97}. 

Reconstructing the bombardment history of the Earth is necessary to associate these impact events with the paleoclimatological and paleontological records. The analysis of lunar and terrestrial craters suggests that the impact rate on the Earth could have been constant for the past $\sim$3\,Gyr and began increasing towards the present rate from about 200\,Myr ago \citep{grieve94,shoemaker98}. The variation of the terrestrial cratering rate can be caused either by the preservation bias (older and smaller craters tend to be strongly eroded by wind and rain), or the disruption of a large asteroid into an asteroid family \citep{bottke07}, or the periodic perturbation of the Oort Cloud by the Galactic tide and stellar encounters (see \citealt{feng14} and chapter \ref{cha:comet}). However, few works have distinguished between these causes by building and comparing models of them. I use a trend to model both the preservation bias and the asteroid impact rate, and model the comet impact rate based on simulations of Oort cloud comets, and compare these models using a Bayesian inference method (see chapter \ref{cha:comet}). 

\subsection{Solar variability}\label{sec:solar}

The Sun has been varied in activity and radiation over different time scales. This variability is most evident in the 11-yr cycle of the number of sunspots, which results in a 0.1\% variation (equivalent to 1\,W/m$^2$) in total solar irradiance. Although the temperature near Earth's surface changes by less than 0.1\,K in response to the variation of the total irradiance, the UV radiation has a variation one order of magnitude higher than the change of total irradiance \citep{lean95,lean05}. UV irradiance increase promotes higher ozone formation, leading to stratospheric heating through absorption of the excess UV radiation by ozone \citep{haigh94,haigh03}. The solar heating of the stratosphere may influence the tropospheric temperature through ``top-down'' mechanisms \citep{shindell99,shindell06}. All these mechanisms can amplify the global average warming by up to 20\% \citep{shindell01,palmer04}, which will influence biodiversity significantly \citep{botkin07}.

Solar activity also generates solar energetic particles (SEPs; mainly protons) which can modulate the heliosphere and thus change the influx of Galactic cosmic rays (GCRs) on the Earth. Both SEPs and GCRs could penetrate the Earth’s geomagnetic field over the poles where they enter the lower atmosphere and catalytically destroy ozone by causing ionization, dissociation, and the production of odd hydrogen and nitrogen species \citep{solomon82, jackman08}. In addition, GCRs could influence cloud formation by increasing atmospheric ion production and modulating the global atmospheric electric circuit \citep{tinsley00}. The above effects could influence the global climate and biosphere through global cooling, damaging the DNA of life forms, destroying the photosynthesis in plants, etc. 

The flux of GCRs is recorded by the concentrations of the isotopes carbon-14 ($^{14}$C), beryllium-10 ($^{10}$Be), and Chlorine-36 ($^{36}$Cl) when they initialize a sequence of nuclear interactions in tree-rings and ice \citep{bard97}. This so-called {\it cosmogenic nuclide production} records the GCR influx, which is dependent on the geomagnetic field, the solar activity, and the GCR flux injected into the solar system. Based on the assumption that the GCR flux from interstellar space is constant over the past million-year timescales, the solar activity (or strength of the heliosphere) can be reconstructed by independently deriving the paleointensity of the geomagnetic field from deep-sea sediment records \citep{sharma02}. However, the results of this approach are not conclusive because of the magnitude of the uncertainties in the data \citep{bard06}. 

\section{The role of Sun's motion around the Galactic center}\label{sec:solar_motion}
{\it This section is adapted from section 1.1 of \cite{feng13}. }\\

Many of the astronomical mechanisms mentioned above are ultimately caused by the presence of nearby stars. Stars turn supernovae, the source of gamma rays, and their remnants are a major source of cosmic rays \citep{koyama95}. Stars perturb the Oort cloud, the main source of comets in the inner solar system \citep{rampino84,sanchez01}. Broadly speaking, when the Sun is in regions of higher stellar density, it is more exposed to extraterrestrial mechanisms of biodiversity change. In its orbit around the Galaxy (once every 200--250\,Myr or so), the Sun's environment changes. For example, it oscillates about the Galactic plane with a (quasi) period of 50--75\,Myr \citep{bahcall85}, and in doing so moves through regions of more intense star formation activity in the Galactic plane. This is particularly true if the Sun crosses spiral arms \citep{gies05, leitch98}, which it may do every 100-200\,Myr or so.

Such changes in the solar environment have been used as the basis for many claims of a causal connection between the solar motion and mass extinctions and/or climate change. Typically, authors have identified a periodicity in the fossil record and then connected this to a plausible periodicity in the solar motion \citep{alvarez84, raup84, davis84, muller88, shaviv03, rohde05, melott10, melott12}. These comparisons are fraught with problems, however, some of which remained unmentioned by the authors. The first is the fact that the solar motion and past environment are poorly constrained by the astronomical data, so a wide range of plausible periods are permissible \citep{overholt09, mishurov11}, yet the coincident one is naturally chosen. Second comes the fact that the solar motion is not strictly periodic even under the best assumptions. Third, many of these studies have not performed a careful model comparison. Typically they identify the best fitting period assuming the periodic model to be true, but fail to accept that a non-periodic model might explain the data even better\citep{kitchell84, stigler87}. In some cases a significance test is introduced to exclude a specific noise model, but this is often misinterpreted, and the resulting significance overestimated.  The reader is referred to \cite{bailer-jones09} for an in-depth review and references.

To clarify the role of the Sun's motion in causing terrestrial events such as extinctions and impacts, I predict the extinction rate of species and the terrestrial impact rate using dynamical (or orbital) models, based on simulations of the solar motion. In these simulations, I take into account the uncertainties in the Sun's current phase space coordinates, and the uncertainties in the parameters of the Galaxy model. I will introduce these dynamical/orbital models in chapter \ref{cha:biodiversity} and \ref{cha:comet}. 

\section{The role of Earth's motion around the Sun}\label{sec:earth_motion}

The Earth's orbit around the Sun and its spin are perturbed by the Moon and the gas giants, leading to variations in the Earth's orbital eccentricity, obliquity or axial tilt, precession (including both axial and apsidal precessions) and orbital inclination relative to the invariant plane. The influence of the orbital variations on the Earth's climate (or ``orbital forcing'') was first studied by \citep{milankovitch41}. According to his theory, the Earth's motion could influence its climate by modulating the summer insolation at 65$^\circ$ N latitude, where the dynamics of sea-ice covers could disintegrate in late spring and summer as a result of high solar radiation \citep{kukla81}. The $\sim$100-kyr eccentricity cycles modulate the amplitude of precession ($\sim$23-kyr), thus changing the total annual/seasonal budget of solar energy, while obliquity ($\sim$41-kyr) varies the latitudinal distribution of insolation \citep{zachos01}. 

Numerous spectral analyses suggest that the 23- and 41-kyr cycles in the isotopic proxies are a linear response to variations in summer insolation at high northern latitudes (classical Milankovitch theory \citep{hays76, imbrie92,imbrie93}). However, the insolation variation caused by eccentricity cycles is not significant enough to drive the 100-kyr climate cycles linearly \citep{hays76, imbrie93}. This and other problems in the orbital theory of 100-kyr cycles could be solved by characterizing the link between eccentricity and the 100-kyr glacial cycles \citep{lisiecki10} and developing a nonlinear-response model of the climate system \citep{imbrie93,abe13}.

Many models \citep{hays76,liu98,huybers05,paillard98,berger99,tziperman06,berger88,saltzman01,ghil94,wunsch03} have been developed to solve the 100-kyr problem but without proper methods to compare them and choose the best \citep{huybers11}. In addition, many of these models either assume a connection between orbital variations and climate change, or assume that climate change is independent of orbital changes. Considering the above problems and the limitation of frequentist approach of model comparison \citep{jeffreys61,kass95,winkler72,mackay03,bailer-jones09}, I combine different orbital/non-orbital forcings with a simple model of climate response, and compare these models using a Bayesian inference method (see chapter \ref{cha:biodiversity}). 

\section{The structure of the thesis}\label{sec:structure}

The structure of this thesis is as follows. First, I introduce the Bayesian inference method and the Galaxy model which have been used in my PhD projects. The research articles (two published and one in preparation) related to the these projects are adapted to form the following three chapters. In the first project, I have modeled the mass extinction rate on the Earth by simulating the solar orbit around the Galactic center and compare this model with other models using the Bayesian inference method. In the second project, I have applied the Bayesian method to study the terrestrial impact rate based on simulations of the Oort cloud comets which are perturbed by the Galactic tide and stellar encounters. In the last project, the Bayesian inference method is applied to explore the role of different astronomical forcings in pacing the glacial-interglacial cycles over the past 2\,Myr on the Earth. Then I assess the potential improvement of these projects from future Gaia data. Finally, I discuss my findings in the last chapter.

\newpage

\chapter{Methods}\label{cha:method}
In this chapter, I introduce the Bayesian inference method proposed by \cite{bailer-jones11,bailer-jones11-err}, and describe a model of the Galatic potential, which is used to simulate the Sun's motion in the Galaxy.

\section{Bayesian model comparison}\label{sec:bayes}

\subsection{Overview}\label{sec:overview}
{\it This section is adapted from section 3.1 in \cite{feng13}. }\\

The goal of this work is to compare how well various models predict the paleontological data sets.  We do this by calculating, for a given data set, the Bayesian evidence for each model. If the models are equally probable a priori, then the one with the highest evidence is the best predictor of the data. This does not exclude the possibility that there exists a better model which we have not yet tested. But it at least allows us to conclude that the lower evidence models are neither appropriate nor sufficient explanations of the phenomenon. (Indeed, we never assume a model is ``true'', just better than the alternatives.) The modeling approach is described in full by \cite{bailer-jones11, bailer-jones11-err}, so will only be outlined here.

Let $D$ denote the paleontolgical time series, and $M$ the model. Examples of $M$ are a periodic model and a trend model, which predict a periodic and a trend variation in the time series, respectively. The probability of a model M with respect to a data set D is given, through the Bayes' rule:
\begin{equation}
  P(M|D) = \frac{P(D|M)P(M)}{P(D)},  
\label{eqn:bayes1}
\end{equation}
where $P(M|D)$ is the posterior of model $M$ given data D, $P(D|M)$ is the probability of obtaining $D$ from model $M$, $P(M)$ is the prior of model M, and $P(D)$ is a normalization factor.

Model M has a set of parameters, $\boldsymbol{\theta}$, which could be the current phase space coordinates of the Sun in the case of the model of extinction rate based on the solar orbit (orbital model). Thus the above formulae can be expanded to explicitly spill out the parameterization as follow:
\begin{equation}
  P(M|D) = \frac{\int P(D|\boldsymbol{\theta},M)P(\boldsymbol{\theta}|M)P(M)d \boldsymbol{\theta}}{P(D)}.  
\label{eqn:bayes2}
\end{equation}
Then we express the posterior distribution of model parameters, $\boldsymbol{\theta}$, as
\begin{equation}
  P(\boldsymbol{\theta}|D,M)=\frac{P(D|\boldsymbol{\theta},M)P(\boldsymbol{\theta}|M)}{P(D|M)},
  \label{eqn:post_par}
\end{equation}
where $P(D|\boldsymbol{\theta},M)$ is the {\it likelihood}, $P(D|M)$ is the so-called {\it evidence} and $P(\boldsymbol{\theta}|M)$ is the {\it prior} distribution of model parameters.

The likelihood of the model, $P(D | \boldsymbol{\theta}, M)$, is the probability of obtaining $D$ from model $M$ with its parameters set to some specific values of $\boldsymbol{\theta}$. Normally we do not know the exact values of these parameters, and the data -- being noisy and imperfectly fit by the model -- do not determine them exactly either. We therefore average the likelihood over all possible values of $\boldsymbol{\theta}$, weighting each by how plausible that value of $\boldsymbol{\theta}$ is. This weighted average is the evidence. This weighting is given by the prior probability distribution, $P(\boldsymbol{\theta} | M)$. In the case of the orbital model, where $\boldsymbol{\theta}$ is the current phase space coordinates of the Sun, $P(\boldsymbol{\theta} | M)$ is determined by the the observational uncertainties in these measurements. Mathematically the evidence is
\begin{equation}
P(D|M)=\int_{\boldsymbol{\theta}} P(D|\boldsymbol{\theta},M) P(\boldsymbol{\theta}|M) d\boldsymbol{\theta} \ \ .
\label{eqn:evidence}
\end{equation}
It gives the probability of getting the data from that the model, regardless of the specific values of the parameters, i.e.\ it measures how well the model explains the data. The absolute value of the evidence is not of interest, so we generally deal with the ratio of two evidences for two models, known as the {\em Bayes factor}. The evidence is a far better measure of the suitability of a model than is the p-value \citep{jeffreys61, kass95, winkler72, mackay03, bailer-jones09}. 

It is worth stressing again that, by averaging over the parameters, the evidence is not sensitive to model complexity per se. This is in contrast to the likelihood at the best fitting parameters (the maximum likelihood): a more complex (flexible) model will always fit the data better, and so will always deliver a higher maximum likelihood. The evidence reports the average likelihood, so it will only increase if the extra complexity gives a net benefit over the plausible parameter space. The model complexity does not then need to be considered separately in some ad hoc way.

\subsection{Likelihood calculation for discrete data sets}\label{sec:discrete}

Based on the above general rule of Bayesian statistics, we move on to specify the principle of likelihood calculation to construct a Bayesian method for model comparison. We first define two types of data sets: discrete data sets and continuous data sets. A discrete data set consists of a series of events, and each event is probabilistically interpreted as a Gaussian distribution with the mean and standard deviation equal to the time and time uncertainty of the event. In contrast a continuous data set is interpreted as one probability distribution rather than a series of Gaussian distributions. We further discuss the definitions of these two types of data sets in section \ref{sec:paleontological_biodiversity}.  

For event $j$ in a discrete data set, a Gaussian distribution gives the probability that the event, such as a mass extinction, an asteroid impact, and a deglaciation\footnote{Hereafter, we use mass extinction to explain the principle of likelihood calculation in this section. }, occurs at time $\tau_j$, given that the true time is $t_j$ and the uncertainty in our measurement is $\sigma_j$. That is,
\begin{equation}
P(\tau_j | \sigma_j, t_j) = \frac{1}{\sqrt{2\pi}\sigma_j} \, e^{-(\tau_j - t_j)^2/2\sigma_j^2} \ \ .
\label{eqn:measurement}
\end{equation}

In order to compare the measurement of this event with the predictions of a model we calculate the likelihood. This is given by an integral over the unknown true time
\begin{eqnarray}
P(\tau_j | \sigma_j, \boldsymbol{\theta}, M)&=&\int_{t_j} P(\tau_j | \sigma_j, t_j,
\boldsymbol{\theta}, M) P(t_j | \sigma_j, \boldsymbol{\theta}, M) \rm{d} t_j \nonumber \\
                            &=&\int_{t_j} P(\tau_j | \sigma_j, t_j) P(t_j | \boldsymbol{\theta}, M) \rm{d}t_j \ .
\label{eqn:likelihood_event}
\end{eqnarray}
The first term in the integral -- Eqn.~\ref{eqn:measurement} -- 
is sometimes called the measurement model. The second term is the prediction
of the time series model, i.e.\ the probability (per unit time) that a mass
extinction occurs at time $t_j$. Our time series models are, therefore,
stochastic in the sense that they do not attempt to predict when mass
extinctions occurred, but rather how the probability of occurrence of a mass
extinction varies over time. Note that the likelihood is just measuring the
degree of overlap between the data and the model predictions, averaged over
all time.

Eqn.~\ref{eqn:likelihood_event} gives the likelihood for a single event. Assuming all events are measured independently\footnote{ More precisely, the events are assumed independent given the model and its parameters. This is probably a reasonable assumption given that the events are distributed quite sparsely over the Phanerozoic, and that the separations between them are generally much longer than their substage durations.}, then the likelihood for all the data is just the product of the event likelihoods
\begin{equation}
P(D|\boldsymbol{\theta}, M) = \prod_j P(\tau_j| \sigma_j,\boldsymbol{\theta}, M)
\label{eqn:likelihood_disc}
\end{equation}
where $D=\{\tau_j\}$. For the sake of the likelihood and evidence calculation we do not consider the $\{\sigma_j\}$ as data, although they are of course measured. That is because $D$ is defined as just those quantities which are predicted by the measurement model.

\subsection{Likelihood calculation for continuous data sets}\label{sec:continuous}

Mathematically the likelihood for the continuous data is very similar as in the discrete case, but the interpretation is different. 

Consider the measurement model, $P(\tau | \sigma, t)$, in Eqn.~\ref{eqn:measurement} (we consider just one event so drop the subscript $j$).  We have interpreted this as the probability of a discrete extinction event being measured at $\tau$, but we could equivalently interpret it as the probability density (i.e.\ probability per unit time) of extinction at time $\tau$. Now, rather than characterizing the probability density as a Gaussian with mean $t$ and standard deviation $\sigma$, we could consider an arbitrary function, characterized by a series of top-hat functions, $\{p_i\}$ (a histogram), each top-hat characterized by a center $t_i$, height $r_i$, and width $\delta_i$.  We can then replace $P(\tau | \sigma, t)$ with $\sum_i p_i(\tau | t_i, \delta_i)$ where
\begin{eqnarray}
   p_i(\tau | t_i, \delta_i) =  
\left\{
     \begin{array}{ll}
          r_i & \mbox{\rm when}~~ t_i - \delta_i/2 < \tau < t_i + \delta_i/2\\
          0   & \mbox{\rm otherwise \ .}
     \end{array}
\right.
\end{eqnarray}
We now see that
\begin{equation}
\lim_{\delta_i \rightarrow 0} \sum_i p_i(\tau | t_i, \delta_i) = E(t)
\end{equation}
i.e.\ we get a continuous function of the variation of the extinction probability with time $t$ (as $\tau$ and $t$ become equivalent in this limit). The likelihood is then
\begin{equation}
P(D|\boldsymbol{\theta}, M) = \int_{t} E(t) P(t | \boldsymbol{\theta}, M) \rm{d}t \ .
\label{eqn:likelihood_cont}
\end{equation}
In practice we characterize $E(t)$ using the extinction rate, $r_j$, tabulated at each time $\tau_j$, which is equivalent to assuming that extinction rate is constant over the substage (or that we have zero uncertainties on the measured times). 

We can actually apply this interpretation to the discrete data sets too. In both cases, the data provide the variation of extinction probability (per unit time) as a function of time, or something proportional to that. The proportionality constant is irrelevant, because we keep the data fixed when comparing different models using the evidence.

\subsection{Numerical calculation of the evidence}\label{sec:calcevi}

The integral in Eqn.~\ref{eqn:evidence} is a multidimensional integral over the parameter space, and cannot be calculated analytically. As in \cite{bailer-jones11}, we estimate it using a Monte Carlo method, by drawing model parameters at random from the prior distribution and calculating the likelihood at each. If the set of $N$ parameter draws is denoted $\{\boldsymbol{\theta}\}$, then Eqn.~\ref{eqn:evidence} can be approximated as the average likelihood
\begin{equation}
P(D | M) \, \simeq \, \frac{1}{N}  \sum_{\boldsymbol{\theta}} P(D | \boldsymbol{\theta}, M) \ \ .
\label{eqn:evidencenum}
\end{equation}
In the following simulations we adopt $N=$\,10\,000 unless noted otherwise. 

\section{Galaxy model}\label{sec:galaxy}
{\it This section is adapted from section 3.1 of \cite{feng14}. }\\

We adopt a Galactic potential with three components, namely an axisymmetric disk and a spherically symmetric halo and bulge
\begin{equation}
  \Phi_{\rm sym}=\Phi_b+\Phi_h+\Phi_d
  \label{eqn:Phi_sym}
\end{equation}
(this is same model as in \cite{feng13, feng14}, and is used in chapter \ref{cha:biodiversity} and \ref{cha:comet}). The components are defined (in cylindrical coordinates) as
\begin{eqnarray}
  \Phi_{b,h}&=&-\frac{GM_{b,h}}{\sqrt{R^2+z^2+b_{b,h}^2}},\\
  \Phi_{d}&=&-\frac{GM_{d}}{\sqrt{R^2+(a_d+\sqrt{(z^2+b_d^2)})^2}},
  \label{eqn:Phi_component}
\end{eqnarray}
where $R$ is the disk-projected galactocentric radius of the Sun and $z$ is its vertical displacement above the midplane of the disk. $M$ is the mass of the component, $b$ and $a$ are scale lengths, and $G$ is the gravitational constant.  We adopt the values of these parameters from \citet{sanchez01}, which are listed in Table \ref{tab:model_par}.

\begin{table}[ht!]
  \centering
    \caption{The parameters of the Galactic potential model for the symmetric component \citep{sanchez01}, the arm \citep{cox02,wainscoat92}, and the bar \citep{dehnen00}.}
    \label{tab:model_par}
    \begin{tabular}{@{}ll@{}}
      \hline
      component& parameter value \\\hline
      Bulge    & $M_b=1.3955 \times 10^{10}~M_\odot$ \\
      & $b_b =0.35$\,kpc \\
      Halo     & $M_h=6.9766\times 10^{11}~M_\odot$\\
      & $b_h=24.0$\,kpc\\
      Disk     & $M_d=7.9080\times 10^{10}~M_\odot$\\
      & $a_d=3.55$\,kpc\\
      & $b_d=0.25$\,kpc\\
      Arm    & $\zeta=15^\circ$\\
      & $R_{\rm min}=3.48$\,kpc\\
      & $\phi_{\rm min}=-20^\circ$\\
      & $\rho_0=2.5 \times 10^7 M_\odot {\rm kpc}^{-3}$\\
      & $r_0=8$\,kpc\\
      & $R_s=7$\,kpc\\
      & $H=0.18$\,kpc\\
      & $\Omega_s=20$\,km$s^{-1}$/kpc\\
      bar   & $R_b/R_{\rm CR}=0.8$\\
      & $\alpha=0.01$\\
      & $R_{\rm CR}=R_\odot(t=0~{\rm Myr})/2$\\
      & $\alpha=0.01$\\
      & $\Omega_b=60$\,km$s^{-1}$/kpc\\
      \hline
    \end{tabular}
\end{table}

In the sections of sensitivity test in chapter \ref{cha:biodiversity} and \ref{cha:comet}, we will add to these non-axisymmetric and time-varying components due to spiral arms and the Galactic bar, to give the new potential
\begin{equation}
  \Phi_{\rm asym}=\Phi_{\rm sym}+\Phi_{\rm arm}+\Phi_{\rm bar} ~,
  \label{eqn:Phi_asym}
\end{equation}
where $\Phi_{\rm arm}$ is a potential of two logarithmic arms from
\citet{wainscoat92} with parameters given in \citet{feng13}, and
$\Phi_{\rm bar}$ is a quadrupole potential of rigid rotating bar from
\citet{dehnen00}. 

The geometry of the arm is
\begin{equation}
  \phi_s(R) = \log(R/R_{\rm min})/\tan(\zeta)+\phi_{\rm min},
  \label{eqn:phi_spiral}
\end{equation}
where $\zeta$ is the pitch angle, $R_{\rm min}$ is the inner radius,
and $\phi_{\rm min}$ is the azimuth at that inner radius. 
A default pattern speed of $\Omega_p=20$ km~s$^{-1}$~kpc$^{-1}$ is adopted
\citep{martos04,drimmel00}. The corresponding potential of this arm model is
\begin{eqnarray}
  \Phi_{arm} &=& -\frac{4 \pi G H}{K_1 D_1}\rho_0 e^{-\frac{R-r_0}{R_s}}\nonumber\\
            &&\times \cos(N[\phi-\phi_s(R,t)])\left[\rm{sech}\left(\frac{K_1z}{\beta_1}\right)\right]^{\beta_1}~,
  \label{eqn:Phi_arm}
\end{eqnarray}
where 
\begin{eqnarray*}
  K_1&=&\frac{N}{R \sin \zeta},\\
  \beta_1 &=& K_1 H (1+0.4 K_1 H),\\
  D_1 &=&\frac{1+K_1 H +0.3 (K_1 H)^2}{1+0.3 K_1 H},
\end{eqnarray*}
and $N$ is the number of spiral arms.
The parameters in equation \ref{eqn:Phi_arm} are given in Table \ref{tab:model_par}.

The bar potential is a 2D quadrupole \citet{dehnen00}. Because the
Sun always lies outside of the bar, we adopt the potential
\begin{equation}
\Phi_{bar}=-A_b \cos[2(\phi-\Omega_b t-\phi_{\rm min})] \left[\left(\frac{R}{R_b}\right)^3 - 2\right]  \ \ \ R \geq R_b
\label{eqn:Phib_geqRb}
\end{equation}
where $R_b$ and $\Omega_b$ are the size and pattern speed of the bar
respectively and $\phi_{\rm min}$ is the bar angle. We assume that the spiral arms start from the ends of the major axis of the bar.
We only consider the barred state and ignore the evolution of the bar, so we
adopt a constant amplitude for the quadrupole potential, i.e.\ $A_b = A_f
$, in equation (3) of \citet{dehnen00}. $A_f$ is determined by the definition of the bar strength
\begin{equation}
\alpha\equiv 3~\frac{A_f}{v^2}\left(\frac{R_b}{R}\right)^3,
\label{eqn:bar_strength}
\end{equation}
where $R$ and $v$ are the current galactocentric distance of the Sun and
the corresponding local circular velocity. The fixed bar strength is given in Table \ref{tab:model_par}, from which we calculate $A_f$ and hence $A_b$.

\newpage

\chapter{Assessing the influence of astronomical phenomena on biodiversity}\label{cha:biodiversity}
{\it This chapter is based on previously published work \citep{feng13}. }
\section{Chapter summary}\label{sec:summary_biodiversity}
The terrestrial fossil record shows a significant variation in the extinction and origination rates of species during the past half billion years. Numerous studies have claimed an association between this variation and the motion of the Sun around the Galaxy, invoking the modulation of cosmic rays, gamma rays and comet impact frequency as a cause of this biodiversity variation. However, some of these studies exhibit methodological problems, or were based on coarse assumptions (such as a strict periodicity of the solar orbit).  Here we investigate this link in more detail, using a model of the Galaxy to reconstruct the solar orbit and thus a predictive model of the temporal variation of the extinction rate due to astronomical mechanisms.  We compare these predictions as well as those of various reference models with paleontological data.  Our approach involves Bayesian model comparison, which takes into account the uncertainties in the paleontological data as well as the distribution of solar orbits consistent with the uncertainties in the astronomical data.  We find that various versions of the orbital model are not favored beyond simpler reference models.  In particular, the distribution of mass extinction events can be explained just as well by a uniform random distribution as by any other model tested.  Although our negative results on the orbital model are robust to changes in the Galaxy model, the Sun's coordinates and the errors in the data, we also find that it would be very difficult to positively identify the orbital model even if it were the true one. (In contrast, we do find evidence against simpler periodic models.)  Thus while we cannot rule out there being some connection between solar motion and biodiversity variations on the Earth, we conclude that it is difficult to give convincing positive conclusions of such a connection using current data.  

\section{\label{sec:introduction_biodiversity}Introduction}

\subsection{Background}

Over the course of Earth's history, evolution has produced a wide variety of life. This is particularly apparent from  around 550\,Myr ago -- the start of the Phanerozoic eon -- when hard-shelled animals first appeared and were preserved in the fossil record. Since then we observe a general increase in the diversity of life, but with significant variation superimposed \citep{sepkoski02, rohde05, alroy08sci}. The largest and most rapid decreases in biodiversity -- defined here are the number of genera extant at any one time -- are referred to as mass extinctions.

The cause of these variations in biodiversity in general, and mass extinctions in particular, have been the subject of intense study and speculation for over a century. Many mechanisms have been proposed for the observed variation, which we can place into four groups.

First, the variations are the result of inter-species interactions. Species
compete for limited resources, and as one species evolves to compete in this
struggle for survival, so other species will evolve too. This idea has been referred to as the ``Red Queen hypothesis'' \citep{van-valen73, benton09} (in reference to the Red Queen's race in Lewis Carroll's {\em Through the Looking Glass}, where Alice must run just to keep still). Recent ecological studies indicate that the interaction between species can minimize competition and enhance biodiversity \citep{sugihara09}, although it is not obvious that these biotic factors are the main cause of large-scale patterns of biodiversity \citep{benton09, alroy08}.

Second, the environment changes with time, and species will evolve in response
to this. This ideas is sometimes called the ``Court Jester hypothesis'' \citep{barnosky01,benton09}.  Some of these (abiotic) geological changes are relatively slow, such as plate tectonics, atmospheric composition, global climate \citep{sigurdsson88, crowley88, wignall01, marti05, feulner09, wignall09}. Others may be more rapid. Large-scale volcanism, for example, would inject dust, sulfate aerosols and carbon dioxide into the atmosphere, resulting in a short-term global cooling, reduced photosynthesis, long periods of acid rain, and resulting ultimately in a long-term global warming (on a timescale of $10^5$ years) \citep{marti05}.

Third, extraterrestrial mechanisms could be involved, either through a direct impact on life or by changing the terrestrial climate. Variations in the Earth's orbit (mostly its eccentricity) over ten to one hundred thousand year time scales are responsible for the ice ages \citep{hays76, muller00}. Extraterrestrial mechanisms on longer time scales could also play a role. These include solar variability \citep{shaviv03,lockwood05,lockwood07}, asteroid or comet impacts \citep{shoemaker83,alvarez80,glen94}, cosmic rays \citep{shaviv05, sloan08}, supernovae (SNe) and gamma-ray burst (GRBs) \citep{ellis95,melott08,domainko13} (for a review see \cite{bailer-jones09}). For example, cosmic rays might influence the Earth's climate if they play a significant role in cloud formation (through the formation of cloud condensation nuclei) \citep{carslaw02,kirkby08}.
Secondary muons resulting from cosmic rays -- as well as high energy gamma rays
from SNe -- could kill organisms directly or damage their DNA
\citep{thorsett95, scalo02, atri12}.

Finally, the apparent variation may be, in part, the result of uneven preservation and sampling bias \citep{raup72, alroy96, peters05}. Fossilization is relatively rare and some animals are more likely to be preserved than others.  Furthermore, the degree of preservation of marine speices (more common in the fossil record) depends on the amount of continental outcrop available at any time, and this depends on the sea level \citep{hallam89, holland12}.  The number of species or genera living at any one time is not observed but must be reconstructed from the times at which species appear and disappear, which implies some kind of sampling or modelling. This can introduce a bias, although it may be diminished to some degree by various techniques \citep{alroy01, alroy10}.

These four types of cause of biodiversity variation are not mutually exclusive. They probably all acted at some point, and will also have interacted. For example, an asteroid impact could release so much carbon dioxide that long-term global warming has the biggest impact on biodiversity. Alternatively, a cool period would lower sea levels, leaving less continental shelf for the preservation of marine fossils, even though biodiversity itself may be unchanged.

Given the limited geological record, untangling the relevance of these different causes in most individual cases is difficult, if not impossible. More promising might be an attempt to identify the overall, long-term significance of these potential causes. The goal of this article is to do that for extraterrestrial phenomena. 

As is mentioned in section \ref{sec:solar_motion}, and also in \cite{feng13}, many of the above astronomical phenomena are ultimately caused by the presence of nearby stars, and thus the rates of these phenomena are modulated by the path of the Sun in the Galaxy. Many studies have assumed a causal connection between the solar motion and mass extinction events, and seek common periodicities in them \citep{alvarez84, raup84, muller88, shaviv03, rohde05,melott12}. These studies have also assumed that the solar motion is strictly periodic, and ignored the uncertainties in the Sun's current space coordinates (i.e.\ ``initial conditions''), and in parameters of Galaxy models. These studies have been reviewed by \cite{bailer-jones09}. 

\subsection{Overview}\label{sec:overview_biodiversity}

Here we attempt a more systematic assessment of the possible role of the solar orbit in modulating extraterrestrial extinction mechanisms. Our approach is new in a number of respects, because we: (1) do a numerical reconstruction of the solar orbit (rather than just assuming it to be periodic); (2) take into account the observational uncertainties in that reconstruction; (3) use models which predict the variation of probability of extinction with time (rather than assuming that extinction events occur deterministically, for instance); (4) do proper model comparison (rather than using p-values in an over-simplified significance test); (5) compare not only the orbital model with the fossil record but also numerous reference (non-orbital) models, such as periodic, quasi-periodic, trend, periodic with constant background, etc.

Our method is as follows. Adopting a model for the distribution of mass in the Galaxy, we reconstruct the solar orbit over the past 550\,Myr by integrating the Sun's trajectory back in time from the current phase space coordinates (position and velocity). This gives us a time series of how the stellar density in the vicinity of the Sun has varied over the Phanerozoic. We then assume that this density is approximately proportional to the terrestrial extinction probability (per unit time). That is, we adopt a non-specific kill mechanism linking the solar motion to terrestrial biodiveristy.  This is naturally a strong and rather simple assumption, but it should be emphasized that we are interested in the overall plausibility of extraterrestrial phenomena rather than trying to identify a specific cause of individual extinction events. The resulting time series is then compared to several different reconstructions of the biodiversity record.

A significant source of uncertainty in the reconstructed solar orbit is the current phase space coordinates (or ``initial'' conditions). We therefore sample over these to build up a set of (thousands of) possible solar orbits, and compare each of these with the data. The comparison is done by calculating the likelihood of the data for each orbit. Rather than finding the single most likely orbit, we calculate the average likelihood over all orbits. This is important, because it properly takes into account the uncertainties (whereas selecting the single most likely orbit would ignore them entirely). Indeed, these initial conditions can be considered as the six parameters of this orbital model (for a fixed Galactic mass distribution). We are, therefore, averaging the likelihood for this model over the prior plausibility of each of its parameters. This average -- or marginal -- likelihood is often called the ``evidence''.  This is just the standard, Bayesian approach to model assessment, which avoids the various flaws of hypothesis testing \citep{jeffreys61, winkler72, mackay03}, but unfortunately it has seen little use in this field of research.

The next step is to compare this evidence with that calculated for various reference models (sometimes called ``noise'' or ``background'' models, depending on the context). One such model is a purely sinusoidal model, parameterized by an amplitude, period and phase. We generate a large number of realizations of the model for different combinations of the parameters, calculate the likelihood of the data for each, and average the results. This averaging plays the crucial role of accommodating the complexity of the model. A complex model with lots of parameters can often be made to fit an arbitrary data set well. That is, it will give a high maximum likelihood. But this does {\em not} make it a good model, precisely because we know that it could have been made to fit any data set well! Such models are highly tuned, so while the maximum likelihood fit may be very good, a small perturbation of the parameters results in poor predictions. Unless supported by the data very well, such models are less plausible. A simpler model, in contrast, may not give such an optimal fit, but it is typically more robust to small perturbations of the model parameters or the data, so gives good fit over a wider portion of the parameter space. The model evidence embodies and quantifies this trade-off, which is why it -- rather than the maximum likelihood -- should be used to compare models.

We have selected four data sets for our study. The first two are compilations of the variation of extinction rate over time, from \cite{rohde05} and \cite{alroy08sci}. In the latter we use the extinction rate standardized to remove the sampling bias. Both report a magnitude as a function of time.
The second two data sets just record the time of mass extinction events. Here we take the times of the ``big 5'' mass extinctions and 18 mass extinctions identified by \cite{bambach06} based on Sepkoski's earlier work \citep{sepkoski86}.  Each mass extinction is represented as a (normalized) Gaussian on the time axis, the mean representing the best estimate of the date of the event and the standard deviation the uncertainty. We refer to these as ``discrete'' data sets, as they just list the discrete dates at which events occur (we do not use any magnitude information). The two rate data sets we therefore refer to as ``continuous'' (even though in practice the rates are also recorded at discrete time points).

This chapter is arranged as follows. We first introduce the data sets.
In Section~\ref{sec:tsmodels_biodiversity} introduce time series models which will be compared using the Bayesian inference method described in section \ref{sec:bayes}. In Section~\ref{sec:solarorbit_biodiversity} we describe how we reconstructed the solar orbit, and also quantify the degree of periodicity typically present (as a strict periodicity has often been assumed in the past).  In Section~\ref{sec:result_biodiversity} we calculate and compare the evidences for the various models and data sets and test the sensitivity of the results to the model parameters and uncertainties in the data.  We conclude in Section~\ref{sec:conclusion_biodiversity}.

\section{\label{sec:paleontological_biodiversity} Paleontological data}

We adopt four data sets: two discrete time series (sequence of time points
with age uncertainties) giving the dates of mass extinctions, and two
continuous time series giving the smoothed and normalized extinction rate as a
function of time.

\subsection{\label{sec:datadiscrete_biodiversity} Discrete data sets}

Sepkoski and others have identified five extinction events to be ``mass extinctions'' \citep{sepkoski86}, often referred to as the ``big five''. Other studies have identified different candidates for these, or have identified a ``big $N$'' for some other value of $N$. For example, Bambach et al.\ identify three mass extinction events as being globally distinct \citep{bambach04}.  Here we adopt a set of 18 mass extinction events (or B18) selected by \cite{bambach06} using an updated Sepkoski genus-level database. They are consistently identifiable in different biodiversity data sets and when using different tabulation methods. The second of our discrete data sets is the ``big five'' as identified from among the B18. Other choices of events are of course possible and our results will, in general, depend on this choice (although as we'll see the results are rather consistent).

\begin{figure}[ht!]
  \centering
  \includegraphics[scale=0.7]{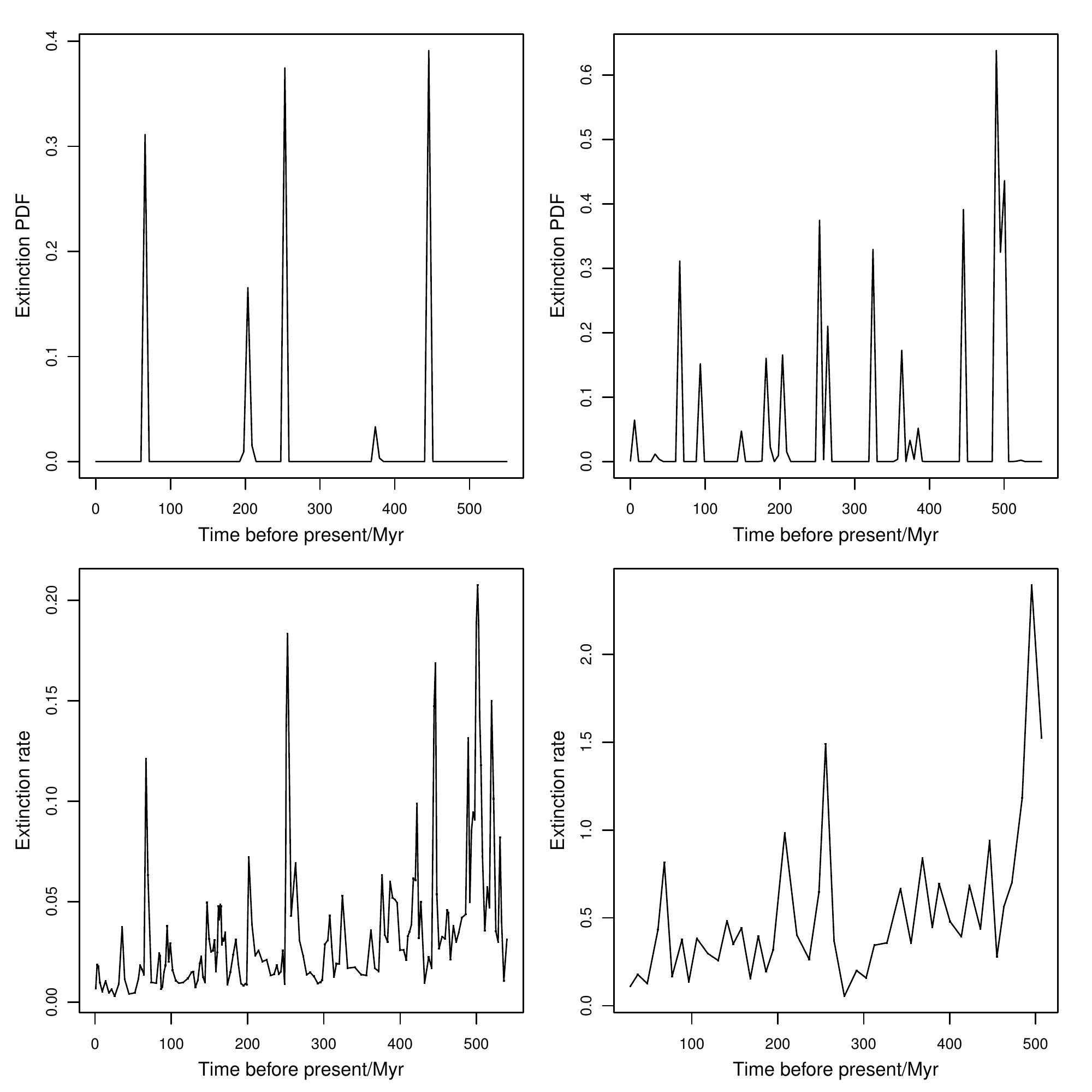}
  \caption{The four data sets used in this study. The top row shows the discrete data sets: the B5 (left) and B18 (right) mass extinction events. These can be interpreted as the extinction probability density function (PDF), which is proportional to the extinction fraction per unit time (i.e.\ a rate).  The bottom row shows the continuous data sets, which give the extinction rate: RM (left) and A08 (right).}
\label{fig:data} 
\end{figure}

The times and durations of the events are listed in Table~\ref{tab:ext-b5-b18}.  The time, $\tau$, is the mid-point between the start age and end age of the substage in which the extinction occurred, and the substage duration, $d$, is the difference between these. The geological record does not resolve the extinction event, so the extinction presumably  took place more rapidly than this substage duration. In that case $\tau$ is our best estimate of the true (but unknown) time, $t$, at which the extinction occurred, and $d$ is a measure of our uncertainty in this estimate. Uncertainty is represented by probability, so we interpret an ``event'' as the probability distribution $P(\tau | t, d)$. This is the probability that we would measure the event time as $\tau$, given $t$ and $d$.  We could represent this as a rectangular (``top-hat'') distribution of mean $t$ and width $d$, but this assigns exactly zero probability outside the substage duration, which implies certainty of the start and end ages.  Even though their (relative) ages have uncertanties which are less the
  event duration, we nonetheless accommodate some uncertainty in these start
  and end ages by considering each event to be a Gaussian distribution with
mean $t$, and standard deviation $\sigma$ equal to the standard deviation of
the rectangular distribution, which is $\sigma = d/\sqrt{12}$.  This
  Gaussian distribution is broader than the corresponding rectangular distribution.  Note that the intensity of the mass extinction is not taken into account. A Gaussian is normalized, so its peak value is determined by its standard deviation (Figure~\ref{fig:data}).\footnote{One might think that a more natural interpretation of an event is $P(t | \tau, d)$, the probability that the true event occurs at $t$ given the measurements. But here we are considering the measurement model (or noise model), that is, given some true time of the event, what possible times might we measure, the discrepancy arising on account of the finite precision of our measurement process.}

\begin{table}
\caption{The B18 mass extinction events, with the B5 shown in bold. BP = before present.}
\centering
\begin{tabular}{l c}
\hline
\hline
Time ($\tau$) / Myr BP & Substage duration ($d$) / Myr\\
\hline
3.565&3.53\\
35.550&3.30\\
{\bf 66.775}&{\bf 2.55}\\
94.515&2.03\\
146.825&2.65\\
182.800&7.00\\
{\bf 203.750}&{\bf 8.30}\\ 
{\bf 252.400}&{\bf 2.80}\\
262.950&5.10\\
324.325&4.15\\
361.600&4.80\\
{\bf 376.300}&{\bf 3.60}\\
386.925&3.25\\
{\bf 445.465}&{\bf 3.53}\\
489.350&2.10\\
495.725&2.15\\
499.950&2.10\\
519.875&2.75\\
\hline 
\end{tabular}
\label{tab:ext-b5-b18}
\end{table}

\subsection{\label{sec:continuous_biodiversity} Continuous data sets}

The discrete data sets are naturally biased in that they only select periods of high extinction rate. It may well be that extraterrestrial phenomena are only relevant in causing (or contributing to) mass extinctions, but a priori it is natural to ask how the overall extinction rate varies.  The extinction rate, $E(t)$, is the fraction of genera which go extinct in a stratigraphic substage divided by its duration. This is directly proportional to the variation of extinction probability per unit time.  For one of our extinction rate data sets we use the linearized and interpolated data set constructed by \cite{rohde05} as reported in \cite{bambach06}. We denote this RM. The other data set is the ``three-timer'' extinction rate from the Paleobiology database\footnote{paleodb.org} \citep{alroy08sci}. The data are binned into 48 intervals averaging 11\,Myr in duration. The counts are derived from 281\,491 occurrences of 18\,541 genera within 42\,627 fossil collections. We use the data set processed using their subsampling method in order to reduce the sampling bias, and denote this A08.  Both data sets are reported as lists of extinction rates at specific times, $\{r_j, \tau_j\}$. These two continuous data sets are plotted in the lower row of Figure~\ref{fig:data}.

\section{Time series models}\label{sec:tsmodels_biodiversity} 

After defining the data sets, we build time series models to predict the extinction rate. The time series models appear in the equation for the likelihood in the form $P(t | \boldsymbol{\theta}, M)$ (see Eqn. \ref{eqn:likelihood_disc} and \ref{eqn:likelihood_cont}), i.e.\ the extinction probability (per unit time) as a function of time predicted by model $M$ at parameters $\boldsymbol{\theta}$. It is important to realize that this probability density function (PDF) over $t$ is normalized, i.e.\ integrates over all time to unity. This is key to model comparison, because a model which assigns a lot of probability to extinctions at some particular time must necessarily assign lower probability elsewhere. This follows because we are not trying to model the absolute value of the extinction rate, but just its relative variations.

In addition to specifying the functional form of the models we must also specify the prior probability distribution of the model parameters, $P(\boldsymbol{\theta} | M)$ (see Eqn. \ref{eqn:post_par}). This describes our prior knowledge of the relative
probability of different parameter settings. For example, given the time scale
in the data, we are not interested in models with time scales less than a few
million years or more than a few hundred million years. It is often difficult to be precise about priors, and the evidence and therefore Bayes factors often depend on the choice. 
The choice of prior must therefore be considered part of the model (e.g.\
``periodic model with permissible periods between 10 and 100\,Myr'' is
distinct from ``periodic model with permissible periods between 50 and
60\,Myr''). We  investigate the sensitivity of the results to
changes in the prior in Section~\ref{sec:sensitivity_biodiversity}.
Except for the orbital model, we adopt a uniform prior over all model parameters over the range specified in Table~\ref{tab:prior}.

\begin{table*}
\centering
\caption{The mathematical form of the time series models and their corresponding parameters. Time $t$ increases into the past and $P_u(t|\boldsymbol{\theta}, M)$ is the unnormalized extinction probability density predicted by the model.}
\begin{tabular}{l c r}
\hline
\hline
model name&$P_u(t|\boldsymbol{\theta},M)$&parameters\\
\hline
Uniform&1&none\\
RNB/RB&$\sum_{n=1}^{N}\mathcal{N}(t; \mu_n,\sigma)$+$B$&$\sigma$, $N$, $B$\\
PNB/PB&$1/2\{\cos[2\pi(t/T+\beta)]+1\}$+$B$&$T$, $\beta$, $B$\\
QPM&$1/2\{\cos[2\pi t/T + A_Q \cos(2 \pi t/T_Q) +\beta]+1\}$&$T$, $\beta$, $A_Q$, $T_Q$\\
SP&$[1+e^{(t-t_0)/\lambda}]^{-1}$&$\lambda$, $t_0$\\
SSP&PNB+SP&$T$, $\beta$, $\lambda$, $t_0$\\
OM(P)/SOM(P)&$n(\overrightarrow{r_\odot}(t),\overrightarrow{v_\odot}(t))$&$\overrightarrow{r_\odot}(t=0),~\overrightarrow{v_\odot}(t=0)$\\
\hline
\end{tabular}
\label{tab:models}
\end{table*}

\begin{table*}
\centering
\caption{Range of parameters adopted in the model prior parameter distributions.
Except for OM(P)/SOM(P), a uniform prior for all parameters for all models is adopted which is constant inside the range shown, and zero outside. The prior PDF of parameters in the OM(P)/SOM(P) model is Gaussian and specified by the uncertainties in the initial conditions.}
\begin{tabular}{l c}
\hline
\hline
model name & range of prior \\
\hline
Uniform& None\\
RNB/RB & $\sigma=10$~Myr, $N\in\{5,18\}$, $B\in\{0,\frac{1}{\sqrt{2\pi}\sigma}\}$\\
PNB/PB & $10<T<100$, $0<\beta<2\pi$, $B\in[0,1]$\\
QPM &$10<T<100$, $0<\beta<2\pi$, $0<A_Q<0.5$, $200<T_Q<500$\\
SP &$-100<\lambda<100$, $100<t_0<500$\\ 
SSP &$10<T<100$, $0<\beta<2\pi$, $-100<\lambda<100$, $100<t_0<500$\\ 
OM(P)/SOM(P) & initial conditions (see Table~\ref{tab:initialconditions})\\
\hline
\end{tabular}
\label{tab:prior}
\end{table*}

\begin{figure}
\centering
\includegraphics[scale=0.7]{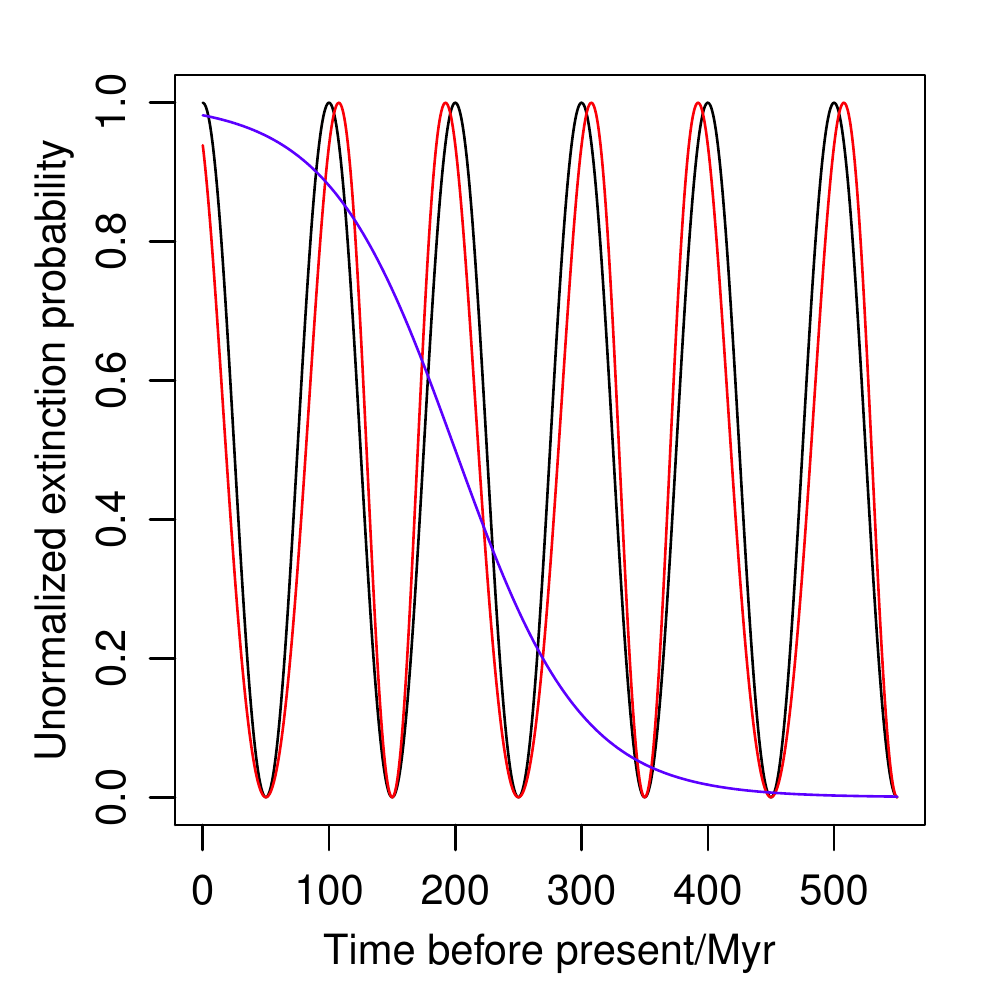}
\caption{Time series models. The black line shows the PNB model with $T=$\,100\,Myr, $\beta=0$. The red line shows the QPM model with $T=$\,100\,Myr, $\beta=0$, $A_Q=0.5$ and $T_Q=$\,200\,Myr; The blue line shows the SP model with $\lambda=$\,50\,Myr and $t_0=$\,200\,Myr.}
\label{fig:models}
\end{figure}

Table~\ref{tab:models} summarizes the functional form of the models, which are now briefly described. 
Figure~\ref{fig:models} plots examples of some of these models.
The range of the data is taken to be 0--550\,Myr BP.
\begin{description}
\item[Uniform] Constant extinction PDF over the range of the data. This has no parameters.

\item[RB/RNB] Random model in which a set of $N$ times are drawn at random from a uniform distribution extending over the range of the data. A Gaussian with standard deviation $\sigma=$\,10\,Myr is assigned to each of these, and then a constant $B$ added before normalizing. This is the RB model.  The RNB (``random no background'') model is just the special case of $B=0$, which produces a model which is similar to our discrete data. 
In practice we fix $N$ and $B$ and calculate the evidence by averaging over a large number of realizations of the model. Specifically, when modelling the B5 and B18 data sets we fix $B=0$, and $N=5,18$ respectively. 

\item[PB/PNB] Periodic model of period $T$ and phase $\beta$ (model PNB). There is no amplitude parameter because the model is normalized over the time span of the data. Adding a background $B$ to this simulates a periodic variation on top of a constant extinction probability (model PB).

\item[QPM] A quasi-periodic model in which the phase is a sinusoid with amplitude $A_Q$, period $T_Q$ and phase $\boldsymbol{\theta}$ (it becomes the same as the PNB model if $A_Q=0$). 

\item[SP] A monotonically increasing or decreasing nonlinear trend in the extinction PDF using a sigmoidal function characterized by the steepness of the slope, $\lambda$ and the center of the slope, $t_0$. In the limit that $\lambda$ becomes zero the model becomes a step function at $t_0$, and in the limit of very large $\lambda$ becomes the uniform model.

\item[SSP] Combination of SP and PNB.

\item[OM(P)/SOM(P)] The orbital/semi-orbital model with/without spiral arms, defined in Section~\ref{sec:OBM_biodiversity}.
\end{description}

\section{\label{sec:solarorbit_biodiversity} Model of the solar orbit}

We now reconstruct the orbit of the Sun around the Galaxy over the past 550\,Myr. This is done by integrating the Sun's path back in time through a fixed gravitational potential, which is expressed in Eqn.~\ref{eqn:Phi_sym}. (The dynamics are reversible because only gravity acts; energy is not dissipated.) It has often been assumed that the solar orbit is periodic with respect to crossings of the Galactic plane and/or spiral arms; we investigate this numerically in Section~\ref{sec:periodicity_biodiversity}.  The stellar mass distribution corresponding to the potential gives the local stellar density which the Sun experiences in its orbit. In Section~\ref{sec:OBM_biodiversity} we use this to derive the variation in the expected extinction rate.

\subsection{Orbit calculation}\label{sec:orbit_biodiversity}

To calculate the motion of a body through the potential from given initial conditions, we solve Newton's equations of motion, which in cylindrical coordinates are
\begin{eqnarray}
\ddot{R} -R\dot\phi^2 &=& -\frac{\partial\Phi}{\partial R} \nonumber \\
R^2 \ddot{\phi}+2R\dot{R}\dot{\phi} &=& -\frac{\partial\Phi}{\partial\phi} \nonumber \\
\ddot{z} &=& -\frac{\partial \Phi}{\partial z} 
\end{eqnarray}
We solve these equations by numerical integration using the {\tt lsoda} method
implemented in the R package {\tt deSolve}, with a time step of 0.1\,Myr. 

The initial conditions are the current phase space coordinates (three spatial
and three velocity coordinates) of the Sun. These are derived from
observations with a finite accuracy, so our initial conditions are Gaussian
distributions, with mean equal to the estimated coordinate and standard
deviation equal to its uncertainty (Table~\ref{tab:initialconditions}). In
order to calculate an orbit we draw the initial conditions at random
from these prior distributions, and a large number of draws gives us a
sampling of orbits which will be used later (e.g.\ in the evidence calculations).

We derive our initial conditions from a number of sources:
The distance to the Galactic centre comes from astrometric and spectroscopic
observations of the stars near the black hole of the Galaxy
\citep{eisenhauer03}. The Sun's displacement from the galactic plane is
calculated from the photometric observations of classical Cepheids by
\cite{majaess09}. The Sun's velocity is calculated from Hipparcos data by \cite{dehnen98b}.

\begin{table*}
\caption{The current phase space coordinates of the Sun, represented as
  Gaussian distributions, and used as the initial conditions in our orbital
  model}
\centering
\begin{tabular}{l*5{c}r}
\hline
\hline
     &$R$/kpc&$V_R$/kpc~Myr$^{-1}$&$\phi$/rad&$\dot\phi$/rad~Myr$^{-1}$&z/kpc&$V_z$
/kpc~Myr$^{-1}$\\
\hline
mean &8.0  &-0.01              & 0        & 0.0275
&0.026&0.00717\\
standard deviation &0.5& 0.00036&0         &0.003
&0.003&0.00038\\
\hline
\end{tabular}
\label{tab:initialconditions}
\end{table*}

\subsection{Geometry of spiral arms}\label{sec:arm_biodiversity}

The model for the spiral arms is described by their geometry and their gravitational potential. However, for the arm crossing periodicity test in the next section we ignore mass of the spiral arms when calculating the solar orbit and only consider their location. Likewise, in one of the class of variants of the orbital model, OM and SOM (defined later), we ignore the arms entirely (for both the orbit and stellar density calculations). This is done so that we can see the additional affect of the arms, the form and mass of which are poorly determined by current observations.

\begin{figure*}
\centering
\includegraphics[scale=0.7]{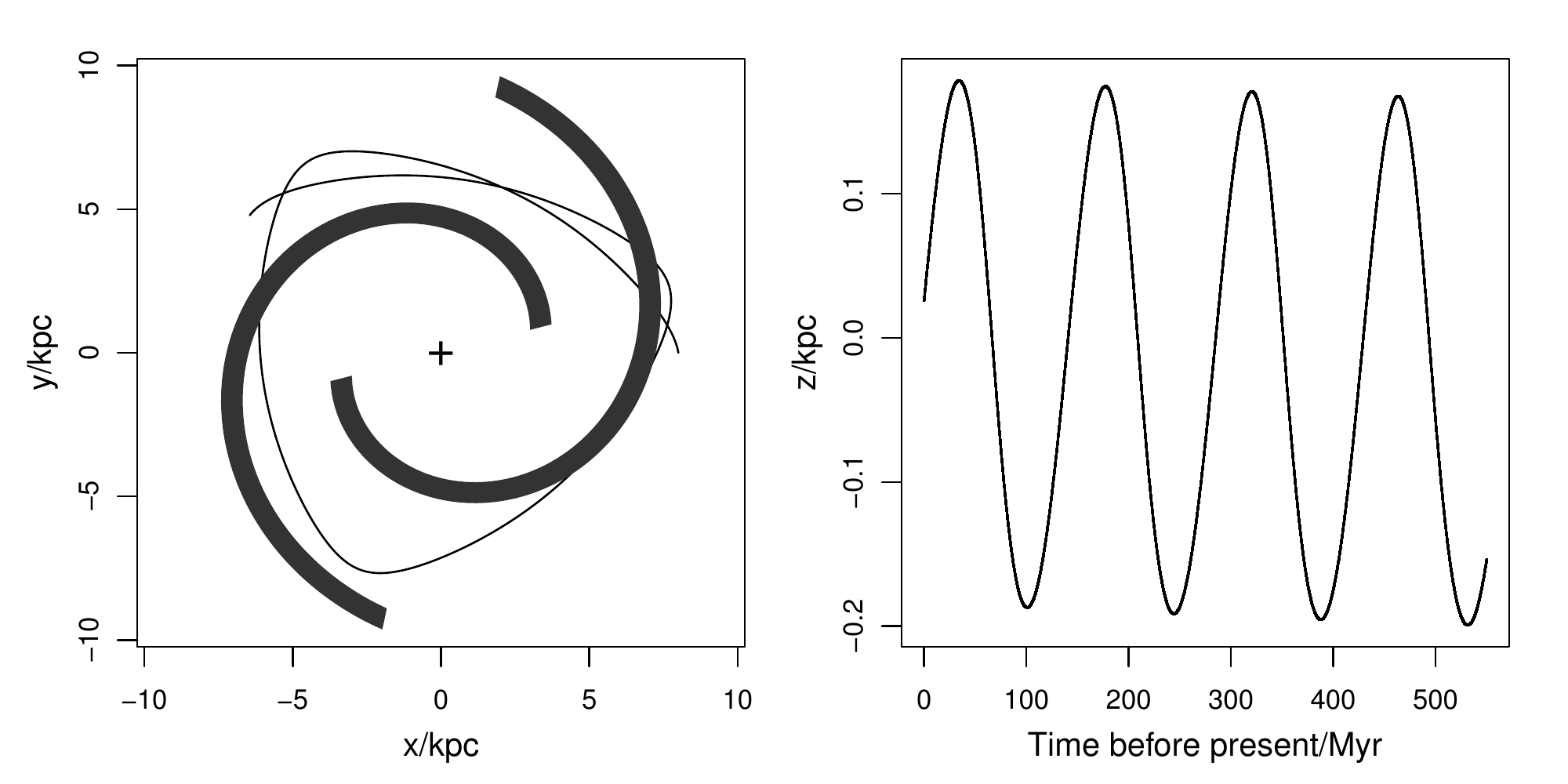}
\caption{The solar orbit in the Galactic plane (left, thin lines) and perpendicular to the
  plane (right). The orbit in the left panel is in a reference frame rotating with the spiral arms (shown in thick lines).} 
\label{fig:spiralzorbit} 
\end{figure*}

The geometric model comprises two logarithmic spiral arms, the positions of which in circular coordinates, $(R, \phi)$, are given by
\begin{equation}
\phi_s(R)=\alpha \log(R/R_{min})+\phi_{min},
\label{eqn:logspiral}
\end{equation}
where $\alpha$ is a winding constant, $R_{min}$ is the inner radius and
$\phi_{min}$ is the azimuth at that inner radius. The radius of the spiral arm
ranges from $R_{min}$ to $R_{max}$. Of the various arm models offered by \cite{wainscoat92}, we selected the main two spiral arms, 1 and $1^\prime$, with $\phi_{min}$ given by \cite{vanhollebeke09} and other parameters given by \cite{wainscoat92} (see Table~\ref{tab:spiralarmpar}).  Their location in the plane of the Galaxy is shown in the left panel of Figure~\ref{fig:spiralzorbit}. The arms rotate rigidly with constant angular velocity (pattern speed) of $\Omega_p=20$\,km\,s$^{-1}$\,kpc$^{-1}$ \citep{martos04, drimmel00}.
Note that we model the geometry of spiral arms with parameters slightly different with those in Eqn. \ref{eqn:Phi_arm} to simplify the geometry of spiral arms. The gravitational potential of the spiral arms are given in Eqn. \ref{eqn:Phi_arm}, and we do not show it explicitely here. 

\begin{table*}
\caption{The parameters of the geometric model for the spiral arms. }
\centering
\begin{tabular}{lcccc}
\hline
\hline
\multicolumn{5}{c}{geometric parameters}\\\hline
arm&$\alpha$&$R_{min}$/kpc&$\phi_{min}$/rad&extent/kpc\\\hline
1        & 4.25 & 3.48 & 0.26 & 6.0\\
$1^\prime$&4.25  & 3.48 & 3.40 & 6.0\\\hline
\hline
\end{tabular}
\label{tab:spiralarmpar}
\end{table*}

\subsection{\label{sec:periodicity_biodiversity} Periodicity test}

In previous studies of the impact of astronomical phenomena on the terrestrial biosphere, it has frequently been assumed that the solar motion shows strict periodicities in its motion perpendicular to the Galactic plane, and sometimes also with respect to spiral arm crossings. We investigate this here using our numerical model.

For each orbit $k$, we calculate the intervals between successive crossings, $\{\Delta t^i\}_k$, (separately for midplane and spiral arm crossings), where $i$ indexes the crossing. We then calculate the sample mean and sample standard deviation of these intervals for each orbit
\begin{eqnarray}
\overline{\Delta t_k} &=& \frac{1}{N_k-1}\sum_{i=1}^{N_k-1} \Delta t_k^i\\
\sigma_k &=& \sqrt{\frac{1}{N_k-2} \sum_{i=1}^{N_k-1} (\Delta t_k^i-\overline{\Delta t_k})^2} \ ,
\end{eqnarray}
where $N_k$ is the number of crossings in the $k^{th}$ orbit. To assess the periodicity of the crossing
intervals, we define the degree of {\em aperiodicity} as
\begin{equation}
a_k=\sigma_k/\overline{\Delta t_k} \ .
\end{equation}
An orbit with $a=0$ is strictly periodic.

We investigated the variation of the aperiodicity of the solar orbit with the six parameters (initial conditions).  This parameter space is too large to report on extensively here, but we find that the aperiodicity is most sensitive to $R(t=0)$ and $\dot\phi(t=0)$.
In the following we vary these initial conditions individually, by drawing $10^4$ samples from the corresponding initial condition distribution. (Larger sample sizes did not alter the results significantly)
We simulate the solar orbits using the arm-free potential, $\Phi_G$.

\subsubsection{\label{sec:midplane_biodiversity} Midplane crossings}

Some earlier studies claimed that Galactic midplane crossings trigger increases in terrestrial extinction due to an enhanced gamma ray or cosmic ray flux or due to larger perturbation of the Oort cloud. These are directly related to 
the increased stellar density and increased occurrence of star forming
regions. The larger tidal forces are postulated to enhance the disruption
of the Oort cloud \citep{rampino00, matese95}, and the higher density of massive stars -- and thus high energy radiation as well as increased supernova rate -- raise the average flux the Earth is exposed to.
The periodicity of the Sun's vertical motion -- not least its period, phase and the assumed stability of this period -- are central to these claims.  We examine these using our model.

\begin{figure}[ht!]
\centering
\includegraphics[scale=0.6]{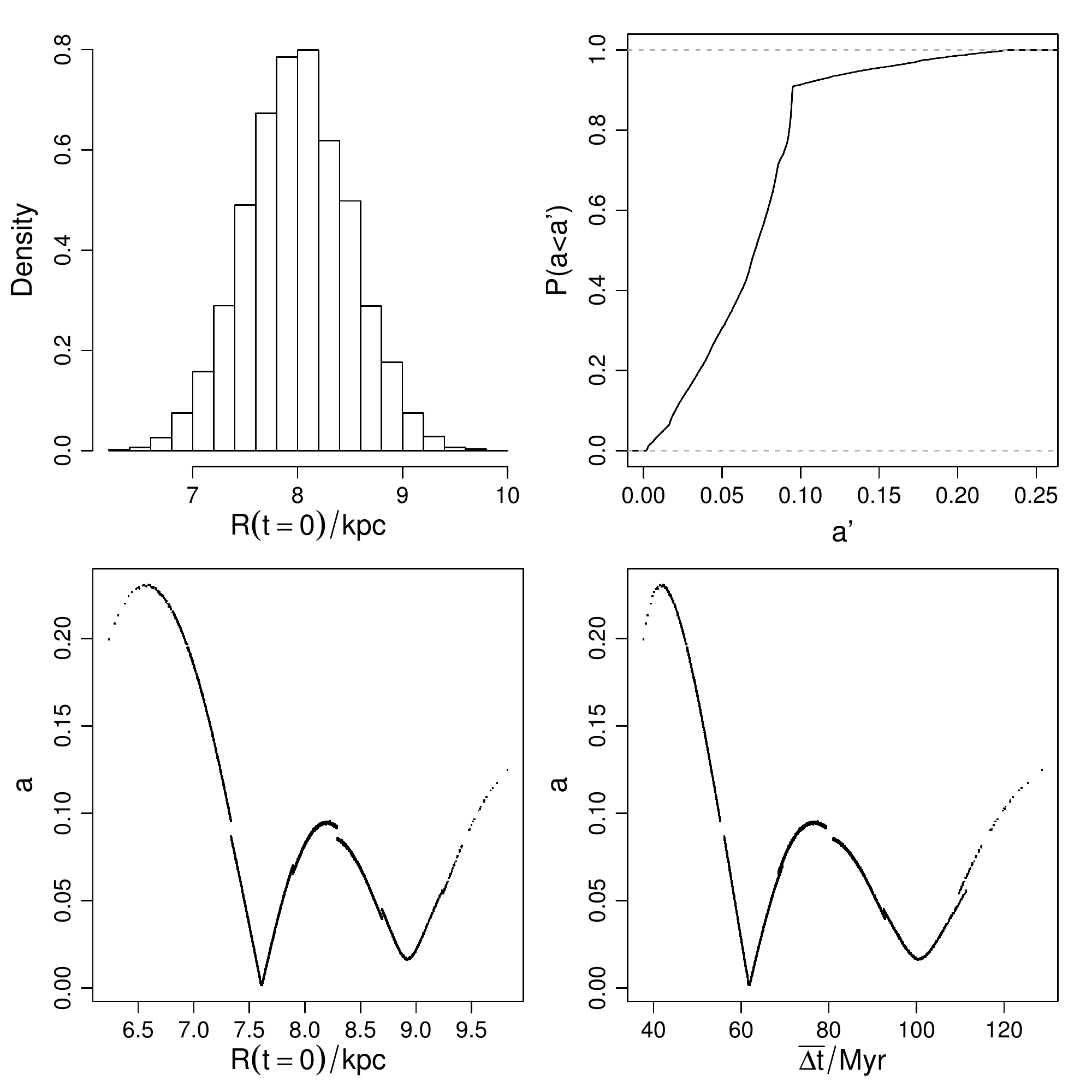}
\caption{Periodicity test of midplane crossings varying just the current galactocentric radius of the Sun, $R(t=0)$.
 Top left: distribution of this initial condition in the simulations (it has a Gaussian distribution with parameters given in Table~\ref{tab:initialconditions}). Top right: cumulative probability of the aperiodicity parameter for the resulting orbits. Bottom left: the variation of aperiodicity with $R(t=0)$. Bottom right: the variation of aperiodicity with the average crossing interval, $\overline{\Delta t_k}$.}
\label{fig:sen-varyR-plane} 
\end{figure}

The results of varying just the initial galactocentric radius of the Sun, $R(t=0)$, are shown in Figure~\ref{fig:sen-varyR-plane}. We see in the top-right panel that about 90\% orbits have an aperiodicity less than 0.1. In the lower two panels we see how $a$ varies with the value of the initial condition and with the average crossing interval.

The aperiodicity is 0.002 (nearly strict periodicity) at $\overline{\Delta t}=61.8$\,Myr.  
This corresponds to a 1:1 resonance between the vertical motion and the radial
motion. Its value is close to a period in the biodiversity data of $62\pm 3$\,Myr claimed by \cite{rohde05}. Little should be read into this coincidence, however, as there is no good (i.e.\ independent) reason to select the specific initial condition that leads to this period over any other. 
Moreover, changing the parameters of the Galactic potential -- which is not very well known -- changes this period. 
(For example, if we increase the mass of the Galactic halo the values of
$\overline{\Delta t}$ are decreased.) The other minimum in the aperiodicity in
the bottom panels is 0.02 at $\overline{\Delta t}=100.2$\,Myr. This
corresponds to an approximately circular orbit in the midplane. If we set $V_R(t=0)=0$, $V_z(t=0)=0$ and $z=0$, this solar orbit would be strictly circular.

The cumulative curve (top-right panel of Figure~\ref{fig:sen-varyR-plane}) makes a sharp turn at $a'=0.1$. This is because of a sudden decrease in the number of orbits with large aperiodicites.  Similarly, the discontinuities in the lower panels are caused by changes in the (small) number of discrete plane crossings which occur for different aperiodicity ranges.

If we now vary the initial condition $\dot\phi(t=0)$ instead, the periodicity test gives very similar results: we find a nearly strict periodicity at $\overline{\Delta t}=60$\,Myr and another minimum in the aperiodicity at about 100\,Myr. That means the nearly strict periodicity is mainly determined by a combination of $R(t=0)$ and $\dot\phi(t=0)$.

In summary, we see that the majority of the simulated orbits (90\%) are quite close to periodic ($a\leq0.1$) in their motion vertical to the midplane, although strict periodicity essentially never occurs.
 
\subsubsection{\label{sec:spiral_biodiversity} Spiral arm crossings}

Regarding spiral arms as regions of increased star formation activity and stellar density, the mechanisms of mass extinction considered for midplane crossing could likewise be applied to spiral arm crossings, and have been by some authors \citep{leitch98, gies05}. However, such studies have over-simplified the solar motion by failing to take into account the considerable uncertainties in the current phase space coordinates of the Sun and thus in its plausible orbits.  Some studies have even claimed a connection between spiral arm crossings and the terrestrial biosphere {\em after} having fit the solar motion to the geological data, but such reasoning is clearly circular.

We examine here the periodicity of spiral arm crossings (although we note that some studies in the literature claiming a spiral arm-extinction link just consider the crossing times and do not claim a periodic crossing).  The crossing intervals are longer than with the midplane, so we only include in our analyses models in which there are at least three arm crossings. We assume that the arms have indefinite vertical extent, so that a crossing on the x--y plane is always a true encounter. In reality the Sun might pass over or under the arms, thus reducing the overall relevance of spiral arm crossings to terrestrial extinction.

\begin{figure}[ht!]
\centering
\includegraphics[scale=0.6]{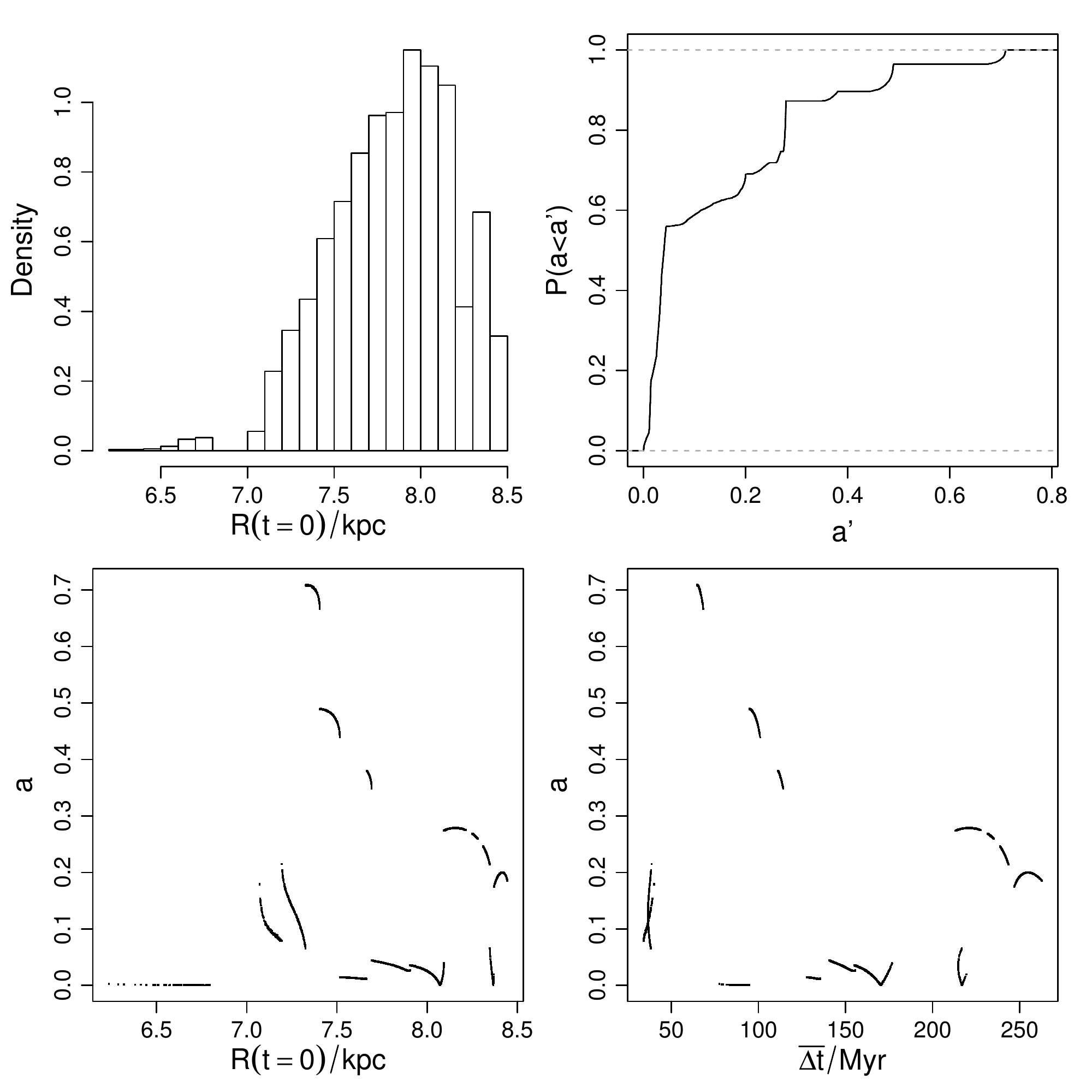}
\caption{As Figure~\ref{fig:sen-varyR-plane}, but now for the spiral arm crossings.}
\label{fig:sen-varyR-arm} 
\end{figure}

Figure~\ref{fig:sen-varyR-arm} shows the result of this analysis for the 7\,407 orbits (out of the original sample of 10\,000) which exhibit at least three arm crossings.  The cumulative probability (top-right panel) shows that about 40\% of the orbits have an aperiodicity larger than 0.2. In other words, it is not very likely that the solar orbit and spiral arms are so tuned to give periodic crossings.  The lower two graphs show how $a$ varies with $R(t=0)$ and $\overline{\Delta t_k}$.  The numerous gaps in these plots are a consequence of the fact that not all orbits for certain ranges of $R(t=0)$ had at least three arm crossings, and so were removed from the analysis. We see, therefore, that the crossing interval is very sensitive to $R(t=0)$.

Note that we have neglected the mass of the arms in the orbital calculations. When we include it the values of aperiodicity increase and there is an even less clear dependence of $a$ on 
$R(t=0)$ or $\overline{\Delta t_k}$.

In summary, we find it unlikely that spiral arm crossings are even close to periodic. If the pattern speed of the spiral arms has not been constant in the past 550\,Myr, or if the pattern itself has not been stable, then this conclusion is strengthened further.

\subsection{\label{sec:OBM_biodiversity}Orbital model}

\subsubsection{Derivation of the extinction rate from the stellar density variation}

As outlined in Section~\ref{sec:introduction_biodiversity}, various astronomical mechanisms for biological extinction have been identified, including comet impacts (from Oort cloud perturbation), gamma rays (from SNe or GRBs), and cosmic rays (from SN remnants) \citep{ellis95,sanchez01,gies05}.  The intensity of all of these depends on the local stellar density. If we consider a general mechanism involving flux from nearby stars, then the flux from a single star is proportional to $f/d^2$, where $f$ is the relevant surface flux and $d$ the distance. The sum of this over the whole relevant volume of space around the Sun is proportional to the total intensity and thus the extinction probability (per unit time). 

Let us assume that the extinction rate, $E$, is linearly proportional to the flux, and that the number density of relevant stars is proportional to the total stellar number density (stars per unit volume), $n$.  Because the density of spiral arms is much less than the density of the other components, we consider at first only the time-independent density arising from halo, disk and bulge.  The density is calculated from the corresponding potential (defined in Eqn. \ref{eqn:Phi_sym}) using Poisson's equation.

In an axisymmetric cylindrical coordinate system, the extinction rate at the Sun is then
\begin{eqnarray}
E(R_\odot, z_\odot) &=& C \int \!\! \int \frac{n(R,z)}{d^2} \,R\,dR\,dz\nonumber\\
                           &=&C \int \!\! \int \frac{n(R,z)}{(R-R_\odot)^2+(z-z_\odot)^2}\,R\,dR\,dz,\nonumber\\
\label{eqn:stellar_density1}
\end{eqnarray}
where $R$ and $z$ are the galactocentric radius and height above the midplane,
respectively, for some star, and $R_\odot$ and $z_\odot$ are the corresponding
(time-varying) coordinates of the Sun, and $C$ is a constant. Notice that the
stellar number density, $n$, is proportional to the corresponding stellar density, $\rho$. Defining the distance from a star to the Sun as
$r\equiv R-R_\odot$ and $Z\equiv z-z_\odot$, the extinction rate is  
\begin{eqnarray}
 E(R_\odot,z_\odot) &=& C \int \!\! \int\frac{n(R_\odot+r,z_\odot+Z)}{r^2+Z^2}(R_\odot+r) drdZ \ .\nonumber\\
\label{eqn:stellar_density2}
\end{eqnarray}
The flux from a star falls off as $1/d^2$, but we can truncate this integral at some upper distance because
at some point the flux is too weak to influence the terrestrial biosphere. We
take $d_{\rm th}=50$\,pc as an upper limit.\footnote{In the case of SNe, \cite{ellis95} conclude that only those which come within 10\,pc of the Sun would have a significant impact on terrestrial life.
GRBs up to 1\,kpc or even more could still have an effect on the
Earth, but we ignore these because the GRB rate (at low redshifts) is comparatively low (e.g.\ \cite{domainko13}).}
This is much smaller than the scale length of the disk and comparable to the scale height of the disk (see Table~\ref{tab:model_par}), so we can approximate $n(R_\odot+r,z_\odot+Z)(R_\odot+r)$ by $n(R_\odot,z_\odot+Z)R_\odot$. The integral then becomes
\begin{equation}
E(R_\odot,z_\odot)\simeq CR_\odot\int_{-d_{\rm th}}^{d_{\rm th}}\int_{-d_{\rm th}}^{d_{\rm th}}\frac{n(R_\odot,z_\odot+Z)}{r^2+Z^2} drdZ \ .
\label{eqn:extinctionPDF1}
\end{equation}
Integrating over $r$ gives
\begin{equation}
E(R_\odot,z_\odot)\simeq 2CR_\odot\int_{-d_{\rm th}}^{d_{\rm th}}n(R_\odot,Z+z_\odot)\frac{\arctan(d_{\rm th}/Z)}{Z}\,dZ \ .
\label{eqn:stellar_density3}
\end{equation}
The geometric factor $\displaystyle \frac{\arctan(d_{\rm th}/Z)}{Z}$ drops close to zero at about $Z=$\,25\,pc, and does so much more rapidly than the stellar density term, which follows the vertical profile of the disk (which has a much larger scale height of 250\,pc).
Thus to a reasonable degree of approximation we can set $n(R_\odot, Z+z_\odot)
\simeq n(R_\odot, z_\odot)$ in this integral. The integral is then just over
the geometric factor, which gives some constant (dependent on $d_{\rm th}$,
but of no further interest).
Thus we are left with 
\begin{equation}
E(R_\odot,z_\odot)\simeq C'R_\odot n(R_\odot,z_\odot),
\label{eqn:stellar_density4}
\end{equation}
for some constant $C'$. For the solar motion, the relative
variation of $R_\odot$ is less than that of $n(R_\odot,z_\odot)$, so we have
\begin{equation}
E(R_\odot,z_\odot) \propto n(R_\odot, z_\odot) \ .
\label{eqn:stellar_density5}
\end{equation}
In other words, the extinction rate is just proportional to the stellar
density at the location of the Sun. 
The approximations in Eqns.~\ref{eqn:extinctionPDF1}--\ref{eqn:stellar_density5} still hold  when we include the low density spiral arms defined in Section~\ref{sec:arm_biodiversity}, in which case we must also introduce the explicit dependence on azimuth and time
\begin{equation}
E(R_\odot, \phi_\odot, z_\odot, t) \propto n(R_\odot,\phi_\odot,z_\odot,t) \ .
\end{equation}

In the above model we assumed that the extinction rate is proportional to
$d^{-2}$, i.e.\ the influence falls off like a flux on the surface of a
sphere. We could generalize this dependence to be $d^{-k/2}$ for $k\geq0$ in
order to reflect other mechanisms, e.g.\ tidal effects. 

\begin{figure}[ht!]
\centering
\psfrag{Rs}[c][c]{\scriptsize{$R_\odot$/kpc}}
\psfrag{Zs}[c][c]{\scriptsize{$z_\odot$/kpc}}
\includegraphics[scale=0.7]{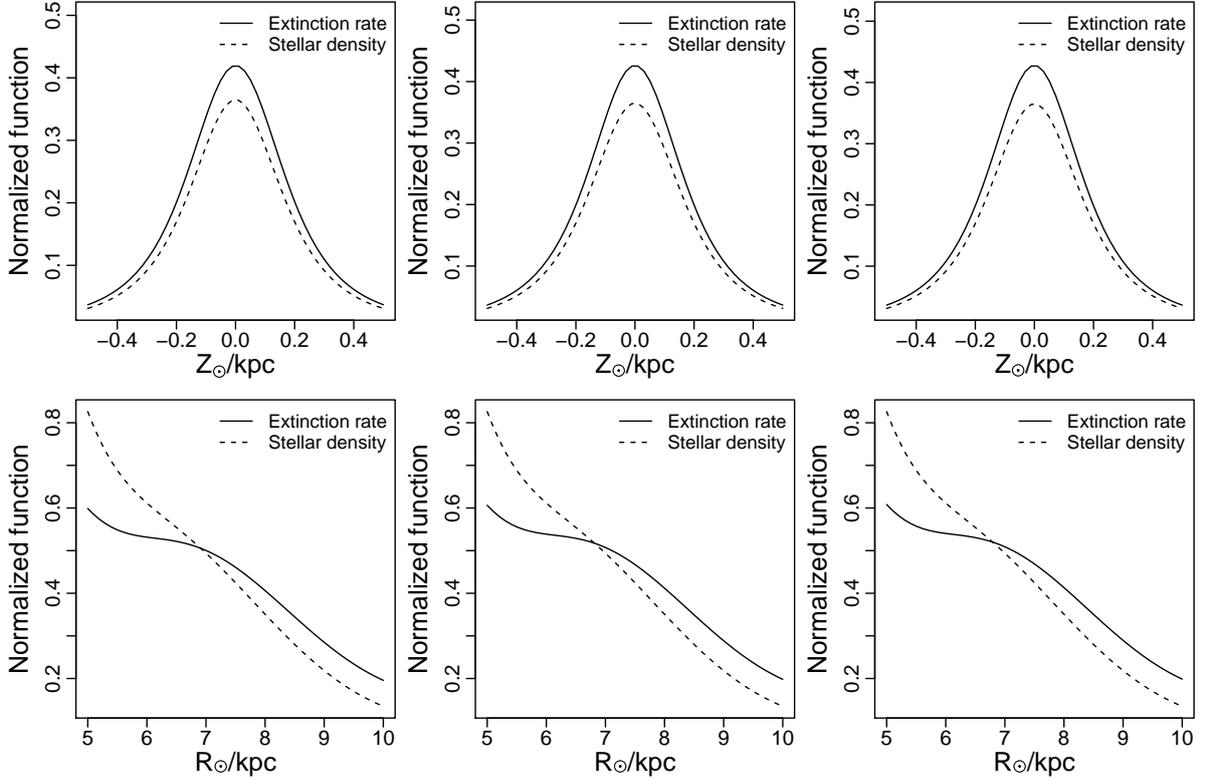}
\caption{Comparison of the extinction rate calculated numerically with the stellar density at the position of the Sun.
The top row shows the variation as a function of $z_\odot$ with $R_\odot$ fixed to 8\,kpc. The bottom row show the variation as a function of $R_\odot$ with $z_\odot$ fixed to 26\,pc. The columns from left to right are for $k=0,2,4$ in the model for the dependence of extinction rate with distance.}
\label{fig:ext-pos10}
\end{figure}

In order to test the validity of the above approximations, we compare in Figure~\ref{fig:ext-pos10}
the extinction rate as given by Eqn.~\ref{eqn:stellar_density1} (by numerical
integration) with the stellar number density $n(R_\odot, z_\odot)$. We plot over
ranges of $R_\odot$ from 5\,kpc to 10\,kpc and $z_\odot$ from $-0.5$\,kpc to
0.5\,kpc, in accordance with the ranges covered by the simulated solar orbits.
We normalize the extinction rate (and the stellar density)
by setting its integral over $R_\odot$ and $z_\odot$ to be unity.
In the upper row of Figure~\ref{fig:ext-pos10}, the difference between the stellar density and the extinction rate reaches a maximum in the midplane ($z_\odot=0$); this is on account of the relatively large density gradient at $z_\odot=0$. The maximum difference is only about 10\% of the peak value of stellar density for all values of $k$. 
In the lower row, the largest difference is at the lower limit of $R_\odot$.
Note that the value of $k$ has very little impact.

In practice, most of the simulated orbits spend most of their time in the region $7<R_\odot/{\rm kpc}<9$ and $-0.3<z_\odot/{\rm kpc}<0.3$, where the differences between local stellar density and extinction rate variation are even smaller. Thus to within a few percent, the stellar density at the Sun is a good predictor of the extinction rate. The time variation of this density is the time series model forms the basis for what we refer to as the ``orbital models'', the forms of which we now define.

\subsubsection{Definition of OM(P) and SOM(P)}

The orbital model ``OM'' is the orbital model which does not include the spiral arm at all, neither in the gravitational potential (for calculation of the orbits) nor in the stellar density (for the extinction rate calculation). The orbital model OMP does include the spiral arm in both senses. Thus both OM and OMP are internally self-consistent. 

Once normalized, $E(t)$ is just the quantity $P(t | \boldsymbol{\theta}, M)$ in Section~\ref{sec:tsmodels_biodiversity} (and it is normalized to give unit integral over the span of the data). The parameters of OM and OMP are the initial conditions of the orbit, and the corresponding priors are the Gaussian distributions summarized in Table~\ref{tab:initialconditions}. Thus one orbit calculated from one draw of the initial conditions allows us to calculate one likelihood for these models (for given data set). Repeating this and averaging the resulting likelihoods gives the evidence for that orbital model (see Section~\ref{sec:calcevi}).

For both of these models we consider four variations, labelled 1--4, according to which initial conditions we vary (and therefore sample over to build up the set of orbits).

\begin{figure*}
\centering
\includegraphics[scale=0.5]{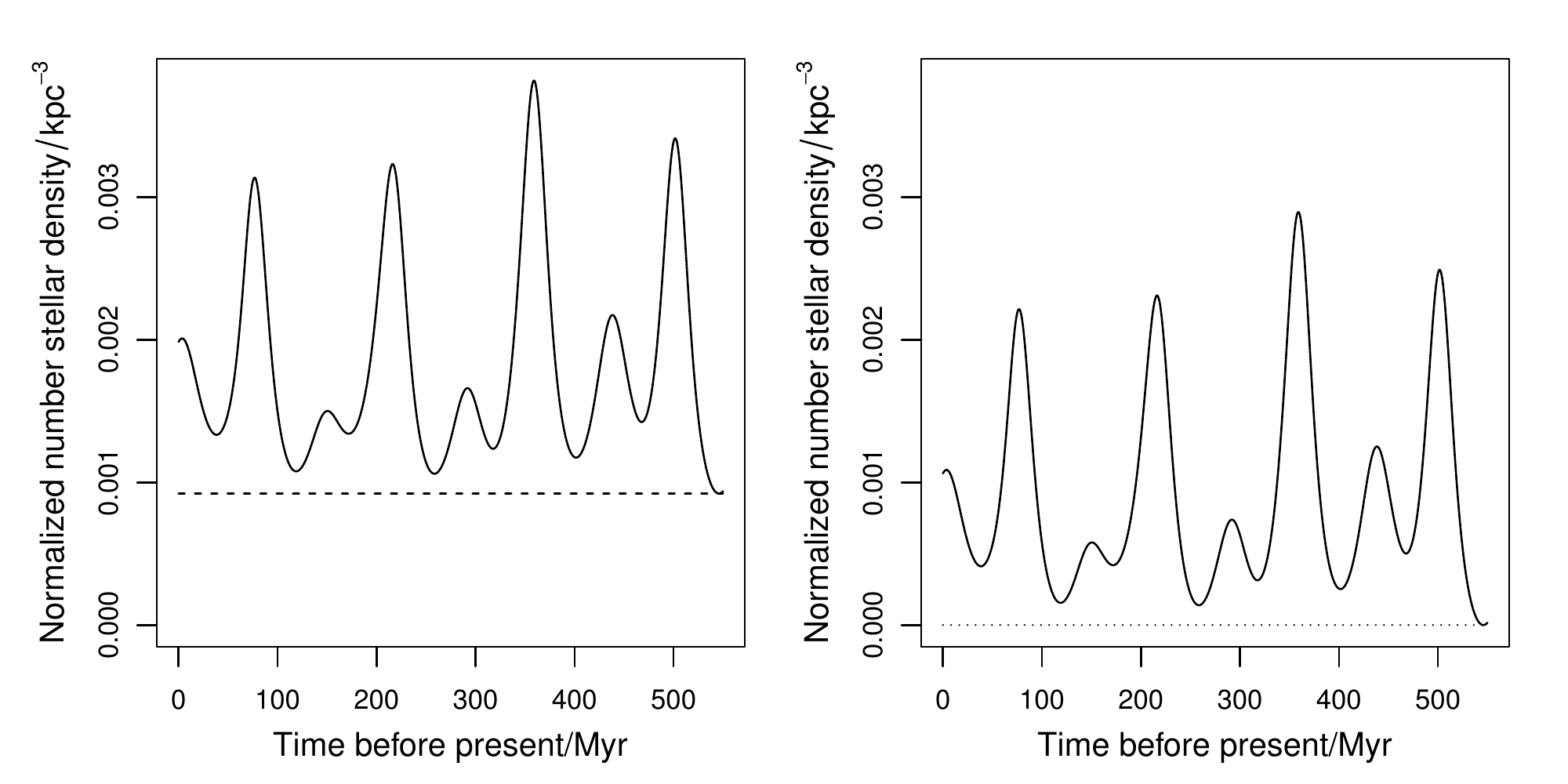}
\caption{The orbital model (OM) and semi-orbital model (SOM). The left
  panel shows the variation of the local stellar density -- and thus the
  extinction probability per unit time -- in one particular orbit calculated
  from OM. The horizontal dashed line is a threshold, $h$, for truncating the stellar density to a minimum level, which gives rise to the extinction probability
  per unit time plotted in the right panel.}
\label{fig:som}
\end{figure*}

In addition to these models, we define the``semi-orbital model'', SOM. This is derived from the OM simply by subtracting from the predicted extinction rate a constant value, $h$, and setting all resulting negative values to zero. Here we simply set $h$ to be the minimum value of the extinction rate (see Figure~\ref{fig:som}). This is intended to model the situation in which the flux causing the extinction must rise above some threshold before it has an effect. (We might consider this as an adaption of life to the extraterrestrial flux background.) In analogy to OM, SOM excludes the spiral arm. SOMP is SOM with the spiral arm potential and density included. Once again we will consider four varieties according to which initial conditions are varied.

\section{\label{sec:result_biodiversity} Results}

\subsection{\label{sec:evidences_biodiversity} Evidences}

\begin{table}[ht!]
\caption{Bayes factors and maximum likelihood ratios of the various time series
  models (rows) relative to the Uniform model for the various data sets
  (columns). OM(P)1--4 refer to the OM(P) model in which different initial
  conditions are varied: $R(t=0)$, $\phi\dot(t=0)$, $\{R(t=0),~\dot\phi(t=0)\}$, $\{R(t=0),~V_R(t=0),~\dot\phi(t=0),~V_z(t=0)\}$, respectively, and likewise for the SOM(P)1--4 models. The other initial conditions are kept fixed. The RB and RNB models are intrinsically discrete, so are not applied to the two continuous data sets.}
\centering
\begin{tabular}{l|llll|llll}
\hline
\hline
     &\multicolumn{4}{c|}{Bayes factor (BF)}&\multicolumn{4}{c}{Maximum
       likelihood ratio (MLR)}\\
\cline{2-9}
Model&B5   &B18   &RM  &A08 &B5 &B18 &RM  &A08\\
\hline
PNB &0.97 &0.62   &0.98&0.87&22 &255 &1.2&1.1\\
PB  &1.0  &0.80   &0.98&0.87&3.5&45  &1.1&0.99\\
QPM &0.99 &0.85   &0.98&0.87&6.8&35  &1.1&0.98\\
RNB &0.041&0.00050&--&--&1153&16 &--&--\\
RB  &0.85 &0.40   &--&--&9.0&8.4 &--&--\\
SP  &0.28 &0.019  &1.02&0.88&2.7&0.36&2.1&1.3\\
SSP &0.73 &0.18   &0.99&0.87&8.6&81  &1.3&1.2\\
OM1 &1.4  &0.74   &0.99&0.88&4.6&2.4 &1.1&1.0\\
OM2 &1.4  &0.72   &0.99&0.89&4.9&2.2 &1.2&1.0\\
OM3 &1.2  &0.63   &0.99&0.88&4.9&2.6 &1.3&1.1\\
OM4 &1.2  &0.65   &0.99&0.88&5.0&3.0 &1.2&1.1\\
OMP1&0.18 &0.014  &0.93&0.88&6.3&0.48&5.9&4.4\\
OMP4&0.14 &0.022  &0.93&0.83&20 &6.1 &6.0&5.0\\
SOM1&1.3  &0.051  &1.0 &0.90&11 &0.34&1.2&1.2\\
SOM2&0.85 &0.037  &1.0 &0.91&5.9&0.33&1.3&1.2\\
SOM3&0.99 &0.032  &1.0 &0.89&24 &0.67&1.4&1.2\\
SOM4&1.0  &0.032  &1.0 &0.89&28 &0.70&1.3&1.3\\
SOMP1&0.11&0.00013&0.94&0.88&3.5&0.0067&5.9&4.4\\
SOMP4&0.10&0.0012 &0.94&0.83&20 &1.8 &6.0 &5.1\\
\hline
\end{tabular}
\label{tab:BF-MLR}
\end{table}

We now calculate the Bayesian evidence (Eqn.~\ref{eqn:evidence}) for the
various models for each data set. This is done by sampling from prior
probability distributions of the model parameters ($P(\boldsymbol{\theta} | M$),
Table~\ref{tab:prior}), calculating the likelihood
(Eqn.~\ref{eqn:likelihood_disc} for discrete time series,
Eqn.~\ref{eqn:likelihood_cont} for continuous time series) and then
averaging these for that model and data set Eqn.~\ref{eqn:evidencenum}. 

To calculate the evidences for RNB and PNB models for the B5 and B18 data sets, we adopt a Monte Carlo sample size of $10^6$. In all other cases we use a sample size of $10^4$. Larger sample sizes did not alter the estimated evidence significantly.\footnote{ For all the data sets, the standard error of the Monte
    Carlo estimates of the evidence is $<1$\% for OM models, $<3$\%
    for SOM and other models, and $<25$\% for RNB and (S)OMP models.}  This sample size is given according to the sensitivity test of the evidence to the sample size for all models and all data sets. This test shows that the evidence estimated from $10^4$ draws in the prior distribution is close to the real evidence when there is a background either
in the model or in the data set. 

As the absolute value of the evidence is not of interest, we report the ratio of evidence, the Bayes factor.  Here we report Bayes factors with respect to the Uniform model. We regard a model as being significantly better than another when its evidence exceeds that of the other by a factor of ten \citep{jeffreys61, kass95}. Note that it is only meaningful to compare evidences -- and therefore Bayes factors -- for a fixed data set.

The results are shown in Table~\ref{tab:BF-MLR}. For the reference models we evaluate the evidence by sampling over all their model parameters, but in the case of OM and SOM we sample over just some of the parameters (initial conditions), keeping the others fixed, in order to investigate the impact of the different parameters.  As shown in Section~\ref{sec:periodicity_biodiversity}, the periodicity of solar orbit is most sensitive to the initial conditions $R(t=0)$ and $\dot\phi(t=0)$. We therefore calculate the evidence for the OM (and SOM) models with four different sets of initial conditions being varied: $R(t=0)$ only; $\phi\dot(t=0)$ only; $\{R(t=0)$ and $\dot\phi(t=0)\}$; $\{R(t=0)$, $V_R(t=0)$, $\dot\phi(t=0)$, and $V_z(t=0)\}$.  In all cases we fix $\phi(t=0)$ and $z(t=0)$, the former because it has no impact on the solar motion in this axisymmetric potential, and the latter because the uncertainty in the current $z$ position of the Sun has a limited impact on the subsequent orbit. To assess the effect of the spiral arm perturbation on the BFs, we have selected four perturbed orbital models, OMP1, OMP4, SOMP1 and SOMP4, to compare with corresponding unperturbed orbital models.

For the B5 data set, the BFs of all time series models relative to the Uniform model are less than 10. 
Thus none of these models are a significantly better explanation of the data. One model, RNB, has a Bayes factor less than 0.1, indicating that we can discount this one as being an unlikely explanation. Given that the Uniform model is the simplest model of the set, the principle of parsimony suggests we should be satisfied with it as explanation. This does not deny the possibility that some other model shows significantly higher evidence. After all, we can only ever make claims about models which we explicitly test.

\begin{figure}
\centering
\includegraphics[scale=0.6]{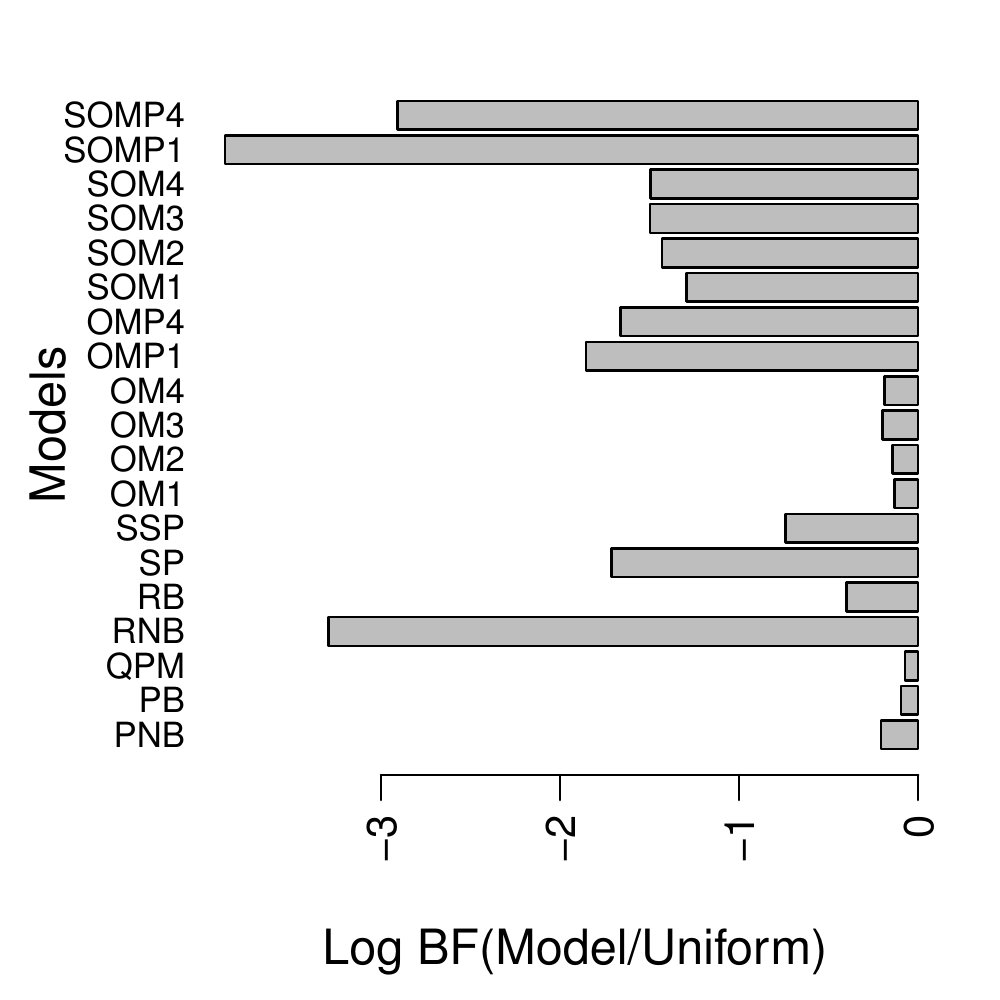}
\includegraphics[scale=0.6]{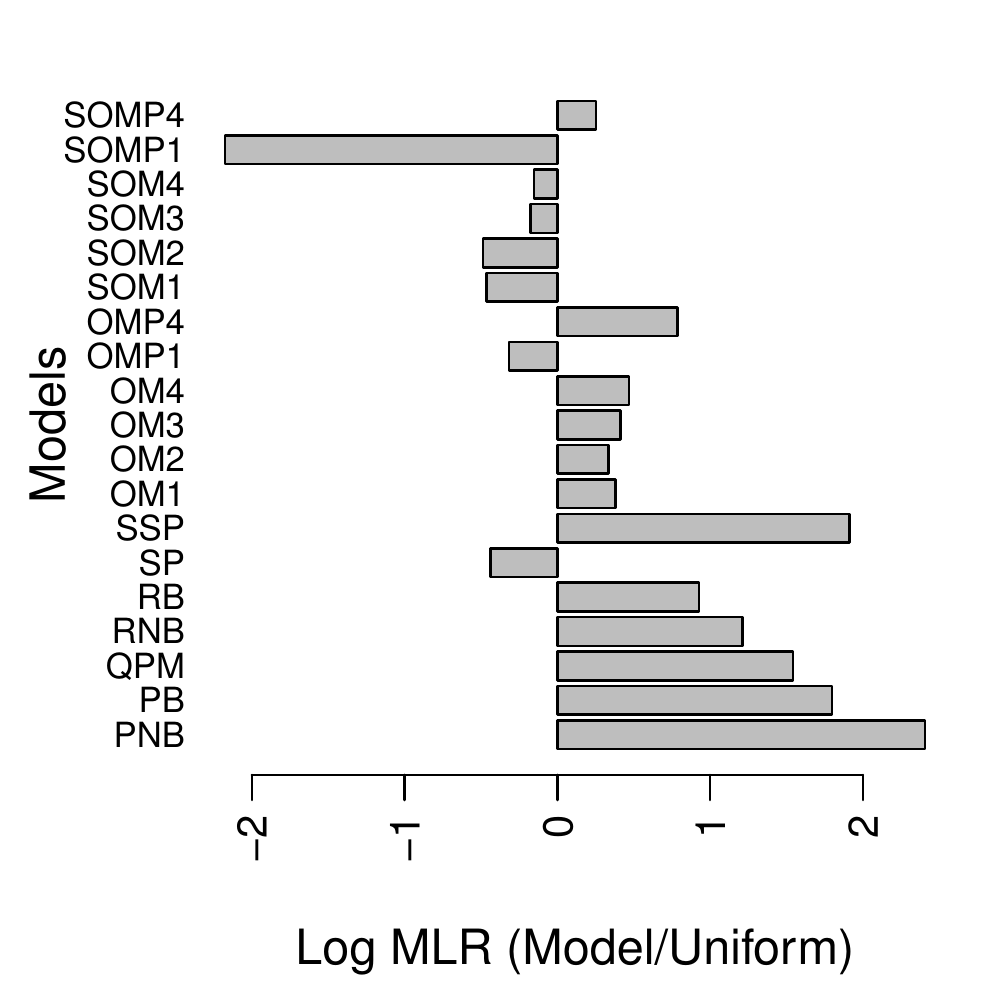}
\caption{Results for the B18 data sets. The upper panel shows the log (base 10)
  Bayes factor of the various models relative to the Uniform model. The lower pannel shows the log (base 10) of the maximum likelihood ratio of various models relative to the Uniform model.
} 
\label{fig:BF-MLR}
\end{figure}

The B18 data set includes more extinction events than the B5 data set, and not
surprisingly it discriminates more between the
models (the Bayes factors show a larger spread). 
(These results are also shown graphically in the upper panel of Figure~\ref{fig:BF-MLR}.)
The OM models are favoured somewhat more than the other models -- e.g.\ the BF
of OM3 to SP is $0.63/0.019=33$ -- although again no model is favoured
significantly more than the Uniform model. In contrast, several models are significantly disfavored (RNB, SP, OMP1, OMP4, SOM1, SOM2, SOMP1, SOMP4). 
In particular, the perturbed orbital models, including OMP1, OMP4, SOMP1 and
SOMP4, are less favored by the data than their corresponding unperturbed orbital
models. All the other perturbed orbital models (not listed in Table
\ref{tab:BF-MLR}) also have lower BFs than the unperturbed orbital models. 

For the two continuous time series, RM and A08, the difference between the
evidences of all of the models is not significant.\footnote{ As the RNB and RB models are obviously conceptually inappropriate models for continuous data sets, we do not apply them to the A08 or RM data sets, so these values are missing from Table~\ref{tab:BF-MLR}.}
Our broad conclusion is that no model significantly outperforms the Uniform model on any of the data sets. On the contrary, a few can be ``rejected'' on the ground of a significantly lower evidence. Recall that all of these models are predicting the extinction probability (per unit time). In terms of the discrete data sets, the Uniform model just means that the mass extinction events occur at random in time. We find this to be no less probable than a periodic or quasi-periodic variation of the probability, or a monotonic trend in the probability, etc.  In terms of the continuous data sets, we obviously do not believe that the Uniform model is a good explanation of the clearly apparent variations in the extinction rate (see Figure~\ref{fig:data}). But the analysis does tell us that this is no worse an explanation than the more complex models of the variation considered, such as periodic, orbital-model based etc. Clearly there must be yet other models which could explain the data even better.  This may explain why previous authors have found an apparent periodicity in the data: the periodic model can explain the data to some degree, but actually no better than simpler models. 

\subsection{\label{sec:likelihood_biodiversity}Likelihood distribution}

We have seen that the evidence hardly discriminates between any of the models on the continuous data sets, and only between some of them on the discrete data sets. (This is by no means inevitable. In other problems the evidence can vary enormously between models.)  This means that, on average over their parameter space, the models differ little in their predictions.  It is nonetheless interesting to see how the likelihood varies over the parameter space.  (We would do this in particular to find the best fitting parameters, although these are only meaningful if the overall model has been identified as the best explanation of the data.)  We focus here mainly on the PNB and OM models for the B18 data. We again normalize the likelihood for a model by dividing it by the likelihood of the Uniform model to form the likelihood ratio.  As the latter model has no parameters, it is likelihood constant and equal to its evidence.
The maximum value of the likelihood ratio we denote as MLR.

\begin{figure}
\centering
\includegraphics[scale=0.6]{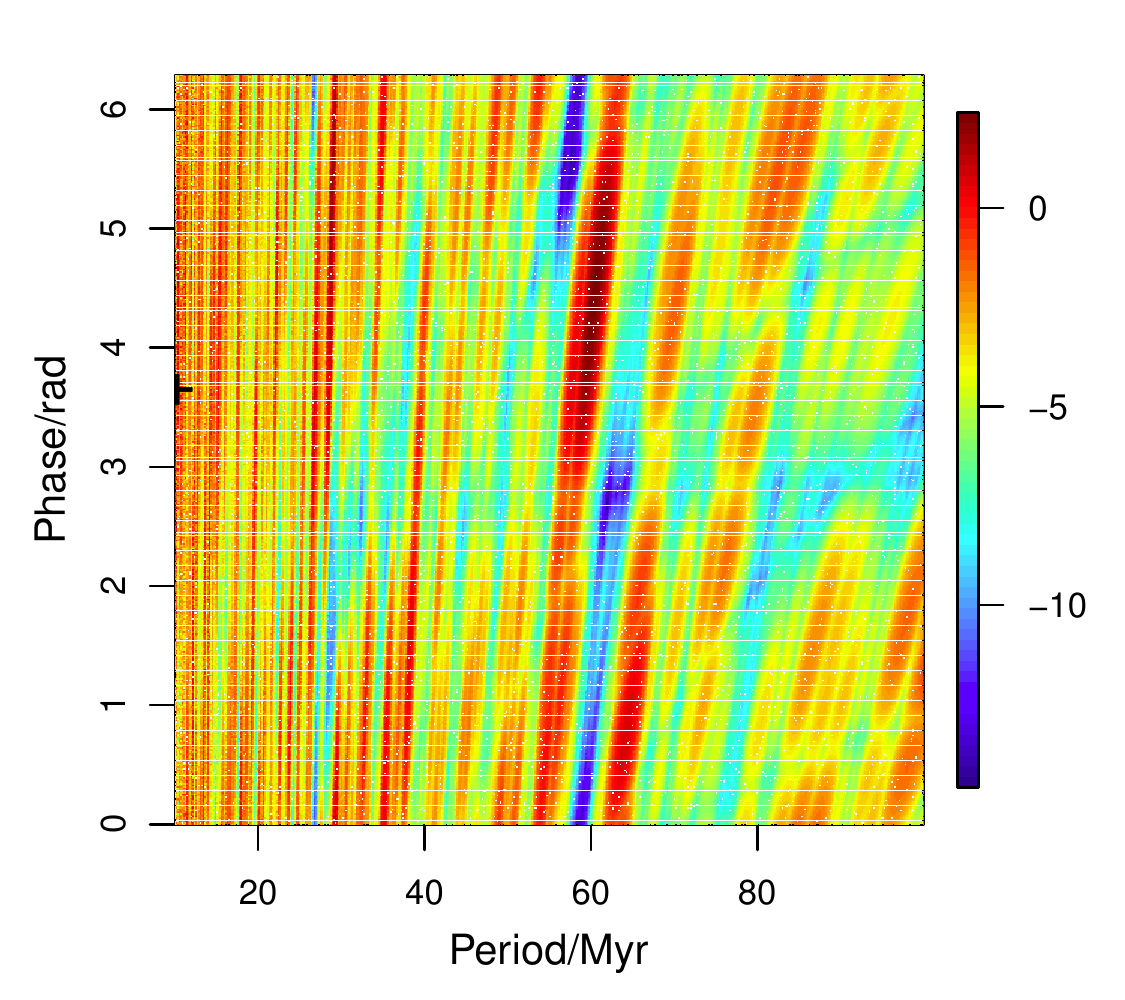}
\includegraphics[scale=0.6]{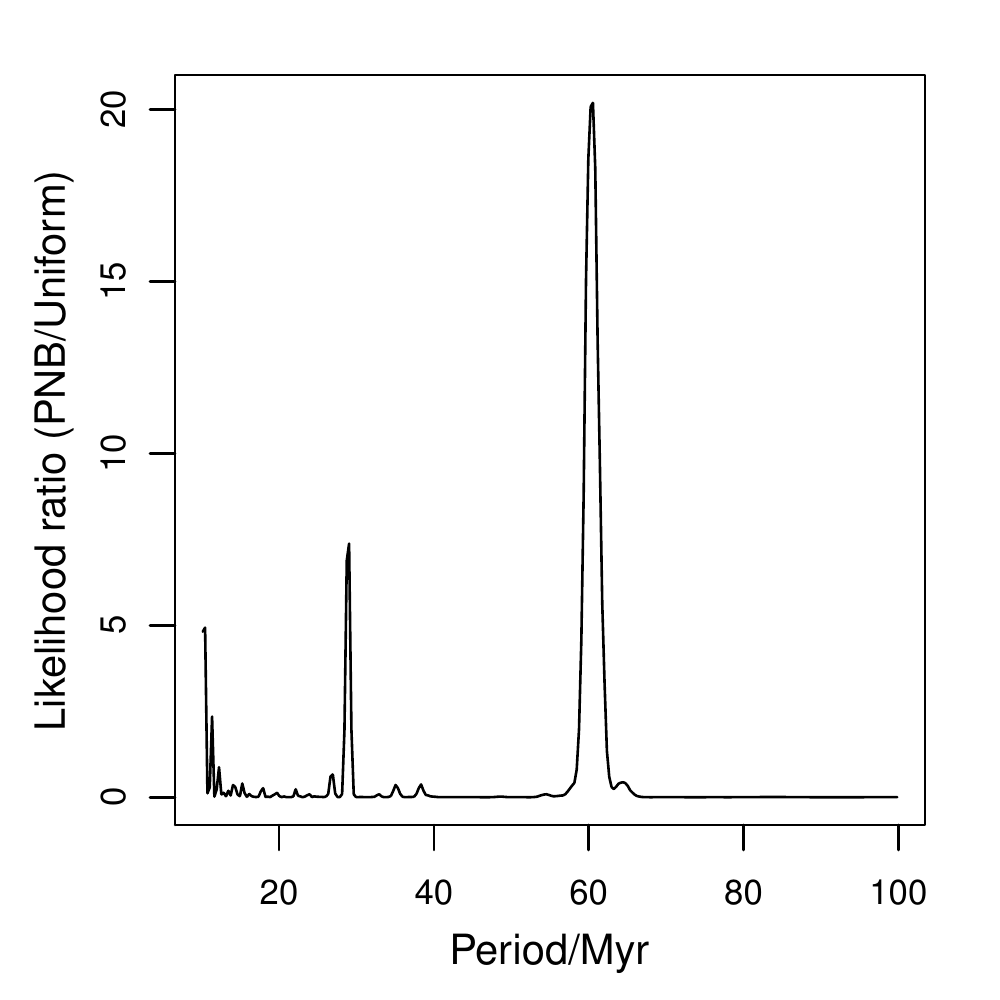}
\caption{Likelihood distribution for the PNB model on the B18 data set as
  a function of period and phase (upper panel) and period only (lower pannel). In both panels we show the
likelihood ratio of the PNB to the Uniform model, on the left as log (base 10)  on a color scale.}
\label{fig:like-2D-pnb}
\end{figure}

Figure~\ref{fig:like-2D-pnb} shows how the likelihood varies over the two-dimensional space formed by the two parameters, period and phase, of the PNB model. There is significant variation. We see numerous local maxima, the largest likelihoods being around $\{T/Myr,~\beta/rad\}=\{60, 4.5\}$.  However, these maxima are rather narrow, so once the (much lower) likelihood in the other (equally plausible) regions are taken into account, the overall evidence for the model is not particularly high.  If we are interested in the variation of likelihood with just the period, then we can marginalize this diagram over phase, and plot with respect to period, thereby forming a (Bayesian) periodogram (lower panel). We see a clear peak around 60\,Myr. This is coincident with the period of $62\pm 3$\,Myr identified by \cite{rohde05}. It is tempting (but incorrect) to associate this peak value of the likelihood with the periodic model as a whole, and use it to claim a larger evidence for the periodic model. Certainly there is a degree of arbitrariness in the prior parameter distribution -- in this case a uniform distribution -- and narrowing this range around this peak would clearly increase the evidence.  For example, if we truncate the period range from its current value of [10,\,100]\,Myr to [50,\,80]\,Myr, then the Bayes factor relative to the Uniform model increases from 0.62 to 1.5. This is a rather modest increase, but we could increase it to a significantly high value with an even narrower prior. However, {\em we may not use the data to find the best fitting parameters and then claim that we should only consider the model near to these}. 
We would need some other reasoning or independent data for making such a
selection. (The \cite{rohde05} time series is not independent of B18, because
both are based on the same paleontological data.) We do not see how, a priori, we could limit the plausible periods of periodicity to something as narrow as 50--80\,Myr, let alone the much narrower range required to favour PNB significantly over other models. In the extreme limit of an infinitesimal region around the maximum likelihood, we end up doing model comparison using the maximum likelihood. Just out of interest, these values are shown in Table~\ref{tab:BF-MLR} and plotted in Figure~\ref{fig:BF-MLR}. If we were to use this (incorrect) metric, then PNB and some of its variants have significantly higher likelihood than the Uniform model and several of the other models (although barely more than a factor of ten above the random model, RB).  Another way of seeing why this is the wrong approach was already discussed in Section~\ref{sec:overview_biodiversity}: by focusing on the {\em best} fits we simply favour the more complex model. We could always define a more flexible model and so fit even better.

We labour this point because many of the claims for a periodicity in biodiversity data have made use of a maximum likelihood approach (of which $\chi^2$ is a special case) or something equivalent.  We must instead use the evidence for model comparison. (Maximum likelihood may be used for estimating the best parameters once we've established we have the best model.) If the periodic model were in fact the true one, then of course only one period and phase would be true. In that case the likelihood around these values would be so high as to result in a large evidence even when averaging over the broader parameter space (see simulations in Section~4.1 of \cite{bailer-jones11} for a demonstration).

Incidentally, the fact that we find a dominant period at all in the B18 data set is actually not that unlikely. The (Bayesian) periodogram of samples drawn at random from a uniform distribution often exhibits a period 
which has a likelihood larger than that of the true Uniform model (see Section~4.2 of \cite{bailer-jones11}). In other words, it is often possible to explain a random data set with {\em some} period, which is just a testament to how flexible the periodic model is.

\begin{figure}[ht!]
\centering
\includegraphics[scale=0.6]{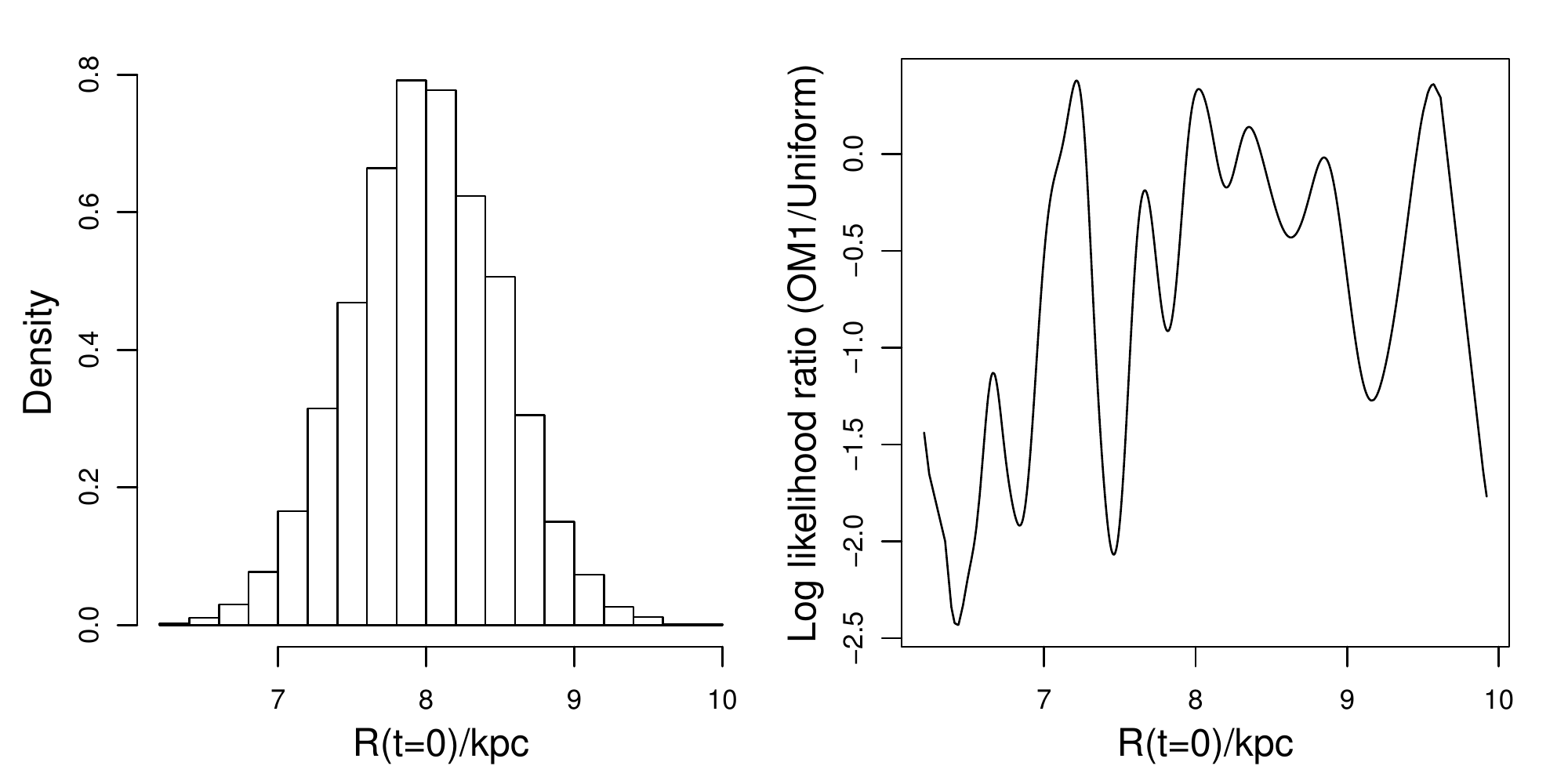}
\includegraphics[scale=0.6]{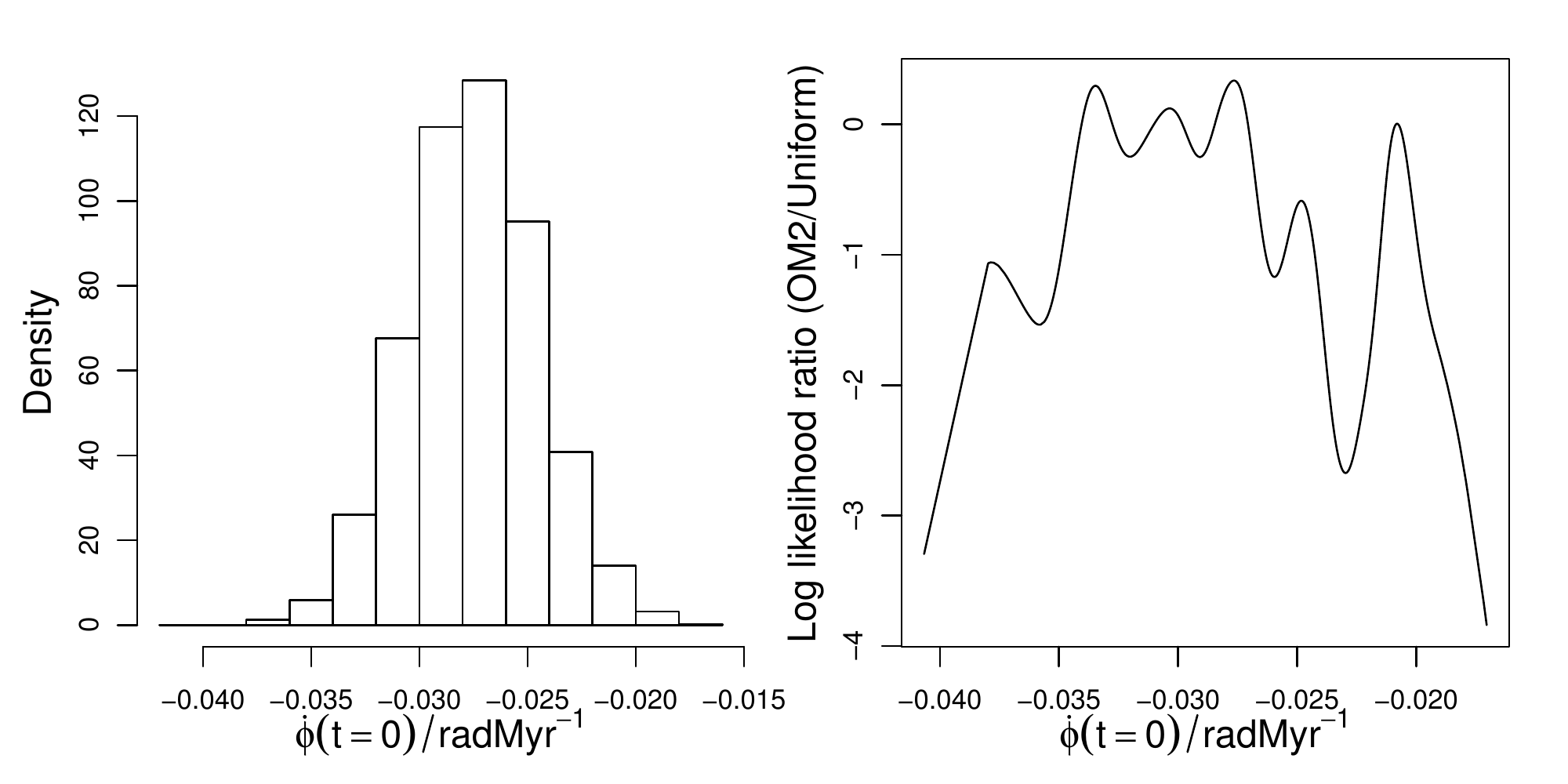}
\caption{The variation of initial conditions (left column) and corresponding
  variation of likelihood ratio relative to the Uniform model (right column) for the OM model on the B18 data set. The top row is for OM1, in which only $R(t=0)$ is varied. The bottom row is for OM2, in which only $\dot\phi(t=0)$ is varied.}
\label{fig:like-r-phidot-B18}
\end{figure}

Moving on from the periodic model, we show in Figure~\ref{fig:like-r-phidot-B18}
the likelihood distribution for OM1 and OM2, i.e.\ where we vary the 
initial conditions $R(t=0)$ and $\dot\phi(t=0)$, respectively.
The likelihood ratio varies by a factor of up to $10^4$, but its obsolute value is never more than about two.  That is, no value of the intial conditions gives a model much more favourable than the Uniform model, whereas as some are far less favourable.  If we had lower uncertainties on these phase space coordinates of the Sun, then we might be able conclude something more definitive. For example, if $R(t=0)=7.5$\,kpc then the OM model would be even less favored.  We see a similar degree of variability for the other OM models and data sets listed in Table~\ref{tab:BF-MLR}.

We performed a similar analysis for the other model, but for the sake of space report only the maximum likelihood ratios in Table~\ref{tab:BF-MLR}.

For the B5 data set, the RNB model has the highest maximum likelihood ratio, yet its evidence was
the lowest. This indicates that while one particular instance of RNB fit the data well, overall it is a poor model.

For the RM and A08 data sets, the evidences are very similar for all models. This means that the data are not able to discriminate between these models very well: they are equally good (or bad).  However, the time series analysis model used here is not best suited to these data sets. These data can be better interpreted as valued time series, ones in which we have an extinction magnitude attached to each time (both, in general, with uncertainties), rather than a time variable probability of extinction.
\cite{bailer-jones12} has extended the present model in order to work with such data sets; the results of its application to RM and A08 will be reported in a future publication.

In summary, we find that for none of the data sets is any model particularly favoured over the simple uniform one.
In particular, there is no evidence to suggest that the orbital model (with or without a background extinction level)
is a particularly good explanation of the data.

\subsection{\label{sec:sensitivity_biodiversity} Sensitivity test}

The evidence is of course sensitive to the choice of prior parameter distribution, and often we have little reason to make a very specific choice. Here we test the sensitivity of the evidence to this, as well as to the parameters of the Galactic potential used in the orbital models, and to the age uncertainties for the discrete data sets.

\begin{figure}[ht!]
\centering
\includegraphics[scale=0.7]{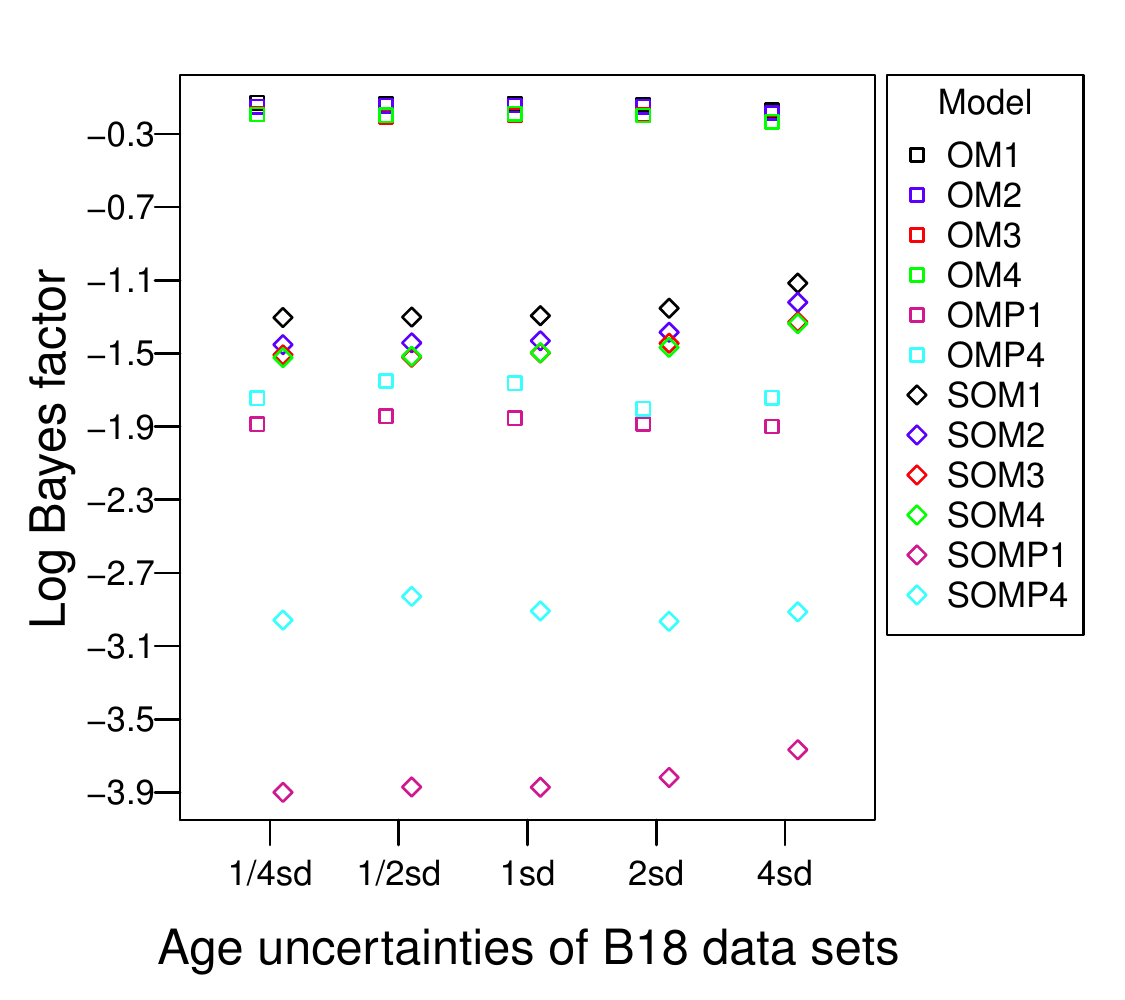}
\includegraphics[scale=0.7]{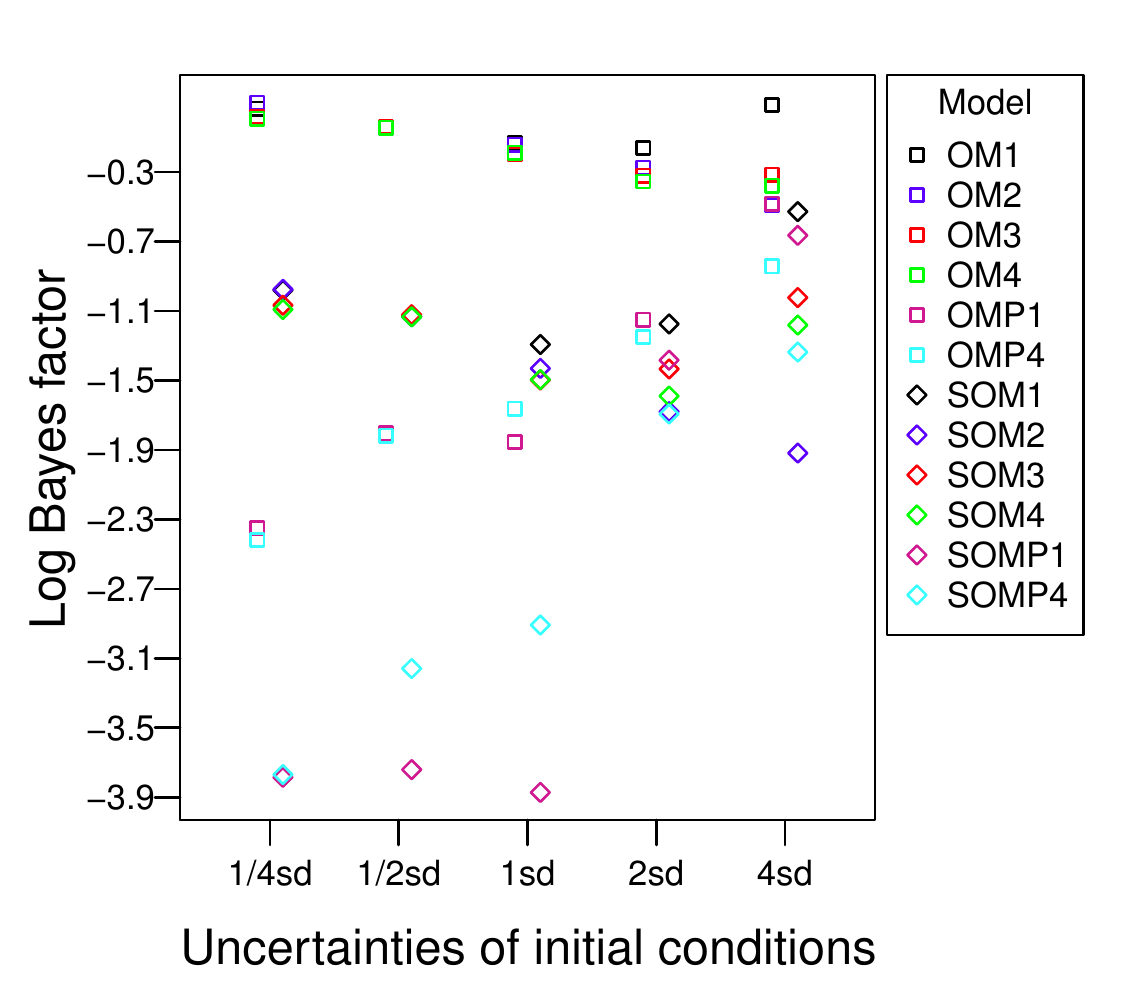}
\caption{Test of the sensitivity of the evidence to variations in the data age uncertainties and model priors. Both panels plot (on the vertical axies) the values of the the Bayes factor of the varied OM(P) and SOM(P) models (relative to the Uniform model) for the B18 data sets. In the left panel, the labels of the horizontal axis (Xsd) denote the varied B18 data sets, where Xsd means that we have multiplied the age uncertainties of the B18 time series by X for X=1/4, 1/2, 1, 2, 4. In the right panel, the labels of the horizontal axis (Xsd) denote variations of the uncertainties in (i.e.\ spread of) the initial conditions, where Xsd means multiplying the uncertainties of the initial conditions by a factor of X (X=1/4, 1/2, 1, 2, 4).  For each value of Xsd we have, for clarity, slightly offset in the horizontal direction the values for the OM(P) models (square points) from those for the SOM(P) models (diamond points).}
\label{fig:priortest}
\end{figure}

The age uncertainties in the discrete data are taken into account by the likelihood function. However, it is often difficult to estimate uncertainties, and we additionally made a plausible, but not unique, translation of the estimated duration of a stratigraphic substage in order to estimate uncertainty (which is the standard deviation of a Gaussian for each event; see Section~\ref{sec:datadiscrete_biodiversity}). To see how this affects our results, we scale the age uncertainties in the B18 data set by a constant factor of $1/4$, $1/2$, $2$, and $4$.  For each of these modified data sets we calculate the evidences for the models (S)OM1--4 and the Uniform model and recalculate the Bayes factors relative to the Uniform model.  These are plotted in the top panel of Figure~\ref{fig:priortest}: The Bayes factors change by just a few percent, so a precise age uncertainty is not necessary.

As a second test, we scale in the same way the uncertainties of the initial conditions of the orbital models (i.e.\ we change the width of the prior parameter distribution). The results are shown in the bottom panel of Figure~\ref{fig:priortest}.
The change in evidence for any particular model is larger than in the previous case, but in most cases less than a factor of 5 (except for the models including the spiral arms). Moreover, the absolute value of the Bayes factor remains below one. 

Our results are therefore also insensitive to considerable imprecision in the uncertainties in the phase space coordinates of the Sun. This (and the previous conclusion) is also true for the B5 data set.

As a third sensitivity test, we allow the number of simulated random events,
$N$, and the standard deviation of each event, $\sigma$, in the RNB and RB
models to vary. (Earlier we fixed $N=5$ when drawing models for the B5 data
set and $N=18$ for the B18 data set.) For the B18 data set, we find that a larger number of peaks or larger standard deviation in the RNB model produces a significantly larger Bayes factor (see Table~\ref{tab:sen-prior}), although it is still below unity. 
The RB model shows much less sensitivity to $N$ and $\sigma$.  We similarly recalculate the evidence for the other models in response to various perturbations of their priors, as also listed in Table~\ref{tab:sen-prior}.  The resulting Bayes factors do not change by more than a factor of 10 in any case, and often by much less.

\begin{table}[ht!]
\caption{The Bayes factors (relative to the Uniform model) 
  on the B18 data set for models with priors varied. Each prior is varied
  individually (listed in the middle column) with the other fixed at their
  canonical values.}
\centering
\begin{tabular}{l|c|r}
\hline
\hline
models&varied prior&BF\\
\hline
\multirow{4}{*}{PNB}
&none&0.62\\
&$50<T<80$&1.5\\
&$10<T<200$&0.31\\
&$10<T<400$&0.15\\
\hline
\multirow{3}{*}{PB}
&none&0.80\\
&$B=1/2$&0.88\\
&$B=2$&0.97\\\hline
\multirow{4}{*}{QPM}
&none&0.85\\
&$0<A_Q<1/4$&0.62\\
&$0<A_Q<1$&0.54\\
&$100<T_Q<300$&0.61\\
&$100<T_Q<500$&0.58\\\hline
\multirow{5}{*}{RNB}
&none&0.00050\\
&$\sigma=5$~Myr&$3.7\times 10^{-11}$ \\ 
&$\sigma=20$~Myr&0.026\\
&$N=9$&0.0085\\
&$N=36$&0.29\\\hline
\multirow{5}{*}{RB}
&none&0.40\\
&$\sigma=5$~Myr&0.73\\
&$\sigma=20$~Myr&0.25\\
&$B=\frac{1}{2\sqrt{2\pi}\sigma}$&0.13\\
&$B=\frac{2}{\sqrt{2\pi}\sigma}$&0.45\\
&N=9&0.84\\
&N=36&0.20\\
\hline
\multirow{4}{*}{SP}
&none&0.019\\
&$-200<\lambda<0$&0.070\\
&$0<\lambda<200$&0.10\\
&$10<t_0<500$&0.047\\\hline
\multirow{3}{*}{SSP}
&none&0.18\\
&$10<T<200$&0.17\\
&$-200<\lambda<200$&0.37\\
&$10<t_0<500$&0.27\\
\hline
\end{tabular}
\label{tab:sen-prior}
\end{table}

Finally, we test the sensitivity to the Bayes factors to changes in the parameters
of the Galaxy model (canonical values listed in Table~\ref{tab:model_par}).
The results are shown in Table~\ref{tab:sen-model-par}) for the B18 data set. If we double the mass of the
halo, for example, then the evidence for the OM and SOM models changes by no more than a factor of three.
Some other changes produce smaller affects, some larger, but not more than by a factor of five (and note that a change in a factor of two of the scale lengths is beyond what is consistent with observed data). 
Changes in the parameters of models which include spiral arms (the OMP and SOMP models) can produce
larger changes in the Bayes factors.
However, most significantly,
none of these changes produce a Bayes factor greater than one. In other words, none of these changes result in the orbital model becoming a better explanation for the paleontological than the Uniform model. 

\begin{table}
\caption{The Bayes factors on the B18 data set for orbital models with varied Galaxy parameters.}
\label{tab:sen-model-par}
\hspace{-3em}{\footnotesize
\begin{tabular}{l*{12}{c}}
\hline
\hline
variation          &OM1  &OM2  &OM3 &OM4 &OMP1   &OMP4  &SOM1  &SOM2 &SOM3  &SOM4   &SOMP1    &SOMP4\\
\hline
none          &0.74 &0.72 &0.63&0.65&0.014  &0.022 &0.051 &0.037 &0.032  &0.032&1.3$\times$10$^{-4}$&1.2$\times$10$^{-3}$\\
$2M_b$          &0.58 &0.45 &0.54&0.55&0.011  &0.018 &0.043 &0.011 &0.027 &0.028 &5.8$\times$10$^{-4}$&3.8$\times$10$^{-4}$\\
$1/2M_b$        &0.78 &0.85 &0.67&0.67&0.0066 &0.021 &0.040 &0.045 &0.030 &0.030 &1.4$\times$10$^{-4}$&1.9$\times$10$^{-3}$\\
$2b_b$          &0.74 &0.72 &0.64&0.63&0.0036 &0.013 &0.055 &0.042 &0.033 &0.032 &4.6$\times$10$^{-5}$&6.0$\times$10$^{-4}$\\
$1/2b_b$        &0.74 &0.73 &0.65&0.63&0.019  &0.025 &0.051 &0.037 &0.032 &0.030 &2.0$\times$10$^{-4}$&1.5$\times$10$^{-3}$\\
$2 M_h$         &0.60 &0.68 &0.57&0.57&0.00014&0.0031&0.080 &0.069 &0.087 &0.083 &4.6$\times$10$^{-6}$&9.4$\times$10$^{-5}$\\
$1/2 M_h$       &0.81 &0.68 &0.66&0.66&0.011  &0.014 &0.20  &0.059 &0.15  &0.15  &2.3$\times$10$^{-4}$&2.8$\times$10$^{-4}$\\
$2 b_h$         &0.81 &0.52 &0.66&0.65&0.011  &0.0061&0.19  &0.027 &0.15  &0.15  &1.1$\times$10$^{-3}$&3.5$\times$10$^{-4}$\\
$1/2 b_h$       &0.073&0.048&0.11&0.11&0.024  &0.062 &0.0046&0.0012&0.010 &0.010 &7.5$\times$10$^{-3}$&1.4$\times$10$^{-2}$\\
$2 M_d$         &0.21 &0.063&0.33&0.32&0.035  &0.058 &0.0047&0.0046&0.027 &0.024 &2.2$\times$10$^{-5}$&7.6$\times$10$^{-4}$\\
$1/2 M_d$       &0.59 &0.56 &0.49&0.48&0.0031 &0.0037&0.0073&0.0018&0.0086&0.0088&3.3$\times$10$^{-4}$&9.6$\times$10$^{-4}$\\
$2 a_d$         &0.86 &0.89 &0.76&0.76&0.13   &0.12  &0.10  &0.088 &0.069 &0.067 &6.3$\times$10$^{-3}$&7.0$\times$10$^{-3}$\\
$1/2 a_d$       &0.63 &0.66 &0.55&0.55&0.021  &0.019 &0.015 &0.011 &0.022 &0.026 &2.5$\times$10$^{-3}$&9.1$\times$10$^{-4}$\\
$2 b_d$         &1.1  &1.1  &0.93&0.93&0.0056 &0.020 &0.028 &0.030 &0.027 &0.025 &1.3$\times$10$^{-4}$&4.2$\times$10$^{-3}$\\
$1/2 b_d$       &0.72 &0.87 &0.60&0.56&0.00085&0.0073&0.15  &0.17  &0.10  &0.090 &1.3$\times$10$^{-4}$&1.5$\times$10$^{-3}$\\
\hline
\end{tabular}
}
\end{table}

In summary, we find that the evidences for most models are not particularly sensitive to the age uncertainties, Galaxy model parameters, or reasonable changes to the prior parameter distributions.

\subsection{\label{sec:power_biodiversity}Testing the discriminative power}

Here we investigate how well our analysis can discriminate between models, by using simulated data which have been drawn from one of these models. Given sufficient data, such a discrimination will always be possible to some threshold Bayes factor, but here we are interested in the case where the data have similar properties (in particular, sparsity) to the real data we have been using.

We investigate this by simulating a number of of time series from the RNB, OM1, and PNB models (which we will refer to as ``generative models'' when used in this way, to distinguish from their use to calculate the evidence on given data). For each generative model, we fix the parameters to certain values, then sample 18 events from the resulting $P(t_j|\boldsymbol{\theta}, M)$ to give a simulated time series, to which we then attach the measured age uncertainties. We generate ten time series in this way (and below we average the Bayes factors over these and report that). For the OM1 and PNB generative models we repeat this at ten different values of the solar initial radius parameter (OM1) or period parameter (PNB) in the generative model.  (RNB has no parameters). We repeat the whole process a second time but using simulated age uncertainties drawn from a log normal distribution with standard deviation and mean calculated from the measured age uncertainties. We finally repeat the process a third time for the OM1 and PNB generative model, but now drawing data to have the same time sampling as our continuous data sets (for which age uncertainties are not used; see section~\ref{sec:continuous_biodiversity}). (We do not do this for the RNB model as it predicts discrete events.)

For each simulated data set we calculate Bayes factors for the RNB, OM1, and PNB models relative to the Uniform model. For the data generated from the RNB model (with age uncertainties taken from the data), the Bayes factors for the three models are as follows: 0.18 for RNB; 0.57 for OM1; 0.52 for PNB. (We get almost identical values when the age uncertainties were drawn at random). Thus no model -- not even the true one -- is favored over the Uniform model (although none is significantly rejected either). This is not that surprising, however, because with only 18 events, and with the evidence effectively averaging the predicted times of events from the RNB model over all time, the Uniform and RNB models end up with similar predictive power. This is unavoidable, because with the RNB model we cannot decide in advance where the events are: we must average over all possibilities.

\begin{figure}
  \centering
  \includegraphics[scale=0.6]{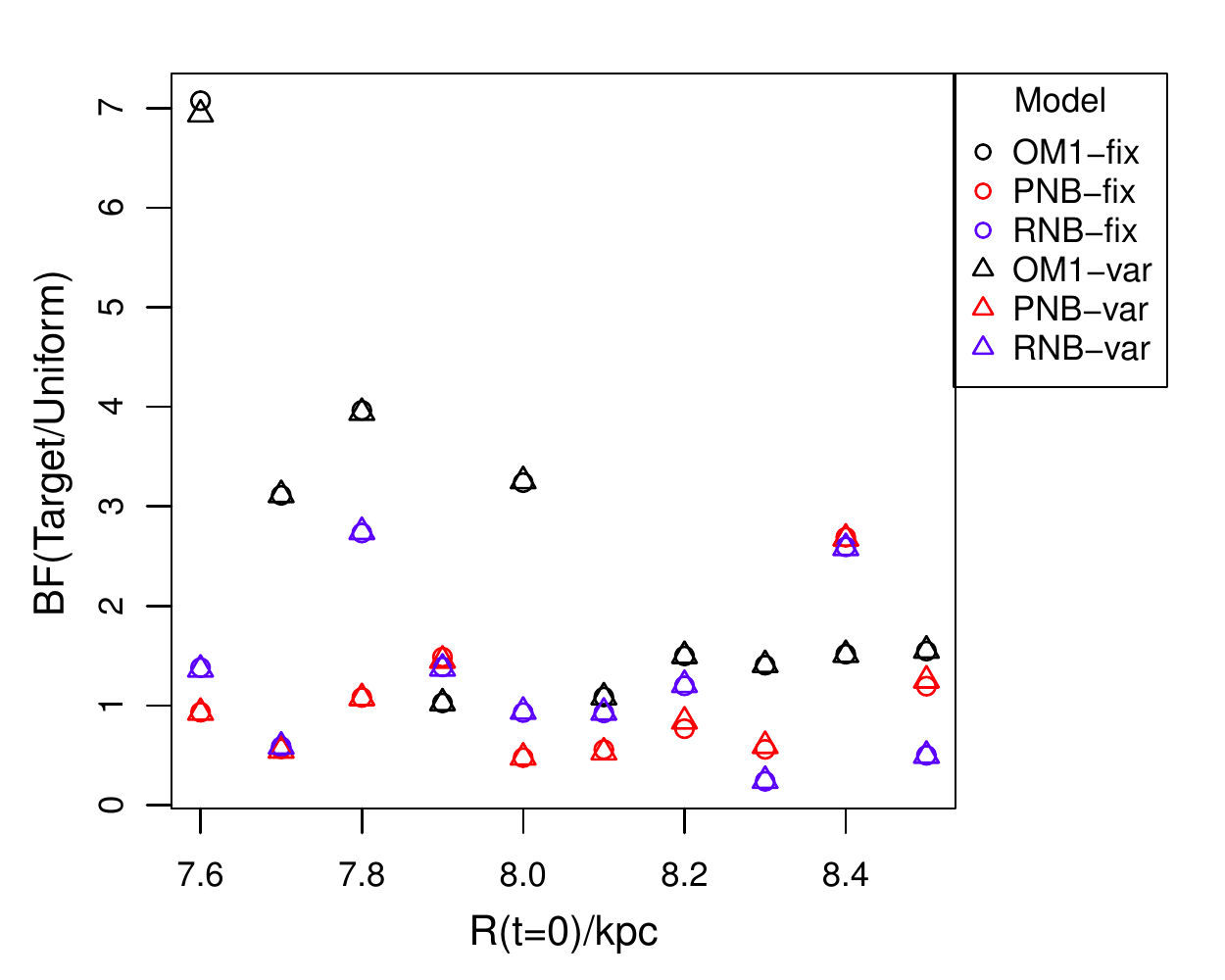}
  \includegraphics[scale=0.6]{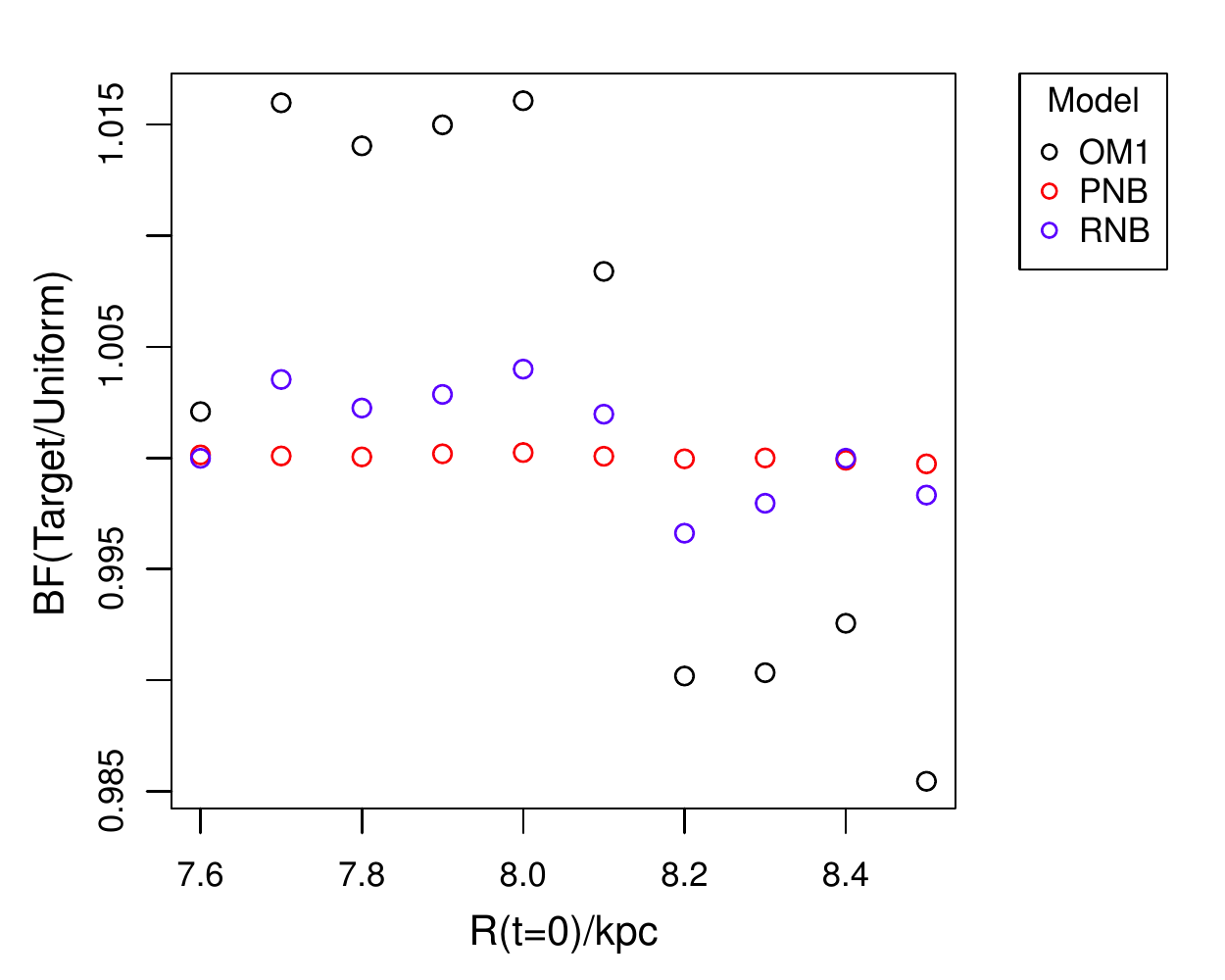}
  \includegraphics[scale=0.6]{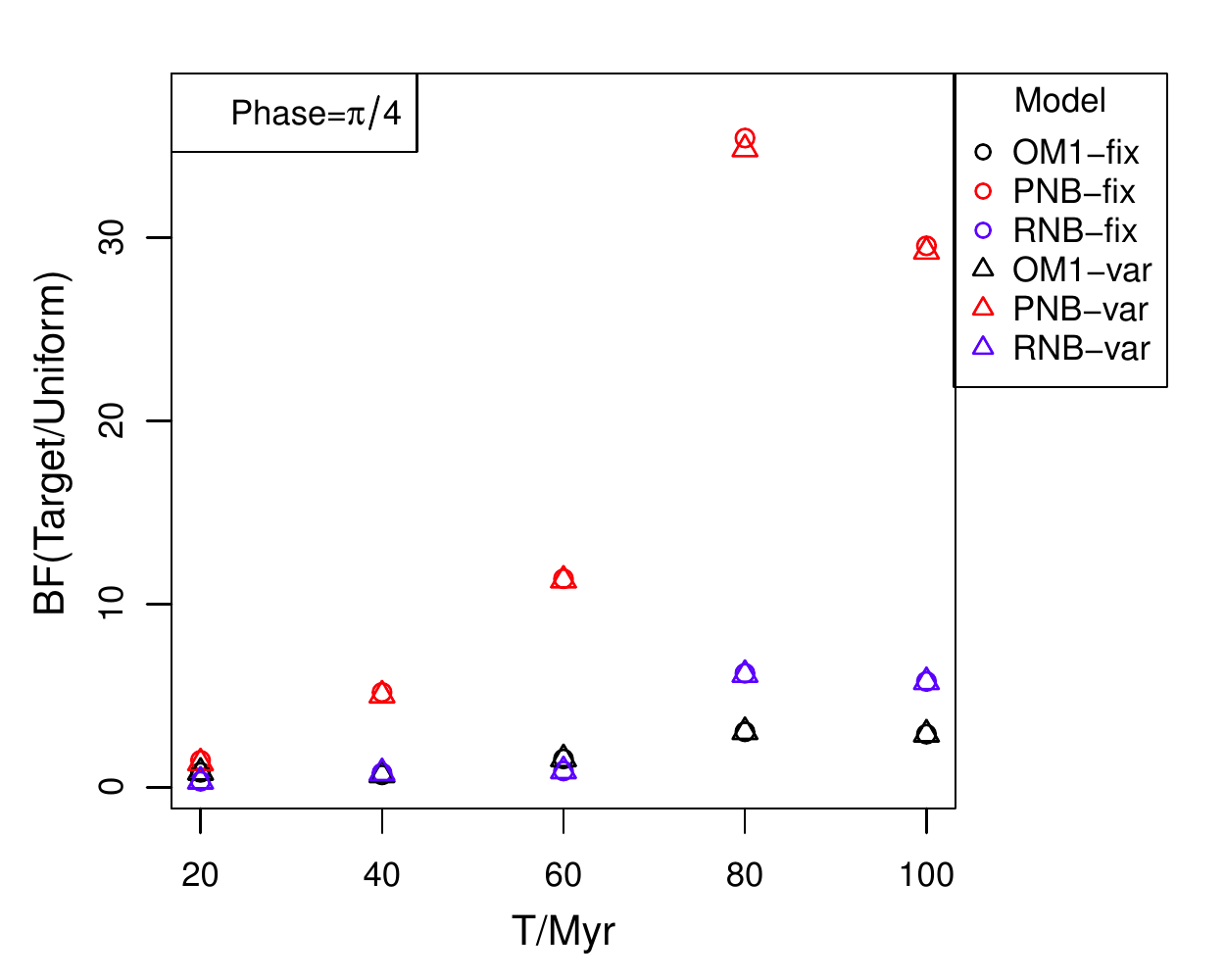}
  \includegraphics[scale=0.6]{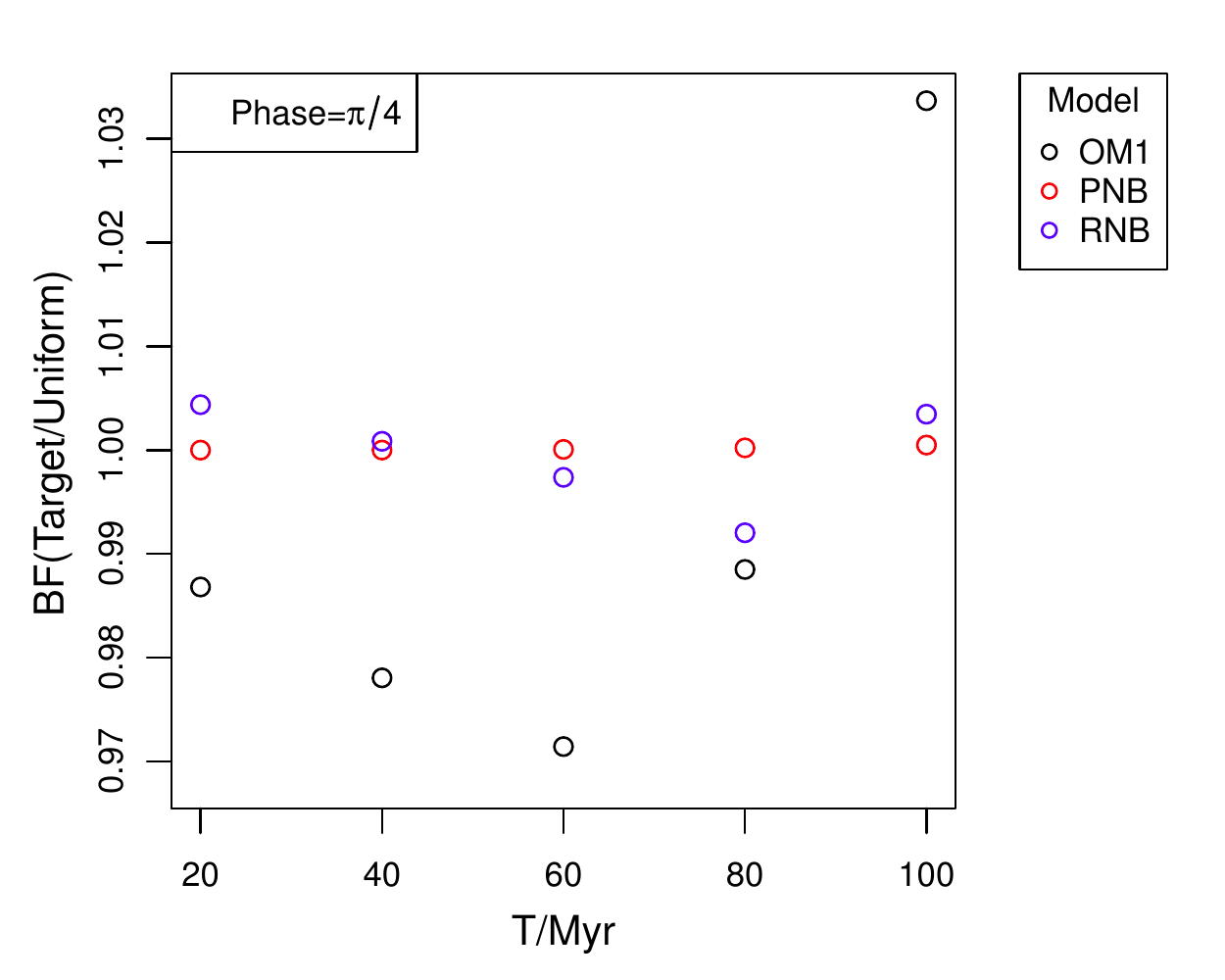}
  \includegraphics[scale=0.6]{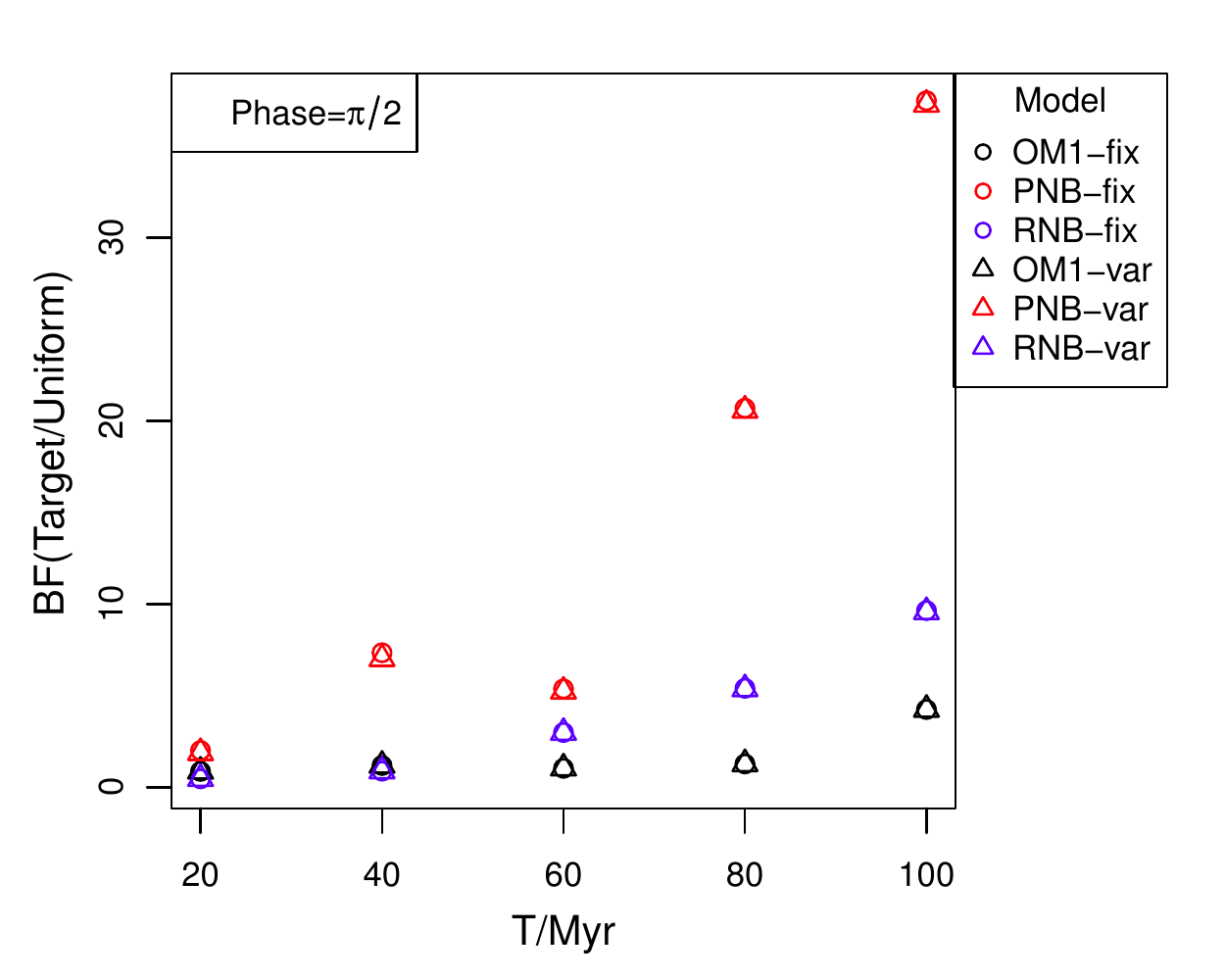}
  \includegraphics[scale=0.6]{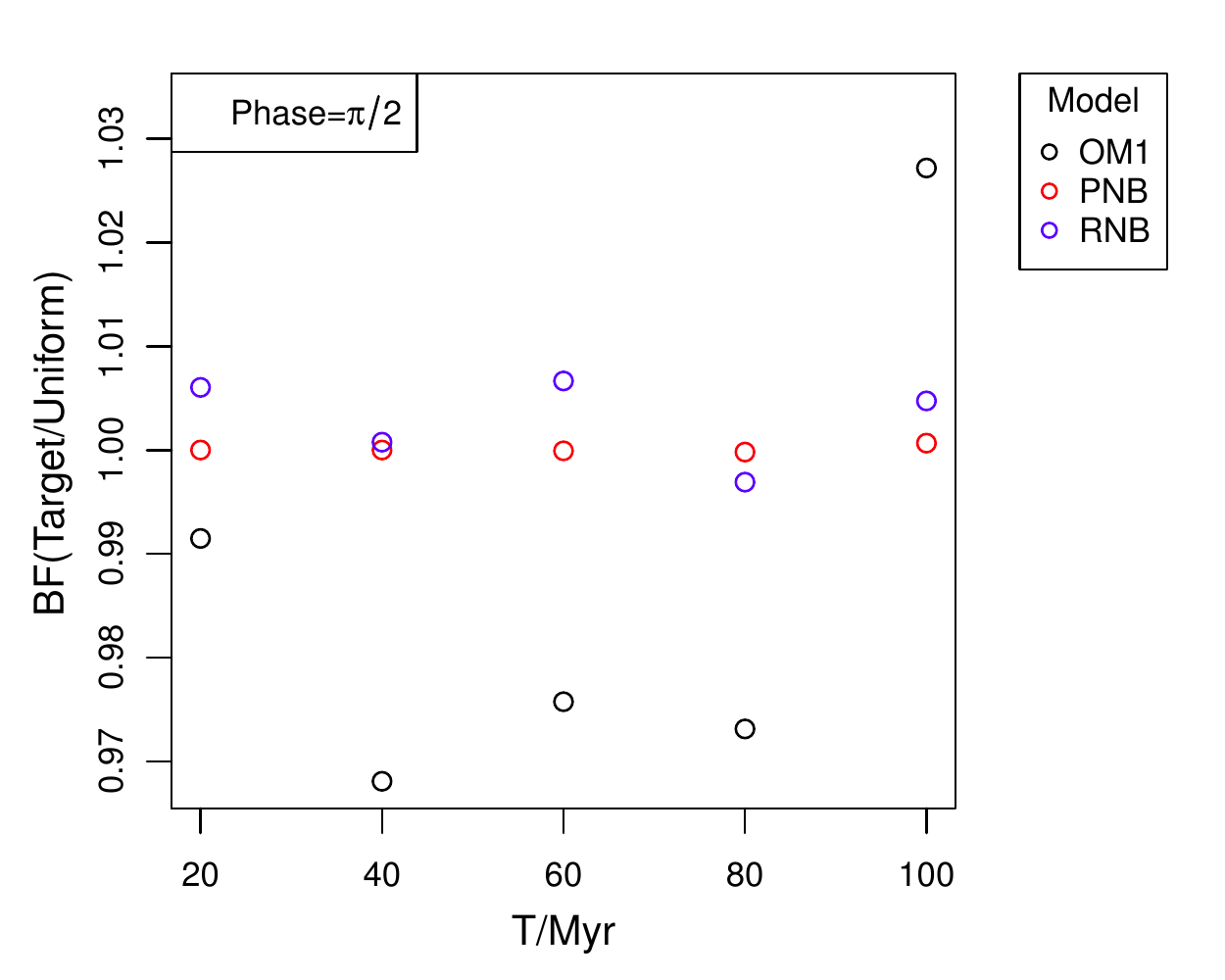}
  \caption{The Bayes factors for various models computed on simulated data sets. The horizontal axis in each panel indicates the value of the parameter in the model used to generate these data sets (all other parameters are kept fixed). The left panels are for discrete data sets and the right ones for continuous data sets.  The top row is for data drawn from the OM1 model, and the bottom two rows for data drawn from the PNB model (for two different values of that generative model's phase parameter, $\pi/4$ and $\pi/2$). The models for which the Bayes factors have been computed are shown in different symbols defined in the plot: OM1, PNB and RNB. The suffices ``fix'' and ``var'' indicate ages uncertainties for the discrete data either taken from the real data or drawn from a log normal distribution (respectively). }
\label{fig:sim-data-comp}
\end{figure}

The results for applying the models to the data generated from the OM1 and PNB models are shown in 
Fig.~\ref{fig:sim-data-comp}, where the horizontal axis shows how the Bayes factor varies with the one parameter which is varied in these generative models.  The top row shows the results for data drawn from the OM1 model, for the discrete (B18-like) data (left) and the continuous data (right). We see that the (true) OM1 model is not significantly favored in either case (Bayes factor always less than ten), although not disfavored either (Bayes factor more than 0.1). In particular, the continuous data show no discriminative power.

The lower two rows show conceptually the same thing, but now for data drawn from the PNB model for two different values of the phase parameter (the two rows). Here we see that, at least for longer periods, the PNB model is generally correctly identified (on the basis of a large Bayes factor), when using the discrete data sets. Yet the continuous data still show no discriminative power.

This difference between discrete and continuous data sets is not unexpected.
In the former, the likelihood is a product of the likelihood for many events, each of which is the convolution
of the event with the model. In the latter, the likelihood is just the result of a single convolution 
of a continuous model over continuous data. This is not the best approach for modeling continuous data. A better choice is the recently developed continuous time series modeling method described in 
\cite{bailer-jones12}, which will be used in future work.

Clearly one could perform many more tests with more simulated time series, varying different parameters in the generative models and with different permutations of the values of the fixed parameters. No doubt there are parts of parameter space where some models are favored over others, in particular if we adopted more informative priors. Thus while these results on simulated data give some check on the discriminative power of the method and data, they should not be over-interpreted to say anything too general. Nonetheless, the tests we have done confirm what we concluded based on the analysis of the real data. Specifically, while our analysis of the real data does not allow us to claim evidence in favor of the orbital-based models, it also cannot rule out these models. This is due partly to the lack of predictive power of the data, and partly to the large flexibility (or broad prior parameter space) of the models. Better constraints on the solar orbit would help reduce the latter.

\section{\label{sec:conclusion_biodiversity} Summary and conclusions}

We have used a Bayesian model comparison method to examine how well different time series models explain the variation of biodiversity over the Phanerozoic eon (the past 550\,Myr). One class of models is derived from the orbit of the Sun around the Galaxy, which we reconstructed from a model of the Galactic mass distribution. Our model comparison takes into account uncertainties in the data as well as uncertainties in the reconstructed path of the Sun. We have compared the evidence for this model with that of various other reference models of no particular causal origin. All models are stochastic in the sense that they predict only the time variation of the extinction probability, rather than the exact magnitude of the extinction rate or the times of mass extinctions.

As part of this analysis we investigated the properties of plausible solar orbits (i.e.\ those consistent with the  accuracy of the present phase space coordinates of the Sun). We find that the majority of orbits have a motion perpendicular to the Galactic plane which is not far from periodic, although a precise period cannot be inferred due to the uncertainties in the present solar phase space coordinates as well as the exact mass distribution (gravitational potential) of the Galaxy. Thus any claims which try to link a variation in geological biodiversity, cratering or climate records to this solar motion must consider the motion as quasi-periodic rather than strictly periodic.

In contrast, only about half of the simulated orbits showed periodic spiral arm crossings, even for a very simple, rigidly rotation arm model. Indeed, many of the orbits did not encounter the spiral arms more than once. It should be noted that the shape and pattern speed of the spiral arms is poorly known (and they may not even be long-lived), so any claims of a causal connection between spiral arm passages and terrestrial conditions should be treated with due skepticism.
 
We have shown how the evidence (marginal likelihood) should be used to do model comparison, as opposed to selecting the model which gives the best single fit. The reason is that an arbitrarily complex model can always be tuned to fit the data arbitrarily well, yet that does not make it a good model.  By averaging the likelihood over the parameter space, the evidence uses the rules of probability to trade off the quality of the fit with the model plausibility in a quantitative fashion. Of the models investigated, we do not claim any one of them to be ``true''. Indeed, no model is exactly true in reality.  All we can hope to do is to find the best of the ones tested so far.

We find that none of the models tested -- including periodic, quasi-periodic and orbital-based -- explain the discrete data sets better than a Uniform model. In other words, the time distribution of mass extinction events is consistent with being randomly distributed in time. There is no need to resort to anything more exotic.

The Uniform model is also no worse than other models for the continuous data sets. This does not mean that we believe the extinction rate has been constant over the Phanerezoic, but rather that none of the other (more complex) models is {\em significantly} better. Assuming the variations in extinction seen in Figure~\ref{fig:data} are true -- something we have no reason to doubt -- then this tells us that there must be other models, not yet tested, which could explain the data better.  This will be investigated in future work using a model more suited for these types of time series.

We found in particular that the orbital-based extinction model is not favored by the data. This conclusion is robust to changes in the parameters of the Galaxy model and to the magnitude of the uncertainties of both the solar phase space coordinates and the ages of the extinction events.  
On the other hand, our analysis of simulated data showed that even if the orbital model were the true one, our analysis could not have identified it with either the discrete or (in particular) the continuous data sets. This ultimately comes down to a combination of a lack discriminative power in the data, plus a large flexibility (or prior parameter space) in the models.
Of course, if the orbit of the Sun could be much better determined then it is possible that this model would then be more -- or less -- favored by our analysis.  We remind the reader that our orbital model adopted an extinction mechanism in which the extinction rate is proportional to the integrated ``flux'' (of a non-specified type) from nearby stars. A radical change in this mechanism would of course correspond to a quite different model, which could give different results.  Thus we do not claim that the solar motion plays no part in terrestrial extinction, nor that astronomical mechanisms are irrelevant.

Indeed, it is quite plausible that the biological extinction rate has been affected by many factors, and so any attempt to connect them solely to the solar motion, or indeed to any simple analytic model, is doomed from the start. We have addressed this to some extent by including compound models and the semi-orbital model, but clearly one could do more. However, given the present uncertainties of the reconstructed solar orbit, it seems unlikely that one could draw a strong conclusion on the positive relevance of the solar orbit on the basis of current geological data.  This, indeed, is the main conclusion of this work, plus the confirmation that periodic models are not a good (or necessary) explanation of the biodiversity variation.  There is some hope that, in the future, results from the Gaia survey of the Galaxy (e.g.\ \cite{lindegren08}) will improve our knowledge of the Galactic potential, spiral arms and inferred solar orbit, to the extent that this study can be repeated to give conclusions of greater certainty.

\cleardoublepage
\newpage

\chapter{Exploring the role of the Sun's motion in terrestrial comet impacts}\label{cha:comet}
{\it This chapter is adapted from my published work \citep{feng14}. }
\section{Chapter summary}\label{sec:abstract_comet}

The cratering record on the Earth and Moon shows that our planet has been exposed to high velocity impacts for much or all of its existence. Some of these craters were produced by the impact of long period comets (LPCs).  These probably originated in the Oort cloud, and were put into their present orbits through gravitational perturbations arising from the Galactic tide and stellar encounters, both of which are modulated by the solar motion about the Galaxy. Here we construct dynamical models of these mechanisms in order to predict the time-varying impact rate of LPCs and the angular distribution of their perihelia (which is observed to be non-uniform). Comparing the predictions of these dynamical models with other models, we conclude that cometary impacts induced by the solar motion contribute only a small fraction of terrestrial impact craters over the past 250\,Myr. Over this time scale the apparent cratering rate is dominated by a secular increase towards the present, which might be the result of the disruption of a large asteroid. Our dynamical models, together with the solar apex motion, predict a non-uniform angular distribution of the perihelia, without needing to invoke the existence of a massive body in the outer Oort cloud. Our results are reasonably robust to changes in the parameters of the Galaxy model, Oort cloud, and stellar encounters.

\section{Introduction}\label{sec:introduction_comet}

\subsection{Background}\label{sec:background_comet} 

As is mentioned in section \ref{sec:geological}, and in \cite{feng14}, comet and asteroid impacts can significantly influence the Earth's biosphere and climate.
Many previous studies have attempted to identify patterns in the temporal distribution of craters and/or mass extinction events (e.g.\ \citealt{alvarez84,raup84,rohde05,melott11}), although other studies have come to other conclusions (e.g.\ \citealt{grieve96,yabushita96,jetsu00}). These controversies are partly caused by not distinguishing different mechanisms which account for the temporal distribution of craters, and partly by unreliable analyses. We try to model the impact rate as a function of time by accounting for different mechanisms, and compare these models in a Bayesian framwork. 

In doing this we should distinguish between asteroid and comet
impacts.  Having smaller relative velocities, asteroid impacts are
generally less energetic. Asteroids originate from within a few AU of
the Sun, so their impact rate is probably not affected much by events
external to the solar system. Comets, on the other hand, originate
from the Oort cloud \citep{oort50}, and so can be affected by the
Galactic environment around the Sun.

As the solar system orbits the Galaxy, it experiences gravitational perturbations from the Galactic tide and from encountering with individual passing stars. These perturbations are strong enough to modify the orbits of Oort cloud comets to inject them into the inner solar system \citep{wickramasinghe08,gardner11}.  The strength of these perturbations is dependent upon the local stellar density, so the orbital motion of the Sun will modulate these influences and thus the rate of comet injection and impact to some degree (e.g.\ \citealt{brasser10,kaib11,levison10}). As the Sun shows a (quasi)-periodic motion perpendicular to the Galactic plane, and assuming that the local stellar density varies in the same way, it has been argued that this could explain a (supposed) periodic signal in the cratering record.  Here we will investigate the connection between the solar motion and the large impact craters (i.e.\ those generated by high energy impacts) more explicitly.  We do this by constructing a dynamical model of the Sun's orbit, the gravitational potential, and the resulting perturbation of comet orbits, from which we will make probabilistic predictions of the time variability of the comet impact rate.

The dates of impact craters are not the only relevant observational
evidence available.  We also know the orbits of numerous long-period
comets (LPCs).  The orbits of dynamically new LPCs -- those which
enter into the inner solar system for the first time -- record the
angular distribution of the cometary flux.  This distribution of their
perihelia is found to be anisotropic. Some studies interpret this as
an imprint of the origination of comets \citet{bogart82, khanna83},
while others believe it results from a perturbation of the Oort Cloud.
Under this perturbation scenario, it has been shown that the Galactic
tide can (only) deplete the pole and equatorial region of the Oort
Cloud \citep{delsemme87} in the Galactic frame, and so cannot account
for all the observed anisotropy in the LPC perihelia. It has been
suggested that the remainder is generated from the perturbation of
either a massive body in the Oort Cloud \citep{matese99, matese11} or
stellar encounters \citep{biermann83,dybczynski02}.

\subsection{Overview}\label{sec:overview_comet} 

Assuming a common origin of both the large terrestrial impact craters
and the LPCs, we will construct dynamical models of the flux and
orbits of injected comets as a function of time based on the solar
motion around the Galaxy. Our approach differs from previous work in
that we (1) simulate the comet flux injected by the Galactic tide and
stellar encounters as they are modulated by the solar motion; (2) use an
accurate numerical method rather than averaged Hamiltonian
\citep{fouchard04} or Impulse Approximation
\citep{oort50,rickman76,rickman05} in the simulation of cometary
orbits; (3) take into account the influence from the Galactic bar and
spiral arms; (4) test the sensitivity of the resulting cometary flux
to varying both the initial conditions of the Sun and the parameters
of the Galaxy potential, Oort Cloud, and stellar encounters.

We build the dynamical models as follows. Adopting models of the
Galactic potential, Oort Cloud and stellar encounters, we integrate
the cometary orbits in the framework of the AMUSE software
environment, developed for performing various kinds of astrophysical
simulations \citep{portegies13, pelupessy13}. The
cometary orbits can be integrated with the perturbation from either
the Galactic tide, or stellar encounters, or both. All three are investigated. In principle, we can
build a three-parameter dynamical model for the variation of the
impacting comet flux as a function of time, Galactic latitude, and
Galactic longitude. In practice we reduce this three-parameter model
to a 1-parameter model of the variation of the comet impact rate over
time, and a 2-parameter model of the angular distribution of the
perihelia of LPCs.  A further simplification is achieved by replacing
the full numerical computations of the perturbations by separating
proxies for the tide-induced comet flux and for the encounter-induced
comet flux.  These are shown to be good approximations which accelerate
considerably the computations.

We combine the predictions of the comet impact history with a (parameterized) component which accounts for the crater preservation bias (i.e.\ older craters are less likely to be discovered) and the asteroid impact rate. We then use Bayesian model comparison to compare the predictions of this model over different ranges of the model parameters to the observed cratering data, using the crater data and statistical method presented in \cite{bailer-jones11}.

We obtain the 2-parameter model for the angular distribution of the perihelia of LPCs by integrating the full 3-parameter model over time. Because we no longer need the time resolution, we actually perform a separate set of numerical simulations to build this model.  We then compare our results with data on 102 new comets with accurately determined semi-major axes (the ``class 1A'' comets of \citealt{marsden08}).

This chapter is organized as follows. We introduce, in section \ref{sec:data_comet}, the data on the craters and LPCs. In section \ref{sec:simulation_comet} we define our models for the Oort cloud, and for stellar encounters, and describe the method for the dynamical simulation of the comet orbits.  In section \ref{sec:bayes} we summarize the Bayesian method of model comparison. In section \ref{sec:impact_comet} we use the dynamical model to construct the 1-parameter model of the cometary impact history. In Section \ref{sec:comparison_comet}, we compare our dynamical time series models of the impact history with other models, to assess how well the data support each. In section \ref{sec:ADP_comet} we use the dynamical model again, but this time to predict the distribution of the perihelia of LPCs (the 2-parameter model), which we compare with the data. A test of the sensitivity of these model comparison results to the model parameters is made in section \ref{sec:sensitivity_comet}. We discuss our results and conclude in section \ref{sec:conclusion_comet}.

The main symbols and acronyms used in this article are summarized in Table \ref{tab:symbol}.
\begin{table}[ht!]
\centering
\caption{Glossary of main acronyms and variables}
\label{tab:symbol}
\begin{tabular}{l l}
  \hline
  \hline
  Symbol & Definition\\
  \hline
  PDF       & probability density function \\
  LSR       & local standard of rest \\
  HRF       & heliocentric rest frame \\
  BP        & before present\\
  LPC       & long-period comet\\
  $s_j$      & crater age\\
  $\sigma_t$ & age uncertainty of crater   \\
  $s^{up}$    & upper limit of the age of crater\\
  ${\vec r}_{\rm enc}$  &impact parameter or perihelion of encounter\\
  $ \vec{v}_\star$&velocity of a star in the LSR\\
  $ \vec{v}_{\rm enc}$&velocity of the stellar encounter relative to the Sun\\
  $ b_\star $ &Galactic latitude of $\vec{v}_\star$   \\
  $ l_\star $ &Galactic longitude of $\vec{v}_\star$   \\
  $ b_{\rm enc} $ & Galactic latitude of $\vec{v}_{\rm enc}$   \\
  $ l_{\rm enc} $ &Galactic longitude of $\vec{v}_{\rm enc}$   \\
  $ b_p$ & Galactic latitude of the perihelion of a stellar encounter \\
  $ l_p$ & Galactic longitude of the perihelion of a stellar encounter \\
  $ b_c $    & Galactic latitude of cometary perihelion \\
  $ l_c $    &Galactic longitude of cometary perihelion\\
  $ q $      & perihelion distance\\
  $ a $      & semi-major axis\\
  $ e $      & eccentricity\\
  $M_{\rm enc}$   &mass of a stellar encounter\\
  $v_{\rm enc}$ &  speed of a star at encounter\\
  $r_{\rm enc}$ & distance of a star at encounter\\
  $f_c$     &injected comet flux relative to the total number of comets\\
  $\bar{f}_c$     &averaged $f_c$ over a time scale\\
  $\gamma$   &parameter of impact intensity$\frac{M_{\rm enc}}{v_{\rm enc}r_{\rm enc}}$\\
  $\gamma_{\rm bin}$ &normalized maximum $\gamma$ in a time bin\\
  $G_1$, $G_2$&coefficients of radial tidal force\\ 
  $G_3$     &coefficient of vertical tidal force\\ 
  $\rho$    &stellar density\\
  $\eta$    & ratio between the trend component and $f_c$\\
  $\xi$     & ratio between the tide-induced flux and encounter-induced flux\\
  $\kappa$  & angle between ${\vec r}_{\rm enc}$ and the solar apex\\
  $M_s$ & mass of the Sun\\
  \hline
\end{tabular}
\end{table}

\section{Data}\label{sec:data_comet} 

\subsection{Terrestrial craters}\label{sec:craters_comet}

\begin{figure}[ht!]
\centering
\includegraphics[scale=0.5]{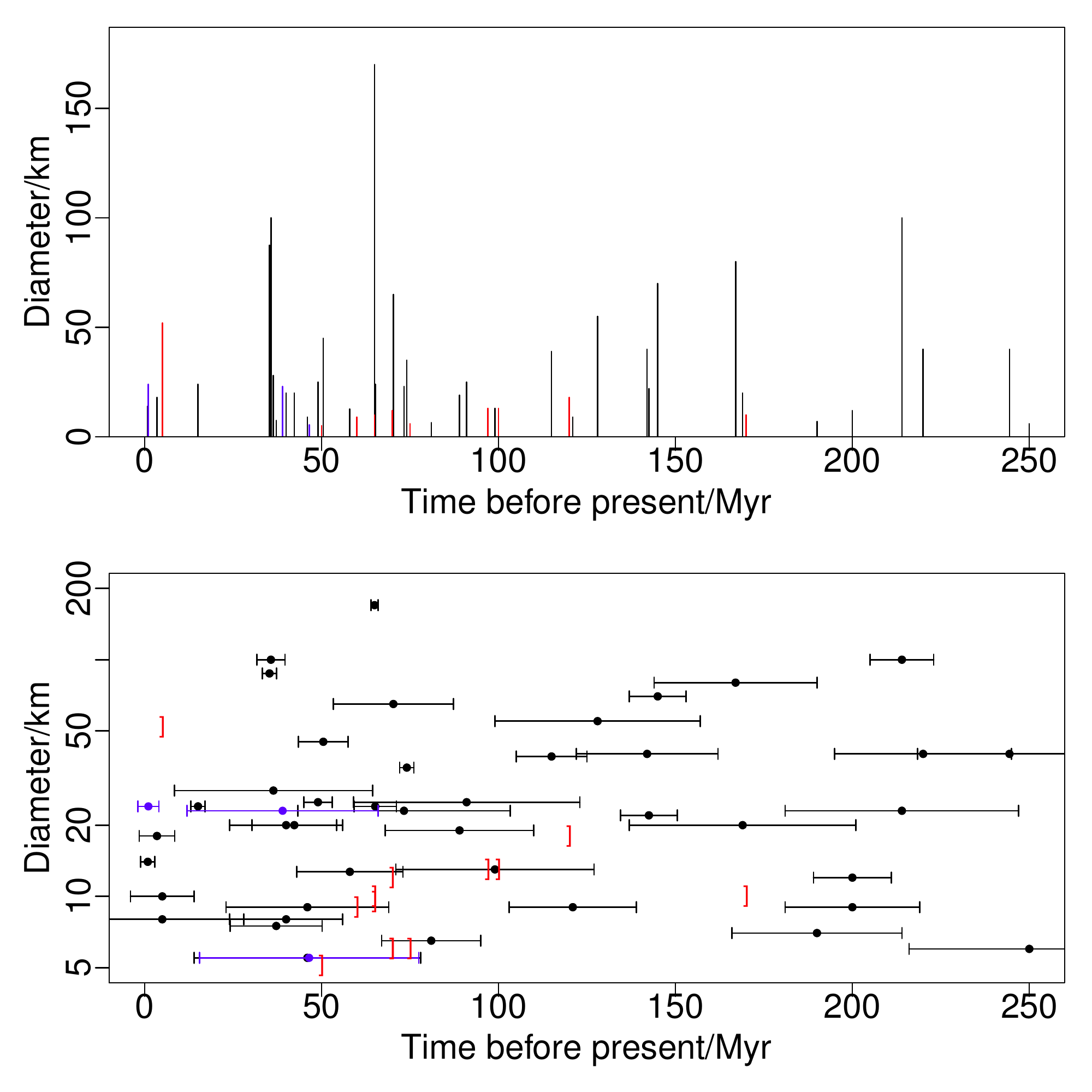}
\caption{The diameters and ages of the 59 craters with (bottom) and
  without (top) age uncertainties plotted. The blue points/lines
  indicate the craters with assigned age uncertainties. The red
  lines/brackets indicate the upper ages of the craters without
  well-defined ages. Adapted from \citet{bailer-jones11}. }
\label{fig:set3lim}
\end{figure}

The data of craters we use in this work is from the {\em Earth Impact
  Database} (EID) maintained by the Planetary and Space Science Center
at the University of New Brunswick.  We restrict our analysis to
craters with diameter $>5$ km and age $<250$\,Myr in order to reduce
the influence of crater erosion (although an erosion effect is
included in our time series models).  We select the following data
sets defined by \citet{bailer-jones11}
\begin{itemize}
\item{\bf basic150} (32 craters) age $\le$ 150\,Myr, $\sigma_t$ original
\item{\bf ext150} (36 craters) age $\le$ 150\,Myr, original or assigned
\item{\bf full150} (48 craters) ext150 plus craters with $s^{up}\le$ 150\,Myr
\item{\bf basic250} (42 craters) age $\le$ 250\,Myr, $\sigma_t$ original
\item{\bf ext250} (46 craters) age $\le$ 250\,Myr, original or assigned
\item{\bf full250} (59 craters) ext250 plus craters with $s^{up}\le$ 250\,Myr
\end{itemize}
The terms ``basic'', ``ext'', and ``full'' refer to the inclusion of
craters with different kinds of age uncertainties.  ``original
$\sigma_t$'' means that just craters with measured age uncertainties
are included. ``original or assigned'' adds to this craters for which
uncertainties have been estimated. The ``full'' data sets further
include craters with just upper age limits (\citealp{bailer-jones11}
explains how these can be used effectively).  As the size
of the existing craters is determined by many factors, e.g.\ the
inclination, velocity and size of the impactor, the impact surface,
and erosion, we only use the time of occurrence ($s_j$) of each impact
crater and its uncertainty ($\sigma_j$). Figure \ref{fig:set3lim}
plots the size and age of the 59 craters we use in the model comparison in Section
\ref{sec:comparison_comet}.

\subsection{Long-period comets}\label{sec:LPCs_comet}

\begin{figure}
  \centering
  \includegraphics[scale=0.5]{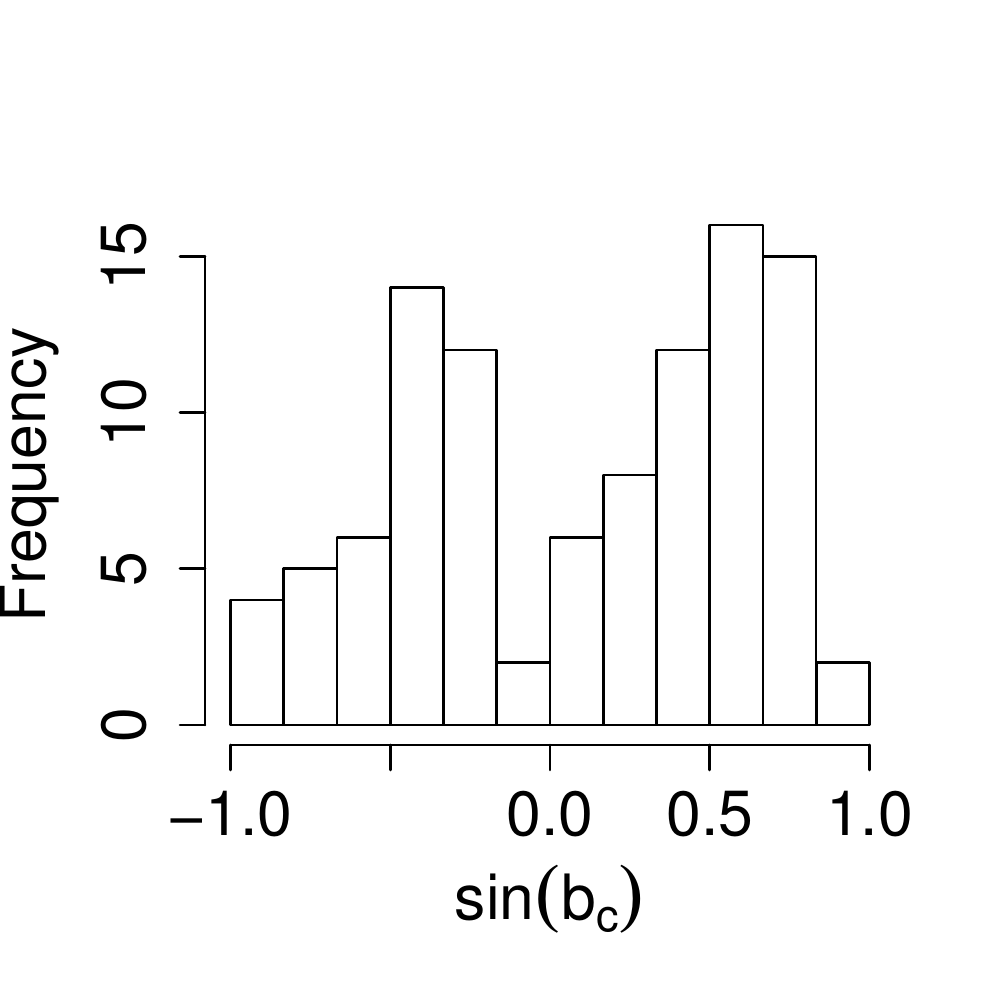}
  \includegraphics[scale=0.5]{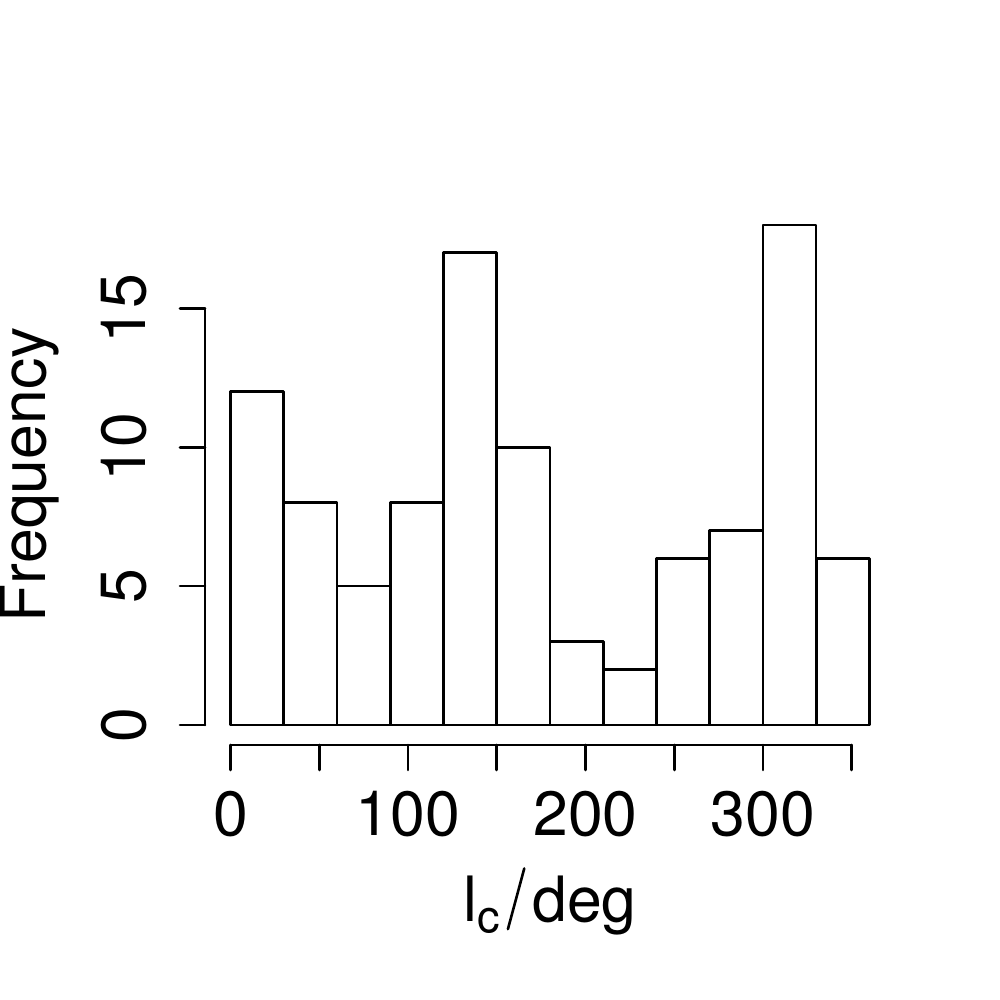}
  \caption{The distribution of $\sin b_c$ (left panel) and $l_c$ (right panel) of perihelia of the 102 LPCs. }
\label{fig:LPC1A_bl}
\end{figure}

The LPCs we use are the 102 dynamically new comets (i.e. class 1A) identified by \citet{marsden08} and discussed by \citet{matese11}. 
Figure \ref{fig:LPC1A_bl} shows the distribution over the Galactic 
latitude ($b_c$) and longitude ($l_c$) of the cometary perihelia. \footnote{Note that our angular distribution is different from the one given in \cite{matese11} because the direction of perihelion is opposite to that of aphelion.}
The two peaks in the longitude distribution suggest a great circle on the sky passing through $l=135^{\circ}$ and $l=315^{\circ}$ \citep{matese99,matese11}. 
We explain this anisotropy in Section \ref{sec:ADP_comet}.

\section{Simulation of cometary orbits}\label{sec:simulation_comet} 

We now build dynamical models of the Oort cloud comets and their perturbation via the Galactic tide and stellar encounters by simulating the passage of the solar system through the Galaxy.  We will use the Galactic potential introduced in section \ref{sec:galaxy}, which yields a tidal gravitational force on the Sun and Oort Cloud comets. Specifically, asymmetric components, i.e. $\Phi_{\rm arm}$ and $\Phi_{\rm bar}$ in Eqn. \ref{eqn:Phi_asym}, are only used in the potential for the calculation of the solar orbit in section \ref{sec:sensitivity_comet}, but not the stellar encounter rate discussed in section~\ref{sec:encmod_comet}. We give the initial conditions of the Oort cloud and the distribution of stellar encounters. Then we outline the numerical methods used to calculate the solar motion and the comet orbits.

\subsection{Oort Cloud}\label{sec:OC_comet}

We generate Oort cloud comets using two different models, one from \citet{duncan87} (hereafter DQT) with the parameters defined in \citet{rickman08}, and another which we have reconstructed from the work of \cite{dones04b} (hereafter DLDW).

In the DQT model, initial semi-major axes ($a_0$) for comets are selected randomly from the interval $[3000,~10^5]$\,AU with a probability
density proportional to $a_0^{-1.5}$. The initial eccentricities ($e_0$) are
selected with a probability density proportional to $e_0$ \citep{hills81}, in such a way that the perihelia ($q_0$) are guaranteed to be larger than 32 AU. We generate the other orbital elements --- $\cos i_0$, $\omega_0$, $\Omega_0$ and $M_0$ --- from uniform distributions.
Because the density profile of comets is proportional to $r^{-3.5}$, where $r$ is the sun-comet distance, about 20\% of the comets lie in the classical Oort Cloud ($a>20\,000$\,AU).

In the DLDW model, the initial semi-major axes, eccentricities, and inclination angles are generated by Monte Carlo sampling from the relevant distributions shown in \cite{dones04b}. This produces semi-major axes in the range 3000 to 100\,000\,AU and ensures that the perihelia are larger than 32\,AU. Unlike the DQT model, there is a dependency of the cometary eccentricity and inclination on the semi-major axis, as can be see in Figures 1 and 2 of \cite{dones04c}. We generate comet positions and velocities relative to the invariant plane and then transform these into vectors relative to the Galactic plane.  In doing so we adopted values for the Galactic longitude and latitude of the north pole of the invariant plane of $98^{\circ}$ and $29^{\circ}$ respectively.

\begin{figure}[ht!]
  \centering
  \includegraphics[scale=0.8]{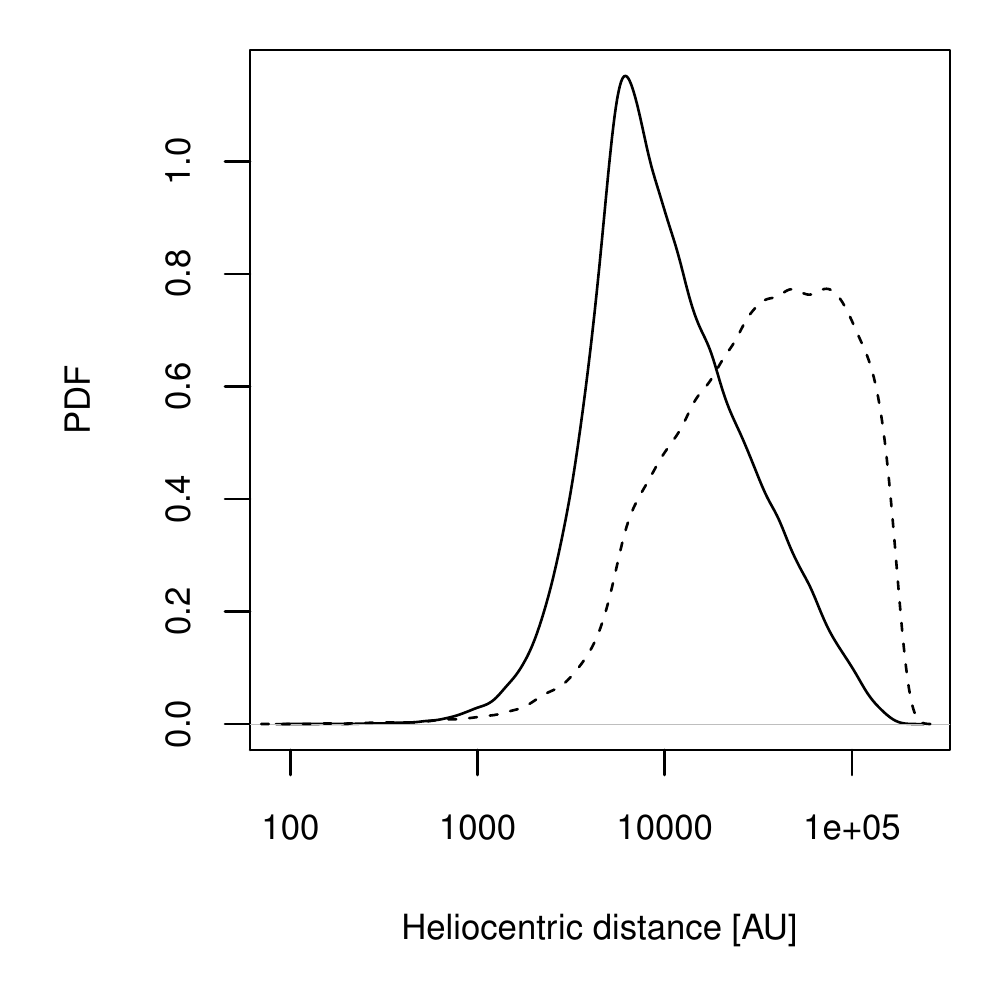}
  \caption{The normalized distributions of initial heliocentric distances of comets generated from the DQT model (solid line) and DLDW model (dashed line) with a sample size of $10^5$.}
  \label{fig:DQT_DLDW}
\end{figure}

The distributions of the cometary heliocentric distances for the DQT and DLDW models are given in Figure \ref{fig:DQT_DLDW}. We see that the DQT model produces more comets in the inner Oort cloud ($<$\,20\,000\,AU) and the DLDW model more in the outer Oort Cloud ($>$\,20\,000\,AU). Our distributions differ slightly from those in Figure 3 of \cite{dybczynski02} because our initial semi-major axes have different boundaries, and because our reconstruction of initial eccentricities and inclination angles is slightly different from the approach used in \cite{dybczynski02}. Many other Oort cloud initial conditions have been constructed numerically \citep{emelyanenko07,kaib11}.
Given the inherent uncertainty of the Oort cloud's true initial conditions, 
we carry out our work using two different Oort cloud models and investigate the sensitivity of our results to this (e.g.\ in section \ref{sec:ADP_comet}).

\subsection{Stellar encounters}\label{sec:encmod_comet} 

The geometry of encounters is complicated by the Sun's motion relative to the
local standard of rest (LSR). This solar apex motion could, by itself, produce
an anisotropic distribution in the directions of stellar encounters in the
heliocentric rest frame (HRF). Any anisotropy must be taken into account when trying to explain the observed anisotropic perihelia of the LPCs. Nonetheless, \citet{rickman08} simulated cometary orbits with an isotropic distribution of stellar encounters which is inconsistent with their method for initializing encounters. Here we use their method to generate encounters, but now initialize stellar encounters self-consistently to have a non-uniform angular distribution.

\subsubsection{Encounter scenario}\label{sec:scenario_enc_comet}

\begin{figure}[ht!]
  \centering
   \includegraphics[height=100mm,width=100mm]{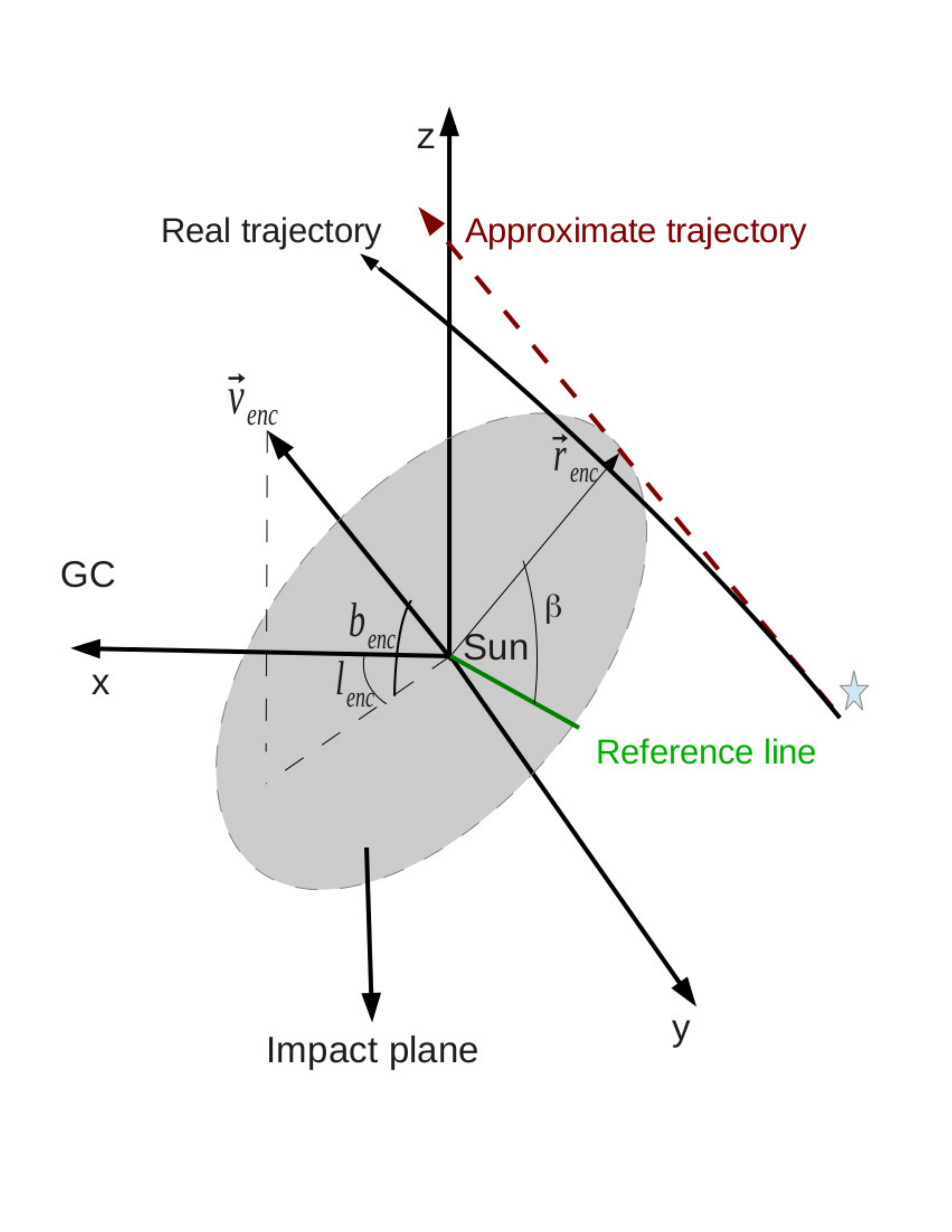}
  \vspace{-6ex}
  \caption{Schematic illustration in the heliocentric rest frame
    of stellar encounters. The circle is the impact plane which is
    defined by its normal, the encounter velocity $\vec{v}_{\rm enc}$. $\beta$
    is the angle in the impact plane measured from the reference axis
    to the stellar perihelion (i.e.\ the encounter). The vector in this plane
    from the Sun to the position of the encounter (i.e. the star’s perihelion)
    is defined as $r_{\rm enc}$. $b_{\rm enc}$ and $l_{\rm enc}$ are the
    Galactic latitude and longitude of $\vec{v}_{\rm enc}$,
    respectively. $(x,y,z)$ is the Galactic coordinate system. $\vec{r}_{\rm enc}$ is defined as the shortest distance from
    the Sun to the approximate trajectory which is a straight line in the
    direction of $\vec{v}_{\rm enc}$. The approximate trajectory of an
    encounter is used for the definition of encounter perihelion $\vec{r}_{\rm enc}$ while the real trajectory is integrated through simulations.} 
  \label{fig:impact_frame}
\end{figure}

The parameters of stellar encounters are generated using a Monte Carlo
sampling method, as follows. We distribute the encounters into different stellar
categories (corresponding to different types of stars) according to their
frequency, $F_i$, as listed in Table 8 of \citet{sanchez01}. In each stellar
category, the stellar mass $M_i$, Maxwellian velocity dispersion $\sigma_{\star i}$, and solar peculiar velocity $v_{\odot i}$, are given. The encounter scenario in the HRF is illustrated in Figure \ref{fig:impact_frame}. The encounter perihelion $\vec{r}_{\rm enc}$ direction (which has Galactic coordinates $b_p$ and $l_p$) is by definition perpendicular to the encounter velocity $\vec{v}_{\rm enc}$. The angle $\beta$ is uniformly distributed in the interval of $[0,2\pi]$. 

In this encounter scenario in the HRF, the trajectory of a stellar encounter is
determined by the encounter velocity $\vec{v}_{\rm enc}$, the encounter perihelion
$\vec{r}_{\rm enc}$, and the encounter time $t_{\rm enc}$. 
In the following paragraphs, we will first find the probability density
function (PDF) of encounters for each stellar category as a function of
$t_{\rm enc}$, $r_{\rm enc}$, and $v_{\rm enc}$, and then sample these parameters from this using the Monte
Carlo method introduced by \cite{rickman08} (hereafter R08). Then we will
sample $b_{\rm enc}$ and $l_{\rm enc}$ using a revised version of R08's method. Finally, $b_p$ and $l_p$ can be easily sampled because $\vec{r}_{\rm enc}$ is perpendicular to $\vec{v}_{\rm enc}$.

\subsubsection{Encounter probability}\label{sec:pdf_enc_comet}

The probability for each category of stars is proportional to the number of stars passing through a ring with a width of $dr_{\rm enc}$ and centered on the Sun. The non-normalized PDF is therefore just
\begin{equation}
  P_u(t_{\rm enc}, r_{\rm enc}, v_{\rm enc}) \,=\, 4 \pi n_i v_{\rm enc} r_{\rm enc} \,\propto\, \rho(t_{\rm enc})v_{\rm enc}r_{\rm enc} ,
  \label{eqn:PDF_enc}
\end{equation}
where $n_i$ is the local stellar number density of the $i^{th}$ category of
stellar encounters, and $\rho(t_{\rm enc})$ is the local stellar mass density, which will change as the Sun orbits the Galaxy.\footnote{We assume that the mass densities of different stellar categories have the same spatial distribution.} Thus the encounter probability is proportional to the local mass density, the encounter velocity and the encounter perihelion. We use a Monte Carlo method to sample $t_{\rm enc}$, $v_{\rm enc}$, and $r_{\rm enc}$ from this.

In different application cases, we sample the encounter time $t_{\rm enc}$
over different time spans according to equation \ref{eqn:PDF_enc}, where the
local mass density is calculated using Poisson's equation with the potentials
expressed in Eqn. \ref{eqn:Phi_sym} and \ref{eqn:Phi_component}. Although we may simulate stellar encounters over a long time scale, we ignore the change of the solar apex velocity and direction when simulating the time-varying comet flux (in section \ref{sec:impact_comet}) and the angular distribution of current LPCs (in section \ref{sec:ADP_comet}). We select $r_{\rm enc}$ with a PDF proportional to $r_{\rm enc}$ with an upper limit of $4\times 10^5$ AU. However, the sampling process of $v_{\rm enc}$ is complicated by the solar apex motion and the stellar velocity in LSR, which we accommodate in the following way.

The encounter velocity in the HRF, $\vec{v}_{\rm enc}$, is the
difference between the velocity of the stellar encounter in the LSR, $\vec{v}_\star$, and the solar apex velocity relative to that type of star (category $i$)
in the LSR, $\vec{v}_{\odot i}$, i.e.\footnote{We
define a symbol without using the subscript $i$ when the symbol is derived
from a combination of symbols belonging and not belonging to certain stellar category.} 
\begin{equation}
  \vec{v}_{\rm enc}=\vec{v}_\star-\vec{v}_{\odot i}~.
  \label{eqn:vector_venc}
\end{equation}
We can consider the above formulae as a transformation of a stellar
velocity from the LSR to the HRF. The magnitude of this velocity in the HRF is
\begin{equation}
  v_{\rm enc}=[v^2_\star+v^2_{\odot i}-2v_{\odot i}v_\star \cos \delta]^{1/2}~,
  \label{eqn:venc}
\end{equation}
where $\delta$ is the angle between $\vec{v}_\star$ and $\vec{v}_{\odot i}$ in
the LSR.

To sample $v_{\rm enc}$, it is necessary to take into account both the encounter probability given in equation \ref{eqn:PDF_enc} and the distribution of $v_\star$. We generate $v_\star$ using
\begin{equation}
  v_\star = \sigma_{\star i} \left[\frac{1}{3}(\eta_u^2+\eta_v^2+\eta_w^2)\right]^{1/2},
  \label{eqn:vstar}
\end{equation}
where $\sigma_{\star i}$ is the stellar velocity dispersion in the $i^{\rm th}$ category, and $\eta_u$, $\eta_v$, $\eta_w$ are random variables, each
following a Gaussian distribution with zero mean and unit variance.

We then realize the PDF of encounters over $v_{\rm enc}$ (i.e.\ $P_u \propto v_{\rm enc}$) using R08's method as follows: (i) we randomly generate $\delta$ to be uniform in the interval $[0,2\pi]$; (ii) adopting $v_{\odot i}$ from table 1 in R08 and generating $v_\star$ from equation \ref{eqn:vstar}, we calculate $v_{\rm enc}$ using equation \ref{eqn:venc}; (iii) we define a large velocity $V_{\rm enc}=v_{\odot i}+3\sigma_{\star i}$ for the relevant star category and randomly draw a velocity $v_{\rm rand}$ from a uniform distribution over $[0, V_{\rm enc}]$. If $v_{\rm rand} < v_{\rm enc}$, we accept $v_{\rm enc}$ and the values of the generated variables $\delta,v_\star$. Otherwise, we reject it and repeat the process until $v_{\rm rand} < v_{\rm enc}$.

\begin{figure}[ht!]
  \centering
  \includegraphics[scale=0.8]{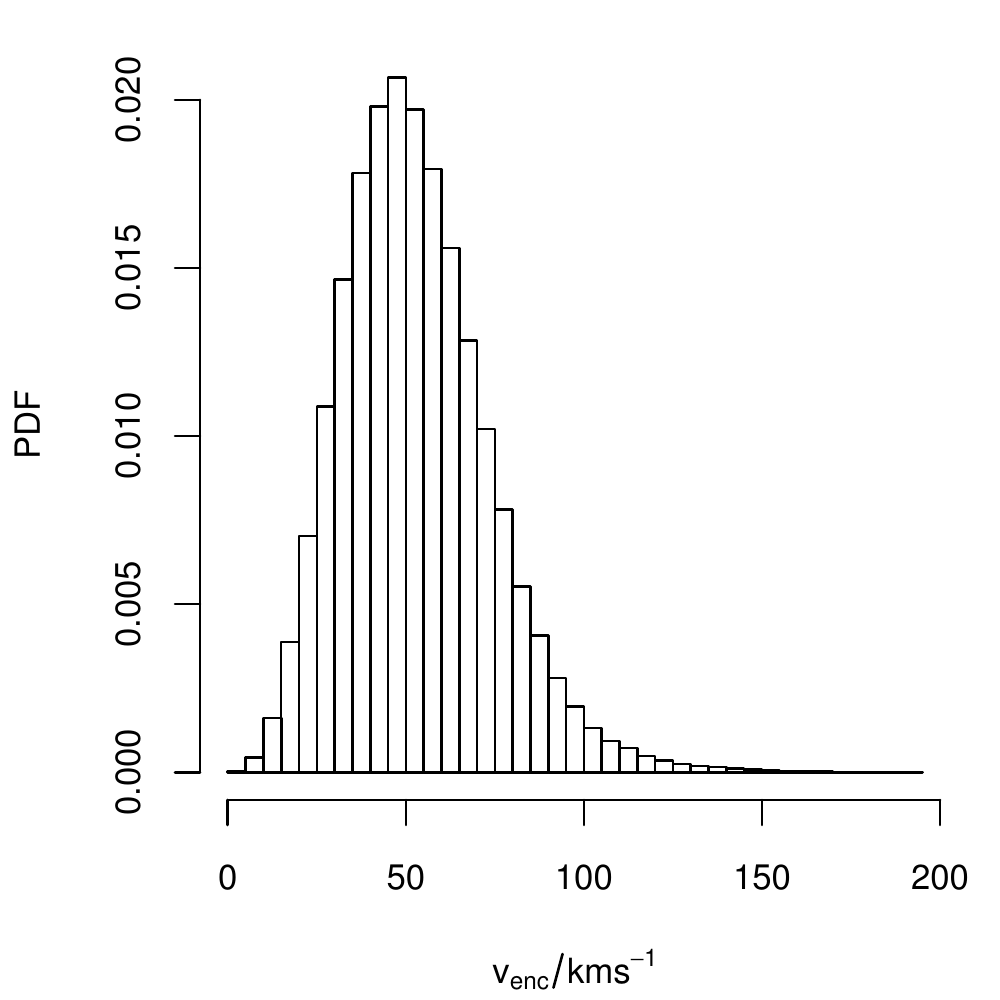}
  \caption{The histogram of the distribution of $v_{\rm enc}$ of all types of stars. 
The total number of encounters is 197\,906, which is the set of simulated encounters over the past 5\,Gyr.} 
  \label{fig:venc}
\end{figure}

We generate $10^5$ encounters in this way. Figure \ref{fig:venc} shows the resulting distribution of $v_{\rm enc}$. It follows a positively-constrained Gaussian-like distribution with mean velocity of 53\,km/s and a dispersion of 21\,km/s, which is consistent with the result in R08. In their modelling, R08 adopt a uniform distribution for $\sin b_{\rm enc}$, and $l_{\rm enc}$. This is not correct, however, because encounters are more common in the direction of the solar antapex where the encounter velocities are larger than those in other directions (equation \ref{eqn:PDF_enc}). We will show how to find the true distribution of $\sin
b_{\rm enc}$, $l_{\rm enc}$, $\sin b_p$ and $l_p$ as follows.

\subsubsection{Anisotropic perihelia of encounters}\label{anisotropic}

To complete the sampling process of encounters, we need to find a 5-variable
PDF, i.e. $P_u(t_{\rm enc}, r_{\rm enc}, v_{\rm enc}, b_{\rm enc}, l_{\rm
  enc})$. We have used R08's original Monte Carlo method to generate $t_{\rm
  enc}$, $r_{\rm enc}$ and $v_{\rm enc}$ according to equation
\ref{eqn:PDF_enc}. However, $b_{\rm enc}$ and $l_{\rm enc}$ are not generated
because R08 only use equation \ref{eqn:venc} to generate the magnitude of
$\vec{v}_{\rm enc}$ rather than the direction of $\vec{v}_{\rm enc}$. To
sample the directions of $\vec{v}_{\rm enc}$, we change the first and second
steps in R08's method introduced in section \ref{sec:pdf_enc_comet} as follows: (i) we
randomly generate $\{b_\star, l_\star\}$ such that $\sin b_\star$ and
$l_\star$ are uniform in the interval of $[-1,1]$ and $[0,2\pi]$,
respectively; (ii) adopting $b_{\rm apex}=58.87^\circ$ and $l_{\rm apex}=17.72^\circ$ for the solar apex direction and generating $v_\star$ according to equation \ref{eqn:vstar}, we calculate $\vec{v}_{\rm enc}$ according to equation \ref{eqn:vector_venc}. 

\begin{figure}[ht!]
  \centering
  \includegraphics[scale=0.5,angle=-90]{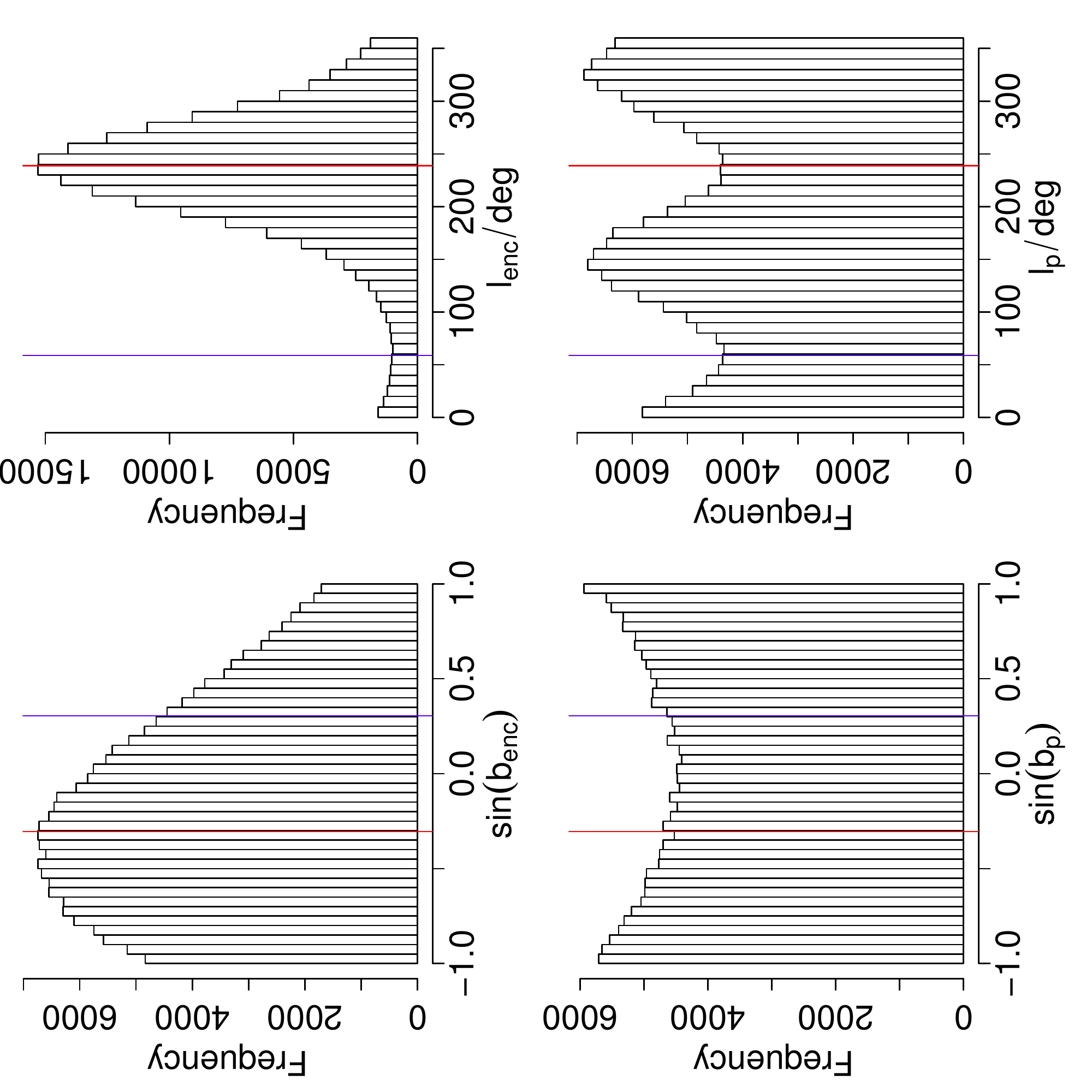}
  \caption{The upper panels show the distributions of the directions of the stellar encounter velocities in our simulations in Galactic coordinates as $\sin b_{\rm enc}$ (upper left)
    and $l_{\rm enc}$ (upper right). The lower panels show the distributions of
the directions of the corresponding perihelia as
    $\sin(b_p)$ (lower left) and $l_p$ (lower right). The blue and red lines
    denote the apex and antapex directions, respectively. The total number of encounters is 197\,906, which is the set of simulated encounters over the past 5 Gyr.} 
  \label{fig:venc_denc_bl}
\end{figure}

Selected in this way, $\sin b_\star$, $l_\star$, $\sin b_{\rm enc}$, and
$l_{\rm enc}$ all have non-uniform distributions. The Galactic latitude $b_p$ and
longitude $l_p$ of the encounter perihelia are also not uniform. 
Like R08, we draw 197\,906 encounters over the past 5 Gyr
from our distribution of encounters. The resulting histograms of $\sin b_{\rm enc}$, $l_{\rm enc}$, $\sin b_p$, and $l_p$ are shown in Figure \ref{fig:venc_denc_bl}. 
We see that the encounter velocity, $\vec{v}_{\rm enc}$, concentrates in the
antapex direction, while the encounter perihelion, $\vec{r}_{\rm enc}$,
concentrates in the plane perpendicular to apex-antapex direction. In
addition, the distribution of $l_p$ is flatter than that of $l_{\rm enc}$
because $\vec{r}_{\rm enc}$ concentrates on a plane rather than along a direction.

\begin{figure}[ht!]
  \centering
  \includegraphics[scale=0.8,angle=-90]{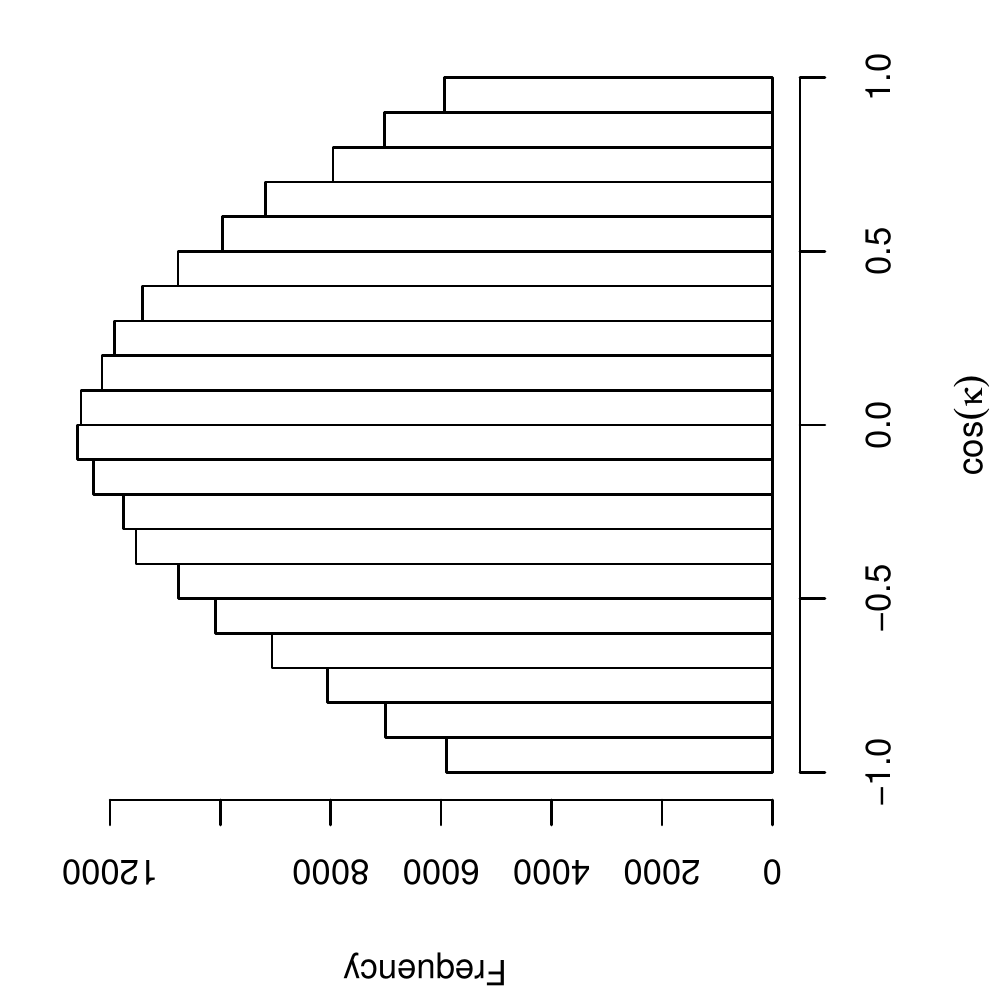}
  \caption{The distribution of the cosine of the angle between the encounter
    perihelion and the solar apex.}
  \label{fig:apex_kappa}
\end{figure}

In order to clarify the effect of the solar apex
motion, we define $\kappa$ as the angle between the encounter perihelion
$\vec{r}_{\rm enc}$ and the solar apex. If there were no solar apex motion, $\cos \kappa$
would be uniform. The effect of solar apex motion is shown in Figure
\ref{fig:apex_kappa}. The solar apex motion would result in the concentration
of encounter perihelia on the plane perpendicular to the apex direction. This
phenomenon is detected by \citet{sanchez01} using Hipparcos data, although the observational incompleteness biases the data. The non-uniform distribution over $\cos \kappa$ results in an anisotropy in the perihelia of LPCs, as we will demonstrate and explain in Section \ref{sec:ADP_comet}.

\subsection{Methods of numerically simulating the comet orbits}\label{sec:method_comet} 

\subsubsection{AMUSE}\label{sec:amuse_comet}

Taking the above models and initial conditions, we
construct an integrator for the orbits of Oort cloud comets via a procedure
similar to that in \citet{wisdom91}, using the Bridge method \citep{fujii07}
in the AMUSE framework\footnote{http://www.amusecode.org} (a platform for coupling existing codes from
different domains; \citealp{pelupessy13,portegies13}). A direct integration of the cometary orbits is  computationally
expensive due to the high eccentricity orbits and the wide
range of timescales involved. We therefore split the dynamics of the comets into
Keplerian and interaction terms (following \citealp{wisdom91}). The Keplerian
part has an analytic solution for arbitrary time steps, while the interaction
terms of the Hamiltonian consist only of impulsive force kicks.
To achieve this we split the Hamiltonian for the system in the following way
\begin{equation}
  H = H_{\rm Kepler} + H_{\rm encounter} + H_{\rm tide}
  \label{eqn:H_split}
\end{equation}
where $H_{\rm Kepler}$, $H_{\rm encounter}$, and $H_{\rm tide}$ describe the
interaction of the comet with the dominant central object (the Sun), a passing
star, and the Galactic tide, respectively. Specifically, the
Keplerian cometary orbits can be integrated analytically according to $H_{\rm
  Kepler}$ while the interactions with the Galactic tide and stellar encounters are taken into account in terms of force kicks. For the time integration a 
second order leapfrog scheme is used, where the Keplerian evolution is 
interleaved with the evolution under the interaction terms. The forces for 
the latter are calculated using direct summation, in which the comet masses are neglected.
Meanwhile, the Sun moves around the
Galactic center under the forces from the Galactic tide and stellar encounters
calculated from $H_{\rm encounter}$ and $H_{\rm tide}$ in the leapfrog scheme.

We first initialize the orbital elements of the Sun and encountering
stars about the Galaxy, and the Oort cloud comets about the Sun.  We treat
the stellar encounters as a N-body system with a varying number of
particles, simulated using the Huayno code \citealt{pelupessy12}. The
interaction between comets and the Sun is simulated with a Keplerian code based
on \citet{bate71}.

At each time step in the orbital integration we calculate the gravitational force from the Galaxy and stellar encounters. The velocities of the comets are changed according to the Hamiltonian in equation \ref{eqn:H_split} at every half time step. Meanwhile, each comet moves in its Keplerian orbit at each time step. All variables are transformed into the HRF in order to take into account the influence of the solar motion and stellar encounters on the cometary orbits.

We use constant time steps in order to preserve the symplectic properties of the integration scheme in AMUSE (although we note that a symplectically corrected adaptive time step is used in some codes, such as SCATR \citep{kaib11b}). We use a time step of 0.1\,Myr for tide-only simulations because we find no difference in the injected flux when simulated using a smaller time step. 
The choice of time step size is a trade-off between computational speed and sample noise in the injected comet sample.  We use a time step of 0.01\,Myr in the encounter-only and in the combined (tide plus encounter) simulations when modelling the angular distribution of the LPCs' perihelia (section \ref{sec:ADP_comet}).  (In section \ref{sec:sensitivity_comet} we repeat some of these simulations with a shorter time step -- 0.001\,Myr -- to confirm that this time step is small enough.)  We use a time step of 0.001\,Myr in all other simulations.

In the following simulations we adopt the initial velocity of the Sun from \citet{schoenrich10} and the initial galactocentric radius from \citet{schoenrich12}. Other initial conditions and their uncertainties are the same as in \cite{feng13}. The circular velocity of the Sun (at $R=8.27$\,kpc), $v=225.06$\,km/s, is calculated based on the axisymmetric Galactic model in section \ref{sec:galaxy}. These values are listed in Table \ref{tab:initial_condition}.
\begin{table*}
\caption{The current phase space coordinates of the Sun, represented as
  Gaussian distributions, and used as the initial conditions in our orbital
  model \citep{schoenrich10,schoenrich12,majaess09,dehnen98b}.}
\centering
\begin{tabular}{l*5{c}r}
\hline
\hline
     &$R$/kpc&$V_R$/kpc~Myr$^{-1}$&$\phi$/rad&$\dot\phi$/rad~Myr$^{-1}$&z/kpc&$V_z$
/kpc~Myr$^{-1}$\\
\hline
mean &8.27  &-0.01135           & 0        & 0.029&0.026&-0.0074\\
standard deviation &0.5& 0.00036&0         &0.003 &0.003&0.00038\\
\hline
\end{tabular}
\label{tab:initial_condition}
\end{table*}

\subsubsection{Numerical accuracy of the AMUSE-based method}\label{sec:accuracy_comet}

\begin{figure}[ht!]
\centering
\includegraphics[scale=0.5,angle=-90]{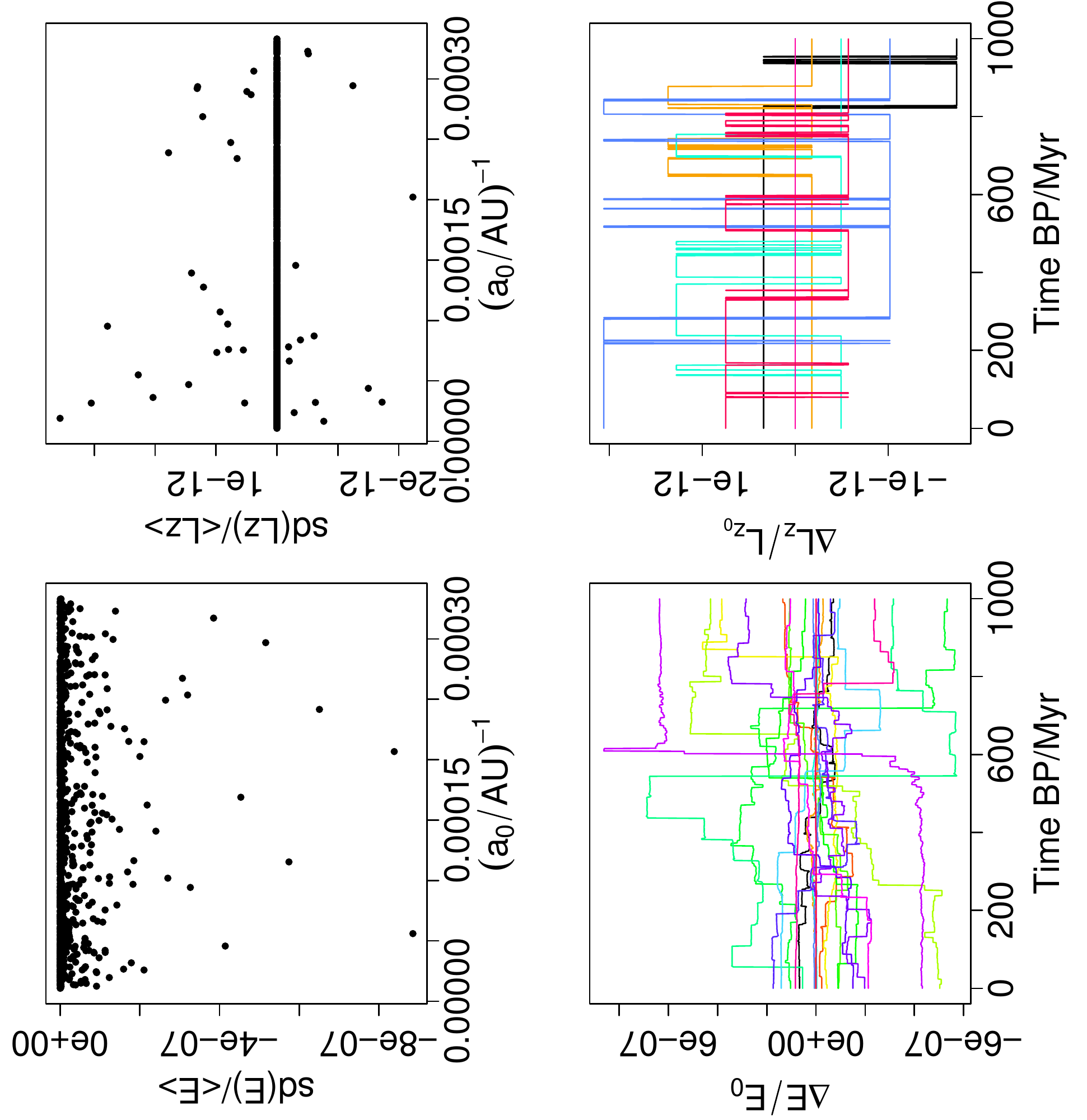}
\caption{Assessment of the numerical accuracy of the AMUSE-based method through monitoring the conservation of energy $E$ and angular momentum $L_z$ for 1000 comets generated from the DLDW Oort cloud model. Upper panels: For each of the 1000 comets, the standard deviation (over its orbit) of $E$ (left) and $L_z$ (right) relative to the average value over the orbit, plotted as a function of the initial energy (which is proportional to $1/a_0$). Lower panels: the fractional change over the orbit of $E$ and $L_z$ for the 20 comets (represented by different colours) with the highest numerical errors. } 
\label{fig:E_Lz_test}
\end{figure}

To test the numerical accuracy of the AMUSE-based method, we generated 1000 comets from the DLDW model and monitored the conservation of orbital energy and angular momentum. As the perturbation from the Galactic potential and stellar encounters used in our work would violate conservation of the third component of angular momentum ($L_z$), we use a simplified Galactic potential for this test, namely a massive and infinite sheet with
  \begin{equation}
    \Phi_{\rm sheet}=2\pi G\sigma |z|,
    \label{eqn:slab_disk}
  \end{equation}
where $G$ is the gravitational constant, $\sigma = 5.0\times 10^6~M_{\odot}/$\,kpc$^2$ is the surface density of the massive sheet and $z$ is the vertical displacement from the sheet. 
Because this potential imposes no tidal force on comets if the Sun does not cross the disk, it enables us to test the accuracy of the bridge method in AMUSE by using the conservation of cometary orbital energy and the angular momentum perpendicular to the sheet. To guarantee that the Sun does not cross the plane during the 1\,Gyr orbital integration (i.e.\ the oscillation period is more than 2\,Gyr), we adopt the following initial conditions of the Sun: $R=0$\,kpc, $\phi=0$, $z=0.001$\,kpc, $V_R=0$\,kpc/Myr, $\dot\phi=0$\,rad/Myr, $V_z=0.0715$\,kpc/Myr.
Integrating the cometary orbits over 1\,Gyr with a constant time step of 0.1\,Myr, we calculate the fractional change of the comets' orbital energies $E$ and the vertical component of their angular momenta $L_z$ during the motion (Figure \ref{fig:E_Lz_test}). Both quantities are conserved to a high tolerance, with fractional changes of less than $10^{-6}$ for $L_z$ and less than
$10^{-12}$ for $E$. The numerical errors are independent of the comet's energy (which is inversely proportional to the semi-major axis). Compared to the magnitude of the perturbations which inject comets from the Oort cloud into the observable zone, these numerical errors can be ignored during a 1\,Gyr and even a 5\,Gyr integration.

\subsubsection{Comparison of the AMUSE-based method with other methods}\label{sec:initial_condition_comet}

Our numerical method calculates perturbations from stellar encounters
and the Galactic tide using dynamical equations directly, instead of
employing an impulse approximation (e.g.\ CIA, DIA, or SIA
\citet{rickman05}) or the Averaged Hamiltonian Method
(AHM)\citep{fouchard04}.  In the latter the Hamiltonian of the
cometary motion is averaged over one orbital period. This can
significantly reduce the calculation time, but is potentially less
accurate. A more explicit method is to integrate the Newtonian
equations of motion directly, e.g.\ via the Cartesian Method (CM) of
\citep{fouchard04}, but this is more time consuming.

\begin{figure}[ht!]
\centering
\includegraphics[scale=0.8]{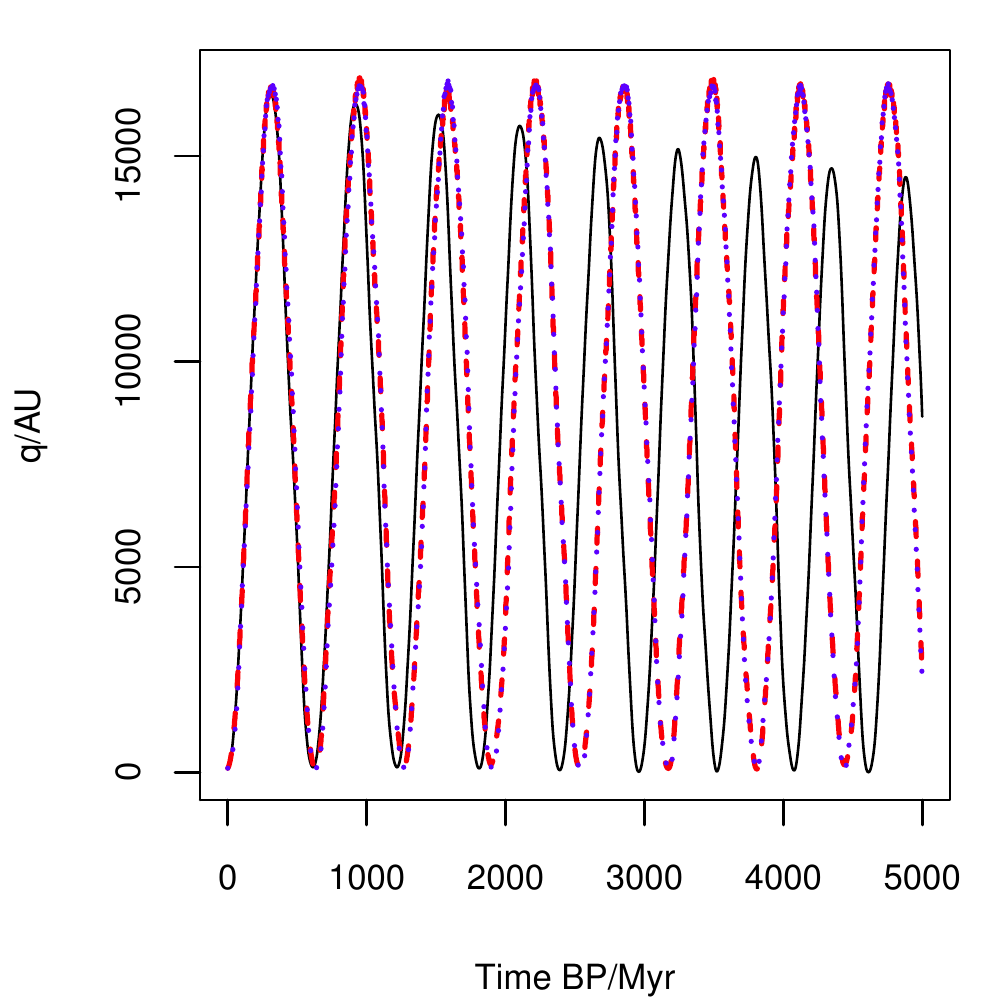}
\caption{The variation of the perihelion of one comet calculated with
  three different integration methods: AHM (black solid), CM (red dashed), and
  AMUSE-based method (blue dotted). } 
\label{fig:amuse_AHM}
\end{figure}

To illustrate the accuracy of the AHM, CM, and AMUSE-based methods in simulating high eccentricity orbits, we integrate the orbit of one comet using all methods. The test comet has a semi-major axis of $a=25\,000$\,AU and an eccentricity of $e=0.996$ (as used in \citet{fouchard04}). Adopting the following initial conditions of the Sun -- $R=8.0$\,kpc, $\phi=0$, $z=0.026$\,kpc, $V_R=-0.01$\,kpc/Myr, $\dot\phi=0.0275$\,rad/Myr, $V_z=0.00717$\,kpc/Myr -- and using the same tide model as described above, the solar orbit under the perturbation from the Galactic tide is integrated over the past 5\,Gyr.  Figure \ref{fig:amuse_AHM} shows that the evolutions of the cometary perihelia calculated using the CM and AMUSE-based methods are very similar, whereas AHM shows an evolution which diverges from these. As CM is the most accurate method, this shows that the
AHM cannot be used to accurately calculate the time-varying, because
it holds the perturbing forces constant during each orbit.
Because the AMUSE-based method computes a large sample of comets more
efficiently than CM does, we have adopted the AMUSE-based method in our work.

\subsubsection{Calculation of the injected comet flux}\label{sec:inject_comet}

A comet which comes too close to the perturbing effects of the giant planets in the solar system will generally have its orbit altered such that it is injected into a much shorter periodic orbit or is ejected from the solar system on an unbound orbit. We regard a comet as having been injected into the inner solar system in this way when it enters into the ``loss cone'' \citep{wiegert99}, i.e.\ that region with a heliocentric radius of 15\,AU or less (the same definition as in \citet{dybczynski05} and R08). These are the comets which can then, following further perturbations from the planets, hit the Earth. If injected comets enter an observable zone within $<5$\,AU then they may be observed as a LPC. Comets which are injected into the loss cone or which are ejected from the solar system (i.e.\ achieve heliocentric distances larger than $4\times 10^5$\,AU) are removed from the simulation. 

The observable comets are only a subset of the injected comets because some injected comets can be ejected again by Saturn and Jupiter. But assuming that this is independent of the orbital elements over long time scales, we assume that the flux of injected comets is proportional to the flux of LPCs. 
Inner Oort cloud comets, in particular comets with $a < 3000$\,AU, may be injected into the loss cone  ($q < 15$\,AU) but not enter the observable zone ($q < 5$\,AU) \citep{kaib09}. 
In our simulations we will examine the properties of comets injected into both types of target zone, and we will refer to such injected comets as LPCs.
Once we have identified the injected comets, we calculate the Galactic latitudes $b_c$ and longitudes $l_c$ of their perihelia. Because the orbital elements of the class 1A LPCs are recorded during their first passage into inner solar system, we can reasonably assume that the direction of the LPC perihelion is unchanged after entering the ``loss cone''. In Section \ref{sec:impact_comet} and \ref{sec:ADP_comet}, we will model the terrestrial cratering time series and the anisotropic perihelion of LPCs based on the injected comet flux. Specifically, in Section \ref{sec:impact_comet}, we will show how we convert the simulations of the perturbations of the cometary orbits into a model for the time variation of the cometary flux entering the inner solar system. 

\section{Time series models}\label{sec:tsmodel_comet}

I have already introduced the Bayesian inference method in section \ref{sec:bayes}. I will introduce time series mdoels below, and calculate their likelihoods using the procedure provided in section \ref{sec:discrete}.

The time series model, $M$, is a model which predicts the variation
of the impact probability with time (the normalized cratering
rate), i.e.\ the term $P(\tau_j|\sigma_j,\boldsymbol{\theta},M)$ in equation
\ref{eqn:likelihood_event}.  The models we use in this work, along with their
parameters, $\boldsymbol{\theta}$, are defined in Table \ref{tab:tsmodels}, and described below 
  \begin{description}
  \item[] {\em Uniform.} Constant impact probability over the range of the data. As any probability distribution must be normalized over this range, this model has no parameters.
  \item[] {\em RandProb, RandBkgProb.} Both models comprise $N$ impact events at random times, with each event modelled as a Gaussian.
$N$ times are drawn at random from a uniform time distribution extending over the range of the data. 
A Gaussian is placed at each of these with a common standard deviation (equal to the average
of the real crater age uncertainties). We then sum the Gaussians, add a constant background, $B$, and normalize.
This is the RandBkgProb (``random with background'') model. RandProb is the special case for $B=0$.
We calculate the evidence by averaging over a large number of realizations of the model (i.e.\ times of the events), and, for RandBkgProb, over $B$. For example, when we later model the basic150 time series, we fix $N=32$ and range $B$ from 0 to $\infty$ (see Table \ref{tab:prior_comet}).
  \item[] {\em SinProb, SinBkgProb.} Periodic model of angular frequency $\omega$ and phase $\phi_0$ (model SinProb). There is no amplitude parameter because the model is normalized over the time span of the data. Adding a background $B$ to this simulates a periodic variation on top of a constant impact rate (model SinBkgProb).
  \item[] {\em SigProb.} A monotonically increasing or decreasing nonlinear trend in the impact PDF using a sigmoidal function, characterized by the steepness of the slope, $\lambda$, and the center of the slope, $t_0$. In the limit that $\lambda$ becomes zero, the model becomes a step function at $t_0$, and in the limit of very large $\lambda$ it becomes the Uniform model. We restrict $\lambda <0$ in our model comparison because the decreasing trend in cratering rate towards the past seems obvious in the time series (see Figure \ref{fig:set3lim}; see also \cite{bailer-jones11}). However, we do include the increasing trend in our sensitivity test in Section \ref{sec:sensitivity_comet}.
  \item[] {\em SinSigProb.} Combination of SinProb and SigProb.
  \item[] {\em TideProb, EncProb, EncTideProb.} Models arising from the dynamical simulation of cometary orbits perturbed by either stellar encounters (EncProb) or the Galactic tide (TideProb) or both (EncTideProb). We describe the modelling approach which produces these distributions in detail in Section \ref{sec:impact_comet}.
  \item[] {\em EncSigProb, TideSigProb, EncTideSigProb.} Combination of EncProb, TideProb, EncTideProb (respectively) with SigProb.
  \end{description}

Some of these models -- those in the first five lines in Table \ref{tab:tsmodels} -- are simple analytic models. The others are models based on dynamical simulations of cometary orbits, which we therefore call dynamical models.
In the next section we will explain how we get from a simulation of the perturbation of the cometary orbits to a prediction of the cratering rate. Table \ref{tab:tsmodels} also lists the parameters of the models, i.e.\ those parameters which we average over in order to calculate the evidence. The prior distributions for these parameters are listed in Table \ref{tab:prior_comet}.
\begin{table}
\centering
\caption{The mathematical form of the time series models and their
  corresponding parameters. Time $t$ increases into the past and
  $P_u(t|\boldsymbol{\theta}, M)$ is the unnormalized cratering rate (probability density)
  predicted by the model. In the dynamical models (EncProb, TideProb,
  EncTideProb, EncSigProb, TideSigProb, and EncTideSigProb), $\vec{r}_\odot(t=0Myr)$ and $\vec{v}_\odot(t=0Myr)$ are Sun's current position and velocity relative to
  the Galactic center. Note that the components in the compound models
  are normalized before being combined. The quantities $\gamma_{\rm bin}(t)$, $G_3(t)$, and $\xi$ are defined in Section
  \ref{sec:impact_comet}. 
  $\eta$ is a parameter which
  describes the relative contribution of the two combined models.}
\begin{tabular}{l c r}
\hline
\hline
model name&$P_u(t|\boldsymbol{\theta},M)$&parameters, $\boldsymbol{\theta}$\\
\hline
Uniform&1&none\\
RandProb/RandBkgProb&$\sum_{n=1}^{N}\mathcal{N}(t; \mu_n,\sigma)$+$B$&$\sigma$, $B$,$N$\\
SinProb/SinBkgProb&$1/2\{\cos[\omega t+\phi_0]+1\}$+$B$&$\omega$, $\beta$, $B$\\
SigProb &$[1+e^{(t-t_0)/\lambda}]^{-1}$&$\lambda$, $t_0$\\
SinSigProb &SinProb+SigProb&$T$, $\beta$, $B$,$\lambda$, $t_0$\\
EncProb &$\gamma_{\rm bin}(t)$&$\vec{r}_\odot(t=0)$, $\vec{v}_\odot(t=0)$\\
TideProb & $G_3(t)$ & $\vec{r}_\odot(t=0)$, $\vec{v}_\odot(t=0)$\\
EncTideProb &$[\gamma_{\rm bin}(t)+\xi G_3(t)]/(1+\xi)$&$\xi$, $\vec{r}_\odot(t=0)$, $\vec{v}_\odot(t=0)$\\
EncSigProb &EncProb + $\eta$ SigProb & $\eta$, $\lambda$, $t_0$, $\vec{r}_\odot(t=0)$, $\vec{v}_\odot(t=0)$\\ 
TideSigProb &TideProb + $\eta$ SigProb & $\eta$, $\lambda$, $t_0$, $\vec{r}_\odot(t=0)$, $\vec{v}_\odot(t=0)$\\ 
EncTideSigProb &EncTideProb + $\eta$ SigProb&$\xi$, $\eta$, $\lambda$, $t_0$, $\vec{r}_\odot(t=0)$, $\vec{v}_\odot(t=0)$\\
\hline
\end{tabular}
\label{tab:tsmodels}
\end{table}

\begin{table}
\caption{The prior distribution and range of parameters for the various time series models. For the non-dynamical models (i.e.\ all except the last five lines), a uniform prior for all the parameters is adopted which is constant inside the range shown and zero outside. $N_{\rm ts}$ and $\tau_{\rm max}$ are the number of events and the earliest time of occurrence of the craters. $\bar{\sigma_i}$ is the averaged age uncertainties of the craters. The prior PDFs over the parameters of the dynamical models (the last five lines) are Gaussian, with means and standard deviations set by the initial conditions as listed in Table \ref{tab:initial_condition}. 
}
\label{tab:prior_comet}
\hspace{-0.6in}{\small
\begin{tabular}{l c}
\hline
\hline
model name & details of the prior over the parameters \\
\hline
Uniform& no parameters\\
RandProb &$\sigma=\bar{\sigma_i}$, $N=N_{\rm ts}$, $B=0$\\
RandBkgProb&$\sigma=\bar{\sigma_i}$, $N=N_{\rm ts}$, $B=\frac{1}{\sqrt{2\pi}\sigma}\frac{b}{(1-b)}$ with $b\in[0,1]$\\
SinProb&$2\pi/100 <\omega<2\pi/10$, $0<\phi_0<2\pi$,$B=0$\\
SinBkgProb&$2\pi/100 <\omega<2\pi/10$, $0<\phi_0<2\pi$, $B=\frac{b}{(1-b)}$ with $b\in[0,1]$\\
SigProb &$-100<\lambda<0$, $0<t_0<0.8\tau_{\rm max}$\\
SinSigProb &Priors from both SinProb and SigProb\\
EncProb & Initial conditions listed in Table \ref{tab:initial_condition} \\
TideProb & Initial conditions listed in Table \ref{tab:initial_condition}\\ 
EncTideProb & $\xi=1$, Initial conditions listed in Table \ref{tab:initial_condition}\\
EncSigProb& $0<\eta<4$, $-100<\lambda<0$, $0<t_0<0.8\tau_{\rm max}$, initial conditions listed in Table \ref{tab:initial_condition}\\
TideSigProb& $0<\eta<4$, $-100<\lambda<0$, $0<t_0<0.8\tau_{\rm max}$, initial conditions listed in Table \ref{tab:initial_condition}\\ 
EncTideSigProb&$\xi=1$, $0<\eta<4$, $-100<\lambda<0$, $0<t_0<0.8\tau_{\rm max}$, initial conditions listed in Table \ref{tab:initial_condition}\\ 
\hline
\end{tabular}
}
\end{table}

\section{Modelling the history of the cometary impact rate}\label{sec:impact_comet} 

The terrestrial impact rate consists of two parts: the asteroid impact
rate and the comet impact rate. We are specifically interested in 
only the latter in the present work. The background asteroid impact rate is
proportional to the number of asteroids in the asteroid belt, which is
depleted by the impact of asteroids on planets and their
satellites. Over a long time scale (longer than 100 Myr), the
background impact rate of asteroids would therefore decrease towards the present.
But we could also see variations in this
due to the disruption of large asteroids into an asteroid family, which would produce phases of enhanced impacting \citep{bottke07}.  
In addition to the actual impact rate, the geological record of all impact craters (comet or asteroid) is contaminated by a selection bias: The older a crater is, the more likely it is to have been eroded and so the less likely it is to be discovered. This preservation bias would lead to an apparent increase in the impact rate towards to the present.  We model the combined contribution of these two components (variable asteroid impact rate and the preservation bias)  to the measured impact rate using a sigmoidal function, which produces a smoothly varying trend with time (model SigProb in Table \ref{tab:tsmodels}). As with the other models, this model has parameters which we average over when computing the model evidence.

The cometary impact rate is determined by the
gravitational perturbations of the Oort cloud due to the Galactic tide
and stellar encounters. Both are modulated by the solar
motion around the Galactic center. Some studies suggest that their
combined effect injects more comets into the inner solar system
 than does each acting alone \citep{heisler87, rickman08}. This so-called synergy
effect is difficult to model, however, and will be ignored in our
statistical approach.

We simulate the effects of the tide and encounters separately (section~\ref{sec:simulation_comet}). The resulting
cometary flux from these is described by
the models TideProb and EncProb respectively. The cometary
flux when both processes operate, the model EncTideProb, is the sum of the
fluxes from each (each being normalized prior to combination). 
To include the contributions from the asteroid
impacts and the crater preservation bias we can add to this the
SigProb model mentioned above. This gives the model EncTideSigProb.
The parameters of all these models and their prior ranges are defined
in Tables \ref{tab:tsmodels} and \ref{tab:prior_comet}.

\subsection{Tide-induced cometary flux}\label{sec:tideflux_comet}

The time variation as the Sun orbits the Galaxy of the tide-induced
cometary flux entering the loss cone is calculated using AMUSE-based method (section \ref{sec:method_comet}).
We define $f_c$ as the relative injected comet flux in a time bin
with width $\Delta t$
\begin{equation}
  f_c=\frac{N_{\rm inj}}{N_{\rm tot}\Delta t},
  \label{eqn:f_tide}
\end{equation}
where $N_{\rm inj}$ is the number of injected comets in this bin and
$N_{\rm tot}$ is the total number of the comets. 

We could use $f_c$ directly as the model prediction of the
comet impact cratering rate, $P_u(t|\boldsymbol{\theta},M)$, for the model TideProb
(section~\ref{sec:tsmodel_comet}) for that particular set of model parameters.
However, as the calculation of the cometary orbits is rather time-consuming,
we instead use a proxy for $f_c$, i.e. the vertical tidal force. 

The tidal force per unit mass experienced by a comet in the Oort Cloud is 
\begin{equation}
  \mathbf{F}=-\frac{G M_\odot \,\mathbf{\hat r}}{r^2} - G_1 x \,\mathbf{\hat x}  -G_2 y \, \mathbf{\hat y} - G_3 z \mathbf{\hat z}
  \label{eqn:tidal_force}
\end{equation}
where $\mathbf{r}$ is the Sun-comet vector of length $r$, $M_\odot$ is the
solar mass, and $G$ is the gravitational constant.\footnote{We don't use this
  equation in simulating cometary orbits in the AMUSE framework.} The three tidal coefficients, $G_1$, $G_2$, and $G_3$ are defined as
\begin{equation}
  \begin{array}{l}
  \displaystyle G_1=-(A-B)(3A+B)\\
  \displaystyle G_2=(A-B)^2\\
  \displaystyle G_3=4\pi G \rho(R,z)-2(B^2 -A^2)
  \end{array}
  \label{eqn:G123}
\end{equation}
where $A$ and $B$ are the two Oort constants, and $\rho(R,z)$ is the local mass
density
which can also be denoted as $\rho(t)$ in the case of using $G_3(t)$ to build models. Because the two components $G_1$ and $G_2$ in the Galactic $(x,y)$ plane are about ten times smaller than the vertical component ($G_3$), it is the vertical tidal force that dominates the perturbation of the Oort Cloud.

\begin{figure}
  \centering
  \includegraphics[scale=0.8]{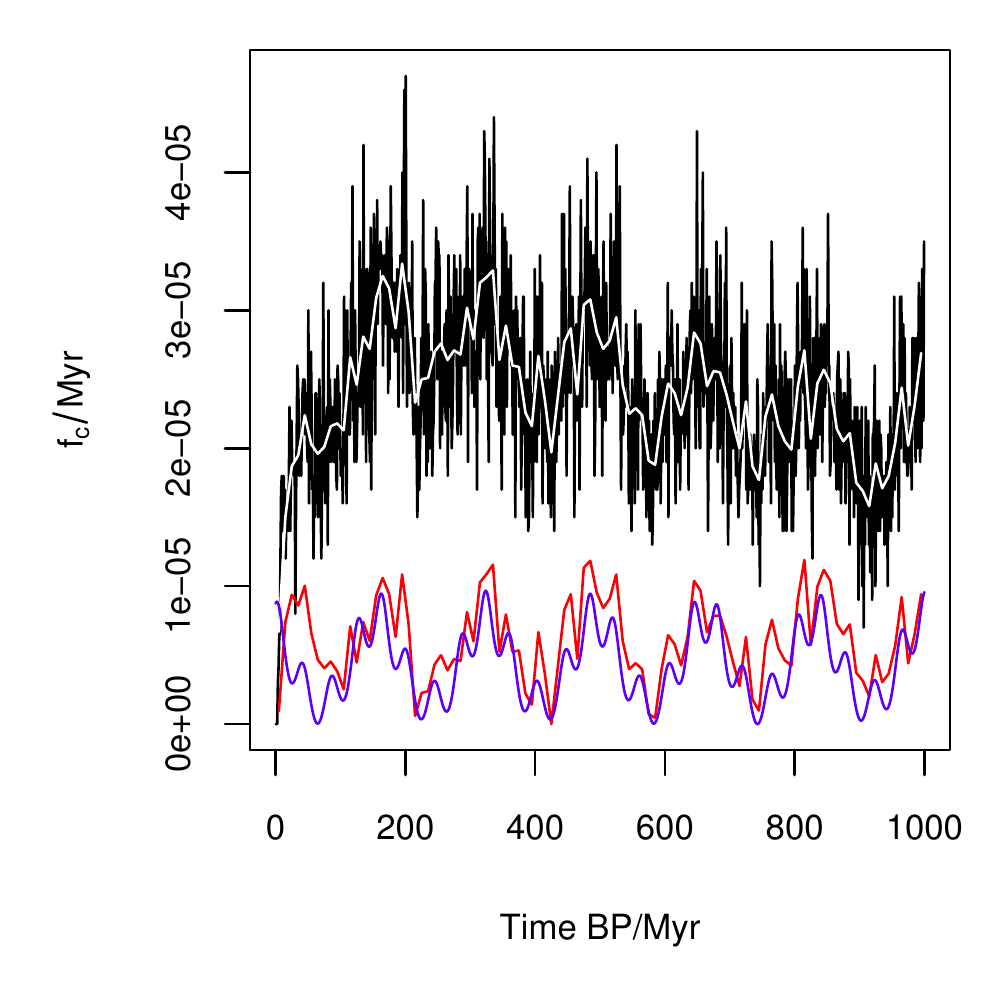}
  \caption{Comparison between the tide-induced injected comet flux ($f_c$) and the vertical Galactic tide ($G_3$). The injected comet flux is shown as a histogram with two different bins sizes: 1\,Myr (black line) and 10\,Myr (white line). The red line is the detrended comet flux with a time bin of 10\,Myr. The blue line shows the variation of $G_3$ (scaled, as it has a different unit to $f_c$). }
\label{fig:Fc_G3}
\end{figure}

To find a relationship between $f_c$ and $G_3$, we simulate the orbits of one million comets generated from the DQT model back to 1\,Gyr in the past under the perturbation of the Galactic tide (stellar encounters are excluded).  We use here the loss cone as the target zone when identifying the injected comets (LPCs).  The two quantities are compared in Figure \ref{fig:Fc_G3}. We see that the detrended comet flux (red line) agrees rather well with $G_3$ (blue line) over the past 1\,Gyr, albeit with an imperfect detrending over the first 100\,Myr.  We made a similar comparison for the DLDW model and also find a very close linear relation. Comparing $G_3$ with the flux of the comets injected into the observable zone (i.e. $q < 5$\,AU) for both the DLDW and DQT models, we find that the result is consistent with what we have found for the loss cone. This confirms the relationship between the tide-induced comet flux and the vertical tidal force, which was also demonstrated by \cite{gardner11} (their Figure 9) with a different approach. We are therefore justified in using $G_3$ as a proxy for the tide-induced comet flux when we build models of cometary impact rate to compare to the crater time series.

\subsection{Encounter-induced cometary flux}\label{sec:encounterflux_comet}

We define the encounter-induced flux entering the loss cone in the same way as $f_c$ in equation \ref{eqn:f_tide}. We now investigate whether we can introduce a proxy for this too. We postulate the use of the quantity \begin{equation}
  \gamma=\frac{M_{\rm enc}}{v_{\rm enc}r_{\rm enc}}
  \label{eqn:gamma}
\end{equation}
which is proportional to the change in velocity of the Sun (or equivalently to the mean change in velocity of the comets) as induced by an encounter according to the classical impulse approximation \citep{oort50,rickman76}. This proxy has also been used in previous studies to approximate the LPC flux injected by stellar encounters (e.g.\ \citet{kaib09,fouchard11}).

\begin{figure}[ht!]
  \centering
  \includegraphics[scale=0.8]{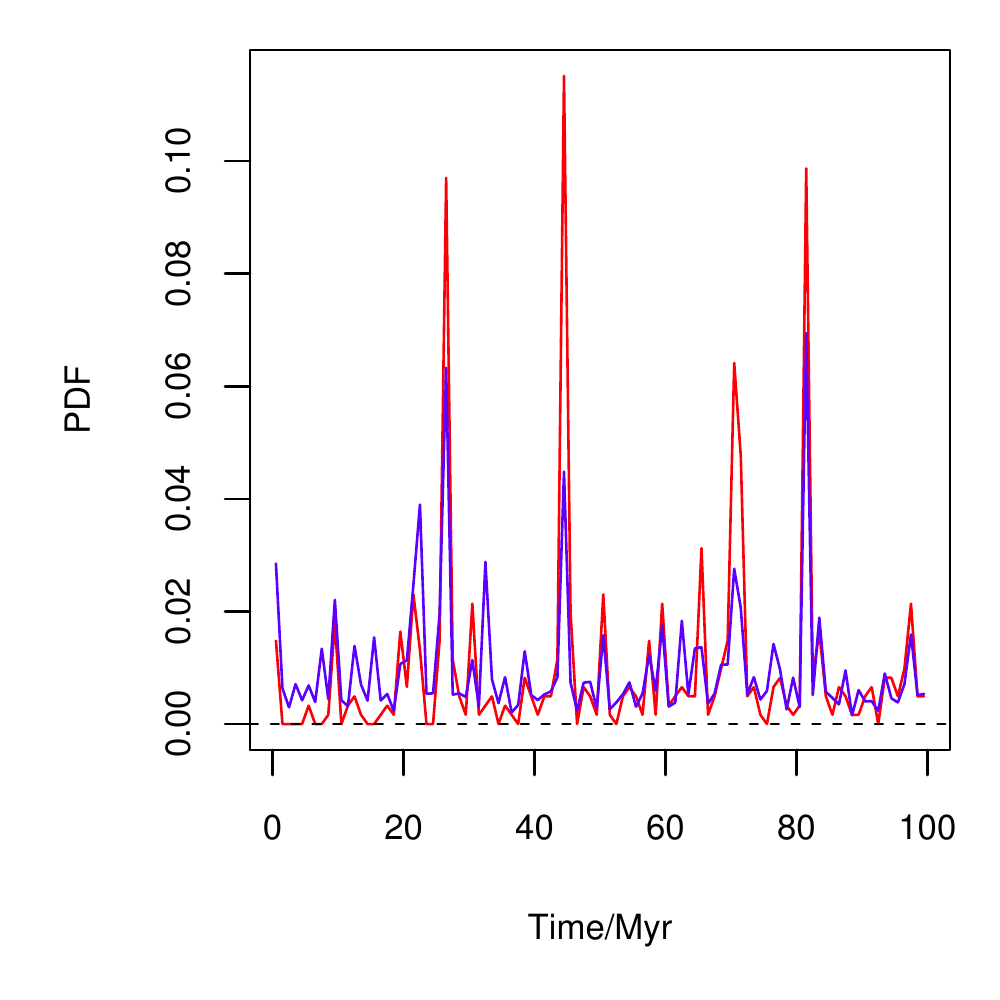}
  \vspace{-2ex}
  \caption{The time-varying probability density of the encounter-induced injected comet flux $f_c$ (red line) and the prediction of proxy $\gamma_{\rm bin}$ (blue line), binned with a time bin of 1\,Myr. }
  \label{fig:gamma_fc_pdf}
\end{figure}

The injected flux is dominated by those encounters which can signifcantly change the velocity and thus the perihelion of the comets
\citep{hills81,heisler87,fouchard11}. Considering the important role of these encounters and the long time scale between them (about 100 Myr according to \citealt{heisler87}), we divide the whole time span of simulated stellar encounters into several time bins and use the (normalized) maximum value of $\gamma$ in each bin to approximate such comet showers. We define this binned proxy as $\gamma_{\rm bin}$, and normalize it over the whole time scale. In Figure \ref{fig:gamma_fc_pdf}, we compare this proxy to the normalized encounter-induced flux which is simulated with a time step of 0.001\,Myr using a sample of $10^5$ comets generated from the DLDW model over 100\,Myr. We find that the main comet showers can be properly predicted by $\gamma_{\rm bin}$, although it may miss small comet showers and predict some non-existent small showers. 

\begin{figure}
  \centering
  \includegraphics[scale=0.8]{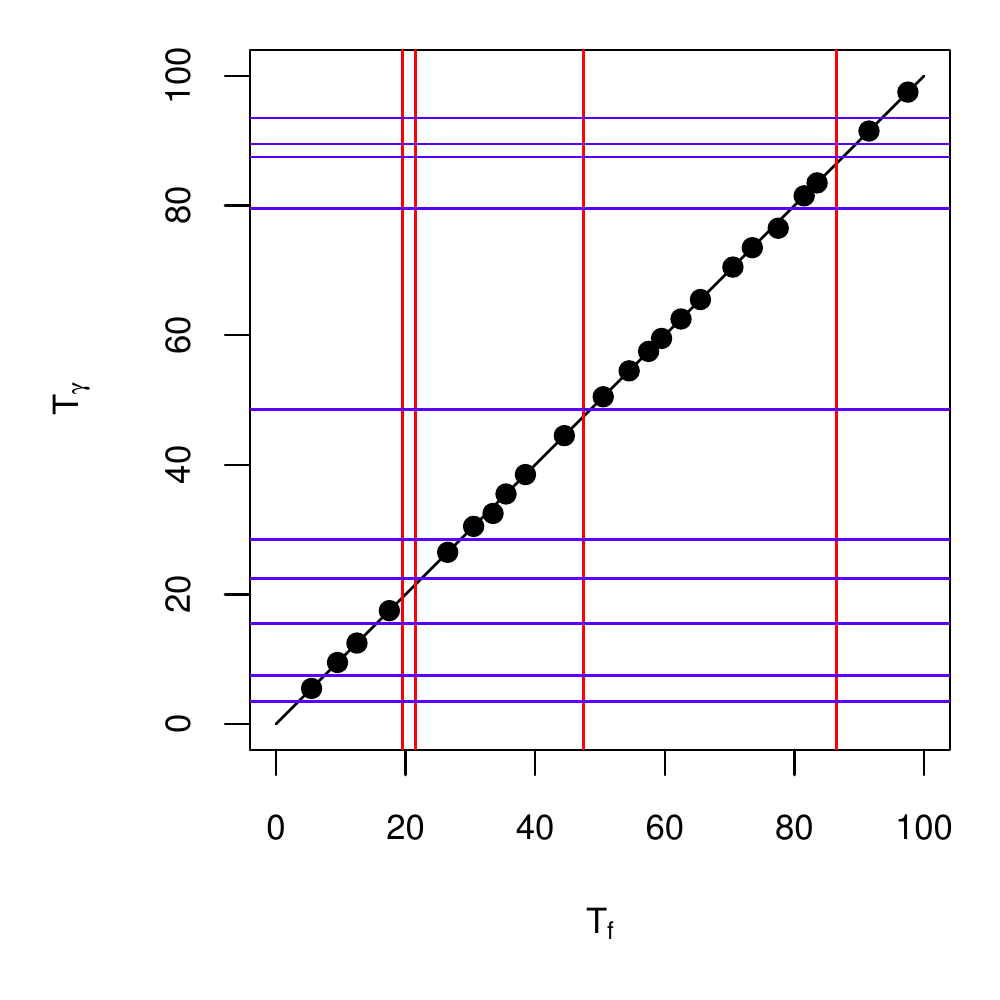}
  \vspace{-2ex}
  \caption{Assessment of the comet shower prediction ability of the proxy $\gamma$. The black points show peaks which are correctly reproduced, by plotting their time of occurrence in the proxy, $T_{\gamma}$, against their true time of occurrence, $T_f$, in $f_c$. Peaks missed by the proxy are shown as vertical red lines and false peaks in the proxy are shown as horizontal blue lines.}
  \label{fig:gamma_fc_peaks}
\end{figure}

To assess the reliability of the shower prediction of the proxy, we evaluate the fraction of
peaks in $f_c$ which are correctly identified by $\gamma_{\rm bin}$, and the fraction of peaks in
$\gamma_{\rm bin}$ which have a corresponding true peak in $f_c$. For the former case, a peak in $f_c$ is counted as correctly predicted by the proxy when it occurs in the same time bin as a peak in $\gamma_{\rm bin}$, or when the $f_c$ peak is one bin earlier (because the shower can occur up to 1\,Myr after the closest approach of the encounter). We find that 23 out of 27 (0.85) flux peaks are correctly predicted by the proxy, 
while 23 out of 33 (0.70) peaks in $\gamma_{\rm bin}$ have corresponding peaks in $f_c$ (Figure \ref{fig:gamma_fc_peaks}).
This simple counting ignores the intensity of the comet showers.
To remedy this use the amplitude of each $\gamma_{\rm bin}$ peak as a weight, 
and count the weighted fractions. We find these to be 0.92 and 0.84 respectively. These results suggests that $\gamma_{\rm bin}$ is a reasonably good proxy for statistical purposes.
Hence we use $\gamma_{\rm bin}$ as the measure of $P_u(t|\boldsymbol{\theta},M)$ for the model EncProb.
The linear relationship between $\rho(t)$ and $G_3(t)$ (equations \ref{eqn:PDF_enc} and \ref{eqn:G123}) indicates that the averaged EncProb model over sequences of $\gamma_{\rm bin}$ is equivalent to the corresponding TideProb model for one solar orbit. We will see in section \ref{sec:comparison_comet} whether there is any significant difference between the evidences for these two models.

\subsection{Combined tide--encounter cometary flux}

\begin{figure}[ht!]
  \centering
  \includegraphics[scale=0.6,angle=-90]{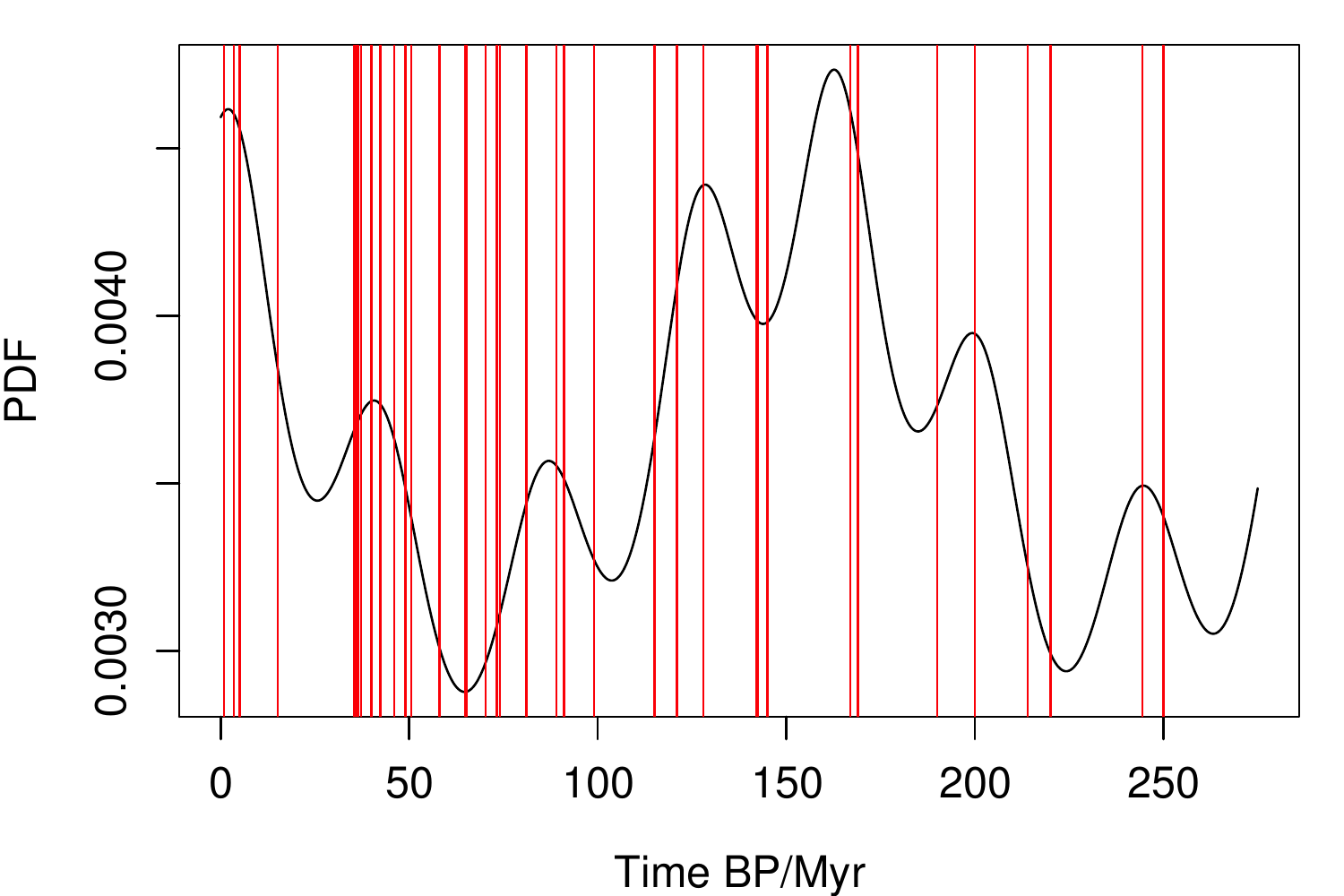}
  \includegraphics[scale=0.6,angle=-90]{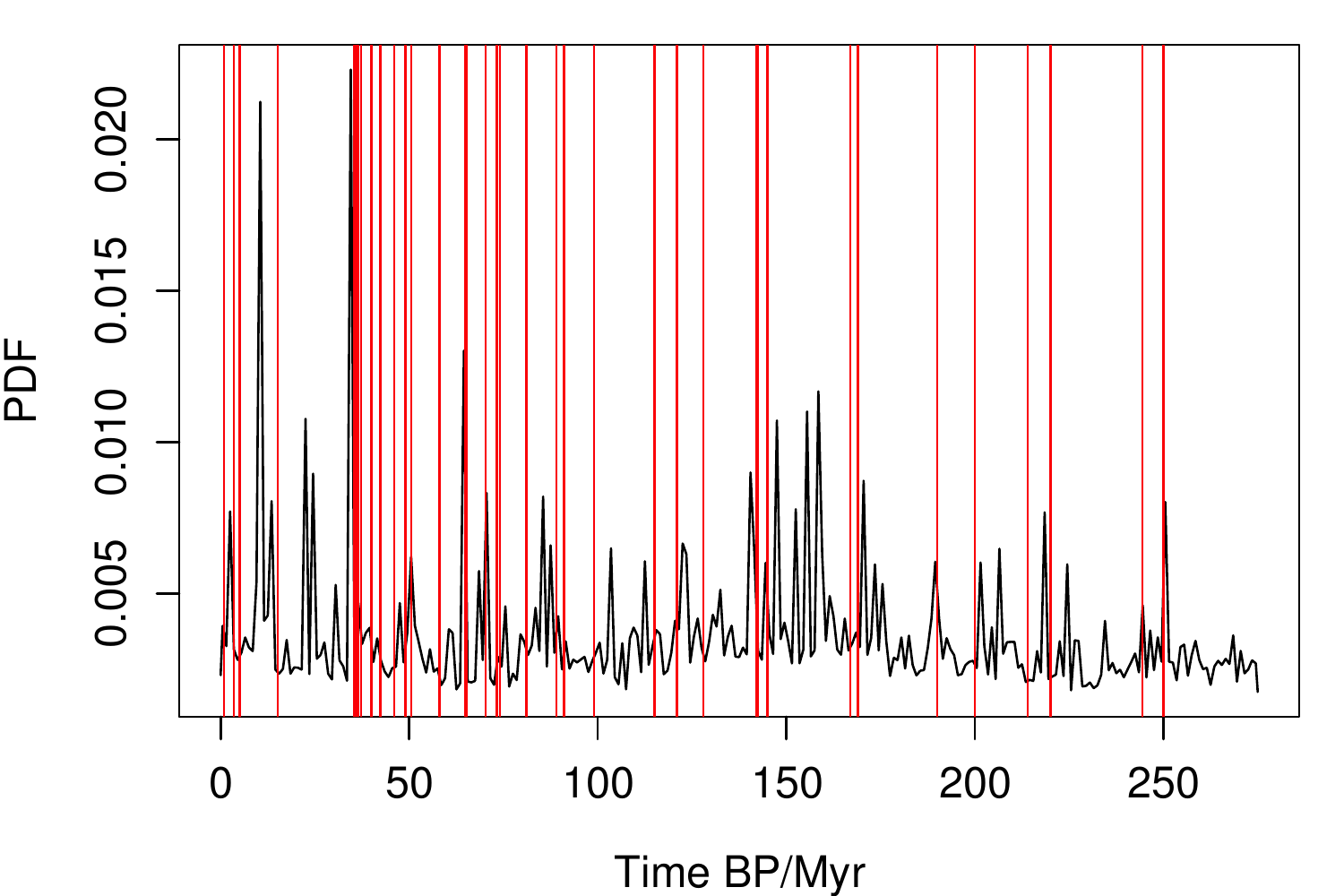}
  \includegraphics[scale=0.6,angle=-90]{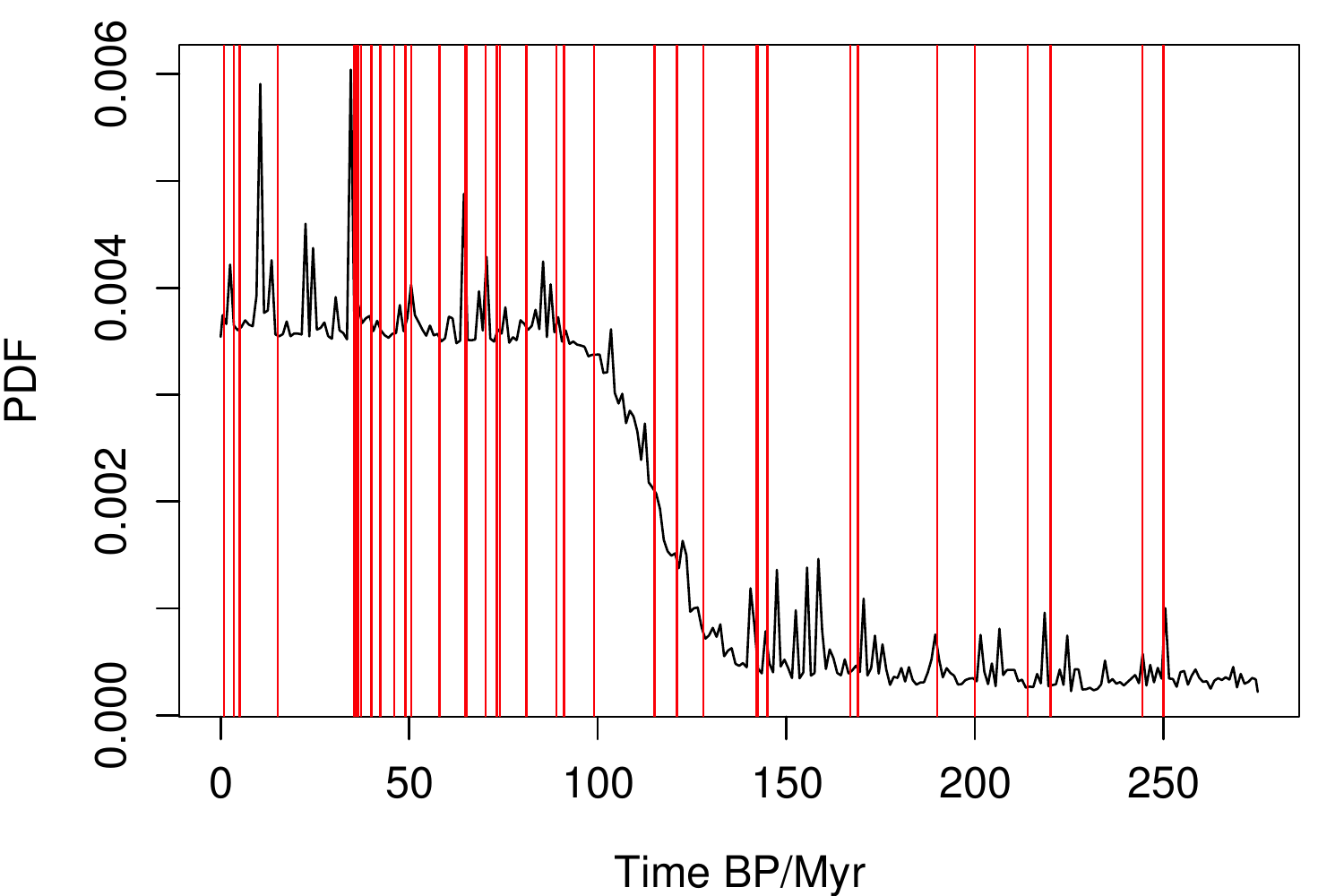}
  \caption{The prediction of the normalized cometary impact rate (i.e.\ probability density function; black line) compared to the actual impacts in the 
    basic250 time series (red lines). The models from top to bottom are TideProb, EncTideProb, and EncTideSigProb. A common solar orbit and encounter
    sample is used in all three cases.}
  \label{fig:dynmodels}
\end{figure}

Having defined TideProb and EncProb, we can combine them to make EncTideProb.
We can further combine this sum with SigProb (scaled by the parameter $\eta$) in order to include a smoothly varying component (see Table \ref{tab:tsmodels}).
Figure \ref{fig:dynmodels} shows examples of the TideProb, EncTideProb and
EncTideSigProb model predictions of the cometary flux for specific values of
their parameters. In the upper panel, we see the TideProb model predicts an
oscillating variation on at least two time scales. In the middle panel, we add
EncProb to TideProb. The amplitude of the background is reduced
due to the normalization effect -- the encounters dominate -- and the high
peaks characterize encounter-induced comet showers. In the bottom panel, the
SigProb model is added onto the EncTideProb model with $\eta=3$. A large value
of $\lambda$ has been used in SigProb here, such that the additional trend is
almost linear. Meanwhile, we also combine TideProb and SigProb to make TideSigProb. This of course does not show the randomly occurring peaks which are characteristic of the encounters model.

In Section \ref{sec:comparison_comet}, we will compare these models with other time
series models defined in Section \ref{sec:tsmodel_comet} using Bayesian method.

\section{Model comparison}\label{sec:comparison_comet} 

Now that we have a way to generate predictions of the comet flux from our dynamical time series models, we use the Bayesian method described in section \ref{sec:bayes} to calculate the evidences for the various time series models defined in section \ref{sec:tsmodel_comet} for different cratering data sets. Because the solar orbit is more sensitive to the Sun's initial galactocentric distance ($R$) and angular velocity ($\dot{\phi}$) than to the other four initial conditions \citep{feng13}, we sample over only those two parameters when calculating the evidences and Bayes factors (ratio of two evidences) for the dynamical models. In order to make our model comparison complete, we will vary all initial conditions individually and simultaneously in section \ref{sec:sensitivity_comet}.

To calculate the evidences we sample the parameter space of the dynamical
models and other time series models with $10^4$ and $10^5$ points
respectively. For the models of EncProb, EncTideProb, EncSigProb and
EncTideSigProb, each point represents an entire simulation of the orbit of the Sun about the Galaxy
and the corresponding simulation of the comet flux as a function of time. For
the latter we use the proxies of $G_3(t)$ and $\gamma(t)$ (i.e.\ the time-varying $\gamma_{\rm bin}$) described in section~\ref{sec:tideflux_comet} and section~\ref{sec:encounterflux_comet} respectively. For each orbit of the Sun we just generate a single sequence $\gamma(t)$ for the comet flux at random. (Because $\gamma(t)$ is modulated by the vertical tide coefficient $G_3(t)$, an average over many sequences of $\gamma(t)$ would be smooth and lack the spikes corresponding to comet showers which we see in the individual sequences.)

The Bayes factors of various models relative to the uniform model are listed
in Table \ref{tab:crater_BF}. We see that the SigProb, EncSigProb, TideSigProb and EncTideSigProb models are favoured by all the data sets, sometimes marginally, sometimes by a significant amount relative to certain models. In these favoured models, the negative trend (a decreasing cratering rate towards the past) is favoured much more than the positive trend. Such a negative trend can be picked out in Figure \ref{fig:set3lim}. As the positive values are so clearly ruled out, we only use negative values of $\lambda$ in all the trend models. This would be consistent with the crater preservation bias or the disruption of a large asteroid dominating over any recent increase in the asteroid impact rate (see section \ref{sec:impact_comet}). 
\begin{table}[ht!]
  \centering
    \caption{Bayes factors of the various time series models (rows) relative to the uniform model for the various data sets (columns). The suffix numbers 1 and 2 in the model names, e.g.\ EncProb1 and EncProb2, refer to which different initial conditions are fixed. 1 means $R(t=0)$ and 2 means $\dot\phi(t=0)$.
}
    \begin{tabular}{@{} l|l l l l l l}
      \hline
      \hline
      Model      &basic150&ext150 &full150&basic250&ext250 &full250 \\
      \hline
      RandProb   &4.4     &9.3    &72    &3.0    &9.4   &4.7$\times 10^2$\\
      RandBkgProb&1.8     &3.8    &31    &2.2    &5.2&1.8$\times 10^2$\\ 
      SinProb    &0.34     &0.62    &1.2     &0.43    &0.76    &1.5     \\
      SinBkgProb &1.0     &1.2    &1.6    &1.0    &1.2    &1.5     \\
      SigProb    &15    &63   &$9.1\times 10^3$&$2.0\times 10^2$ &$1.8\times 10^3$&$5.8\times 10^6$\\
      SinSigProb &10     &36   &$1.6\times 10^2$&$1.0\times 10^2$&$6.0\times 10^2$&$2.6\times 10^5$\\
      EncProb1   &1.5     &3.9    &26     &1.7   &5.2&$1.1\times 10^2$\\
      EncProb2   &1.7     &3.3    &77    &1.6    &8.5&$2.7\times 10^2$\\
      TideProb1   &0.73   &0.87   &6.7   &0.81    &0.91&1.1\\
      TideProb2   &0.79   &0.86   &10    &0.69    &0.76&0.94\\
      EncTideProb1&1.0    &1.6   &18 &1.3 &2.1&10\\
      EncTideProb2&1.2    &1.8  &25   &1.2    &2.1&24\\
      EncSigProb1 &11     &41 &$4.6\times 10^3$&$1.5\times 10^2$&$1.5\times 10^3$&$5.9\times 10^6$\\
      EncSigProb2 &12     &52 &$8.7\times 10^3$&$1.7\times 10^2$&$1.5\times 10^3$&$6.6\times 10^6$\\
      TideSigProb1&11   & 38  &$4.6\times 10^3$&$1.6\times 10^2$&$1.4\times 10^3$&$6.2\times 10^6$\\
      TideSigProb2&10   & 37  &$4.5\times 10^3$&$1.6\times 10^2$&$1.4\times 10^3$&$6.1\times 10^6$\\
      EncTideSigProb1&11&40   &$5.0\times 10^3$&$1.6\times 10^2$&$1.4\times 10^3$&$6.0\times 10^6$\\
      EncTideSigProb2&11&40  &$4.7\times 10^3$&$1.6\times 10^2$&$1.5\times 10^3$&$6.1\times 10^6$\\
      \hline
    \end{tabular}
    \label{tab:crater_BF}
\end{table}

The SinSigProb model is not favoured more than SigProb, which means the periodic component is not necessary in explaining cratering time series. This is consistent with the conclusion in \citet{bailer-jones11}. Moreover, the pure periodic model is actually slightly less favoured than the uniform model for the ``basic'' and ``ext'' data sets.  The pure random model (RandProb) is slightly more favoured than the random model with background (RandBkgProb). Both are more favoured than the uniform model, but with relatively low Bayes factors compared to the models with trend components.

EncProb is slightly more favoured than the TideProb model.
This suggests that the stochastic component of EncProb is slightly preferable to the smooth tidal component of TideProb in predicting the cratering data, although the difference is small.
Combining them to make the EncTideProb models does not increase the evidence.

The best overall model for explaining the data is SigProb, the pure trend model. Adding the tide or encounters or both does not increase the evidence by a significant amount for any of the data sets. This suggests that the solar motion has little influence on the total observed impact rate (i.e.\ comets plus asteroids and the preservation bias) either through the Galactic tide or through stellar encounters, at least not in the way in which we have modelled them here. This minor role of the solar motion in generating terrestrial craters weakens the hypothesis that the (semi-)periodic solar motion triggers mass extinctions on the Earth through modulating the impact rate, as some have suggested \citep{alvarez84,raup84}. We note that a low cometary impact rate relative to the asteroid impact rate has been found by other studies \citep{francis05,weissman07}.

The evidence is the prior-weighted average of the likelihood over the parameter space.  It is therefore possible that some parts of the parameter space are much more favoured than others (i.e.\ there is a large variation of the likelihood), and that this is not seen due to the averaging. In that case changing the prior, e.g.\ the range of the parameter space, could change the evidence. (We investigate this systematically in section~\ref{sec:sensitivity_comet}).  In other words, the tide or encounter models may play a more (or less) significant role if we had good reason to narrow the parameter space. This would be appropriate if we had more accurate determinations of some of the model parameters, for example. We now investigate this by examining how the likelihood varies as a function of individual model parameters (but still be averaged over the other model parameters).

\begin{figure}
  \centering
  \includegraphics[scale=0.6]{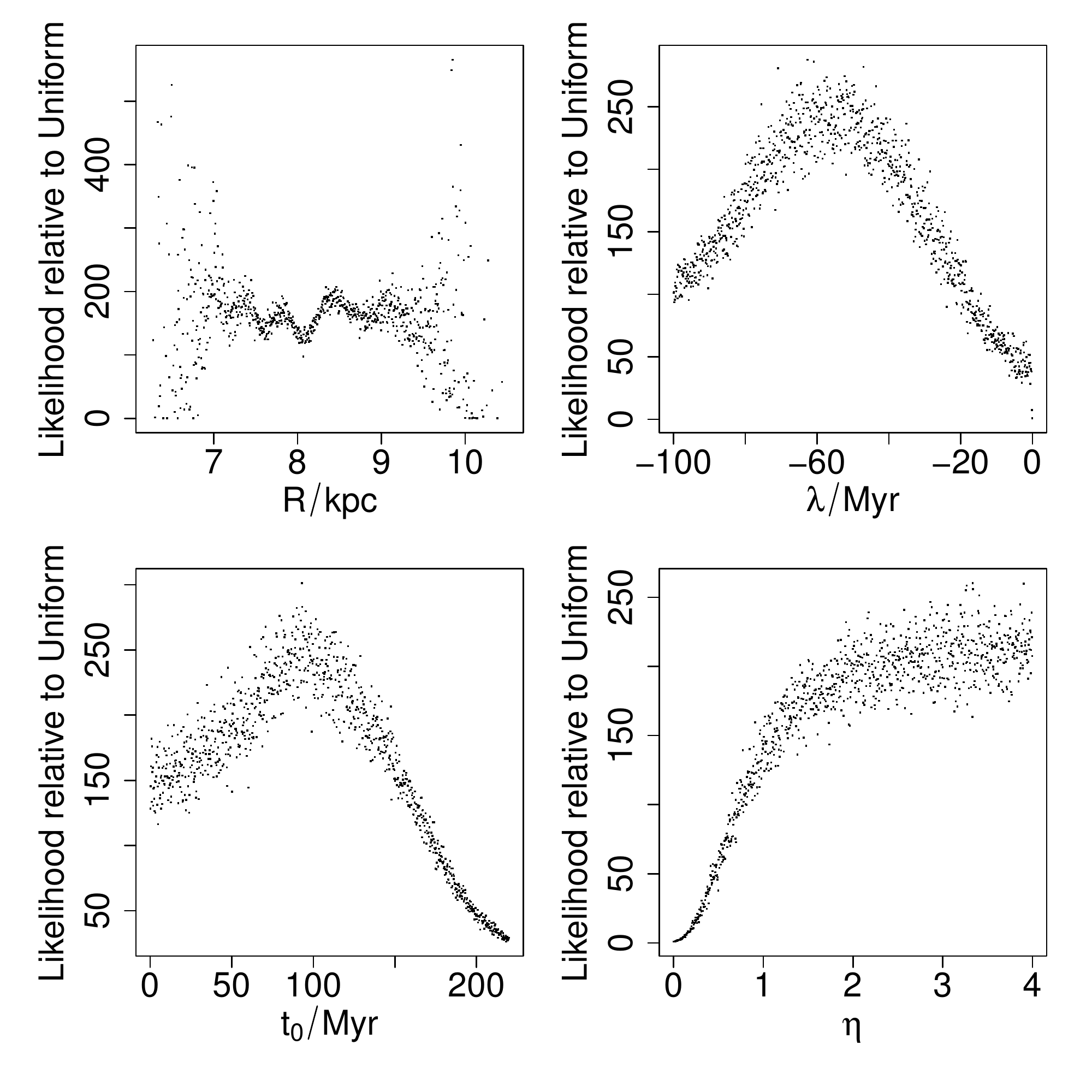}
  \vspace{-2ex}
  \caption{The distribution of the likelihood over each of the parameters in the TideSigProb1 model for the basic250 data set, sampling over all other parameters in each case. The parameters are divided into 1000 bins. For each bin, the likelihoods are averaged to reduce the noise generated by the randomly selected sequence of stellar encounters. There are 100\,000 samples in the parameter space. }
  \label{fig:like_1D}
\end{figure}

Figure \ref{fig:like_1D} shows how the resulting likelihood varies as a function of the
four parameters in the TideSigProb1 model.  The most favoured
parameters of the trend component are $\lambda \approx -60$\,Myr and $t_0
\approx 100$\,Myr. This trend component represents an increasing cratering
rate towards the present over the past 100 Myr
\citep{shoemaker98,gehrels94,mcewen97}, either real or a result of preservation bias. In the upper left graph, the
likelihood varies with $R$ slightly and varies a lot in the region where
$R<8$\,kpc and $R>9$\,kpc.  In the lower right panel, the likelihood increase with $\eta$,
which means that the trend component is important in increasing the likelihood
for the TideSigProb model. 

To find the relationship between the likelihood for TideSigProb and the
Sun's initial galactocentric distance $R$ and the scale parameter $\eta$, we
fix the parameters of the trend component to $\lambda=-60$\,Myr and
$t_0=100$\,Myr.
\begin{figure}[ht!]
  \centering
  \includegraphics[scale=0.9]{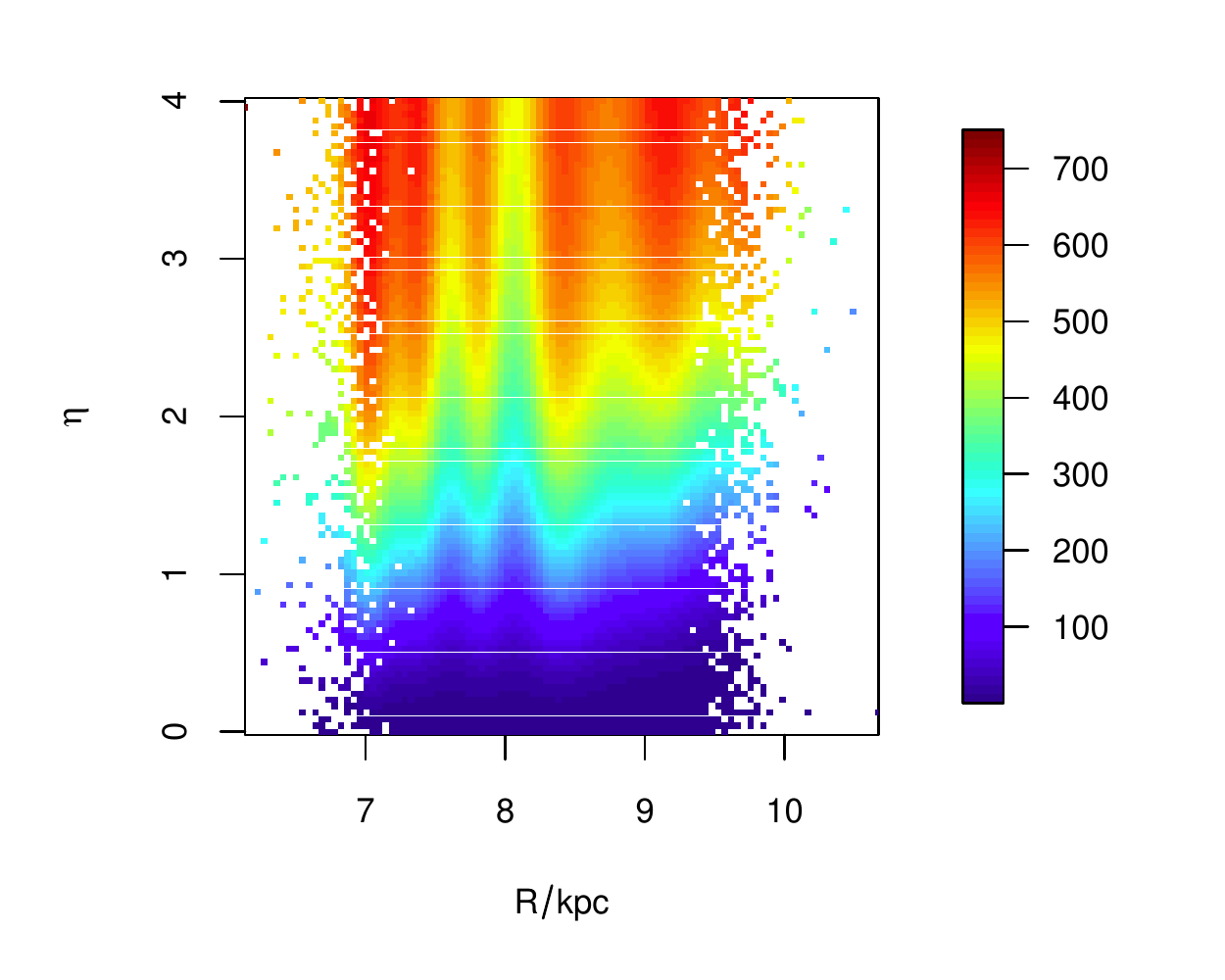}
  \vspace{-2ex}
  \caption{The distribution of the likelihood over the parameters $R$ and $\eta$ in the TideSigProb1 model relative to the Uniform model for the  basic250 data set. The relative likelihood is shown as the colour scale indicated in the legend. There are 100\,000 samples in the parameter space. }
  \label{fig:like_2D}
\end{figure}
In Figure \ref{fig:like_2D} we see that the likelihood for TideSigProb
increases monotonically with $\eta$ over this range, but has a more complex dependence on $R$. 
The likelihood is highest at around $R=7.0$ and $R=9.5$\,kpc.
In Figure \ref{fig:tideprob_R70} we compare the dates of the craters in the basic250 data set
with the prediction of the cratering rate from TideProb with $R=7.0$\,kpc.
There are 7 craters within the first 30\,Myr compared to 16 and 13 craters in
the intervals [30,60]\,Myr and [60,90]\,Myr respectively. This lack of craters
in the first 30\,Myr can be better predicted by TideSigProb than by
the SigProb model with a negative $\lambda$. 
While this is small number statistics, it may suggest that even though we have little evidence for
the effect of the tide on cometary impacts in the overall cratering data, it may have had more of an effect in selected time periods. Other explanations are also possible, of course: we cannot say anything about models we have not actually tested, such as a more complex model for the asteroid impact rate variation.
\begin{figure}
  \centering
  \includegraphics[scale=0.6,angle=-90]{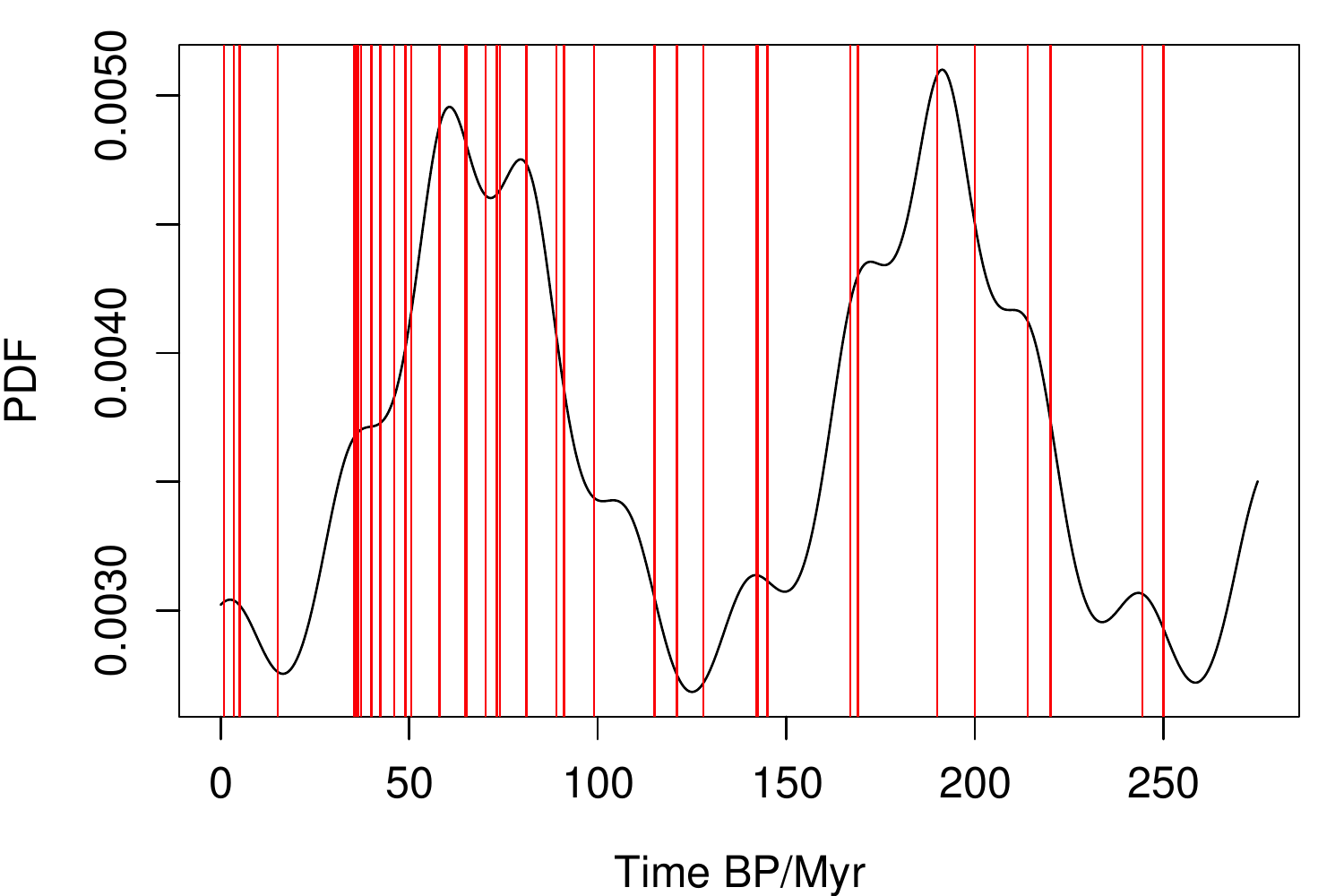}
  \caption{Comparison between the prediction of TideProb with $R= 7.0$\,kpc (shown as a probability distribution function in black) and the times of the impact craters in the basic250 data set (shows as vertical red lines). } 
  \label{fig:tideprob_R70}
\end{figure}

\section{Modelling the angular distribution of cometary perihelia}\label{sec:ADP_comet}

In this section we predict the 2D angular distribution (latitude, longitude) of the perihelia of LPCs, the observed data for which are shown in Figure \ref{fig:LPC1A_bl}.  To do this we need to identify from the simulations comets injected over an appropriate time scale.  Figure \ref{fig:gamma_fc_pdf} shows that a comet shower usually has a duration of less than 10\,Myr, something which was also demonstrated by \cite{dybczynski02} in detailed simulations of individual encounters.  The Galactic tide varies little over such a time scale, because the vertical component of the tide, which dominates the total Galactic tide, varies over the period of the orbit of the Sun about the Galaxy, which is of order 200\,Myr.  We may therefore assume that the solar apex is also more or less fixed during the past 10\,Myr, which is then an appropriate time scale for constructing our sample. 

We simulate cometary orbits over the past 10\,Myr as follows: (1) generate one million comets from the Oort cloud model (DLDW or DQT), as well as a set of stellar encounters (about 400 over 10\,Myr); (2) integrate the cometary orbits under the perturbations of only the Galactic tide (tide-only simulations with a time step of 0.1\,Myr), only stellar encounters (encounter-only simulations with a time step of 0.01\,Myr), and both of them (combined simulations with a time step of 0.01\,Myr) back to 10\,Myr ago; (3) identify the injected comets and their longitudes and latitudes. We then repeat steps (1)--(3) ten times (i.e.\ resample the Oort cloud and the set of stellar encounters) and combine the results in order to increase the number statistics.

\subsection{Latitude distribution}\label{sec:latitude_comet}

\begin{figure}
  \centering
  \includegraphics[scale=0.7]{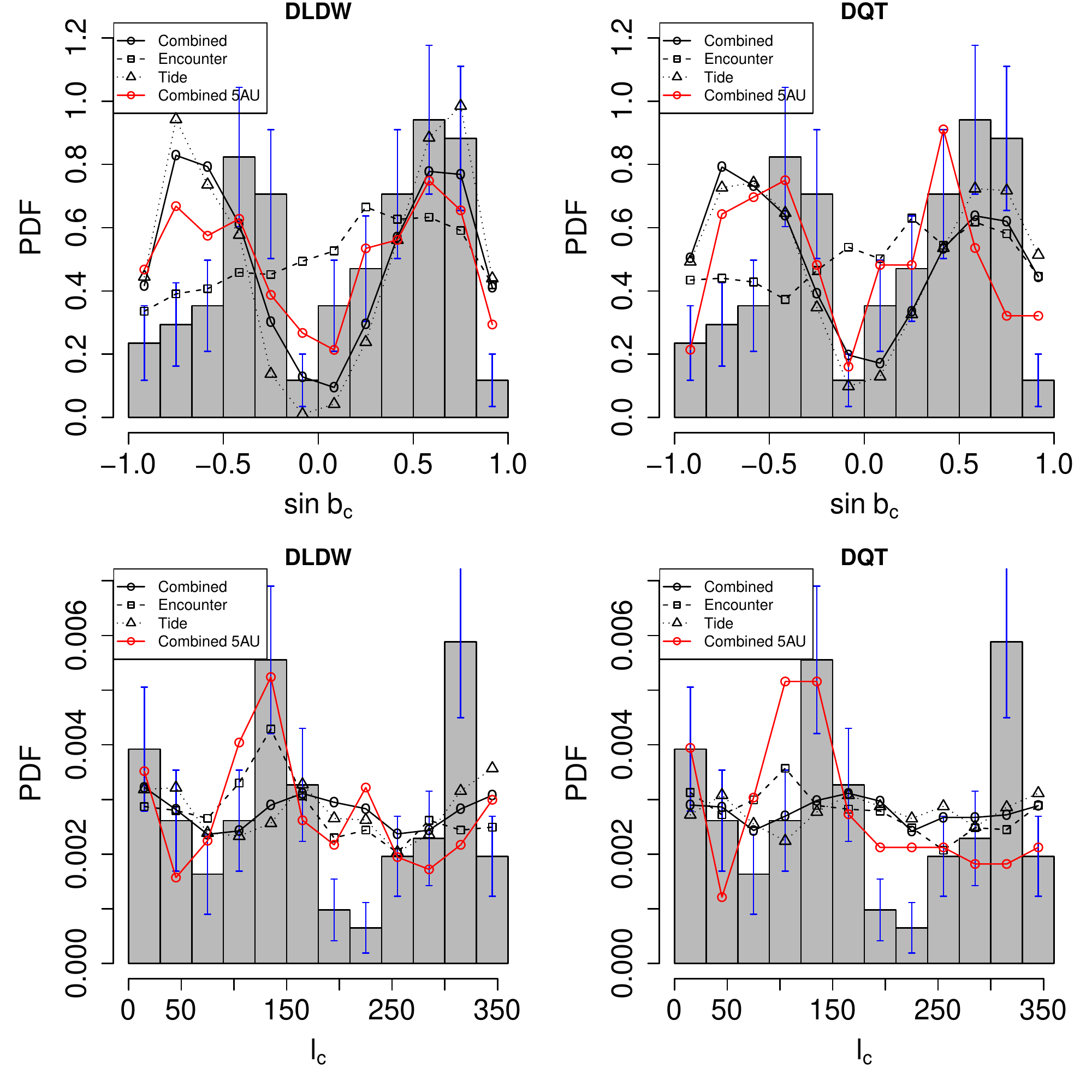}
  \vspace*{-2ex}
\caption{Comparison between the observed distribution (histogram blocks) and model-predicted distributions (points/lines) of the perihelia of long-period comets (LPCs) with Galactic latitude (upper panels) and longitude (lower panels) for the DLDW (left panels) and the DQT (right panels) Oort cloud initial conditions.  All distributions are normalized. The error bars on the data have been calculated using a Poisson noise model (arising from the binning) with a total of 102 class 1A LPCs. The model-predicted distributions show the comets injected into the loss cone for three modes of simulations, namely including only the Galactic tide (triangles), only stellar encounters (squares), and both (circles). The number of injected comets in these simulations for the DLDW (DQT) models are 1858 (981), 1133 (1976), and 12\,751 (2796), respectively.  The red circles connected by red lines show the number of comets injected into the observable zone ($q<5$\,AU), and comprise 449 comets for the DLDW model, and 112 for the DQT model. }
  \label{fig:dyn_bl}
\end{figure}

The upper panels of Figure \ref{fig:dyn_bl} compare the Galactic latitudes of the LPC perihelia with our model predictions. In addition to showing the model predictions for the comets injected into the loss cone, we also show the predicted distributions for comets injected into the observable zone ($q<5$\,AU). The former contains more comets, but the latter is of course closer to the observed sample.
The small sample of comets within the observable zone have significant sample noise in their angular distributions, so we will only compare model predictions of the angular distribution of comets in the (larger) loss cone.

The upper panels show that the injected LPCs in the pole and equatorial regions are depleted for both DLDW and DQT models, as also found by \cite{delsemme87}.  According to theoretical prediction, the tide-induced flux should be proportional to $|\sin b \cos b|$ \citep{matese99},
in very good agreement with our tide-only simulations.  The observed data broadly agree with this, the main difference being that for negative latitudes the peak is at around -0.4 rather than the model-predicted value of -0.7. This discrepancy was also noticed by \citet{matese11}, for example, and could be a consequence of the small size of the data set (note the errors bars in the figure).

We see in the figure that the PDF of the latitude distribution predicted by the combined simulation always lies between those predicted by the single perturbation simulations.
Although the combined simulation of comets injected into the loss cone predicts a flatter distribution than the tide-only simulation does, the stellar encounters cannot entirely smooth out the peaks in the latitude distribution. This is consistent with the results in \cite{rickman08}. Thus the observed non-uniform latitude distribution does not indicate that the Galactic tide dominates at the present epoch, as was claimed by \cite{matese11}.

We can attempt to make a more quantitative assessment of how well our models predict the observed distribution. Using model comparison techniques we can ask whether our dynamical models (the combined tide plus encounters model) explain the data better than a uniform distribution. 
We can do this crudely on the binned data/simulations shown in the figure via a likelihood test. The act of binning means that the model-predicted number of events per bin is determined by the Poisson distribution, thus defining our likelihood. However, such a test is dependent on the choice of binning, and we have tried out a range of bin widths and centres. While we find that the combined model for the DQT Oort cloud model is always more favoured than a uniform distribution, the significance is marginal.

An alternative approach is to use the unbinned data and unbinned model predictions, and to apply a kernel density estimate (KDE) to each. This produces a non-parametric density function for the data and for the model, the difference between which we quantify using the (symmetrized) Kullback-Leibler divergence (KLD). A value of zero divergence means that the two distributions are identical; larger (positive/negative) values indicate larger differences.  We find that our dynamical models give smaller KLD values than do the uniform model (i.e.\ the former predict the data better), for both the DLDW and DQT. Although the distributions formed by the KDE are sensitive to size of the kernel adopted,\footnote{This is analogous to the size of the histogram bins. A histogram is just a particular type of kernel.}  we find that the KLD values are quite insensitive to this, and consistently favour the dynamical models.  This suggests that the dynamical models explain the data better than a flat distribution in latitude (although because calibrating KLD ratios into formal significances is not easy, we leave this as a qualitative statement).

\subsection{Longitude distribution}\label{sec:longitude_comet}

The perihelia of LPCs are not distributed uniformly on the celestial sphere. It has been suggested \citep{matese99, matese11} that they lie preferentially on a great circle, as evidenced by two peaks at $l_c\simeq135$\deg and $l_c\simeq315$\deg seen in Figure \ref{fig:LPC1A_bl}.  The comets on this great circle could be induced by stellar encounters with preferred directions, thereby producing the apparent anisotropy.  In the lower two panels in Figure \ref{fig:dyn_bl}, we see that the model predictions do not produce any very large peaks, although one around $l_c\simeq135$\deg\ is discernable.  We also observe a peak around $l_c =$\,0--60\deg\ which is proposed as a signal of the ``Biermann comet shower'' \citep{biermann83,matese99}. In our model, this peak is probably the result of accumulated perturbations from several stellar encounters with preferred directions.
  
The peak around $l_c=135$\deg\ is more prominent in the model prediction for the comets injected into the observable zone (red points/line in the figure). This peak is generated primarily by one or more massive stellar encounters. Hence, stellar encounters play a more significant role in injecting comets into the observable zone than just into the loss cone. This is consistent with the ``synergy effect'' investigated by \cite{rickman08}. 

As with the latitude distribution, we also measured the KLD for the model predictions (for the loss cone) and for a uniform distribution. The dynamical models predict the data little better than a uniform distribution. (The likelihood test gives a similar result.) One reason for this lack of support for our dynamical (combined) model could be the fact that we are averaging the predicted distribution from the encounters over ten different realizations of the stellar encounters. This will tend to smooth out individual peaks, which are probably produced by just a few encounters with massive stars.\footnote{Such massive stars (or stars with relatively high $\gamma$) move slowly relative to the Sun, and so would generate a relatively narrow peak in comet flux with $l_c$.}
If we instead only used a single random realization of encounters, we are unlikely to reproduce exactly the showers which occurred. This is an inherent problem of modelling stellar encounters in a stochastic way.  This does not affect our model prediction of the latitude distribution nearly as much, however, because its shape is dominated by the non-stochastic tide.

In order to investigate this we again use our encounter model via the proxy $\gamma$ (a proxy of comet flux) defined in equation \ref{eqn:gamma}, but now as a function of $b_{\rm p}$ and $l_{\rm p}$, the direction toward the perihelion of the stellar encounter. Moreover, we now impose a minimum threshold, $\gamma_{\rm lim}$, on the proxy: The larger the value of $\gamma_{\rm lim}$, the larger the encounter perturbation must be for it to be included in the model.

\begin{figure}[h!]
  \centering
  \includegraphics[scale=0.8,angle=-90]{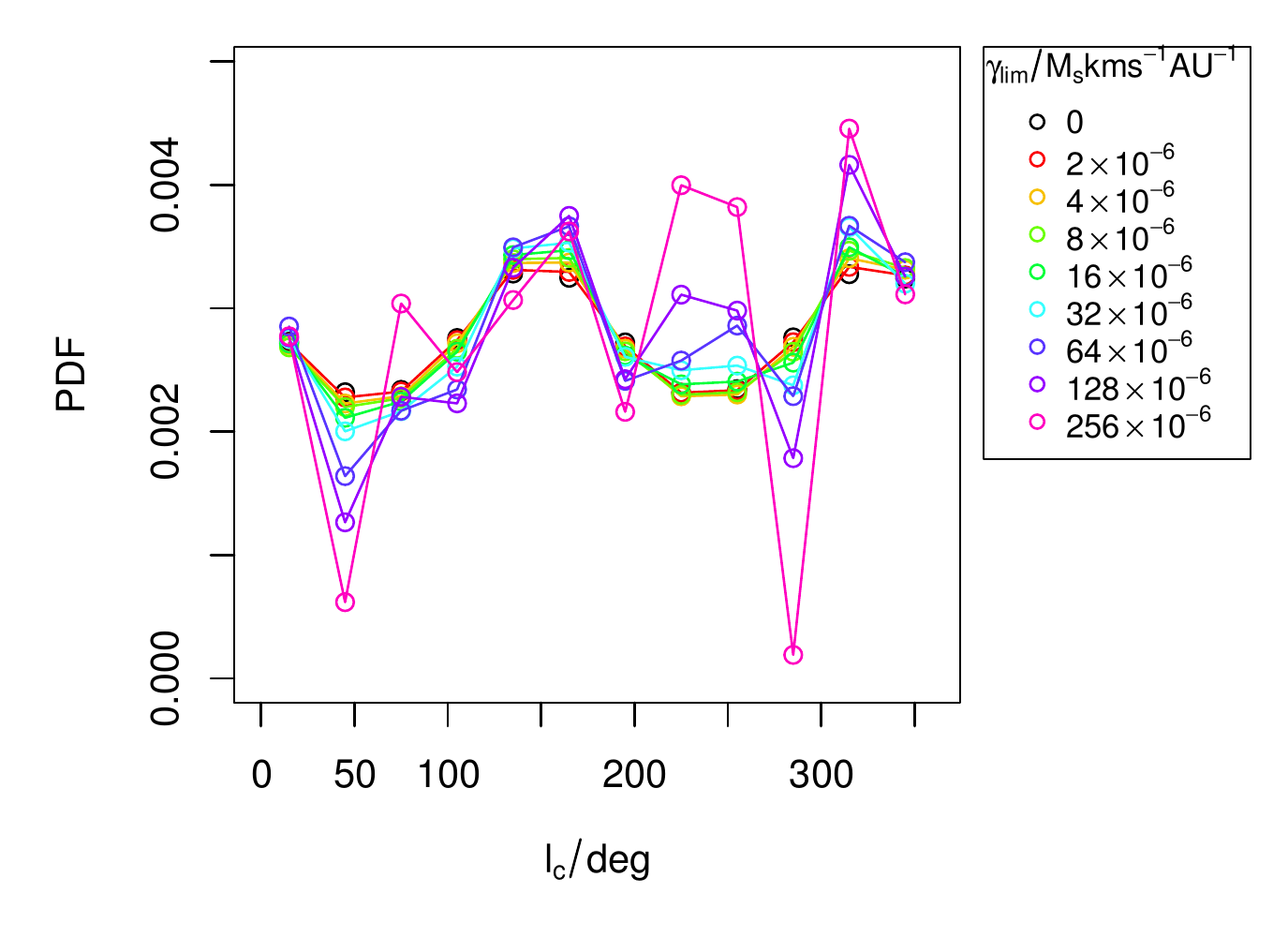}
  \caption{Predictions of the enounter-induced cometary flux when adopting different lower limits, $\gamma_{\rm lim}$, on the value of $\gamma$ required for an event to have an influence on the Oort cloud. There are $10^7$ and $10^8$ encounters generated for the model predictions with $\gamma_{\rm lim}=0$ and $\gamma_{\rm lim}\neq0$ respectively. 
}
  \label{fig:l_gamma_lim}
\end{figure}

Using the encounter model described in section \ref{sec:encmod_comet}, we simulate 10 million encounters and calculate $\gamma$, $b_{\rm p}$, and $l_{\rm p}$ for each. The predicted direction of an LPC's perihelion is opposite on the sky to the direction of the encounter perihelion. Thus we can calculate $b_c$ and $l_c$ accordingly and use $\gamma(b_c,~l_c)$ to predict the PDF of $b_c$ and $l_c$.  Then we divide the range of the Galactic longitude into 12 bins and sum $\gamma$ in each bin including only those encounters with $\gamma>\gamma_{\rm lim}$. Normalizing this gives the angular PDF of the encounter-induced flux, as shown in Figure \ref{fig:l_gamma_lim}. For larger values of $\gamma_{\rm lim}$ we observe a larger variation in the flux with longitude, as expected, because then fewer encounters contribute to the distribution. As we can see from equation \ref{eqn:gamma}, these are the more massive and/or slower stars. These encounters may induce a series of weak comet showers rather than a single strong comet shower. Because strong encounters are rare and extremely weak encounters cannot induce enough anisotropic LPCs, the spikes in the longitude distribution can be caused by at least two weak encounters rather than one strong or many extremely weak encounters. From Figure \ref{fig:dyn_bl}, we see that the tide cannot completely wash out the anisotropy in the longitude distribution induced by these encounters. 

Consistent with our results, \cite{matese11} found that the two spikes in the longitude distribution result from weak impulsive perturbations by analyzing the energy and angular momentum of dynamically new LPCs. Similar to the definition of weak comet showers in \cite{matese02} and \cite{dybczynski02}, we define encounters with $\gamma$ in the interval $\lbrack 1\times 10^{-7}, 5\times 10^{-6}\rbrack M_{\odot} ~km~s^{-1}~AU^{-1}$ as weak encounters. We do not find strong peaks in the longitude distribution of $\gamma$ for these encounters in Figure \ref{fig:l_gamma_lim},  because we know that $\gamma$ can underestimate the intensity of the shower (see Figure \ref{fig:gamma_fc_pdf}). Thus a small enhancement of the two peaks in Figure \ref{fig:l_gamma_lim} may correspond to a large enhancement of the peaks in the longitude distribution as predicted by our dynamical model in Figure \ref{fig:dyn_bl}.

Inspecting the catalogue of the frequencies of different types of stellar encounters in table 8 of \cite{sanchez01}, we see that there were at least eight encounters with masses equal to or larger than one solar mass encountering the solar system in the past 10\,Myr with perihelia less than 1\,pc. These encounters can move to a heliocentric distance much larger than 50\,pc over that time, which is the upper limit for their unbiased sample of stellar encounters with $M_V < 5$ -- see Figure 13 of \cite{sanchez01}. 

We also point out that GL 710 will have a close approach with the solar system in about 1.4\,Myr at a perihelion longitude of around 135$^\circ$. According to studies, it will induce a weak comet shower which is expected to increase the cometary flux by 40\%-50\% \citep{sanchez99,matese02}. This supports the suggestion that the solar apex motion induces the non-uniform longitude distribution of the LPCs' perihelia (see Figure \ref{fig:venc_denc_bl} and \ref{fig:dyn_bl}). In addition, Algol, a triple-star system with a total mass of 5.8\,$M_{\odot}$, encountered the solar system with a closest distance of 2.5\,pc 6.9\,Myr ago \citep{sanchez01}. The Galactic longitude of Algol was also close to $135^\circ$.

Based on the above plausible scenario, we conclude that the peaks in the longitude distribution of LPC perihelia could arise from the perturbations of a few strong stellar encounters, the encounter directions of which depend on the solar apex motion. Considering the important role of the Galactic tide in generating a non-uniform latitude distribution, and the role of stellar encounters in generating a non-uniform longitude distribution, the synergy effect plays a role in maintaining -- rather than smoothing out -- the anisotropy in the observed LPCs. In other words, we can explain the anisotropy of the LPC perihelia based only on the solar apex motion and the Galactic tide, without needing to invoke the Jupiter-mass solar companion as proposed by \cite{matese11}. To date there is no observational evidence for such a companion. We note that a recent analysis of data from the WISE satellite has excluded the existence of a Jupiter-mass solar companion with a heliocentric distance less than 1\,pc \citep{luhman14}.

\section{Sensitivity test}\label{sec:sensitivity_comet}

\subsection{Spiral arms and Galactic bar }\label{sec:armbar_comet}

The spiral arms and Galactic bar are non-axisymmetric, time-varying components of the Galactic potential. These make only a small contribution to the tidal force acting on the Sun and Oort cloud (\citet{binney08_book,cox02}). However, if their contribution is always in the same direction, the effect of their perturbation could accumulate.  This can occur when the Sun is near to the co-rotation resonance, when the rotation velocities of the disk and of the spiral pattern coincide.  To test this hypothesis, we simulate the solar and cometary motion adopting various constant pattern speeds of the spiral arms and the bar with fixed Galactic density distributions (specified in section \ref{sec:galaxy}).

\begin{figure}[h!]
  \centering
  \includegraphics[scale=0.5]{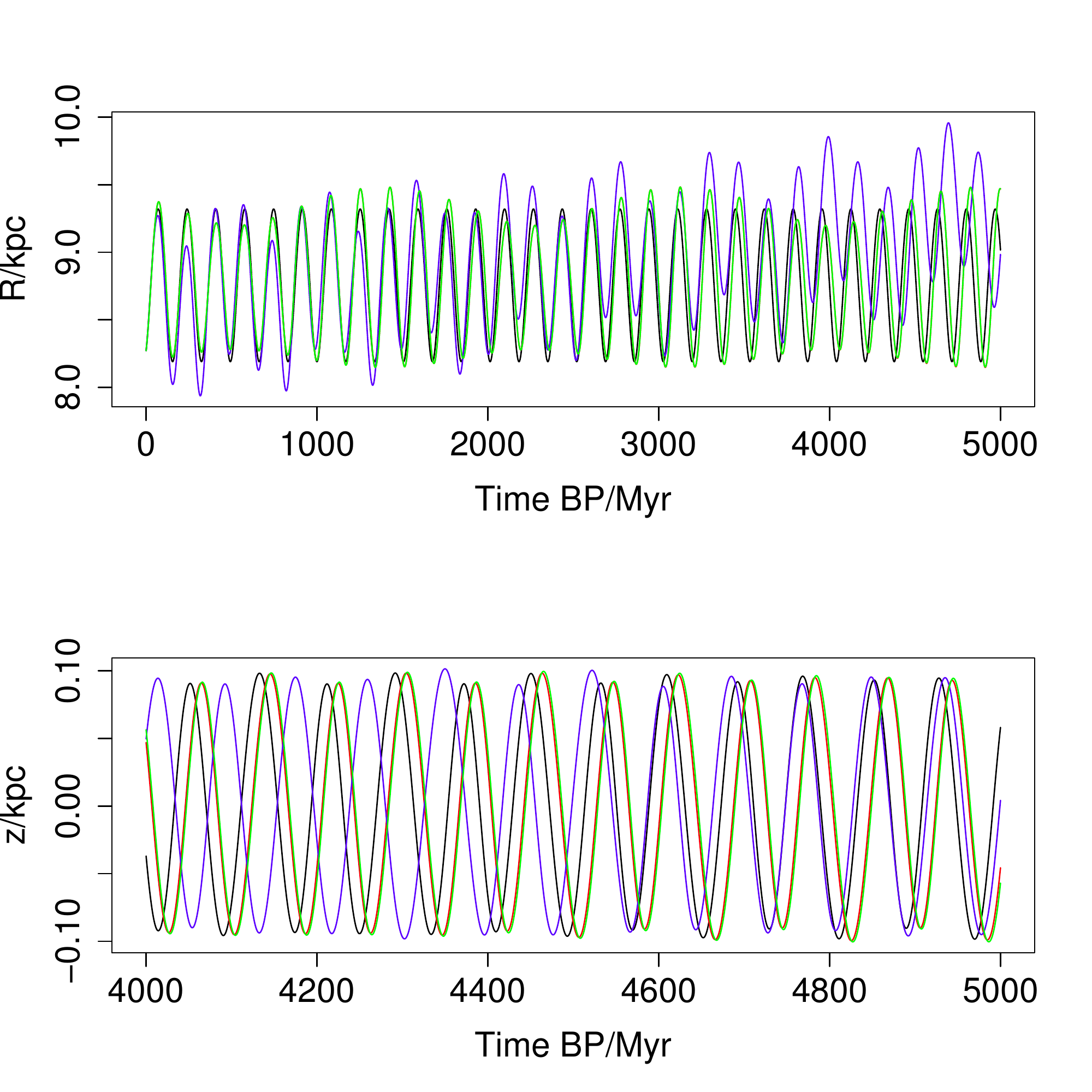}
  \caption{The variation of Sun's galactocentric radius (upper panel) and
    vertical displacement from the disk (lower panel) as calculate for
    different potentials: axisymmetric potential (black); potential including
    Galactic bar (red); potential including spiral arm (blue); potential
    including both bar and arm (green). To show different lines in the lower
    panel better, we plot the variation of the Sun's vertical displacement
    over a shorter time scale.} 
  \label{fig:solar_orbit}
\end{figure}

We integrate the solar orbit in the Galactic potential both including and excluding the non-axisymmetric components. The initial conditions of the Sun and potential parameters are given in Table \ref{tab:model_par}. We find that the gravitational force from the bar is always much larger than that from the spiral arms. However, the difference between the pattern speed of the Galactic bar $\Omega_b$ and solar angular velocity is much larger than the difference between the pattern speed of the spiral arms $\Omega_s$ and solar angular velocity, which results in a much lower accumulated perturbation due to the bar. To see this effect, we integrate the solar orbit back to 5 Gyr in the past.  The variations of galactocentric radius and vertical displacement of the Sun are shown in Figure \ref{fig:solar_orbit}.  The arms have a stronger effect on the solar orbit than does the bar.  The spiral arms tend to increase the galactocentric radius of the Sun as the integration proceeds (back in time), while the bar modulates the galactocentric radius by a comparatively small amount. Neither the bar nor the arms significantly affect the vertical displacement amplitude of the Sun. Here the combined perturbation from the potential including both the Galactic bar and spiral arms changes the solar motion the same way as the perturbation from the bar alone.  

\begin{figure}[ht!]
  \centering
  \vspace{-2ex}
  \includegraphics[scale=0.9,angle=-90]{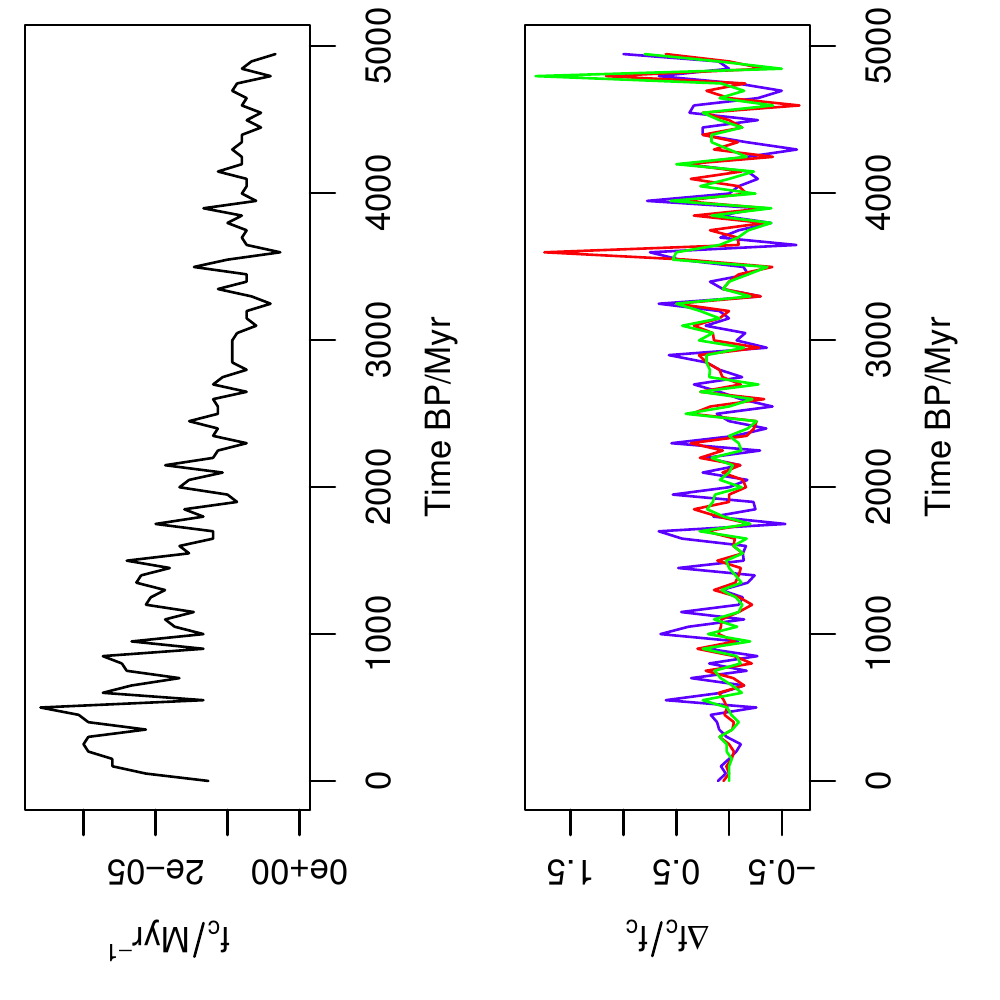}
  \caption{The magnitude of the tide-induced flux, $f_c$, generated by the
    axisymmetric potential model (upper panel) and the relative flux difference,
    $\Delta f_c/f_c$, generated by asymmetric Galactic potential models (lower panel) over the past 5 Gyr with a sample of 3$\times 10^4$ comets. The potentials are: axisymmetric potential only (black); including the arms (blue); including the Galactic bar (red); including both the arms and the Galactic bar (green).
} 
  \label{fig:flux_nonsym}
\end{figure}

We now simulate the tide-induced flux corresponding to these different
potential models. The lower panel in Figure \ref{fig:flux_nonsym} shows that the non-axisymmetric components do not alter the flux very much. Although the
perturbation from the arms can change the solar orbit slightly, the resulting change in the perturbation of the Oort cloud is minimal. The changed tidal force may change some individual cometary orbits, but has little effect on the overall injected comet flux, because the effect of the tide depends also on the distribution of the comets, which is nearly isotropic. 
We also see that the arms modify the cometary flux more than the bar,
consistent with its larger impact on the stellar density. (The limited number
of injected comets contributes to the sharp peaks in the relative flux difference, $\Delta f_c/f_c$, after 3\,Gyr.)

We also investigated the sensitivity of the solar motion and comet flux to the pattern speed of the asymmetric components.  We find that the closer the pattern speed of the arms is to the angular velocity of the Sun, the larger the perturbation from the arms is. (We can understand this in terms of a resonance.) Meanwhile, the perturbation from the bar is not sensitive to the bar's pattern speed.

Finally, we also find that the distribution of $b_c$ and $l_c$ of the comet flux does not change very much for different non-axisymmetric components of the Galactic potential.

In summary, we find that the model predictions of the tide-induced cometary flux are generally insensitive to changes in the non-axisymmetric components of the Galactic potential, except when a resonance between the arms and the solar orbit occurs, which increases the variation in the cometary flux.

\subsection{Variations of the prior}\label{sec:components_comet} 

As discussed earlier, the evidence depends on the prior distribution adopted
for the model parameters. As this prior frequently cannot be determined with
any certainty, it is important to investigate the sensitivity of the evidence
to changes in the prior.\footnote{A more robust -- but also more
  time-consuming -- way of calculating the evidence is presented in
  \cite{bailer-jones12}.}  To complete the calculation of evidences for
dynamical models, we also vary the other three initial conditions,
$V_R(t=0~{\rm Myr})$, $z(t=0~{\rm Myr})$, and $V_z(t=0~{\rm Myr})$, in the
EncTideSigProb models, which we previously kept constant.  Together with
SigProb, EncSigProb and TideSigProb, this was previously the best favoured model (Table \ref{tab:crater_BF}).  We made numerous changes in the priors by altering their parameter ranges, and re-did all necessary Monte Carlo samplings, numerical simulations, and likelihood calculations and recomputed the Bayes factors. Some of our results are shown in Table \ref{tab:prior_change}.

\begin{table}[ht!]
\caption{The Bayes factors for various time series models (rows) relative to the uniform model for two different data sets (cf.\ Table \ref{tab:crater_BF}). The second column describes what change has been made to the range of which parameter in the prior. The other priors are kept fixed.
TideSigProb3--6 refer to the TideSigProb model in which different initial conditions are varied:
  $V_R(t=0~{\rm Myr})$; $z(t=0~{\rm Myr})$; $V_z(t=0~{\rm Myr})$; all three (respectively)}
\hspace{-0.2in}\begin{tabular}{l|l|c|c}
\hline
\hline
models&varied prior&Bayes factor for basic150&Bayes factor for basic250\\
\hline
\multirow{5}{*}{RandProb}
&none&4.4&3.0\\
&$\sigma=2\bar{\sigma_i}$&2.0&4.8\\
&$\sigma=1/2\bar{\sigma_i}$&2.2&4.7\\
&$N=2N_{\rm ts}$&1.9&1.8\\
&$N=1/2N_{\rm ts}$&2.4&7.6\\
\hline
\multirow{5}{*}{RandBkgProb}
&none&1.8&2.2\\
&$\sigma=2\bar{\sigma_i}$&1.6&3.7\\
&$\sigma=1/2\bar{\sigma_i}$&1.8&2.6\\
&$N=2N_{\rm ts}$&1.5&1.5\\
&$N=1/2N_{\rm ts}$&2.4&2.9\\
\hline
\multirow{4}{*}{SinProb}
&none&0.34&0.43\\
&$10<T<100$&0.12&0.14\\
&$2\pi/300<\omega<2\pi/10$&0.34&0.39\\
&$10<T<300$&0.88&$5.4\times 10^{-2}$\\
\hline
\multirow{4}{*}{SinBkgProb}
&none&1.0&1.0\\
&$10<T<100$&0.90&0.88\\
&$2\pi/300<\omega<2\pi/10$&1.0&1.0\\
&$10<T<300$&1.8&1.4\\
\hline
\multirow{4}{*}{SigProb}
&none&15&$2.0\times 10^2$\\
&$0<t_0<1.2\tau_{\rm max}$&13&$1.4\times 10^2$\\
&$-100<\lambda<100$&7.7&$1.0\times 10^2$\\
&$0<\lambda<100$&$1.3\times 10^{-2}$&$1.8\times 10^{-3}$\\
\hline
\multirow{3}{*}{SinSigProb}
&none&6.4&80\\
&$0<t_0<1.2\tau_{\rm max}$&8.3&71\\
&$2\pi/300<\omega<2\pi/10$&9.9&97\\
\hline
TideSigProb3&none&9.0&$1.7\times 10^2$\\
\hline
TideSigProb4&none&9.1&$1.7\times 10^2$\\
\hline
TideSigProb5&none&9.0&$1.7\times 10^2$\\
\hline
TideSigProb6&none&11&$1.6\times 10^2$\\
\hline
\end{tabular}
\label{tab:prior_change}
\end{table}

The difference in Bayes factors for random models (RandProb, RandBkgProb) and periodic models (SinProb, SinBkgProb) with different prior distributions is less than five. The Bayes factors also remain less than ten so they remain no better explanations of the cratering data than the Uniform model. Thus our former conclusions about these models are not very sensitive to plausible changes in the priors. 

The TideSigProb models in which other parameters are varied have nearly the same evidences as the TideSigProb models listed in Table \ref{tab:crater_BF}, so these too are insensitive to these changes in the priors.
We also see that the SigProb model with positive $\lambda$ has Bayes factors much lower than
SigProb with negative $\lambda$ for both the basic150 and basic250 data sets. 

The dynamical models have parameters of the Galaxy potential,
Sun's initial conditions and combination ratio parameters ($\eta$ and $\xi$) which are listed in Table \ref{tab:prior_comet}). To keep things simple, we change the fixed parameters and the ranges of the
varying parameters individually, and then calculate the evidence by sampling
the prior defined by the changed parameter and other parameters shown in Table \ref{tab:prior_comet}. We calculate evidences for dynamical models with double or half the disk mass ($M_d$), halo mass ($M_h$), standard deviation of the initial value $R$ ($\sigma_R$), and the range of the varying ratio between the EncTideProb (or TideProb) and SigProb models ($\eta$). In addition, previous studies suggest that the number of tide-induced LPCs is not identical to the encounter-induced LPCs, i.e.\ $\xi\neq1$ \citep{heisler87,rickman08}. Thus we multiply the ratio between the tide-induced flux and the encounter-induced flux ($\xi$) by a factor of 4 or 1/4 for the sensitivity test. 

The resulting Bayes factors calculated for the basic150 data set are
shown in Table \ref{tab:dyn_prior_change}. In each row we see little variation: the Bayes factors
are relatively insensitive to these parameters. This means that either the parameter space of the EncTideSigProb1 model is evenly favoured by the
basic150 data set, or the data are unable to discriminate between the compound dynamical models.
\begin{table*}
\caption{The Bayes factors for EncProb1, EncTideProb1 and EncTideProb1 for basic150 with different Galaxy parameters.}
\label{tab:dyn_prior_change}
\hspace{-1in}{\small
  \begin{tabular}{l|*{11}{c}}
\hline
\hline
models &none&$2M_d$&$1/2M_d$&$2M_h$&$1/2M_h$&$2\sigma_{R}$&$1/2\sigma_{R}$&$\xi=4$&$\xi=1/4$&$0<\eta<8$&$0<\eta<2$\\
\hline
\multirow{1}{*}{EncProb1}
&1.5&2.5&3.4&2.5&4.1&2.3&2.6&---&---&---&---\\
\hline
\multirow{1}{*}{EncTideProb1}
&1.0&2.1&2.3&2.6&3.5&1.8&1.0&1.5&0.73&---&---\\
\hline
\multirow{1}{*}{EncTideSigProb1}
&11&15&11&13&12&12&11&12&10&13&8.8\\
\hline
\end{tabular}
}
\end{table*}

The model prediction of the anisotropic LPCs (see Figure \ref{fig:dyn_bl}) depends to a greater or lesser extent on the Galactic potential, the Sun's initial condition, the Oort Cloud model, and the model of encounters. We vary the model parameters in the same way as we did in Table \ref{tab:dyn_prior_change} and simulate ten million orbits of DLDW comets perturbed by the tide and ten samples of stellar encounters backwards to 10\,Myr ago. We find that the latitude distribution of the LPC perihelia is not sensitive to the change of the Galactic halo mass, the initial conditions of the Sun, or the direction of the solar apex. The amplitudes of the peaks in the latitude distribution are reduced if we decrease the mass of the Galactic disk or increase the stellar masses, which make the stellar encounters play a more important role in injecting comets into the loss cone.
However, the overall profile of the peaks is not changed in the latitude distribution.

The peaks in the longitude distribution shift slightly if we change the solar apex direction, the masses of the encounters, or the mass of the Galactic disk. The longitude distribution is not sensitive to changes in the other model parameters.

Finally, we also tested the effect of changing the time step in the (combined) simulations. We simulated four million comets generated from the DLDW model perturbed by the tide and ten samples of stellar encounters backwards to 10\,Myr ago using a time step of 0.001\,Myr (as opposed to 0.01\,Myr). We find little change in either the latitude or longitude distributions. In addition, we see only 4\% more comets injected when using this smaller time step.

In summary, we find that the overall shape of the angular distribution of LPC perihelia in both longitude and latitude is not very sensitive to changes in the model parameters, in particular not to the initial distribution of Oort Cloud comets, not to the masses of Galactic halo and disk, and not to the initial conditions of the Sun.

\section{Discussion and Conclusion}\label{sec:conclusion_comet}

We have built dynamical models for the impact rate and angular distribution of comets induced by the Galactic tide and stellar encounters, as modulated by the solar motion around the Galaxy.  Without using the approximate methods (the averaged Hamiltonian or impulse approximation), we numerically simulate the tide-induced flux and encounter-induced flux separately. We use these to validate the use of proxies for tide-induced flux, $G_3$, and for the encounter-induced flux, $\gamma_{\rm bin}$, in our models.  

Using the Bayesian evidence framework, we find that the pure trend model
(SigProb) together with the dynamical models including a trend component
(EncSigProb, TideSigProb and EncTideSigProb) for the cratering record are better favoured than other models we have tested. The trend component indicates a decreasing cratering rate ($\lambda<0$) towards the past over the past 100 Myr \citep{shoemaker98,gehrels94,mcewen97,bailer-jones11}. This suggests that either the asteroid impact rate or the preservation bias or both dominates the cratering record. Because the craters in our data sets are larger than 5\,km, the preservation bias may not be very significant over this time scale. The disruption of a single large asteroid could explain the trend in the data, as suggested by \citep{bottke07}. In addition, our models, which include the solar apex motion, can properly predict the anisotropic perihelia of LPCs without assuming a massive body in the outer Oort Cloud or an anisotropic Oort Cloud.

The EncTideSigProb, EncSigProb and TideSigProb models have Bayes factors of the
same magnitude as the SigProb model, which indicates that either the tide and encounter components are unnecessary in modelling the temporal distribution of craters, or the data cannot effectively discriminate between the models.

The stochastic component in the comet flux arising from encounters -- as represented by the term $\gamma$ -- in the EncProb and EncTideProb models can slightly increase their evidence relative to the TideProb model. We have performed a sensitivity test by changing the prior PDF over the parameters in the dynamical models and other time series models, and find only small changes of the Bayes factors. 

The asymmetrical components in the Galactic potential could, in principle, increase the time-variation of the comet flux and hence impact rate predicted by the dynamical models, by inducing larger deviations of the Sun's motion from a circular orbit and thus larger changes in the local stellar density.
It turns out that the non-axisymmetric component has relatively little impact on the predicted cometary flux, with the exception of when the Sun is in co-rotation with the spiral arms. In that case the transient resonance can produce large variations in the flux.

By including the solar apex motion, our dynamical models for anisotropic LPCs can predict reasonably well the distribution of Galactic latitude and longitude in a set of 102 dynamically new comets. In this model, the asymmetry in the distribution of Galactic latitudes caused by the Sun's current location and its motion over the past 10\,Myr (comparable with the time scale
of a comet shower). 

The two narrow peaks in the cometary perihelia at $l_c=135^\circ$ and $l_c=315^\circ$ could be caused by a handful of strong stellar encounters encountering the Sun with their encountering velocities in the direction of antapex in the HRF. On the other hand, we might also see something similar due to the periodic orbital motion about the Sun of a massive body (such as a brown dwarf) residing within the Oort cloud \citep{matese99, matese11}. However, our dynamical model, which takes into account the solar apex motion, can predict the longitudinal asymmetry without assuming the existence of such a body. In addition, the latitude distribution of LPC perihelia predicted by our simulations is consistent with the theoretical prediction, although one peak in the observed distribution is not properly predicted by our simulations. The synergy effect between the encounters and the tide cannot entirely eliminate the anisotropy induced by either the tide or the encounters.

A non-uniform distribution in the perihelion direction of encounters was found by \citet{sanchez01}, although the signal is of questionable significance due to the incompleteness, i.e.\ faint stars which high velocities being too faint after 10\,Myr for Hipparcos to have observed.

An anisotropy in the longitude of LPCs will not correspond to an anisotropy in longitudes of impacts on the Earth's surface due to the rotation of the Earth and its orbit about the Sun. Some latitude variation may be expected, despite the long-term variation in inclination and obliquity of the Earth's orbit \citep{feuvre08,werner10}. Disrupted comets generally retain their original orbital plane \citep{bottke02}, so the resulting asteroids would tend to impact in the plane perpendicular to solar apex. Yet these are all higher order effects which would be difficult to convincingly detect and relate to the solar orbit in the analysis of terrestrial impact craters.

Our modelling approach has, like any other, introduced various assumptions and approximations. We have ignored the synergy effect between the Galactic tide and stellar encounters highlighted by \citet{rickman08}. We instead simply sum the
tide-induced flux and the encounter-induced flux in the ratio $\xi$ to 1.
Because the cometary impact rate modulated by the solar motion around the
Galactic center seems to be unnecessary in order to explain the data, the
synergy effect, which is also influenced by the solar motion, may not change the
result significantly. In addition, we use a decreasing impact rate towards the past (negative trend component) to model the combined effect of preservation bias and asteroid impact rate. In modelling the angular distribution of the LPC perihelia, the sample noise in the comets injected into the observable zone prevent us from building a more robust model, especially for the longitude distribution. This problem could be resolved by calculating perturbations based on a more accurately measured Galactic tide and using an actual catalogue of encountering stars in the solar neighborhood as opposed to our stochastic model of plausible encounters. 

In common with some other studies (e.g.\ \citet{rickman08, gardner11, fouchard11, wickramasinghe08}), 
we have ignored the perturbing effect on comets from the giant planets, although we acknowledge that the giants planets could influence the predicted LPC flux in particular \citep{kaib09}. 
The planetary perturbations can also change the fraction of the inner Oort cloud comets among the injected LPCs \citep{kaib09}, which in turn could change the angular distribution of the LPC perihelia. However, these perturbations should not have a significant effect over the relatively short time scale of 10\,Myr which we use in the simulations to generate the LPC distribution.
As the main goal of our work is to study the variable effect of the solar orbit on the LPC flux and angular distribution, rather than to predict the absolute LPC flux precisely, our conclusions should not be overly affected by neglecting the giant planets in this way.

In the future, the Gaia survey allow us to detect many more recent stellar encounters down to fainter magnitude limits and larger distances than Hipparcos, thereby allowing us to extend the time scale over which we can get a complete sample of recent stellar encounters. The Gaia magnitude limit of G=20 which is low enough to cover the high velocity stars in a time scale of 10 Myr. For example, a star with absolute magnitude of 10 and a velocity of 80\,km/s in the  HRF would move 800 pc in 10 Myr and so have an apparent magnitude of 19.5. Thus Gaia will be able to observe all stars more massive than early M dwarfs (and thus essentially all relevant stars) encountering the solar system over the past 10 Myr. For more recent timescales Gaia can observe even less massive objects. Moreover, the Gaia catalogue of more massive stellar encounters (stars with absolute magnitudes larger than that of the Sun) may shed light on the study of terrestrial craters over since the beginning of the Phanerozoic era, some 550\,Myr ago. Gaia can further improve the measurement of Sun's initial conditions and the potential of the Galaxy \citep{lindegren08,koposov09}. After including planetary perturbations, this would make the simulation of cometary orbits accurate enough to trace the stellar encounter back to the time when it generated comet showers and corresponding terrestrial craters \citep{rickman12}.

\newpage

\chapter{Obliquity or precession paces deglaciations over the last 2 million years: a Bayesian approach}\label{cha:climate}
\section{Chapter summary}\label{sec:abstract_climate}

Milankovitch proposed that the orbital eccentricity, precession and obliquity of the Earth can influence the climate by modulating the summer insolation at high latitudes in the Northern hemisphere. Despite a great success of the Milankovitch theory in explaining the climate change over the Pleistocene, it is inconclusive with regard to which combination of orbital elements drove or paced the sawtooth $\sim$100-kyr glacial-interglacial cycles over the late Pleistocene. To explore the roles of different orbital elements in pacing the Pleistocene deglaciations, we model the ice-volume variations over the Pleistocene by combining simple ice-volume models with different orbital elements. The Bayesian formalism allows us to compare these models and we find that obliquity plays a dominant role in pacing the glacial cycles over the whole Pleistocene while precession only becomes important in pacing major deglaciations after the transition from the 100-kyr dominant to the 41-kyr dominant glacial-interglacial cycles (mid-Pleistocene transition). Unlike the traditional Milankovitch theory that the insolation at the summer solstice at $65^\circ$N drives the climate change, our results confirm previous studies that the climate response to the insolation is interconnected in multiple spatial and temporal scales. We also conclude that the mid-Pleistocene transition was with a time scale of about 130\,kyr and the mid-point of the transition is about 700\,kyr. We find that the geomagnetic field and orbital inclination variations are unlikely to pace the Pleistocene deglaciations. Our results are consistent with Milankovitch's theory but also indicate a rather rapid change of the response of the climate system to the Northern hemisphere summer insolation during the mid-Pleistocene transition.

\section{Introduction}\label{sec:introduction_climate}

Roughly over the past 1\,Myr (the Pleistocene), the ice sheet gradually grew (glaciation) and abruptly retreated to relatively ice-free conditions (deglaciation or glacial termination) with an interval of $\sim$100-kyr. This semi-periodic glacial-interglacial cycles dominate terrestrial climate change, mainly documented by climate records in the deep sea sediments and ice cores. The glacial cycles are recorded by paleoclimatic proxies such as $\delta^{18}$O (i.e. a measure of $\delta^{18}{\rm O}/\delta^{16}{\rm O}$) of foraminiferal calcite, which is sensitive to global ice volume and ocean temperature. Milankovitch proposed that the climate change is driven by the summer insolation (i.e. incoming solar radiation) at $65^{\circ}$N, which is the so-called Milankovitch's theory \citep{milankovitch41}. Milankovitch also claimed that the climate change over different time scales are caused by climate responses to different orbital elements, including eccentricity, obliquity and precession. These orbital variations can potentially influence the climate and thus are one type of climate forcing, i.e. orbital or Milankovitch forcing. Many previous studies have confirmed Milankovitch's theory and the role of Milankovitch forcing in driving the Pleistocene climate change by spectral analyses of paleoclimatic time series derived from deep-sea sediments \citep{hays76, shackleton73,kominz79}. These studies have demonstrated that the climate variance is concentrated in spectral peaks at periods of around 19\,kyr, 23\,kyr, 42\,kyr and 100\,kyr which are close to periods of precession ($\sim$23 and 19\,kyr), obliquity ($\sim$41\,kyr) and precession ($\sim$100 and 400\,kyr). 

\begin{figure}[h!]
  \centering
  \includegraphics[scale=0.8]{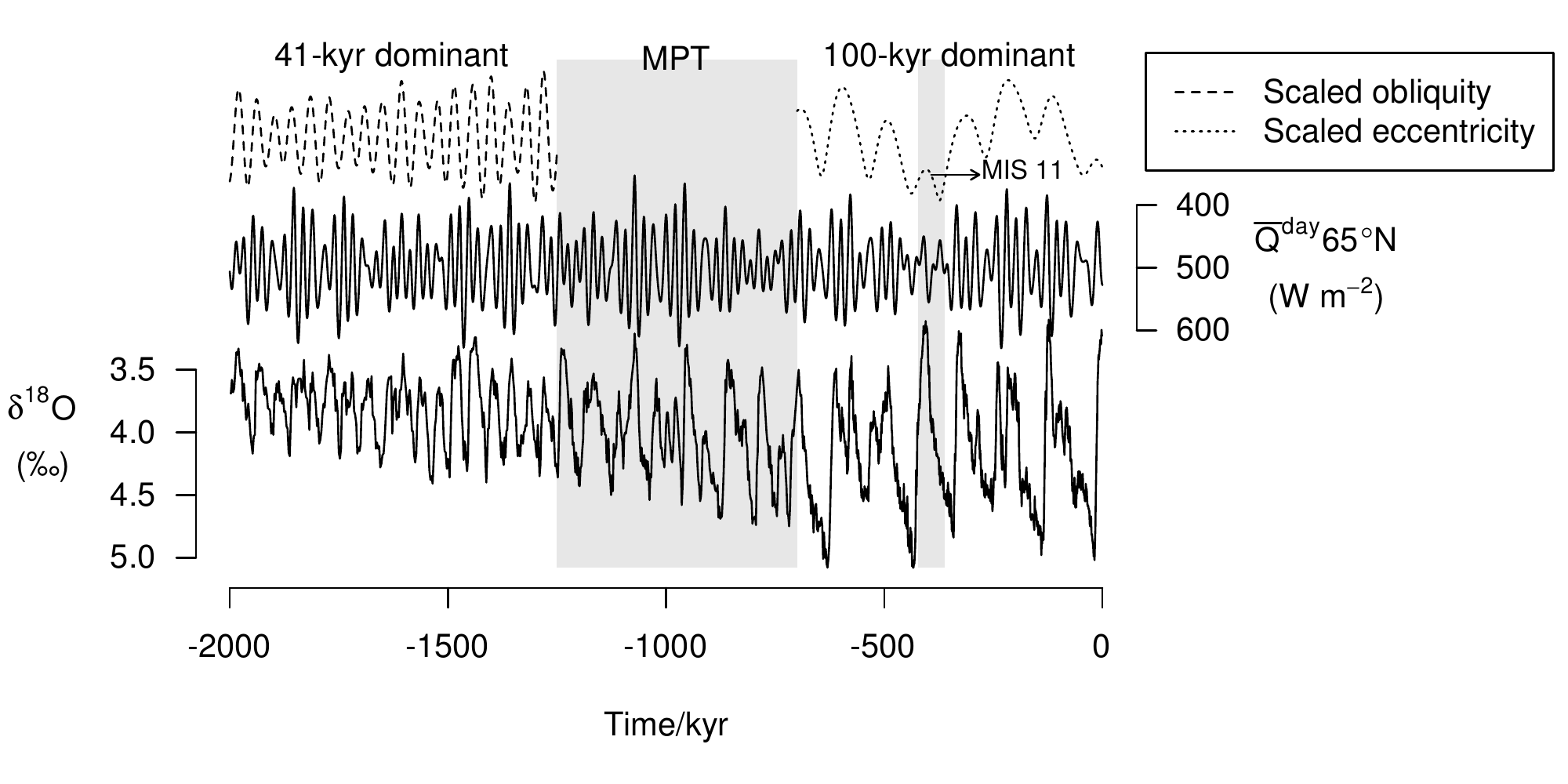}
  \caption{The $\delta^{18}$O record (lower solid line) stacked by \cite{lisiecki05} compared with the daily-averaged insolation at the summer solstice at $65^{\circ}N$, $\bar{Q}^{\rm day}65^{\circ}N$ (upper solid line), scaled obliquity (dashed line) and eccentricity (dotted line) calculated by \cite{laskar04} over the past 2\,Myr. The grey region at around 1\,Mya represents the MPT extending from 1.25\,Mya to 0.7\,Mya \citep{clark06}. The grey bar extending from 423 to 362\,kya (i.e. kyr ago) represent stage 11 or MIS 11. The $\delta^{18}$O variations are dominated by 41-kyr and 100-kyr cycles before and after the MPT, respectively. }
  \label{fig:milankovitch_diagram}
\end{figure}

However, there are several difficulties to reconcile theory and observations. In particular, there are mainly two types of difficulties in explaining the 100-kyr cycles based on Milankovitch's theory: the transition from the 41-kyr dominant to the 100-kyr dominant climate variations at the mid-Pleistocene (around 1\,Myr ago or 1\,Mya) and the difficulties in generating 100-kyr sawtooth variations with orbital forcings and climate response mechanisms (see \citealt{imbrie93}, \citealt{huybers07} and \citealt{lisiecki10} for details). On the one hand, as is shown in Fig. \ref{fig:milankovitch_diagram}, the onset of 100-kyr power at the mid-Pleistocene transition (MPT) is without a corresponding change in the summer insolation at high Northern latitudes (represented by the daily-averaged insolation at June 21 at $65^{\circ}$N). On the other hand, the $\sim$100-kyr eccentricity cycle only produces negligible 100-kyr power in seasonal or mean annual insolation variations despite its modulation of the precession amplitude. In addition, the 400-kyr cycles in eccentricity variations does not appear in paleoclimatic time series \citep{imbrie80}. Further, the eccentricity cycles and the 100-kyr climatic variations are anti-correlated, notably in marine isotope stage (MIS) 11 (see Fig. \ref{fig:milankovitch_diagram} and \citealt{imbrie80,howard97}). These problems are related to the 100-kyr cycles over the past one million years and thus are called ``100-kyr problem'', which is illustrated in Fig. \ref{fig:milankovitch_diagram}. 

Models with different climate forcings and response mechanisms are proposed to solve the ``100-kyr problem''. Most hypotheses are proposed based on either deterministic climate forcing or stochastic internal climate variation. The former hypothesis proposes that the 100-kyr cycles are attributed to being driven by orbital variations, particularly precession and eccentricity. This hypothesis has inspired the development of deterministic models which depend on orbital forcings \citep{imbrie80,paillard98,gildor00}. Many models treat the insolation variation as a pacemaker which sets the phase of the glacial-interglacial oscillation by directly controlling summer melting of ice sheets \citep{gildor00}. According to the latter hypothesis, stochastic internal climate variability plays a main role in generating the 100-kyr glacial cycles \citep{saltzman82,pelletier03,wunsch03}. Apart from these hypotheses, there are also other hypotheses that the glacial cycles resulted from the accretion of interplanetary dust when the Earth crosses the invariant plane \citep{muller97} or from the cosmic ray influx modulated by the geomagnetic paleointensity \citep{christl04,courtillot07}. Based on the above mentioned hypotheses, some models also explain the MPT with \citep{raymo97, paillard98,honisch09,clark06} or without \citep{saltzman93,huybers09,lisiecki10,imbrie11} an internal change in the climate system. For example, the erosion of a continental regolith allow larger ice to grow, with an attendant change in the climate response to the orbital forcing \citep{clark06}. 

The above models always consist of climate forcings and responses. According to current studies, climate forcings usually determine the time of occurrence of a certain climate feature such as deglaciations when the climate system reaches a threshold such as a maximum ice volume, which is the so-called pacing model. As is mentioned by \cite{huybers11}, dozens of pacing models are proposed but with a lack of means to choose among them. Our current work aims to compare different forcings based on a simple ice volume model for the Pleistocene glacial-interglacial cycles. We adopt the simple pacing model given by \cite{huybers05} and combine it with different forcings to predict the glacial terminations which are identified from different $\delta^{18}$O records. Unlike most conceptual models, these models do not try to describe the physical mechanisms of the climate response to external forcings. They aim instead to investigate the roles of different forcings in determining the time of deglaciations which are insensitive to dating uncertainties of $\delta^{18}$O records because of the large magnitude and abruptness of deglaciations \citep{huybers05}. These conceptual models are also called statistical models \citep{crucifix12}. 

Instead of using p-value to reject null hypotheses as in \cite{huybers05,huybers11}, we propose to compare all models/hypotheses on an equal footing in a Bayesian framework. The complexity of a model is properly taken into account by the prior of a model, and the Bayes factor is a metrics developed to compare models with different flexibilities (see \citealt{kass95,spiegelhalter02,bailer-jones09} for details). 

This chapter is organized as follows. We introduce the 100-kyr problem and our approach to compare different climate forcings in section \ref{sec:introduction_climate}. Then we choose different stacked $\delta^{18}O$ records and derive glacial terminations from them in section \ref{sec:data_climate}. In section \ref{sec:model_climate}, we build models based on orbital elements and geomanetic paleointensity (GPI) proxies to predict the Pleistocene glacial terminations. In section \ref{sec:comparison_climate}, these models are compared for different data sets and time scales. We perform a test of sensitivity of the results to model parameters and time scales in section \ref{sec:sensitivity_climate}. Finally, we discuss the results and conclude in section \ref{sec:conclusion_climate}. 

\section{Data}\label{sec:data_climate}

\subsection{$\delta^{18}$O with depth-derived age model}\label{sec:delta18O_climate}

The climate of the past can be reconstructed from proxies such as isotopes which are recorded in ice cores, deep-sea sediments, etc.. For example, air bubbles in ice cores are atmosphere samples from the past and can be analyzed for CO$_2$ concentrations. These are sensitive to the temperature of the atmosphere, so the history of Earth's surface temperature can be reconstructed from ice core records. The longest ice core can trace the climate history back to about 800\,kyr \citep{augustin04}. In order to reconstruct the climate change over the last 2\,Myr, the $\delta^{18}$O proxy recorded in foraminifera fossils (including species of benthos and plankton) in ocean sediment cores is used due to its sensitivity to the deep ocean temperature and ice volume in the past. During a glaciation period, the lighter isotope $^{16}$O evaporates from the ocean and is captured in ice sheets, leading to a high concentration of $^{18}$O in the oceans. This $^{18}$O concentration also depend on the salinity of seawater. Foraminifera in the oceans absorb more $^{18}$O into their skeletons when the water temperature is lower and the concentration of $^{18}$O is higher in the water.


To calibrate $\delta^{18}$O measurements and assign ages to depths (or age model) of sediment cores, researchers either assume a constant globally averaged sedimentation rate or a constant phase relationship between $\delta^{18}$O measurements and an insolation forcing based on Milankovitch theory (see \citealt{huybers04} for details). The former is called depth-derived age model \citep{huybers04, huybers07}, which is always used to test Milankovitch theory \citep{huybers05,lisiecki10,huybers11}. On the contrary, the latter calibration method is named ``orbital tuning'' \citep{imbrie84,martinson87,shackleton90}. However this method is not appropriate for testing theories related to Milankovitch forcings because it intrinsically assumes a link between $\delta^{18}$O variations and orbital forcings.


\cite{huybers07} (hereafter H07) have stacked and averaged twelve benthic (benthos related) and five planktic (plankton related) $\delta^{18}$O records to generate three $\delta^{18}$O global records: average of all $\delta^{18}$O records (``HA'' data set); average of the benthic records (``HB'' data set) and average of the planktic records (``HP'' data set)\footnote{The planktic $\delta^{18}$O records may not produce a stack as good as benthic records because surface water is less uniform in temperature and salinity than the deep ocean \citep{lisiecki05}. }. Apart from these three data sets, we also use the orbital-tuned benthic $\delta^{18}O$ stacked by \cite{lisiecki05} (defined as ``LR04'' data set) despite its orbital assumptions. In addition, the LR04 record is re-calibrated by H07 to generate a tuning-independent LR04 data set (defined as ``LRH'' data set and refer to the supplementary material of H07 for details).

\begin{figure}[h!]
  \centering
  \includegraphics[scale=0.8]{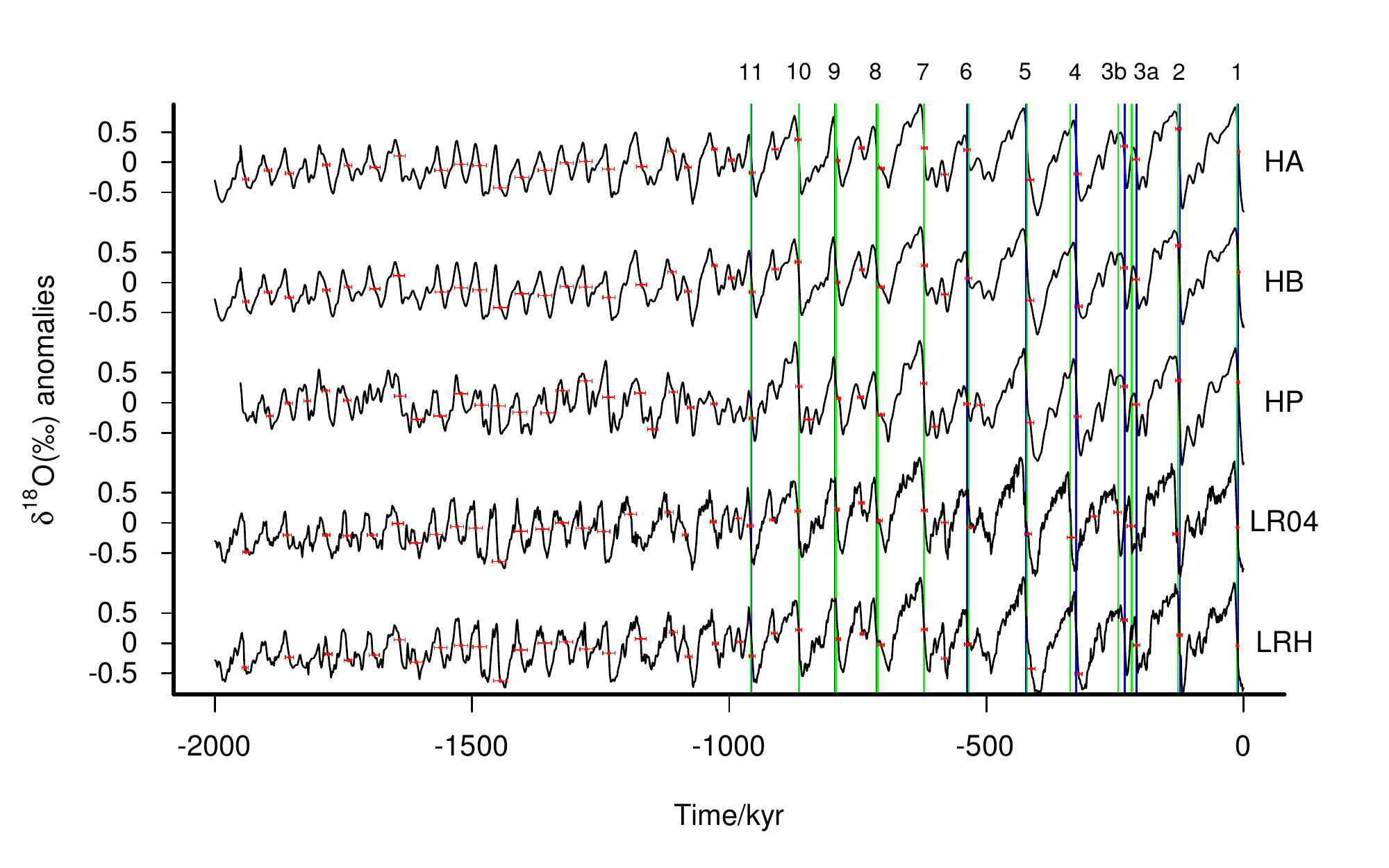}
  \caption{The $\delta^{18}$O anomalies with ages determined by extended depth-derived age-model (HA, HB, HP and LRH) and orbital-tuning model (LR04). The past 2000\,kyr, i.e. Pleistocene, is divided into two parts: the early Pleistocene extending from 2\,Mya to 1\,Mya and the late Pleistocene extending from 1\,Mya to the present. The deglaciations for each data set are identified, and the timing uncertainties are shown in red error bars. The DD terminations are denoted by blue lines while the ML/MS terminations are denoted by green lines. The 11 terminations are denoted by numbers and termination 3 is split into two events. }
  \label{fig:huybers_data}
\end{figure}

All the above $\delta^{18}$O records over the past 2\,Myr are normalized to the mean and unit variance (i.e. $\delta^{18}$O anomalies) and shown in Figure \ref{fig:huybers_data}. The terminations shown in this figure will be identified from $\delta^{18}$O records in the following section. We can see that the sawtooth 100-kyr glacial-interglacial cycles become significant over the late Pleistocene while 41-kyr cycles dominate the climate change over the early Pleistocene. All records show gradual glaciations and abrupt deglaciations over the late Pleistocene. Hereafter, in the context without mentioning the MPT, the late Pleistocene means a period ranging from 1\,Mya to 0\,Mya, and the early Pleistocene means a period ranging from 2\,Mya to 1\,Mya. 

\subsection{Identification}\label{sec:identification_climate}

Recent paleoclimatic studies has gradually achieved a consensus on the main role of orbital forcing in pacing the 100-kyr glacial-interglacial cycles which are mainly generated by internal climate variability. In other word, orbital forcing determines the time of deglaciations when the internal climate system reaches a certain threshold. Because our goal is to select out the true pacemaker rather than the true climate response mechanism, we will not build a complex model to predict all variations in the $\delta^{18}$O records. Thus we only identify the {\it deglaciations} or {\it glacial terminations} from the full $\delta^{18}$O times series and use them to test our models. 

Following H07, a deglaciation is identified when a local maximum and the following minimum (defined as a maximum-minimum pair) have a difference in $\delta^{18}$O larger than one standard deviation of the whole $\delta^{18}$O record. The time and time uncertainty of a deglaciation is the mid-point of the maximum-minimum pair and the age uncertainty of this mid-point calculated by running a stochastic sediment accumulation rate model \citep{huybers07}. In order to identify sustained events in all data sets, different $\delta^{18}$O records are filtered with different moving-average (or ``Hamming'') filters to remove the potential noise in records. The data sets of HA and HB are filtered with a 7\,kyr filter, HP and LRH are filtered with a 11\,kyr filter, and LR04 is filtered with a 9\,kyr filter, respectively. As a result, we identify 20 deglaciations during the early Pleistocene, and 16 deglaciations during the late Pleistocene for HA, 20 and 16 for HB, 21 and 16 for HP, 20 and 17 for LR04, 20 and 16 for LRH. The age uncertainties of these deglaciations are denoted by error bars in Fig. \ref{fig:huybers_data}. 

However, H07's method identifies extra terminations apart from the 11 established late-Pleistocene terminations which are characterized by a rapid and abrupt shift from extreme glacial to extreme interglacial conditions \cite{broecker84,raymo97b}. This may be caused by applying a single filter to identify deglaciations over both the early and late Pleistocene, which may have rather different termination features.
To distinguish the deglaciations identified using H07's method and these 11 terminations, we define {\it major terminations} as the 11 deglaciations which are frequently studied in the literature. In contrast, the terminations identified using H07's method contain many terminations do not have the feature of major terminations and thus are defined as {\it minor terminations}. Among the major terminations, termination 3 is usually split into two events: 3a and 3b, and thus we actually use 12 events for our model comparison (see Fig. \ref{fig:huybers_data}). 

The time of the major terminations from different papers in the literature are collected by \cite{huybers11} and are given in his supplementary material. Based on his Table S2, we define another three data sets of terminations:
\begin{itemize}
\item DD: termination times and corresponding uncertainties estimated from the depth-derived timescale in H07,
\item MS: each termination with time and time uncertainty respectively equal to the median and standard deviation of different termination times for each event given in the literature,
\item ML: termination times the same as those in the MS data set but with larger uncertainties by adding the time uncertainties of the depth-derived time scales in quadrature with the corresponding uncertainties in the MS data set.
\end{itemize}
Because MS and ML have same termination times, we only show the termination times of DD and MS with vertical lines in Figure \ref{fig:huybers_data}. 

Finally, we define three hybrid data sets particularly for climate models which predict the climate change over the last 2\,Myr and the MPT. Considering that the HA data set is a stack of both benthic and planktic records, we combine the early-Pleistocene deglaciation events identified from the HA data set and late-Pleistocene terminations from the DD, ML and MS data sets to generate HADD, HAML, HAMS data sets, respectively. The time and time uncertainties of all terminations identified from various $\delta^{18}$O records are listed in Table \ref{tab:terminations}.

\begin{table}[ht!]
  \caption{Deglaciations identified from different $\delta^{18}$O records using H07's method (HA, HB, HP, LR04 and LRH) and deglaciations extensively studied in the literature (DD, MS and ML). Combining the early Pleistocene deglaciations of HA with the DD, MS and ML data sets, we obtain the hybrid data sets of HADD, HAMS and HAML. For each column, the deglaciation ages are listed on the left side and the age uncertainties are listed on the right side. All quantities are in unit of kyr. }
  \label{tab:terminations}
  \hspace{-0.3in}
  \footnotesize{
    \begin{tabular}{|cc|cc|cc|cc|cc|cc|cc|cc|}
      \hline
    \multicolumn{2}{|c|}{HA}&\multicolumn{2}{c|}{HB}&\multicolumn{2}{c|}{HP}&\multicolumn{2}{c|}{LR04}&\multicolumn{2}{c|}{LRH}&\multicolumn{2}{c|}{DD}&\multicolumn{2}{c|}{MS}&\multicolumn{2}{c|}{ML}\\\hline
    -10&0.81&-10&0.81&-11&1.9&-12&2.2&-12&2.2&-11&1.9&-13&1.8&-13&3.1\\
    -127&5.3&-127&5.3&-127&5.3&-131&6.3&-125&5&-124&5&-128&3.6&-128&6.6\\
    -209&6.6&-209&6.6&-209&6.6&-219&7.5&-208&6.4&-208&6.4&-218&4.3&-218&8.7\\
    -233&6.4&-233&6.4&-233&6.4&-245&7&-233&6.4&-231&6.3&-244&4.8&-244&8.6\\
    -323&6.8&-321&7&-323&6.8&-290&7.5&-321&7&-326&7&-337&4.5&-337&9.8\\
    -415&7.4&-415&7.4&-415&7.4&-335&8.4&-413&7.6&-423&7.1&-421&4.4&-421&8.2\\
    -537&6.5&-535&6.6&-537&6.5&-531&7.3&-581&6.9&-622&5.8&-621&2.7&-621&6.4\\
    -581&6.9&-581&6.9&-537&6.5&-531&7.3&-581&6.9&-622&5.8&-621&2.7&-621&6.4\\
    -621&5.8&-621&5.8&-601&6.4&-581&6.9&-621&5.8&-714&4.5&-712&7.5&-712&8.8\\
    -705&5.9&-705&5.9&-622&5.8&-621&5.8&-705&5.9&-794&3.7&-793&1.8&-793&1.8\\
    -743&5&-742&4.8&-705&5.9&-708&5.4&-741&4.5&-864&5.7&-864&0.84&-864&5.8\\
    -789&4.2&-789&4.2&-745&5.5&-743&5&-788&4.2&-957&5.8&-958&1.7&-958&6.0\\
    -866&5.8&-866&5.8&-787&4.1&-791&4.1&-865&5.7&&&&&&\\
    -911&6&-911&6&-845&8&-867&5.7&-912&6&&&&&&\\
    -955&5.9&-955&5.9&-865&5.7&-915&5.9&-955&5.9&&&&&&\\
    -996&5.5&-996&5.5&-955&5.9&-959&5.7&-978&7&&&&&&\\
    -1029&5.6&-1029&5.6&-1030&5.6&-983&6.5&-1027&5.5&&&&&&\\
    -1080&6.6&-1080&6.6&-1075&6.1&-1031&5.5&-1079&6.5&&&&&&\\
    -1111&8.1&-1111&8.1&-1109&8&-1085&6.5&-1109&8&&&&&&\\
    -1170&10.4&-1171&10.5&-1149&9.9&-1117&8&-1172&10.5&&&&&&\\
    -1235&11.7&-1234&11.7&-1173&10.5&-1192&11.4&-1234&11.7&&&&&&\\
    -1279&12.3&-1279&12.3&-1235&11.7&-1244&12&-1278&12.3&&&&&&\\
    -1316&12.9&-1316&12.9&-1279&12.3&-1285&12.3&-1317&13&&&&&&\\
    -1358&13.2&-1358&13.2&-1324&12.7&-1325&12.7&-1359&13.2&&&&&&\\
    -1403&13.3&-1403&13.3&-1353&13&-1363&13.1&-1405&13.2&&&&&&\\
    -1445&13.4&-1445&13.4&-1407&13.2&-1405&13.2&-1445&13.4&&&&&&\\
    -1485&13.2&-1485&13.2&-1449&13.2&-1447&13.3&-1485&13.2&&&&&&\\
    -1521&12.9&-1521&12.9&-1481&13.1&-1493&12.9&-1521&12.9&&&&&&\\
    -1560&12.9&-1559&12.4&-1521&12.9&-1529&12.5&-1561&12.3&&&&&&\\
    -1641&10.8&-1642&10.8&-1562&12.3&-1569&12&-1608&11.5&&&&&&\\
    -1688&9.8&-1689&9.8&-1607&11.5&-1609&11.5&-1641&10.8&&&&&&\\
    -1741&7.4&-1741&7.4&-1640&10.8&-1644&10.7&-1690&9.7&&&&&&\\
    -1783&6.9&-1783&6.9&-1742&7.4&-1694&9.4&-1741&7.4&&&&&&\\
    -1855&7.7&-1855&7.7&-1784&7&-1743&7.3&-1855&7.7&&&&&&\\
    -1897&7.3&-1897&7.3&-1820&6.9&-1783&6.9&-1855&7.7&&&&&&\\
    -1940&5.8&-1940&5.8&-1856&7.7&-1859&7.6&-1941&5.9&&&&&&\\
    &&&&-1893&7.1&-1940&5.8&&&&&&&&\\
    \hline
  \end{tabular}}
\end{table}

Because there are dating errors and identification uncertainties, we don't know exactly when a deglaciation occurs. To take into account these uncertainties, we treat the time of each deglaciation probabilistically by generating a Gaussian distribution with the mean and standard deviation equal to the time and time uncertainty of the termination respectively. Thus the terminations identified from a $\delta^{18}$O record are actually a discrete data set, which is described within a Bayesian framework in section \ref{sec:discrete}, and also in section \ref{sec:datadiscrete_biodiversity} and \cite{feng13}. 

\section{Models}\label{sec:model_climate}

The ``sawtooth'' variations of the global ice volume (also present in other climate proxies) and the significant 100-kyr cycles over the late Pleistocene requires a non-linear response of the climate system to Milankovitch forcings which include precession, obliquity and eccentricity. This can be modeled by simple conceptual models which combine different feedback mechanisms such as ice-albedo feedback \citep{tziperman03}, or CO$_2$ feedback \citep{saltzman90}. Another modeling approach is to construct a differential model of ice volume with some parameters changing with some thresholds of the ice volume \citep{gildor00,tziperman03,ashkenazy04} or parameters changing with the Milankovitch forcing \citep{paillard98,parrenin03}. All these models can explain the sawtooth structure and even MPT to various degrees of success through transitions between different equilibria or bifurcations generated from the non-linear climate system.

\begin{figure}
  \centering
  \includegraphics[scale=0.6]{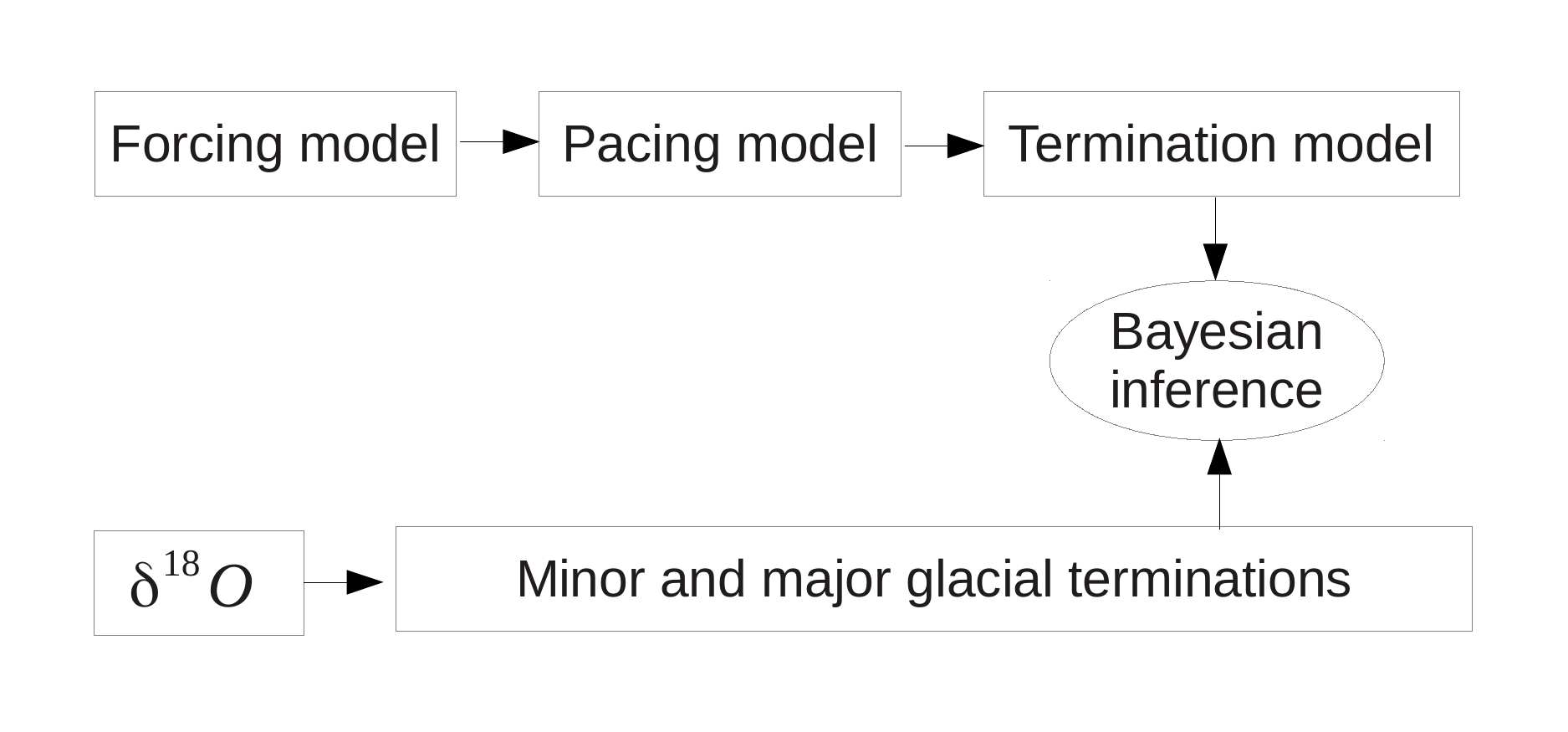}
  \caption{A schematic description of the climate modeling and model inference process.}
  \label{fig:climate_model}
\end{figure}

However, the models differ in their interpretation of non-linear climate response to forcings. For instance, they assume a dependence or independence on Milankovitch forcing, either the summer insolation at 65$^\circ$ N (i.e. classical Milankovitch theory) or combinations of orbital elements, over the Pleistocene. In this present work, we assess these assumptions as \cite{huybers05} and \cite{huybers11} did, but within a Bayesian framework. Therefore, we (1) introduce different climate forcings (section \ref{sec:forcing_climate}); (2) build pacing models with thresholds depending on different forcings (section \ref{sec:pacing_climate}); (3) identify glacial terminations from these pacing models and compare them with different data sets of deglaciations (section \ref{sec:termination_climate}). This procedure is illustrated by Fig. \ref{fig:climate_model}. In sum, we probabilistically compare termination predictions from various models that in particular include forcing and pacing prescriptions.  We develop these prescriptions below. 

\subsection{Forcing models}\label{sec:forcing_climate}

The solar insolation influences the climate by heating the lower atmosphere, changing the ice volume through modifying the ice accumulation rate, and modulating the CO$_2$ concentration in the atmosphere, and by altering the rate of CO$_2$ dissolution in the ocean \citep{saltzman90}. Studies claim that these climate changes are more sensitive to the Northern summer insolation at high latitudes because the temperature in continental areas is critical for ice melting or sublimation in the Northern Hemisphere \citep{milankovitch41}. The summer insolation at high latitudes depends on the geometry of the Earth's orbit and the inclination of Earth's spin axis, and thus depends on eccentricity, precession and obliquity.

However, the climate response to these three semi-periodic orbital variables has different time scales, and insolation variations at different latitudes and seasons vary differently with respect to these orbital elements. It is therefore necessary to combine these orbital variables to form a compound forcing model \citep{imbrie80,huybers11,crucifix13}.
The forcing models based on normalized time-varying eccentricity, $f_{\rm E}(t)$, precession, $f_{\rm P}(t)$, obliquity, $f_{\rm T}(t)$, and combinations thereof are described as follows:
\begin{equation}
  \begin{array}{l}

    \displaystyle f_{\rm E}(t) = e(t)\\
    \displaystyle f_{\rm P}(t) = e(t) \sin(\omega(t)-\phi)\\
    \displaystyle f_{\rm T}(t) = \epsilon(t)\\
    \displaystyle f_{\rm EP}(t) = \alpha^{1/2}f_{\rm E}(t) + (1-\alpha)^{1/2}f_{\rm P}(t)\\
    \displaystyle f_{\rm ET}(t) = \alpha^{1/2}f_{\rm E}(t) + (1-\alpha)^{1/2}f_{\rm T}(t)\\
    \displaystyle f_{\rm PT}(t) = \alpha^{1/2}f_{\rm P}(t) + (1-\alpha)^{1/2}f_{\rm T}(t)\\
    \displaystyle f_{\rm EPT}(t) = \alpha^{1/2}f_{\rm E}(t) + \beta^{1/2}f_{\rm P}(t) + (1-\alpha-\beta)^{1/2}f_{\rm T}(t),
    \label{eqn:ts_function}
  \end{array}
\end{equation}

where $e(t)$, $\epsilon(t)$ and $e(t)\sin(\omega(t)-\phi)$ are the time-varying eccentricity, obliquity and precession index, respectively. The precession index is also called climatic precession because it directly relates to insolation. In the precession index, $\omega(t)$ is the angle between perihelion and the moving vernal equinox, and $\phi$ is a free parameter controlling the phase of the precession. We adopt the variations of these three orbital elements over the past 2\,Myr as calculated by \cite{laskar04}.
Finally, all variables in compound models are normalized to zero-mean and unit variance before combination. Therefore, $\alpha$ and $\beta$ are contribution factors, they determine the relative contribution of each component in the compound models, where $0\leq\alpha\leq 1$ and $0\leq\beta\leq 1$. In addition to these models, we also consider the classical Milankovitch theory (i.e. daily-averaged insolation at 65$^\circ$N on July 21) as
\begin{equation}
  f_{\rm Ins}(t) = \bar{Q}_{\rm day}(e(t), e(t)sin(\omega(t)-\phi),\epsilon(t)).
  \label{eqn:insolation}
\end{equation}
$\bar{Q}_{\rm day}$ is calculated from the values of orbital elements given by \cite{laskar04}. 

Although conceptual models based on Milankovitch forcing achieve a great success in explaining the precession and obliquity related cycles in the climate change over the Pleistocene \citep{hays76,imbrie84,imbrie93}, the 100-kyr problem (see section \ref{sec:introduction_climate}) has motivated scientists to propose other climate forcings, such as cosmic rays \citep{svensmark97,kirkby04}, Earth's orbital inclination with respect to the invariant plane \citep{muller97}, the geomagnetic field \citep{courtillot07,knudsen08} and solar activity \citep{sharma02}. Therefore we also consider 2 models based on the variations of Earth's orbital inclination and geomagnetic field paleointensity (GPI). However, we ignore the cosmic ray forcing and solar activity forcing because the history of cosmic ray influx and the solar activity cannot be accurately reconstructed from the concentrations of cosmogenic isotopes such as $^{10}$Be over a time scale longer than 1\,Myr \citep{bard06}.

Although the orbital inclination relative to the invariant plane also has a $\sim$100-kyr variance, it is not included into Milankovitch theory because it cannot directly change the insolation at the top of Earth's atmosphere. However, \cite{muller97} proposed that the insolation at the Earth's surface can be modulated by an increase in the amount of interplanetary dust or meteoroids when the Earth crosses the invariant plane. To test this hypothesis, we build a inclination-based forcing model, $f_{\rm Inc}(t)$, using the orbital inclination calculated by \cite{muller97}. 

The geomagnetic field on the Earth can influence the climate through changing cosmic-ray induced nucleation of clouds \citep{courtillot07}. We model this forcing as a geomagnetic paleointensity (GPI) time series normalized to the mean and unit variance, $f_{\rm G}(t)$, which is collected by \cite{channell09}. 

All forcing models and corresponding prior distributions over their parameters (defined as ``forcing parameters'') are shown in Table \ref{tab:ts_models}. In Table \ref{tab:ts_models} and following sections, all parameters are treated as dimensionless variables after setting the time unit as 1\,kyr and the ice volume unit as an unspecified volume of ice in unit of ${\rm km}^3$. For the precession model, we set $\phi=0$ to treat precession according to the classical Milankovitch theory, i.e. a high summer insolation in the Northern Hemisphere tend to trigger a deglaciation. Because we don't have any prior knowledge about the value of contribution factors, compound models have uniform prior distributions over the interval of $[0,1]$ for these contribution factors. Following \cite{huybers05} and \cite{huybers11}, we adopt positive contribution factors because eccentricity, precession and tilt contribute positively to the daily average insolation at summer solstice. However, we will test this by reversing the sign of the variations of these orbital elements and assigning time lags to different forcing models in section \ref{sec:sensitivity_climate}.

\begin{figure}[h]
  \centering
  \includegraphics[scale=0.8]{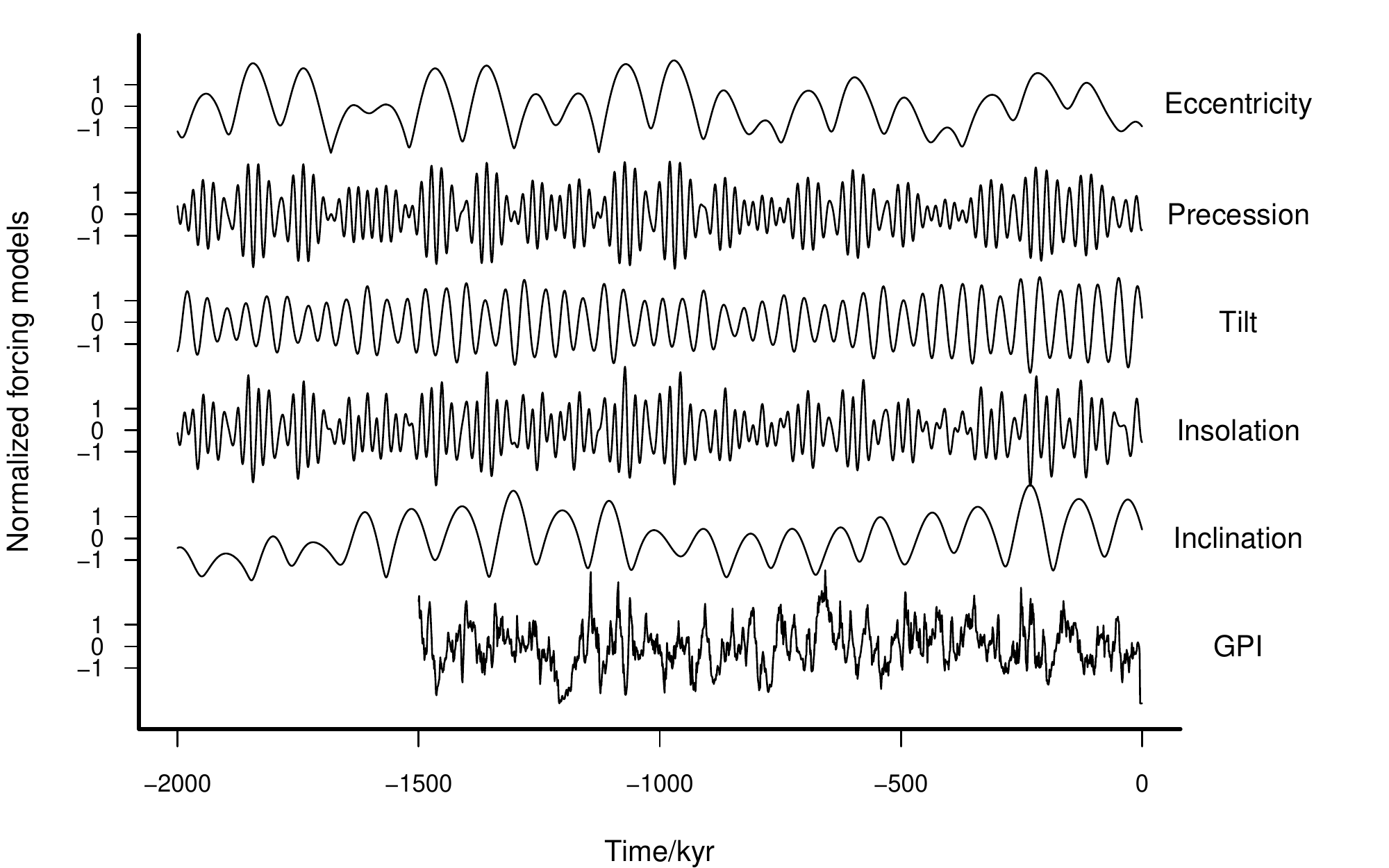}
  \caption{Normalized single-component forcing models. For each model, a deglaciation is likely to be triggered by a forcing peak. The values of eccentricity, precession and obliquity are simulated by \cite{laskar04}, the orbital inclination relative to the invariant plane is from \cite{muller97}, and the GPI record is from \cite{channell09}.}
  \label{fig:forcing_models}
  \end{figure}

Fig. \ref{fig:forcing_models} shows the single-component forcing models normalized to the mean and unit variance \footnote{These forcing models don't have any adjustable forcing parameter. Despite being a optimized combination of eccentricity, precession and obliquity, the Inclination model also contains a single component.}. All forcing models will be included in pacing models and corresponding termination models in the following sections. Hereafter, for each forcing model, the corresponding pacing and termination models share the same name which is shown in the first column of Table \ref{tab:ts_models}. 

\begin{table}[h]
  \begin{adjustwidth}{-1.2in}{-1.2in}
  \centering
  \caption{The termination models and corresponding forcing models and the uniform prior distributions over forcing parameters. Common priors are prior distributions of parameters in pacing and termination models, which are given in section \ref{sec:pacing_climate} and \ref{sec:termination_climate}.}
  \label{tab:ts_models}
  \scalebox{0.8}{\begin{tabular}{llll}
  \hline
  \hline
  Termination models & Description &Forcing models & Uniform prior distribution\\
  \hline
    Periodic &100-kyr pure periodic model & None & common priors\\
    Eccentricity &Eccentricity& $f_{\rm E}(t)$ & common priors\\
    Precession &Precession& $f_{\rm P}(t)$ & common priors and $\phi=0$\\
    Tilt &Tilt or obliquity& $f_{\rm T}(t)$ & common priors\\
    EP &Eccentricity plus Precession& $f_{\rm EP}(t)$ &common priors and $0\leq \alpha\leq 1$, $\phi={\rm 0}$\\
    ET &Eccentricity plus Tilt& $f_{\rm ET}(t)$ &common priors and $0\leq \alpha\leq 1$\\
    PT &Precession plus Tilt& $f_{\rm PT}(t)$ &common priors and $0\leq \alpha\leq 1$, $\phi=0$\\
    EPT &Eccentricity plus Precession plus Tilt&$f_{\rm EPT}(t)$&common priors and $0\leq \alpha, \beta\leq 1$ and $\alpha+\beta\leq 1$, $\phi=0$\\
    Insolation &Insolation& $f_{\rm Ins}(t)$ &common priors\\
    Inclination &Inclination& $f_{\rm Inc}(t)$ &common priors\\
    GPI &Geomagnetic paleointensity& $f_{\rm G}(t)$ &common priors\\
    \hline
  \end{tabular}}
\end{adjustwidth}
\end{table}

\subsection{Pacing models}\label{sec:pacing_climate}

\begin{figure}[ht!]
  \centering
  \includegraphics[scale=0.9]{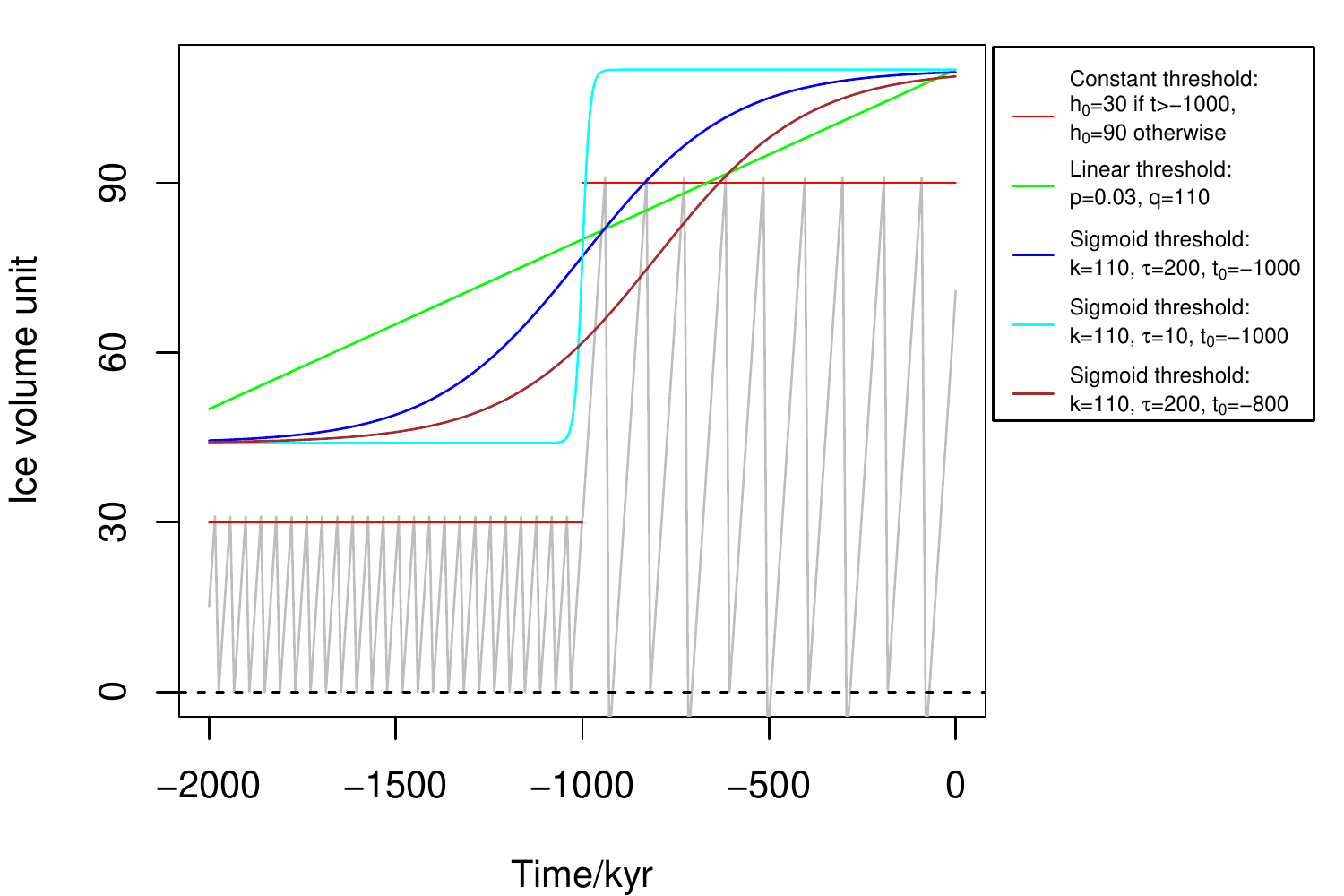}
  \caption{Examples of background thresholds, $h_0$: constant (red), linear (green) and sigmoid thresholds (other colors). The legend shows the values of parameters of linear and sigmoid background thresholds which are expressed in equation \ref{eqn:linear} and equation \ref{eqn:sigmoid}, respectively. The Periodic model, i.e. a pacing model with a constant threshold, with $h_0=30$ over the early Pleistocene and $h_0=90$ over the late Pleistocene are shown by grey lines. }
  \label{fig:icevolume}
\end{figure}

As is mentioned in section \ref{sec:introduction_climate}, pacing means that the variation of a climate system is independent with external forcings until the climate system reaches a threshold which is modulated by climate forcing. We model this pacing effect on ice volume variations using the deterministic model adapted from the stochastic model introduced by \cite{huybers05}. The model of ice volume with time $t$ is
\begin{equation}
  v(t)=v(t-1)+\eta(t) \quad \quad \text{and if } v(t)>h(t) \text{ then terminate},
\label{eqn:deterministic}
\end{equation}
with
\begin{equation}
  h(t)=h_0-af(t),
  \label{eqn:threshold}
\end{equation}
where $v(t)$ is the ice volume which increases by a value of $\eta(t)$ until it passes a threshold $h(t)$ which is modulated by a climate forcing $f(t)$ with a contribution factor, $a$. The initial ice volume is $v_0$, and the background threshold, $h_0$, is either a constant or a trend varying with time. In the stochastic model of ice volume defined by \cite{huybers05}, $\eta(t)$ is a random length drawn from a Gaussian distribution with mean and standard deviation equal to 1 and 2, respectively. We modify this here to build a deterministic model. The increment $\eta(t)$ is 1 ice volume unit, when the threshold is not passed; otherwise, the increment has such a value that the ice volume can linearly decrease to 0 over 10\,kyr\footnote{Note that the ice volume sometimes goes below 0 slightly due to numerical errors.}. If the contribution factor $a=0$, the ice volume will vary with a period modulated by the background threshold, $h_0$. We define this model as the Periodic model. If $h_0$ is a constant, this model predicts a pure periodic glacial-interglacial cycles with a single period of $h_0+10$\,kyr. We further explain this using a realization of the Periodic model shown by Fig. \ref{fig:icevolume}. We see that the Period model with $h_0=30$ and $h_0=90$ can well predict the $\sim$41\,kyr and $\sim$100\,kyr saw-tooth cycles over the early and late Pleistocene, respectively. 

Because the background threshold, $h_0$, controls the period of ice volume variations, different values of $h_0$ are required to model the 100\,kyr-dominated glacial cycles over the late Pleistocene and the 40\,kyr-dominated glacial cycles over the early Pleistocene. Considering this complexity in the climate change, we will first build pacing models to separately predict the deglaciations over the two Pleistocene periods, and then predict all deglaciations over the Pleistocene. We use a constant background threshold for the former models and a trend background threshold for the latter models (examples are shown in Fig. \ref{fig:icevolume}). 

\subsubsection{Threshold with a constant background}\label{sec:pacing_constant_climate}

\begin{figure}
  \centering
  \includegraphics[scale=0.6]{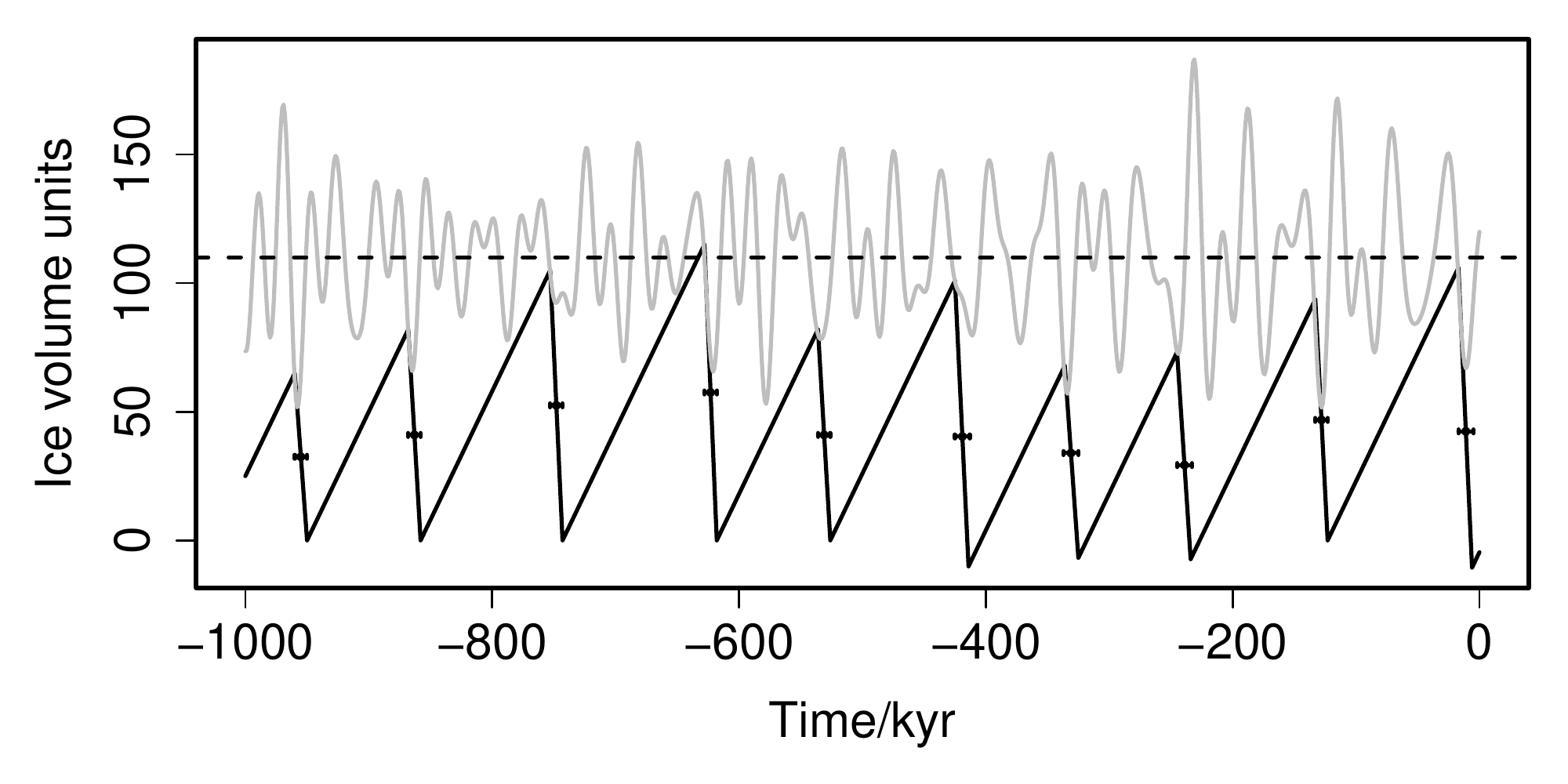}
  \caption{A pacing model with threshold $h(t)$ modulated by the PT forcing model with $\alpha=0.5$ and $\phi=0$, i.e. $f(t)=f_{\rm PT}(t; \alpha=0.5,\phi=0)$. This model has parameters as follows: the background threshold, $h_0=90$, the initial ice volume, $v_0=25$ and the contribution factor of forcing, $a=25$. The dashed line denote the constant threshold, and the grey line represents the threshold modulated by the Insolation forcing model, i.e. $h(t)=h_0-af_{\rm PT}(t;\alpha=0.5,\phi=0)$. Terminations are identified from this pacing model, and are shown with error bars. }
  \label{fig:pacing_model}
\end{figure}

The pacing model with a constant background threshold is appropriate for modeling glacial-interglacial cycles without a transition such as the MPT. 
A realization of this pacing model with threshold modulated by a PT forcing model is shown in Fig. \ref{fig:pacing_model}. The ice volume grows until it passes the forcing-modulated threshold. After it passes the threshold, the ice volume decreases rapidly to zero within the next 10\,kyr. We see that a deglaciation tends to occur when the insolation is around a peak. From this example, we observe that the pacing model expressed in equations \ref{eqn:deterministic} and \ref{eqn:threshold} can generate $\sim$100\,kyr saw-tooth cycles and allow climate forcing to pace the phase of these cycles, enabling us to test which forcing determines the exact time of terminations. This pacing model has many free parameters that have prior distributions described below.

The predicted time of glaciation termination is determined by the initial ice volume, $v_0$, and the threshold, $h(t)$. A pure periodic pacing model is generated by adopting a constant threshold, $h(t)=h_0$. Other pacing models are generated by modulating the threshold by different forcing models (see Fig. \ref{fig:forcing_models}). These pacing models have common pacing parameters: initial ice volume, $v_0$, threshold constant, $h_0$, and contribution factor of forcing, $a$.

As is shown in Fig. 4, the threshold background, $h_0$, control the period of ice volume variations. When forcings are added onto the constant threshold, the average period of ice volume variations would decrease by an amount of $\sim a$ ice volume units, due to the effect that the ice volume accumulation tend to terminate at forcing maxima. To be specific, the average period of ice volume variation is $\sim (h_0+10-a)$\,kyr. To model the 41-kyr cycles and 100-kyr cycles separately, we define different prior distributions of $h_0$ using a parameter, $\gamma$, for the early and late Pleistocene.

For pacing models with a constant background threshold, we define uniform prior distributions of $v_0$, $h_0$ and $a$ over the following intervals: $0<v_0<90\gamma$, $90\gamma<h_0<130\gamma$ and $15\gamma<a<35\gamma$, where $\gamma=0.4$ when we model $\sim$41-kyr cycles and $\gamma=1$ when we model $\sim$100-kyr cycles. The range of $v_0$ is just the range of the ice volume variation while the mean values of the prior distributions of $h_0$ and $a$ with $\gamma=1$ are the fitted values obtained by \cite{huybers11}. For the periodic model, $a$ is zero and $h_0$ has a uniform prior distribution over $70\gamma <h_0< 110\gamma$. In section \ref{sec:sensitivity_climate}, we will check whether our results are sensitive to our priors by varying our assumptions. 


\subsubsection{Threshold with a linear trend}\label{sec:pacing_linear_climate}

As we have mentioned, the above pacing model with a constant background threshold is not capable to model the transition from the 41\,kyr world to the 100\,kyr world. If we treat $h_0$ as a step function as is shown in Fig. \ref{fig:icevolume} by red lines, the corresponding pacing model does predict an abrupt MPT but with an additional parameter. To predict the MPT and the whole Pleistocene climate change, we will introduce another two versions of the pacing model by modeling the background threshold as a trend varying time. We will not introduce new names because each pacing model with a specific forcing has shared the same name with the relevant forcing model. But we will mention different versions by specifying the changes we have made in the background threshold, $h_0$.

Many studies have suggested various mechanisms which may be involved in climate change before and after the MPT (about 0.8$\sim$1\,Mya) \citep{saltzman84,maasch90,ghil94,raymo97,paillard98,clark99,tziperman03,ashkenazy04}. However, H07 suggests that a simple model with a threshold modulated by obliquity and a linear trend can explain changes in glacial variability over the last 2\,Myr without invoking complex mechanisms. To investigate this scenario and assess the role of different orbital elements in triggering the MPT, we build one more version of pacing models (equation \ref{eqn:deterministic}) in which we replace the threshold constant $h_0$ with a linear trend with time, i.e.
\begin{equation}
  h_0=pt+q,
  \label{eqn:linear}
\end{equation}
where $p$ and $q$ are the slope and intercept of the trend respectively. The prior distributions of the pacing parameters in this new model are uniform over the following intervals: $0<v_0<36$, $0<p<0.1$, $106<q<146$ and $10<a<30$. For the periodic model, two prior distributions of pacing parameters are changed: $a=0$ and $86<q<126$.

An example of the linear trend is shown by the green line in Fig. \ref{fig:icevolume}. We see that the linear trend starting from 50 ice volume units and linearly increases to 110 ice volume units. If the threshold in the pacing model is not modulated by forcing, i.e. $h(t)=h_0$, the pacing model will predict a gradual transition from 50\,kyr cycles 2\,Mya to 110\,kyr cycles at the present. 

\subsubsection{Threshold with a sigmoid trend}\label{sec:pacing_sigmoid_climate}

To enable the occurrence of an abrupt or rapid MPT, we introduce another version of pacing models with a sigmoid trend which is
\begin{equation}
  h_0=0.6k/(1+e^{-(t-t_0)/\tau})+0.4k,
  \label{eqn:sigmoid}
\end{equation}
where $k$ is a scaling factor, $t_0$ denotes the transition time and $\tau$ represents the time scale of the MPT. The uniform priors of the parameters of this version of pacing models are: $0<v_0< 36$, $90<k<130$, $10<\tau<500$, $10<a<30$ and $700<t_0<1250$ according to the range of MPT time given by \cite{clark06}. For the periodic model, two prior distributions are changed: $a=0$ and $70<k<110$.

In the above equation, the variables $0.6k$ and $0.4k$ are used to rescale the trend model such that the ice volume threshold including a sigmoid trend allows both $\sim$41\,kyr and $\sim$100\,kyr ice volume variations. From Fig. \ref{fig:icevolume}, we see that the three example sigmoid thresholds have shapes varying with $\tau$ and $t_0$. A late transition time, $t_0$, moves the trend to the present, and a small transition time scale, $\tau$, generate a rapid transition. The sigmoid trend becomes a step function and a constant trend when $\tau=0$ and $\infty$, respectively. 

\subsection{Termination models}\label{sec:termination_climate}

Because we aim to select out the forcing that paces the deglaciations, we will only build termination models to predict the deglaciations identified from $\delta^{18}$O data sets. Modeling the ice volume variations using termination models has several strengths: i) it predicts the significant events -- glacial terminations -- in $\delta^{18}$O with few parameters; ii) it is specially designed for statistical analysis of glacial terminations; iii) it efficiently accounts for the pacing effect of different forcings with only one free parameter, i.e. the contribution factor of forcing, $a$. The plausibility of a termination model is strongly related to the plausibility of the corresponding forcing model and pacing model. The procedure of generating termination models from pacing models is described below. 

We identify terminations from pacing models by finding the start and end time of a deglaciation and assigning the median and half width of the deglaciation period as the termination time and time uncertainty, respectively (see Fig. \ref{fig:pacing_model}). Like we did in section \ref{sec:data_climate}, we interpret the time of each termination probabilistically using a Gaussian distribution with the mean and standard deviations equal to the termination time and time uncertainty, respectively. Thus a termination model, which predicts a sequence of terminations, is actually a Gaussian sequence. This termination model aims to predict the probability of finding a termination at a given time, i.e. $P(\tau_j|\boldsymbol{\theta},M)$ in equation \ref{eqn:post_par}.

\begin{figure}
  \centering
  \includegraphics[scale=0.6]{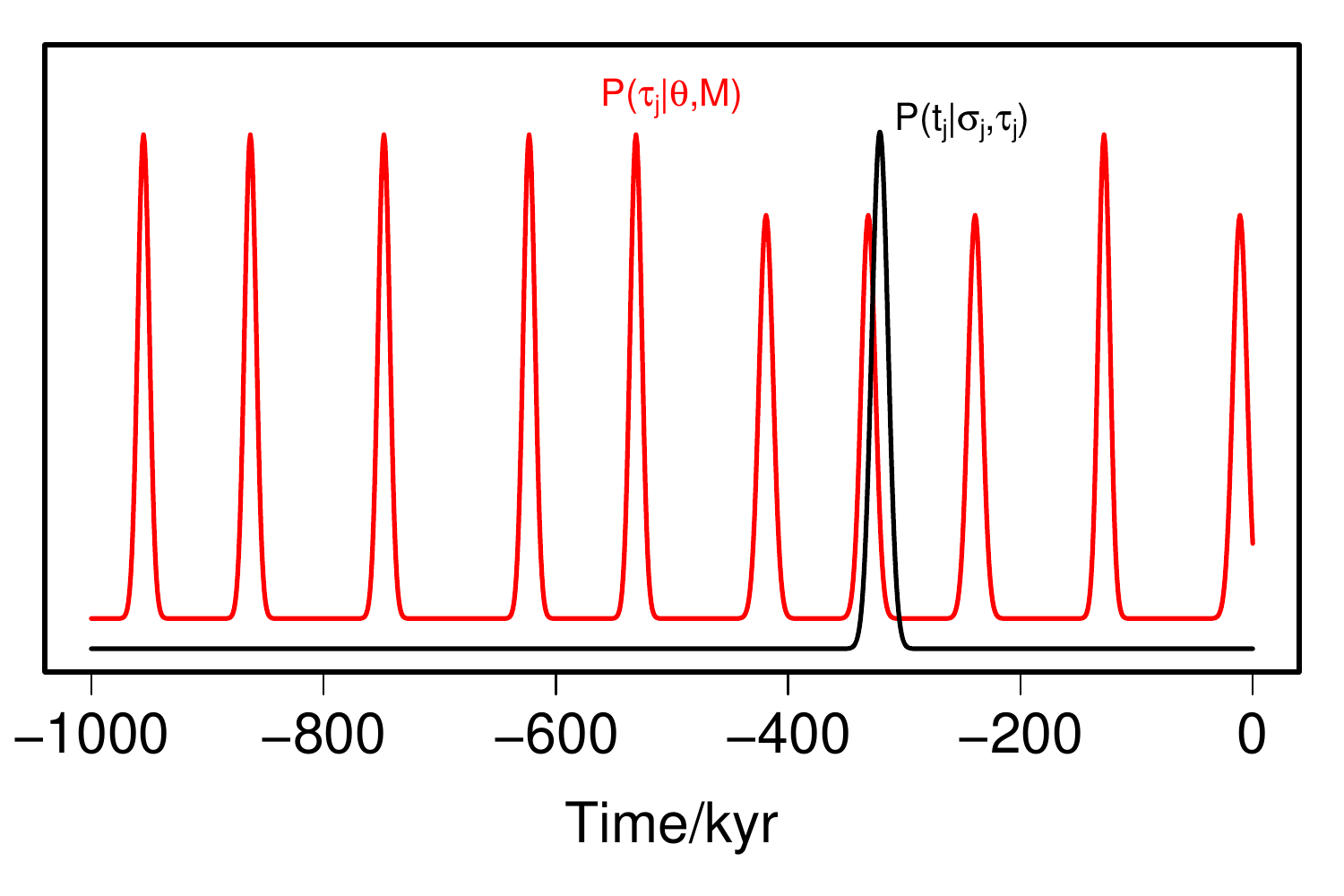}
  \caption{Principle of likelihood calculation (equation \ref{eqn:post_par}). The red line is the termination model generated from the pacing model shown in Fig. \ref{fig:pacing_model}. The black line represents termination $j$ which is interpreted probabilistically using a Gaussian distribution over time.}
  \label{fig:termination_model}
\end{figure}

A termination model is derived from the pacing model in Fig. \ref{fig:pacing_model}, and is shown by the red line in Fig. \ref{fig:termination_model}. Considering possible random contributions from the climate system to the timing of a termination, we add a background to this termination model. The background is represented by the background fraction $b=H_b/(H_b+H_g)$, where $H_b$ is the amplitude of the background and $H_g$ is the difference between the maximum and minimum of the Gaussian sequence. We adopt a uniform prior for the background fraction: $0<b<0.1$. The event likelihood for a termination is calculated by integrating the product of a measurement model for this termination, $P(t_j|\sigma_j,\tau_j)$, and the model prediction of the true termination time, $P(\tau_j|\boldsymbol{\theta},M)$. The product of event likelihoods for all terminations in a data set is the likelihood for a termination model (equation \ref{eqn:post_par}).

Based on this example, the whole procedure of model comparison is described as follows: a forcing model (see Fig. \ref{fig:forcing_models}) modulates the ice volume threshold (equation \ref{eqn:threshold}) of the pacing model (equation \ref{eqn:deterministic}) from which a termination model (see Fig. \ref{fig:termination_model}) is derived and compared with a Gaussian sequence of terminations identified from a $\delta^{18}$O data set. With the above procedure, we generate termination models from pacing models with their thresholds depending on various forcing models (see Fig. \ref{fig:climate_model}). In addition, we define a simple reference model, i.e. the uniform model, which predicts a uniform probability distribution over the termination time. In section \ref{sec:comparison_climate}, we will calculate Bayes factors with respect to the uniform model. 

\section{Model comparison}\label{sec:comparison_climate}

\subsection{Bayes factor}\label{sec:BF_climate}

We calculate the evidence for the termination models listed in Table \ref{tab:ts_models} for the terminations listed in Table \ref{tab:terminations} using the Monte Carlo method described in section \ref{sec:bayes}. Considering that different pacing models predict terminations over different Pleistocene periods, we calculate Bayes factors for terminations extending over three different time spans: 1\,Mya to 0\,Myr, 2\,Mya to 1\,Mya and 2\,Mya to 0\,Mya. The first time span is chosen to model the $\delta^{18}$O variations over the same time span with \cite{huybers11}. However, many previous studies claim that the onset of strong 100-kyr power in glacial cycles occurred around 0.8\,Mya. We will check if our results are sensitive to the change of the duration of the late Pleistocene in Section \ref{sec:sensitivity_climate}. In the following sections, we will choose terminations (see Table \ref{tab:terminations}) over different time spans and decide which version of pacing models will be used for model inference.  

\subsubsection{Late Pleistocene}\label{sec:LP_climate}

Although the $\delta^{18}$O responses to forcings over the late Pleistocene (-1 to 0\,Mya) are dominated by 100-kyr cycles, the deglaciations identified using H07's method (in the data sets HA, HB, HP, LR04 and LRH) contain many minor terminations which may be better explained by models which predict $\sim$40\,kyr cycles. Thus, we choose both $\gamma=1$ and $\gamma=0.4$ for all termination models to predict 100-kyr and 41-kyr cycles in $\delta^{18}$O variations over the past 1\,Mya, respectively. Note that we set different $\gamma$ to avoid unreasonable prior distributions rather than to fit our models to data. 

Using the method described in section \ref{sec:bayes}, for each termination model, we calculate and show the Bayes factor (BF) and maximum likelihood (ML) relative to the uniform model in Table \ref{tab:BF_deterministic-1Myr}. Comparing BFs in each column, we find that the HA, HB, LR04 and LRH data sets favor the models with tilt component and with $\gamma=0.4$. Although compound models such as EPT and Insolation sometimes have BFs slightly higher than the Tilt model, precession and eccentricity may not be necessary to explain the terminations identified from these data sets according to the Occam's razor which increases the prior of simple models, i.e. $P(M)$ (see equation \ref{eqn:bayes1}). 

In addition, the HP data set favors the PT model with $\gamma=1$. This can be caused by a mismatch between the terminations identified in HP and the terminations identified in other data sets. For example, nearby the time of termination 6 shown in Figure \ref{fig:huybers_data}, two terminations are identified in HP while only one termination is identified in other data sets. In particular, the discrepancy between HP and other data sets becomes larger before 0.8\,Mya, which indicates a more ambiguous definition of terminations before the late Pleistocene particularly for planktic $\delta^{18}$O records. Considering this problem, we will choose terminations which occur only over the last 0.8\,Myr (a more conservative time scale of late Pleistocene) and calculate BFs for our models again in section \ref{sec:sensitivity_climate}. Despite this discrepancy, for all the data sets containing minor terminations, tilt is a common factor in the preferred models.  

For terminations identified from the DD, ML and MS data sets, the PT and Insolation models with $\gamma=1$ are best favored. This means precession can be combined with tilt to pace the major terminations better than tilt or precession alone. Because the EPT model and the Insolation model, i.e. a fitted EPT model, does not have BFs as high as the PT model has, the eccentricity component seems to be unlikely to pace the glacial terminations directly. But eccentricity can determine the glacial terminations indirectly through modulating the amplitude of the precession maxima ({\it i.e.,} $e\sin{\omega}$). A similar conclusion has been drawn using the p-value to reject null hypothesis by \cite{huybers11}. However, the rejection of null hypothesis may not validate the alternative hypothesis because there may be yet other hypotheses which fit the data better. Bayesian inference is more appropriate for model comparison not only because it treats all models equally, but also because it accounts for model complexity using a marginalized likelihood, i.e. the evidence (see equation \ref{eqn:evidence}). We conclude that the combination of precession and tilt paces the major glacial terminations while only tilt is necessary to pace both the major and minor terminations over the past 1\,Myr. 

\begin{table}
  \begin{adjustwidth}{-1.2in}{-1.2in}
    \centering
    \caption{The BFs and MLs relative to the uniform model for terminations occurring over the late Pleistocene (1 to 0\,Mya). For each model and data set, the BF is shown on the left and the ML is shown on the right. The prior of each model (see Table \ref{tab:ts_models}) is sampled with $10^5$ points. The priors of the first 11 termination models are determined by $\gamma=1$ while the priors of the last 11 models are determined by $\gamma=0.4$. For each data set, all models are compared in the corresponding column. Note that the data sets on the left side of the double vertical line are terminations identified using H07's method while the data sets on the right are major terminations extensively studied in the literature.}
    \label{tab:BF_deterministic-1Myr}
    \scalebox{0.5}{\begin{tabular}{c|c|ll|ll|ll|ll|ll||ll|ll|ll}
      \hline
      \hline
      &Termination model & \multicolumn{2}{|c|}{HA} & \multicolumn{2}{c|}{HB} & \multicolumn{2}{c|}{HP} & \multicolumn{2}{c|}{LR04} & \multicolumn{2}{c||}{LRH} & \multicolumn{2}{c|}{DD} &\multicolumn{2}{c|}{ML} & \multicolumn{2}{c}{MS}\\
      \hline
      \multirow{11}{*}{$\gamma=1$}&Periodic &0.066& 21&0.072 &32&0.52&400&0.03 &12&0.10 &90&1.4&500&0.64&160&0.30&140\\
      &Eccentricity &0.061&4.4&0.067&5.6&0.090&17&0.11&7.9&0.040&16&0.31&88&1.2&67&0.55&83\\
      &Precession &1.4&380&1.4&410&2.9&$2.1\times 10^3$&0.42&74&0.57&210&9.6&$9.0\times 10^3$&14&$8.3\times 10^3$&12&$4.3\times 10^3$\\
      &Tilt &1.6&$1.2\times 10^3$&1.5&$1.1\times 10^3$&1.3&$1.9\times 10^3$&0.34&160&0.43&470&3.0&$5.1\times 10^3$&2.7&$1.4\times 10^3$&7.4&$1.2\times 10^4$\\
      &EP &1.3&480&1.3&550&2.0&$2.0\times 10^3$&0.42&81&1.3&340&2.7&$7.7\times 10^3$&5.2&$1.1\times 10^3$&4.7&$5.3\times 10^3$\\
      &ET &2.5&$1.0\times 10^3$&2.2&$1.0\times 10^3$&2.6&$2.6\times 10^3$&1.2&380&0.71&840&6.4&$1.3\times 10^4$&21&$8.6\times 10^4$&69&$2.2\times 10^5$\\
      &PT &20&$3.2\times 10^3$&17&$2.7\times 10^3$&100&$6.4\times 10^3$&10&810&15&$1.1\times 10^3$&120&$1.3\times 10^4$&220&$6.4\times 10^4$&740&$2.0\times 10^5$\\
      &EPT &16&$4.4\times 10^3$&13&$3.1\times 10^3$&13&$5.7\times 10^3$&5.0&910&5.2&$1.1\times 10^3$&19&$1.5\times 10^3$&47&$7.4\times 10^4$&170&$3.9\times 10^5$\\
      &Insolation &32&$4.7\times 10^3$&27&$3.6\times 10^3$&69&$4.6\times 10^3$&10&680&17&$2.6\times 10^3$&130&$1.9\times 10^4$&450&$1.1\times 10^5$&$1.2\times 10^3$&$2.3\times 10^5$\\
      &Inclination &0.0047&2.0&0.0051&2.2&0.018&10&0.012&7.7&0.0094&3.4&0.035&13&0.018&4.1&0.022&11\\
      &GPI &0.12&37&0.12&36&0.19&62&0.019&3.6&0.090&25&0.38&120&0.16&31&0.073&11\\\hline
      \multirow{11}{*}{$\gamma=0.4$} &Periodic&14&$3.1\times 10^3$&10&$2.1\times 10^3$&0.32&26&17&$3.9\times 10^3$&2.5&340&0.49&30&1.1&150&3.4&690\\
      &Eccentricity &0.67&270&0.98&240&0.37&240&0.72&400&0.84&180&1.0&550&3.0&750&2.0&$2.5\times 10^3$\\
      &Precession &1.5&570&1.9&980&0.18&50&2.4&920&0.79&300&1.5&96&1.2&43&2.5&300\\
      &Tilt &220&$1.3\times 10^4$&170&$1.4\times 10^4$&3.8&100&220&$9.9\times 10^3$&59&$5.2\times 10^3$&10&480&22&960&79&$9.1\times 10^3$\\
      &EP &1.3&280&1.7&430&0.67&93&2.4&$2.8\times 10^3$&0.94&180&2.0&180&2.7&560&5.1&$1.7\times 10^3$\\
      &ET &150&$2.5\times 10^4$&130&$1.3\times 10^4$&1.6&87&240&$2.9\times 10^4$&22&$4.3\times 10^3$&7.0&$8.0\times 10^3$&26&$5.1\times 10^3$&100&$2.2\times 10^4$\\
      &PT &170&$3.9\times 10^4$&140&$3.7\times 10^4$&3.4&510&400&$8.9\times 10^4$&94&$1.3\times 10^4$&14&880&38&$2.3\times 10^3$&210&$2.5\times 10^4$\\
      &EPT &240&$3.2\times 10^4$&230&$4.2\times 10^4$&2.5&120&730&$9.3\times 10^4$&83&$1.5\times 10^4$&18&$1.4\times 10^3$&71&$4.7\times 10^3$&540&$2.5\times 10^5$\\
      &Insolation &170&$1.6\times 10^4$&152&$1.4\times 10^4$&5.7&570&410&$2.8\times 10^4$&97&$8.7\times 10^3$&21&$1.6\times 10^3$&29&$1.5\times 10^3$&160&$2.2\times 10^4$\\
      &Inclination &0.72&$1.3\times 10^3$&0.81&560&0.61&100&1.7&430&1.4&350&4.2&670&2.7&410&2.6&$1.4\times 10^3$\\
      &GPI &0.022&3.7&0.019&6.8&0.039&12&0.082&24&0.026&7.4&0.20&71&0.29&48&0.16&31\\\hline
   \end{tabular}}
  \end{adjustwidth}
\end{table}

\subsubsection{Early Pleistocene}\label{sec:EP_climate}

Because the terminations of the DD, ML and MS data sets are only within the late Pleistocene, we will only use the terminations identified from the HA, HP, HB, L04 and LRH data sets. Moreover, we do not calculate evidences for models with $\gamma=1$ because the $\sim$40\,kyr cycles are significant in all data sets (see Fig. \ref{fig:huybers_data}). Thus we use $\gamma=0.4$ to define prior distributions for each pacing model such that the corresponding termination model can predict $\sim$40\,kyr cycles in the early-Pleistocene deglaciations. We don't use the GPI model because the GPI record has a time span less than 2\,Myr. The BFs and MLs for other termination models are shown in Table \ref{tab:BF_deterministic-2-1Myr}.

We find that the Tilt model is best favored by all data sets. Given that the combination of tilt with other orbital elements does not give a higher evidence, the other orbital elements must not play a main role in pacing the deglaciations over the early Pleistocene. However, this does not indicate a priori penalization of complex models in a Bayesian framework because a more complex (multi-component) model could in principle get a higher evidence if supported by the data.

\begin{table}[ht!]
    \begin{adjustwidth}{-1.0in}{-1.0in}
  \centering
  \caption{Same as Table \ref{tab:BF_deterministic-1Myr} but for data extending from 2 to 1\,Mya. The DD, ML and MS data sets are not included because they do not exist over this time scale. For all models, only $\gamma=0.4$ is used due to the obvious 41-kyr dominant $\delta^{18}$O variations over this time scale (see Fig. \ref{fig:huybers_data}). }
  \label{tab:BF_deterministic-2-1Myr}
  \scalebox{1.0}{\begin{tabular}{c|ll|ll|ll|ll|ll}
    \hline
    \hline
    Termination model & \multicolumn{2}{|c|}{HA} & \multicolumn{2}{c|}{HB} & \multicolumn{2}{c|}{HP} & \multicolumn{2}{c|}{LR04} & \multicolumn{2}{c}{LRH} \\
    \hline
    Periodic &2.5&200&2.6&190&1.4&92&3.2&390&2.1&150\\
    Eccentricity &0.53&120&0.49&210&0.16&26&0.49&110&0.65&61\\
    Precession &1.0&270&1.0&260&0.61&160&0.44&150&1.0&240\\
    Tilt &22&$1.3\times 10^3$&21&$1.3\times 10^3$&11&170&14&590&18&870\\
    EP&0.55&340&0.51&280&0.24&91&0.18&67&0.49&140\\
    ET&5.9&$1.2\times 10^3$&5.5&$1.7\times 10^3$&3.9&260&5.1&370&6.2&930\\
    PT &9.3&$2.9\times 10^3$&9.0&$6.3\times 10^3$&4.3&390&6.1&420&9.7&920\\
    EPT &5.0&$1.0\times 10^3$&4.8&$1.1\times 10^3$&2.3&400&3.3&450&6.0&640\\
    Insolation &10&910&10&960&4.7&480&3.9&340&9.1&690\\
    Inclination&0.30&59&0.29&55&0.041&8.1&0.12&11&0.14&13\\\hline
  \end{tabular}}
\end{adjustwidth}
\end{table}

\subsubsection{Whole Pleistocene}\label{sec:WP_climate}

For the time scale of the last 2\,Myr, we use the data sets of HA, HB, HP, LR04, LRH and the hybrid data sets, HADD, HAML and HAMS. We use pacing models with and without a trend threshold to model the terminations over the whole Pleistocene. The BFs and MLs for the above models and data sets are shown in Table \ref{tab:BF_trend-2-0Myr}.

\begin{table}[ht!]
    \begin{adjustwidth}{-1.2in}{-1.2in}
  \centering
  \caption{Same as Table \ref{tab:BF_deterministic-1Myr} but for data extending over the last 2\,Myr. Three sets of models are given: termination models with a linear trend defined in equation \ref{eqn:linear} (upper 10 models), with a sigmoid trend defined in equation \ref{eqn:sigmoid} (middle 10 models), and without any trend but with $\gamma=0.4$ (lower 10 models). The GPI model is not included because the corresponding GPI record only has a length of 1.5\,Myr. Note that the data sets on the left side of the double lines are terminations identified with H07's method while the data sets on the right side are hybrid deglaciation events combining events in HA from 2 to 1\,Mya and the well-studied terminations in the last 1\,Myr.}
  \label{tab:BF_trend-2-0Myr}
  \scalebox{0.45}{\begin{tabular}{c|c|ll|ll|ll|ll|ll||ll|ll|ll}
    \hline
    \hline
    &Termination model & \multicolumn{2}{|c|}{HA} & \multicolumn{2}{c|}{HB} & \multicolumn{2}{c|}{HP} & \multicolumn{2}{c|}{LR04} & \multicolumn{2}{c||}{LRH} &\multicolumn{2}{c|}{HADD} & \multicolumn{2}{c|}{HAML} & \multicolumn{2}{c}{HAMS}\\
    \hline
    \multirow{10}{*}{{\parbox[t]{3cm}{Linear trend\\No $\gamma$}}} &Periodic &$8.9\times 10^{-4}$&7.1&$1.0\times 10^3$&6.5&$1.1\times 10^{-3}$&12&$1.2\times 10^{-3}$&2.3&$9.7\times 10^{-4}$&2.5&0.050&$2.5\times 10^3$&0.013&920&$3.3\times 10^{-3}$&81\\
    &Eccentricity &$1.6\times 10^{-3}$&1.2&$1.6\times 10^{-3}$&1.7&$9.1\times 10^{-4}$&0.74&$4.9\times 10^{-3}$&3.4&$2.7\times 10^{-3}$&2.0&0.0051&2.5&0.012&7.7&$8.2\times 10^{-3}$&5.9\\
    &Precession &0.17&$1.4\times 10^3$&0.17&860&0.14&$2.1\times 10^3$&0.057&400&0.10&430&8.8&$1.6\times 10^5$&4.1&$1.5\times 10^4$&6.8&$3.8\times 10^4$\\
    &Tilt &33&$8.9\times 10^4$&33&$1.2\times 10^5$&54&$5.8\times 10^4$&6.8&$1.5\times 10^4$&21&$3.5\times 10^4$&$2.2\times 10^3$&$3.5\times 10^7$&$4.9\times 10^3$&$8.7\times 10^6$&$2.4\times 10^4$&$3.4\times 10^7$\\
    &EP&0.026&350&0.040&$1.1\times 10^3$&0.013&92&$7.2\times 10^{-3}$&28&0.013 &50&0.17&$2.4\times 10^3$&0.35&$1.5\times 10^4$&0.31&$7.0\times 10^3$\\
    &ET&40&$2.5\times 10^5$&27&$1.3\times 10^5$&17&$1.4\times 10^5$&2.8&$1.6\times 10^4$&11&$8.2\times 10^4$&910&$1.5\times 10^7$&840&$7.8\times 10^6$&$3.3\times 10^3$&$7.2\times 10^7$\\
    &PT &380&$4.7\times 10^5$&310&$3.3\times 10^{-5}$&130&$2.2\times 10^5$&11&$5.7\times 10^4$&78 &$7.2\times 10^4$&$3.8\times 10^3$&$9.2\times 10^6$&$9.9\times 10^3$&$4.9\times 10^7$&$4.1\times 10^4$&$3.2\times 10^8$\\
    &EPT &89&$5.1\times 10^5$&74&$3.9\times 10^5$&30&$1.3\times 10^5$&3.1&$1.8\times 10^4$&18&$8.9\times 10^4$&446&$2.6\times 10^6$&$1.5\times 10^3$&$3.2\times 10^7$&$5.3\times 10^3$&$2.7\times 10^7$\\
    &Insolation &9.4&$3.0\times 10^4$&10&$6.0\times 10^4$&7.9&$3.1\times 10^4$&1.3&$6.1\times 10^3$&3.0&$6.7\times 10^3$&260&$7.1\times 10^6$&460&$2.3\times 10^6$&$1.4\times 10^3$&$1.5\times 10^7$\\
    &Inclination&$4.5\times 10^{-4}$&1.5&$5.0\times 10^{-4}$&5.1&$5.5\times 10^{-4}$&0.79&$4.5\times 10^{-4}$&2.5&$7.3\times 10^{-4}$&7.9&0.0023&5.3&$1.8\times 10^{-3}$&6.0&0.0018&3.3\\\hline
    \multirow{10}{*}{{\parbox[t]{3cm}{Sigmoid trend\\No $\gamma$}}} &Periodic&0.15&$2.8\times 10^3$&0.15&$1.7\times 10^3$&0.37&$4.3\times 10^3$&0.031&550&0.048&270&27&$5.0\times 10^5$&110&$7.8\times 10^6$&13&$1.7\times 10^5$\\
    &Eccentricity &0.060&920&0.069&890&0.096&710&0.015&74&0.036&120&0.44&$5.2\times 10^3$&0.63&$4.3\times 10^3$&0.34&$5.1\times 10^3$\\
    &Precession &0.73&740&1.1&$1.0\times 10^4$&0.68&$3.1\times 10^3$&0.12&980&0.68&$1.7\times 10^4$&36&$3.7\times 10^5$&38&$7.3\times 10^5$&30&$2.0\times 10^5$\\
    &Tilt &160&$1.2\times 10^6$&160&$2.5\times 10^6$&29&$1.5\times 10^5$&21&$9.1\times 10^4$&48&$6.6\times 10^5$&580&$4.9\times 10^6$&590&$2.2\times 10^7$&$1.8\times 10^3$&$2.5\times 10^7$\\
    &EP&0.18&940&0.24&$2.1\times 10^3$&0.23&500&0.038&960&0.15&680&12&$7.8\times 10^5$&2.5&$3.6\times 10^4$&2.0&$3.6\times 10^4$\\
    &ET&98&$3.8\times 10^5$&170&$4.9\times 10^6$&32&$3.2\times 10^5$&41&$2.2\times 10^5$&70&$1.3\times 10^6$&$1.6\times 10^3$&$2.8\times 10^7$&$2.3\times 10^3$&$6.0\times 10^7$&$6.1\times 10^3$&$2.3\times 10^8$\\
    &PT &$3.1\times 10^3$&$1.0\times 10^7$&$2.8\times 10^3$ &$8.9\times 10^6$&300&$6.6\times 10^5$&320&$8.6\times 10^5$&$4.1\times 10^4$&$2.0\times 10^8$&$1.7\times 10^4$&$8.1\times 10^7$&$4.1\times 10^4$&$2.0\times 10^8$&$2.2\times 10^5$&$4.8\times 10^8$\\
    &EPT &550&$8.6\times 10^6$&510&$5.3\times 10^6$&61&$4.7\times 10^5$&110&$1.9\times 10^6$&220&$1.2\times 10^6$&$8.9\times 10^3$&$3.5\times 10^8$&$1.0\times 10^4$&$1.2\times 10^8$&$4.7\times 10^4$&$3.2\times 10^8$\\
    &Insolation &190&$6.2\times 10^6$&144&$2.1\times 10^6$&47&$4.7\times 10^5$&58&$2.3\times 10^5$&230&$1.7\times 10^6$&$1.0\times 10^4$&$6.2\times 10^8$&$9.7\times 10^3$&$1.2\times 10^8$&$2.7\times 10^4$&$8.7\times 10^8$\\
    &Inclination&$2.3\times 10^{-3}$&6.1&$2.9\times 10^{-3}$&6.5&$6.8\times 10^{-4}$&3.7&$5.8\times 10^{-4}$&1.7&$1.6\times 10^{-3}$&22&0.027&220&0.011&85&$8.4\times 10^{-3}$&210\\\hline
    \multirow{10}{*}{{\parbox[t]{3cm}{No trend\\$\gamma=0.4$}}} &Periodic&560&$1.8\times 10^5$&320&$9.6\times 10^4$&7.1&920&990&$3.9\times 10^5$&70&$2.2\times 10^4$&18&$4.8\times 10^3$&0.83&$2.5\times 10^3$&240&$1.0\times 10^5$\\
    &Eccentricity &0.21&880&0.27&470&0.081&170&0.30&830&0.27&800&0.25&580&0.83&$2.5\times 10^3$&0.48&730\\
    &Precession &0.54&$2.0\times 10^3$&0.55&$1.4\times 10^3$&0.14&100&1.6&$2.3\times 10^4$&0.81&520&2.3&$1.4\times 10^3$&1.8&$1.7\times 10^3$&5.2&$6.6\times 10^3$\\
    &Tilt &$1.3\times 10^4$&$2.1\times 10^6$&$9.2\times 10^3$&$1.6\times 10^6$&75&$1.7\times 10^3$&$1.6\times 10^4$&$1.7\times 10^6$&$1.8\times 10^3$&$2.5\times 10^5$&160&$1.3\times 10^4$&720&$4.8\times 10^4$&$3.7\times 10^3$&$3.2\times 10^5$\\
    &EP&0.26&600&0.27&570&0.16&220&0.71&$4.1\times 10^3$&0.42&$1.3\times 10^3$&0.84&$1.5\times 10^3$&0.65&$1.2\times 10^3$&1.7&$8.0\times 10^3$\\
    &ET&$2.6\times 10^3$&$1.8\times 10^6$&$1.9\times 10^3$&$5.4\times 10^5$&18&$3.1\times 10^3$&$4.2\times 10^3$&$2.0\times 10^6$&400&$1.0\times 10^5$&44&$1.8\times 10^4$&170&$9.2\times 10^4$&820&$5.8\times 10^5$\\
    &PT &$6.7\times 10^3$&$3.5\times 10^6$&$5.0\times 10^3$&$3.9\times 10^6$&28&$8.8\times 10^3$&$6.3\times 10^3$&$3.3\times 10^6$&$1.3\times 10^3$&$8.0\times 10^5$&110&$1.7\times 10^5$&430&$2.4\times 10^5$&$3.5\times 10^3$&$4.1\times 10^6$\\
    &EPT &$3.5\times 10^3$&$2.4\times 10^6$&$2.9\times 10^3$&$4.5\times 10^6$&13&$6.4\times 10^3$&$5.1\times 10^3$&$5.6\times 10^6$&690&$5.5\times 10^5$&140&$6.1\times 10^5$&470&$6.4\times 10^5$&$4.0\times 10^3$&$8.9\times 10^6$\\
    &Insolation &$7.2\times 10^3$&$7.9\times 10^6$&$6.5\times 10^3$&$6.0\times 10^6$&69&$4.0\times 10^4$&$3.9\times 10^3$&$3.6\times 10^6$&$1.6\times 10^3$&$1.7\times 10^6$&260&$4.0\times 10^5$&390&$4.3\times 10^5$&$3.3\times 10^3$&$2.5\times 10^6$\\
    &Inclination&0.21&240&0.23&450&0.027&34&0.2&570&0.21&220&0.85&620&0.77&600&0.74&$1.0\times 10^3$\\\hline
  \end{tabular}}
  \end{adjustwidth}
\end{table}

For the HA, HB and LR04 data sets, the Tilt model with $\gamma=0.4$ is best favored, and other combinations with the tilt component and with $\gamma=0.4$ also give comparative evidences. However, the PT model with a sigmoid trend is the best favored for the HP and LRH data sets and also gives high evidences for the HA, HB and LR04 data sets. In addition, for all of the above data sets, the Precession and Eccentricity models have rather low evidences and the Periodic model has evidences not as high as models with tilt component. All of these results indicate a major role of tilt and a minor role of precession in pacing the Pleistocene deglaciations comprising both major and minor late-Pleistocene terminations. Additionally, for all the above data sets, the Insolation model with $\gamma=0.4$ has high evidences but not higher than other models with tilt component, which means that the deglaciations may not be paced by a daily-averaged insolation at a specific day and latitude as Milankovitch suggested. We will investigate this further in section \ref{sec:sensitivity_climate}. 

For the HADD, HAML and HAMS data sets, the PT model with a threshold modulated by a sigmoid trend are best favored and those compound models with tilt component also have high evidences. Considering that the Tilt model has higher evidences than the Precession model, the whole Pleistocene deglaciations may be mainly paced by tilt while precession only plays a minor role. This is consistent with the results for the data sets with minor late-Pleistocene deglaciations. Thus the role of precession in pacing major deglaciations is probable to intensify the late-Pleistocene glaciations which are resonant with the $\sim$100 eccentricity cycles in precession. Because the EPT and Insolation models have evidences around 10 times lower than the PT model with a linear or a sigmoid trend, eccentricity may not directly pace terminations over the whole Pleistocene. In addition, the PT model with a sigmoid trend is more favored than the PT model with a linear trend, which indicates that the MPT may not be as gradual as claimed by \citep{huybers07}. We will discuss this in details in section \ref{sec:sensitivity_climate}.

According to the evidences shown in Table \ref{tab:BF_deterministic-1Myr}, \ref{tab:BF_deterministic-2-1Myr} and \ref{tab:BF_trend-2-0Myr}, the Inclination and GPI models are not favored and even less favored than the uniform model. That means the geomagnetic paleointensity does not pace glacial cycles over the last 2\,Myr despite a possible link between the GPI and climate changes \citep{courtillot07}. In contrast to the conclusion of \cite{muller97}, there is no evidence for a cause-effect link between the orbital inclination and climate changes, particularly the ice volume change. 

\subsection{Discrimination power}\label{sec:discrimination_climate}

To validate our method as an effective inference tool to select out the true model, we check the discrimination power for each model by generating data from the models and then applying the full analysis (all models) to these data. The data are simulated from all models with common parameters: $h_0=110\gamma$, $a=25\gamma$, $b=0$ and $v_0=45\gamma$, where $\gamma=1$ over the last 1\,Myr and $\gamma=0.4$ from 2 to 1\,Mya. For the Periodic model, the values of $h_0$ and $a$ are different (recall that period $\sim h_0+a+10$), namely $90\gamma$ and 0 respectively. Other parameters in corresponding forcing models are fixed as: $\alpha=0.5$ for compound models with two components, $\alpha=0.3$ and $\beta=0.2$ for the EPT model and $\phi=0$ for models with the precession component.

BFs and MLs for simulated data over the last 1\,Myr are shown in Table \ref{tab:BF_simulation_-1Myr}. We see that all models based on a single orbital element are correctly selected,\footnote{According to Occam's razor, a model with fewer components or free parameters, which has comparative evidence with a model with more components, has fewer assumptions and thus is better favored by the data.}, although those models combining the correct single orbital element with other elements may also give comparative evidences. Incorrect models, in contrast, generally receive much lower Bayes factors. For the PT-simulated data set, the PT model is correctly discriminated from the Insolation model, which is actually a fitted EPT model. In addition, although the ET model may not be corrected selected out when its evidence is close to the evidences for the EP, PT, EPT and Insolation models, the ratios of the Bayes factors are small. The much larger ratios between them for the real data validate our inference of the ET model. We see that, the EP model is not more favored than the Eccentricity model even though it is the true model. However the Eccentricity model is never found to be better favored than others for all real data sets and time scales. Thus this test of discrimination power does support the validity of our conclusion that the PT model is best favored by the 11 major terminations over the late Pleistocene.  

\begin{table}
  \begin{adjustwidth}{-1.2in}{-1.2in}
    \centering
    \caption{Same as Table \ref{tab:BF_deterministic-1Myr} but for simulated data extending over the last 1\,Myr. The data sets simulated from termination models with $\gamma=1$ are represented by the column names. The BF and ML for each true model are underlined. }
    \label{tab:BF_simulation_-1Myr}
    \scalebox{0.4}{\begin{tabular}{c|*{10}{ll|}*{1}{ll}}
        \hline
        \hline
        & \multicolumn{2}{|c|}{Periodic} & \multicolumn{2}{c|}{Eccentricity} & \multicolumn{2}{c|}{Precession} & \multicolumn{2}{c|}{Tilt} & \multicolumn{2}{c|}{EP} & \multicolumn{2}{c|}{ET} & \multicolumn{2}{c|}{PT} & \multicolumn{2}{c|}{EPT} & \multicolumn{2}{c|}{Insolation} & \multicolumn{2}{c|}{Inclination} & \multicolumn{2}{c}{GPI}\\
        \hline
    Periodic &\underline{$2.8\times 10^3$}&\underline{$4.1\times 10^6$}&2.6&$5.7\times 10^3$&58&$6.8\times 10^4$&14&$1.3\times 10^5$&1.3&$4.6\times 10^3$&66&$1.4\times 10^5$&3.0&$3.9\times 10^3$ &320&$1.4\times 10^6$&1.5&690&130&$1.9\times 10^5$&14&$1.8\times 10^4$\\
    Eccentricity &0.052&23&\underline{$3.3\times 10^4$}&\underline{$3.4\times 10^6$}&7.2&$9.0\times 10^4$&0.29&6.9&$1.5\times 10^4$&$2.1\times 10^6$&910&$2.0\times 10^6$&1.1&130&$2.9\times 10^3$&$3.0\times 10^6$&1.0&61&0.011&24&0.33&$8.9\times 10^3$\\
    Precession &0.29&600&$1.5\times 10^3$&$4.4\times 10^5$&\underline{$3.4\times 10^4$}&\underline{$2.0\times 10^7$}&0.82&$1.3\times 10^3$&290&$8.8\times 10^5$&$1.2\times 10^3$&$4.8\times 10^5$&13&$4.2\times 10^5$&$7.4\times 10^3$&$2.1\times 10^6$&6.4&$3.9\times 10^3$&0.41&850&5.1&$4.8\times 10^4$\\
    Tilt &18&$8.5\times 10^4$&5.2&$1.1\times 10^3$&0.034&26&\underline{$1.7\times 10^4$}&\underline{$1.1\times 10^7$}&2.6&480&350&$3.8\times 10^4$&140&$6.9\times 10^4$&$1.3\times 10^3$&$3.6\times 10^5$&2.2&$8.5\times 10^3$&0.067&140&5.5&$4.5\times 10^4$\\
    EP &0.086&230&$3.0\times 10^4$&$2.5\times 10^6$&$1.3\times 10^4$&$1.8\times 10^7$&0.42&640&\underline{$1.3\times 10^4$}&\underline{$8.1\times 10^6$}&$3.0\times 10^3$&$1.8\times 10^6$&7.7&$6.9\times 10^4$&$6.9\times 10^3$&$3.0\times 10^6$&3.6&$2.0\times 10^3$&0.089&49&0.85&$1.3\times 10^4$\\
    ET &0.61&$1.3\times 10^3$&$3.3\times 10^3$&$2.5\times 10^6$&3.0&$5.7\times 10^3$&310&$6.0\times 10^6$&$2.2\times 10^3$&$1.3\times 10^6$&\underline{$6.4\times 10^3$}&\underline{$4.7\times 10^6$}&390&$4.7\times 10^6$&$1.1\times 10^4$&$4.5\times 10^6$&91&$1.6\times 10^5$&0.031&18&12&$8.5\times 10^4$\\
    PT &0.77&$3.2\times 10^3$&180&$1.9\times 10^6$&270&$1.9\times 10^6$&$1.1\times 10^3$&$1.0\times 10^7$&37&$4.4\times 10^4$&$1.0\times 10^4$&$2.0\times 10^6$&\underline{$7.6\times 10^4$}&\underline{$1.6\times 10^7$}&$1.7\times 10^4$&$3.8\times 10^6$&$5.9\times 10^3$&$1.4\times 10^7$&1.5&$7.0\times 10^3$&13&$1.7\times 10^4$\\
    EPT &0.20&130&$6.7\times 10^3$&$1.9\times 10^6$&340&$2.6\times 10^6$&110&$4.3\times 10^6$&$1.6\times 10^3$&$6.1\times 10^6$&$1.4\times 10^4$&$5.5\times 10^6$&$5.1\times 10^3$&$1.2\times 10^7$&\underline{$5.1\times  10^4$}&\underline{$6.6\times 10^6$}&$2.9\times 10^3$&$1.6\times 10^7$&0.29&450&3.5&$1.4\times 10^4$\\
    Insolation &0.32&590&310&$3.3\times 10^5$&51&$1.4\times 10^5$&9.6&$2.2\times 10^3$&67&$5.0\times 10^3$&$1.4\times 10^4$&$1.5\times 10^6$&$2.5\times 10^4$&$1.2\times 10^7$&$2.9\times 10^4$&$3.6\times 10^6$&\underline{$7.9\times 10^3$}&\underline{$1.6\times 10^7$}&0.66&200&4.2&$3.0\times 10^3$\\
    Inclination &0.027&3.1&0.075&52&3.7&86&$3.0\times 10^{-3}$&1.8&0.046&19&0.057&56&0.031&18&0.063&110&0.27&230&\underline{$4.0\times 10^4$}&\underline{$1.0\times 10^7$}&0.18&460\\
    GPI &19&$1.3\times 10^5$&0.084&$1.8\times 10^3$&0.97&440&0.59&$1.5\times 10^4$&0.017&60&$7.6\times 10^{-3}$&30&0.026&38&0.016&240&0.011&9.3&2.6&$4.8\times 10^3$&\underline{$8.0\times 10^3$}&\underline{$3.5\times 10^6$}\\\hline
  \end{tabular}
}
\end{adjustwidth}
\end{table}

The BFs and MLs relative to the uniform model applied to simulated data from 2 to 1\,Mya are shown in Table \ref{tab:BF_simulation_-2-1Myr}. We find that the correct model is always identified with the largest Bayes factor. Yet we do see, for example, that for data from the PT model, the Insolation model and EPT models have similar evidences. However, as the PT model is not as fine tuned as the Insolation model and has fewer adjustable parameters than the EPT model, we would invoke the principle of parsimony (Occam's razor) to select the PT model. 

\begin{table}
    \begin{adjustwidth}{-1.2in}{-1.2in}
    \centering
    \caption{Same as Table \ref{tab:BF_deterministic-1Myr} but for simulated data extending from 2 to 1\,Mya. The data sets simulated from termination models with $\gamma=0.4$ are represented by the column names. The BF and ML are underlined for each true model. }
    \label{tab:BF_simulation_-2-1Myr}
    \scalebox{0.4}{\begin{tabular}{c|*{10}{ll|}*{1}{ll}}
        \hline
        \hline
        & \multicolumn{2}{|c|}{Periodic} & \multicolumn{2}{c|}{Eccentricity} & \multicolumn{2}{c|}{Precession} & \multicolumn{2}{c|}{Tilt} & \multicolumn{2}{c|}{EP} & \multicolumn{2}{c|}{ET} & \multicolumn{2}{c|}{PT} & \multicolumn{2}{c|}{EPT} & \multicolumn{2}{c|}{Insolation} & \multicolumn{2}{c|}{Inclination} & \multicolumn{2}{c}{GPI}\\
        \hline
        Periodic &\underline{$2.4\times 10^5$}&\underline{$1.1\times 10^8$}&1.0&460&1.0&$8.7\times 10^3$&120&$3.5\times 10^4$&2.6&$1.6\times 10^3$&7.4&$1.6\times 10^3$&4.8&$5.3\times 10^3$&11&$4.5\times 10^3$&4.9&680&0.39&96&29&$1.0\times 10^4$\\
        Eccentricity &8.2&$7.3\times 10^4$&\underline{$2.3\times 10^4$}&\underline{$1.4\times 10^8$}&460&$2.5\times 10^6$&0.44&370&330&$1.7\times 10^6$&220&$3.4\times 10^5$&1.2&$1.4\times 10^3$&38&$1.6\times 10^5$&3.7&$1.4\times 10^4$&0.78&570&1.7&$2.1\times 10^3$\\
        Precession &160&$3.6\times 10^5$&$6.3\times 10^3$&$1.2\times 10^7$&\underline{$2.6\times 10^5$}&\underline{$3.1\times 10^8$}&1.8&$2.9\times 10^3$&830&$1.4\times 10^6$&4.3&$8.4\times 10^3$&240&$2.6\times 10^5$&7.1&$1.9\times 10^4$&$7.8\times 10^3$&$7.6\times 10^7$&6.8&$1.5\times 10^3$&120&$1.3\times 10^5$\\
        Tilt &1.4&$3.3\times 10^4$&7.2&$2.7\times 10^4$&45&$1.3\times 10^5$&\underline{$6.8\times 10^4$}&\underline{$1.7\times 10^8$}&0.14&$1.4\times 10^3$&41&$4.4\times 10^4$&640&$8.5\times 10^5$&280&$1.7\times 10^5$&60&$2.8\times 10^5$&0.35&220&0.68&$4.8\times 10^3$\\
        EP &32&$3.2\times 10^5$&$2.7\times 10^4$&$5.4\times 10^7$&$2.2\times 10^4$&$1.5\times 10^8$&0.70&$1.0\times 10^3$&\underline{$6.8\times 10^4$}&\underline{$1.5\times 10^8$}&23&$1.7\times 10^5$&71&$2.1\times 10^5$&190&$1.6\times 10^5$&$1.6\times 10^3$&$6.2\times 10^6$&2.0&$2.1\times 10^3$&45&$2.4\times 10^5$\\
        ET &12&$3.3\times 10^5$&$7.3\times 10^3$&$4.2\times 10^7$&$1.8\times 10^3$&$4.5\times 10^6$&$4.2\times 10^3$&$3.7\times 10^7$&4.1&$3.8\times 10^4$&\underline{$1.3\times 10^4$}&\underline{$9.2\times 10^7$}&$1.8\times 10^3$&$1.1\times 10^6$&$3.0\times 10^3$&$1.6\times 10^5$&320&$1.7\times 10^6$&2.1&810&0.24&$1.7\times 10^3$\\
        PT &110&$2.1\times 10^5$&110&$4.8\times 10^5$&$4.4\times 10^3$&$6.4\times 10^7$&760&$1.8\times 10^7$&22&$6.0\times 10^5$&52&$5.4\times 10^4$&\underline{$3.1\times 10^5$}&\underline{$2.1\times 10^8$}&$1.7\times 10^3$&$3.0\times 10^6$&$1.3\times 10^4$&$6.3\times 10^7$&0.54&$2.3\times 10^3$&0.26&$2.2\times 10^3$\\
        EPT &29&$1.4\times 10^5$&$4.7\times 10^3$&$1.6\times 10^7$&$1.7\times 10^4$&$4.9\times 10^7$&120&$5.1\times 10^5$&$1.3\times 10^3$&$4.4\times 10^7$&$2.2\times 10^3$&$3.3\times 10^7$&$1.4\times 10^5$&$6.0\times 10^7$&\underline{$5.0\times 10^4$}&\underline{$8.6\times 10^7$}&$8.6\times 10^3$&$4.6\times 10^7$&2.5&$6.9\times 10^3$&1.4&$2.5\times 10^4$\\
        Insolation &540&$4.7\times 10^5$&140&$7.3\times 10^5$&$5.8\times 10^3$&$3.7\times 10^7$&6.2&$9.3\times 10^3$&1.9&$4.6\times 10^4$&67&$1.2\times 10^5$&$1.4\times 10^5$&$5.5\times 10^7$&$4.8\times 10^3$&$3.3\times 10^7$&\underline{$4.8\times 10^4$}&\underline{$1.4\times 10^8$}&0.53&$1.2\times 10^3$&0.24&840\\
        Inclination &14&$2.4\times 10^4$&0.40&$1.4\times 10^3$&0.23&190&0.018&22&0.22&$1.2\times 10^3$&0.88&$3.4\times 10^3$&1.1&$1.4\times 10^3$&2.0&$1.5\times 10^3$&0.057&28&\underline{250}&\underline{$2.5\times 10^6$}&6.9&$8.6\times 10^4$\\
        GPI &1.5&$1.8\times 10^4$&0.15&83&$1.5\times 10^{-3}$&31&0.043&260&1.8&660&0.24&140&0.033&150&0.078&43&0.18&170&0.027&59&\underline{$8.4\times 10^5$}&\underline{$2.0\times 10^8$}\\\hline
      \end{tabular}
    }
  \end{adjustwidth}
\end{table}

Given the ability of the Bayesian inference method for model comparison, we conclude that tilt (or obliquity) is the main ``pace-maker'' of the deglaciations over the last 2\,Myr while precession may pace the deglaciations over the late Pleistocene. This indicates that precession becomes important in pacing terminations after the MPT. Other climate forcings, including GPI and inclination forcing, are very unlikely to pace the deglaciations over the Pleistocene. 

\section{Sensitivity test}\label{sec:sensitivity_climate}

We perform a sensitivity test to check how sensitive the evidences of the models are to the choices of time scales and model priors. 

First, we change the time of the onset of the 100-kyr cycles. We calculate the BFs and MLs for all termination models for all data sets over the last 0.8\,Myr (as opposed to the last 1\,Myr as before) and show them in Table \ref{tab:BF_deterministic-0.8Myr}. In this new list of models, we have added another GPI model based on a GPI data set from \cite{guyodo99} (G99), using the method of modeling the GPI record from \cite{channell09} (C09). We find that the combination of obliquity and precession still pace the well-studied or main terminations (DD, ML and MS) better than obliquity alone. Thus our conclusion is robust to the change of the late-Pleistocene time span. 

\begin{table}
  \begin{adjustwidth}{-1.2in}{-1.2in}
    \centering
    \caption{Same with Table \ref{tab:BF_deterministic-1Myr}, but for terminations over the last 0.8\,Myr. }
    \label{tab:BF_deterministic-0.8Myr}
    \scalebox{0.5}{\begin{tabular}{c|c|ll|ll|ll|ll|ll||ll|ll|ll}
      \hline
      \hline
      &Termination model & \multicolumn{2}{|c|}{HA} & \multicolumn{2}{c|}{HB} & \multicolumn{2}{c|}{HP} & \multicolumn{2}{c|}{LR04} & \multicolumn{2}{c||}{LRH} & \multicolumn{2}{c|}{DD} &\multicolumn{2}{c|}{ML} & \multicolumn{2}{c}{MS}\\
      \hline
      \multirow{11}{*}{$\gamma=1$}&Periodic &0.33&130 &0.38&150&0.98&110&0.10&29&0.41&130&2.8&$1.9\times 10^3$&1.5&380&1.0&400\\
      &Eccentricity &0.17&26 &0.18&30&0.61&99&0.24&27&0.16&19&0.40&56&2.9&380&1.7&420\\
      &Precession &1.3&250 &1.5&310&3.8&580&0.62&140&1.6&280&9.0&$1.7\times 10^4$&11&$1.2\times 10^4$&12&$6.9\times 10^3$\\
      &Tilt &2.0&540&1.7&510&1.5&500&0.70&120&1.3&370&4.5&960&1.8&820&4.6&590\\
      &EP &2.0&460&2.1&430&5.1&$1.4\times 10^3$&0.76&120&2.3&510&3.8&$2.1\times 10^3$&6.6&$8.8\times 10^3$&6.4&$1.0\times 10^4$\\
      &ET &2.5&480&2.2&430&3.4&840&2.4&450&1.6&330&6.5&$4.0\times 10^3$&26&$2.9\times 10^4$&87&$2.4\times 10^5$\\
      &PT &11&720&10&710&12&920&6.6&510&8.3&580&26&$1.1\times 10^4$&58&$1.7\times 10^4$&120&$4.0\times 10^4$\\
      &EPT &8.4&720&7.4&590&5.3&790&6.0&770&7.1&720&7.6&$9.5\times 10^3$&22&$9.7\times 10^3$ &50&$4.8\times 10^4$\\
      &Insolation &14&710&12&560&14&770&8.4&560&13&610&27&$3.6\times 10^3$&83&$3.0\times 10^4$&150&$4.0\times 10^4$\\
      &Inclination &0.012&1.7&0.012&2.1&0.018&2.5&0.018&4.0&0.013&1.9&$9.8\times 10^{-3}$&1.9&$7.2\times 10^{-3}$&1.2&$6.2\times 10^{-3}$&1.3\\
      &GPI(G99) &0.16&18&0.17&23&0.097&18&0.020&3.0&0.21&26&0.094&28&0.065&7.8&0.059&15\\
      &GPI(C09) &2.5&120&2.5&110&3.3&260&0.20&14&2.9&170&7.3&880&2.6&290&1.1&150\\\hline
      \multirow{11}{*}{$\gamma=0.4$} &Periodic&1.7&200&1.3&140&0.30&17&6.4&$1.2\times 10^3$&1.2&92&0.44&26&0.98&84&2.8&520\\
      &Eccentricity &0.58&94&0.82&140&0.47&77&0.84&140&0.78&140&0.98&300&2.4&230&1.7&520\\
      &Precession &0.86&250&1.1&260&0.51&270&1.6&520&0.99&300&0.82&130&0.80&110&1.3&320\\
      &Tilt &30&$1.1\times 10^3$&24&990&2.9&71&78&$2.5\times 10^3$&17&640&2.7&220&6.9&130&20&540\\
      &EP &1.6&100&2.5&280&1.2&120&2.3&300&2.1&180&1.4&320&4.2&760&6.5&$4.1\times 10^3$\\
      &ET &12&$1.1\times 10^3$&11&760&1.4&78&60&$7.7\times 10^3$&7.0&390&2.3&710&9.2&$1.6\times 10^3$&23&$8.8\times 10^3$\\
      &PT &23&$1.1\times 10^3$&21&860&1.9&270&71&$7.5\times 10^3$&18&880&3.1&280&7.9&320&29&$4.0\times 10^3$\\
      &EPT &17&$1.3\times 10^3$&17&$1.2\times 10^3$&2.1&200&110&$9.5\times 10^3$&13&840&4.3&470&17&$1.4\times 10^3$&63&$8.6\times 10^3$\\
      &Insolation &18&970&17&900&2.9&370&52&$6.1\times 10^3$&15&670&4.2&240&8.4&390&31&$3.3\times 10^3$\\
      &Inclination &1.7&360&2.0&340&1.1&110&1.7&330&1.9&300&3.0&290&1.6&90&1.4&310\\
      &GPI(G99) &0.17&23&0.15&22&0.040&2.9&0.32&62&0.16&24&0.38&120&0.67&130&0.52&110\\
      &GPI(C09) &0.38&34&0.30&35&0.13&9.1&0.83&67&0.37&45&1.4&260&1.6&160&0.85&79\\\hline
   \end{tabular}}
  \end{adjustwidth}
\end{table}

We also change the prior distributions over some model parameters and keep others fixed. We apply this sensitivity test to the ML, HA and HAML data sets with time spans of 1 to 0\,Mya, 2 to 1\,Mya and 2 to 0\,Mya, respectively. These three data sets are representative and conservative because they contain the major terminations with large time uncertainties over the late Pleistocene and minor terminations identified in the HA data set, which is stacked from both benthic and planktic data sets. The values of $\gamma$ and the types of trend are fixed for different data sets such that the corresponding models can better explain the periodicity in the data. The above three data sets, time spans and corresponding $\gamma$ are listed in the first column of Table \ref{tab:sensitivity}. For all models, we change priors as follows:
\begin{itemize}
\item $\lambda=0 \rightarrow -10 \leq \lambda\leq 10$: Accounting for the possible time lags between the cause and effect. Here, $\lambda$ represents the time lag(s) of any model listed in Table \ref{tab:ts_models}, and $\lambda$ can be integers varying from -10 to 10 with a time unit of 1\,kyr. For models with a single component, a single time lag is assigned by shifting the corresponding time series to the past (positive lag) or to the future (negative lag). For compound models, each component is shifted independently. A positive lag means that the model leads the data while a negative lag means the model lags the data. The motivation for this is that the orbital inclination is claimed to force the climate with a time lag \citep{muller97}. 
\item $90\gamma<h_0<130\gamma \rightarrow 80\gamma<h_0<140\gamma \text{ and } 100\gamma<h_0<120\gamma$: Changing the prior distribution of $h_0$ is equivalent to changing the prior distribution of the period of a pacing model because the average period is about $h_0+10-a$ (see section \ref{sec:pacing_constant_climate}). However, the above changes only apply to models with $a\neq 0$ while the prior distribution of the Periodic model ($a=0$) is changed from $70\gamma<h_0<110\gamma$ to $60\gamma<h_0<120\gamma$ and $80\gamma<h_0<100\gamma$. For models with a sigmoid trend, the prior distribution of $k$, rather than $h_0$, is changed from $90<k<130$ to $80<k<140$ and $100<k<120$
\item $15\gamma<a<35\gamma \rightarrow 5\gamma<a<45\gamma \text{ and } 20\gamma<a<30\gamma$: these changes do not apply to the Periodic model, for which $a=0$.
\item $0<b<0.1 \rightarrow 0<b<0.2 \text{ and } 0<b<0.05$: Varying the prior distribution over the contribution fraction of the background, $b$, in the termination models. 
\item $\phi=0 \rightarrow -\pi<\phi<\pi$: The phase of the precession is closely related to the season of the insolation that forces the climate change. A varying phase enables a contribution of other season's insolation to the evidence of a termination model. 
\end{itemize}

The BFs and MLs for models with the above changes of priors are shown in Table \ref{tab:sensitivity}. For the ML data set over the last 1\,Myr, the Insolation and PT model are always better favored than the Tilt model for all changed priors. In addition, the PT and Insolation model without time lags are more favored than corresponding models with lags. This indicates that the Tilt and Precession pace the climate change without significant time lags. Over the early Pleistocene, the Tilt model is always best favored by the HA data set. The evidences of the EPT model vary a lot but are never higher than the Tilt model. For the HAML data set over the last 2\,Myr, the model combining a sigmoid trend and the PT forcing is the most favored for all changed priors. Moreover, the evidence for the PT model increases after shrinking the range of the background fraction, $b$, which suggests a high signal to noise ratio for the obliquity signal in the climate change over the past 2\,Myr.

We also see that the MLs for some compound models decrease a little after adding one free parameter. For example, the MLs for the ET model with $\gamma=1$ for the ML data set are $8.6\times 10^4$ and $1.1\times 10^4$ before and after adding a time lag to the model, respectively. This is caused by the sample noise in the prior distributions of models. Unlike MLs, BFs would not be sensitive to the sample noise because they measure the overall plausibility rather than the fitting ability of a model. 

\begin{table}[ht!]
\begin{adjustwidth}{-1.30in}{-1.30in}
  \caption{The BFs and MLs relative to the uniform model for changed prior distributions and different data sets. The priors are sampled using 100,000 points for each model. Each column name denotes a changed prior distribution over a common parameter. Note that the column with the name ``None'' gives the reference BFs and MLs for models without any change in priors. The first column gives the value of $\gamma$ (or threshold with a trend), the time scale and the data set for each set of models. Some models are not relevant for certain prior changes and thus the corresponding positions in the table are denoted by ``---''. }
  \label{tab:sensitivity}
  \centering
  \scalebox{0.45}{
  \begin{tabular}{c|c|ll|*{7}{ll|}ll}
    \hline
    \hline
    &Models& \multicolumn{2}{|c|}{None} & \multicolumn{2}{c|}{$\lambda\neq 0$} & \multicolumn{2}{c|}{$80\gamma<h_0<140\gamma$} & \multicolumn{2}{c|}{$100\gamma<h_0<120\gamma$} & \multicolumn{2}{c|}{$10\gamma<a<40\gamma$} & \multicolumn{2}{c|}{$20\gamma<a<30\gamma$} & \multicolumn{2}{c|}{$0<b<0.2$} & \multicolumn{2}{c|}{$0<b<0.05$} & \multicolumn{2}{c}{$-\pi<\phi<\pi$}\\
    \hline
    \multirow{11}{*}{\parbox[t]{3cm}{$\gamma=1$\\age: 1$\sim$0\,Mya\\data set: ML}}&Periodic&0.64&160&---&---&0.50&180&0.99&160 &0.65&170&0.65&180&0.98&170&0.30&160&---&---\\
    &Eccentricity&1.2&67&1.4&150&1.1&78&1.3&75 &1.2&121&1.1&50&2.0&50&0.36&44&---&---\\
    &Precession&14&$8.3\times 10^3$&13&$2.0\times 10^4$&13&$6.8\times 10^3$&6.5&$1.0\times 10^4$ &12&$7.5\times 10^3$&13&660&11&$8.7\times 10^3$&16&$9.0\times 10^3$&7.8&$1.1\times 10^4$\\
    &Tilt&2.7&$1.4\times 10^3$&1.4&$2.1\times 10^3$&11&$1.9\times 10^4$&1.7&$1.2\times 10^3$ &9.1&$1.2\times 10^4$&1.1&190&3.1&$1.1\times 10^3$&1.7&$1.5\times 10^3$&---&---\\
    &EP&5.2&$1.1\times 10^3$&5.3&$1.1\times 10^4$&9.5&$9.3\times 10^3$&3.9&780 &6.0&$1.6\times 10^3$&4.2&$2.0\times 10^3$&6.2&$8.4\times 10^3$&3.8&$4.1\times 10^3$&4.0&990\\
    &ET&21&$8.6\times 10^4$&8.9&$1.1\times 10^4$&48&$1.2\times 10^5$&9.7&$4.1\times 10^3$ &26&$1.1\times 10^5$&14&$5.4\times 10^3$&17&$7.9\times 10^4$&22&$9.7\times 10^4$&---&---\\
    &PT&220&$6.4\times 10^4$&22&$8.7\times 10^4$&170&$7.8\times 10^4$&210&$4.9\times 10^4$ &170&$8.1\times 10^4$&250&$7.8\times 10^4$&140&$7.0\times 10^4$&250&$6.8\times 10^4$&57&$1.4\times 10^5$\\
    &EPT&47&$7.4\times 10^4$&17&$2.9\times 10^4$&86&$6.3\times 10^4$&25&$1.1\times 10^4$ &50&$1.3\times 10^5$&48&$1.4\times 10^5$&36&$7.1\times 10^4$&51&$2.2\times 10^5$&20&$2.0\times 10^5$\\
    &Insolation&450&$1.1\times 10^5$&150&$1.4\times 10^5$&330&$1.1\times 10^5$&190&$1.4\times 10^5$&330&$1.1\times 10^5$&680&$1.3\times 10^5$&260&$1.2\times 10^5$&680&$1.4\times 10^5$&---&---\\
    &Inclination&0.018&4.1&0.015&7.1&0.032&30&0.0067&3.0&0.016&4.0&0.019&4.3&0.084&7.8&0.0017&0.73&---&---\\
    &GPI&0.16&31&0.14&74&0.48&560&0.041&12&0.16&87&0.11&28&0.36&26&0.057&70&---&---\\\hline
    \multirow{10}{*}{\parbox[t]{3cm}{$\gamma=0.4$\\age: 2$\sim$1\,Mya\\data set: HA}}&Periodic &2.5&200&---&---&2.0&190&1.5&91&2.5&200&2.5&200&2.1&200&2.9&200&---&---\\
    &Eccentricity &0.53&120&0.65&130&0.53&160&0.50&150&0.54&210&0.54&110&0.60&120&0.48&130&---&---\\
    &Precession &1.0&270&0.86&280&0.93&270&1.2&240&1.2&240&0.86&250&1.0&230&1.0&270&1.1&560\\
    &Tilt &22&$1.3\times 10^3$&9.0&$1.5\times 10^3$&31&$1.1\times 10^3$&6.3&$2.2\times 10^3$&23&$1.6\times 10^3$&21&$1.1\times 10^3$&16&$1.1\times 10^3$&27&$1.6\times 10^3$&---&---\\
    &EP &0.55&340&0.71&360&0.83&490&0.33&200&0.52&260&0.71&230&0.59&210&0.56&250&0.62&330\\
    &ET &5.9&$1.2\times 10^3$&3.0&$2.1\times 10^3$&15&$1.3\times 10^3$&1.5&$1.2\times 10^3$&6.2&$2.7\times 10^3$&5.5&$1.7\times 10^3$&4.6&$5.8\times 10^3$&6.8&$2.4\times 10^3$&---&---\\
    &PT &9.3&$2.9\times 10^3$&6.0&$2.2\times 10^3$&20&$1.2\times 10^3$&2.4&$1.5\times 10^3$&12&$3.5\times 10^3$&7.3&$1.6\times 10^3$&6.9&$1.4\times 10^3$&11&$2.2\times 10^3$&11&$2.0\times 10^3$\\
    &EPT &5.0&$1.0\times 10^3$&2.7&$1.1\times 10^3$&16&810&0.74&840&5.2&$1.1\times 10^3$&3.9&760&3.8&820&6.4&850&4.4&790\\
    &Insolation &10&910&4.1&830&13&880&3.5&780&11&980&7.8&890&7.4&910&13&950&---&---\\
    &Inclination &0.30&59&0.31&54&0.23&60&0.48&59&0.27&62&0.32&41&0.49&57&0.18&59&---&---\\\hline
    \multirow{10}{*}{\parbox[t]{3cm}{Sigmoid trend\\age: 2$\sim$0\,Mya\\data set: HAML}}&Periodic&110&$7.8\times 10^6$&---&---&29&$3.4\times 10^5$&6.1&$1.2\times 10^5$&33&$1.2\times 10^6$&23&$2.3\times 10^5$&25&$6.2\times 10^5$&69&$1.3\times 10^6$&---&---\\
    &Eccentricity &0.63&$4.3\times 10^3$&1.2&$1.7\times 10^4$&0.71&$4.3\times 10^3$&0.13&700&0.55&$1.6\times 10^3$&0.66&$4.3\times 10^3$&1.0&$3.6\times 10^3$&0.46&$3.7\times 10^3$&---&---\\
    &Precession &38&$7.3\times 10^5$&18&$8.7\times 10^4$&15&$1.7\times 10^5$&36&$2.5\times 10^5$&33&$4.9\times 10^5$&27&$1.6\times 10^5$&22&$1.2\times 10^5$&35&$3.6\times 10^5$&14&$1.3\times 10^5$\\
    &Tilt &590&$2.2\times 10^7$&220&$5.2\times 10^6$&350&$1.4\times 10^6$&390&$5.1\times 10^6$&460&$3.4\times 10^6$&560&$8.0\times 10^6$&300&$1.8\times 10^6$&480&$8.8\times 10^6$&---&---\\
    &EP &2.5&$3.6\times 10^4$&2.6&$9.7\times 10^4$&2.6&$5.6\times 10^4$&1.3&$3.9\times 10^4$&2.2&$2.5\times 10^4$&16&$1.3\times 10^6$&2.5&$1.1\times 10^4$&3.5&$9.9\times 10^4$&2.8&$6.8\times 10^4$\\
    &ET &$2.3\times 10^3$&$6.0\times 10^7$&660&$1.2\times 10^7$&$1.8\times 10^3$&$1.4\times 10^7$&700&$3.7\times 10^6$&$1.6\times 10^3$&$4.2\times 10^7$&$1.7\times 10^3$&$2.2\times 10^7$&600&$2.5\times 10^6$&$2.4\times 10^3$&$4.7\times 10^7$&---&---\\
    &PT &$4.1\times 10^4$&$2.0\times 10^8$&$3.2\times 10^3$&$3.4\times 10^7$&$3.0\times 10^4$&$1.3\times 10^8$&$6.7\times 10^4$&$1.5\times 10^8$&$4.6\times 10^4$&$2.3\times 10^8$&$4.0\times 10^4$&$1.4\times 10^8$&$2.1\times 10^4$&$1.0\times 10^8$&$7.7\times 10^4$&$2.2\times 10^8$&$1.1\times 10^4$&$8.6\times 10^7$\\
    &EPT &$1.0\times 10^4$&$1.2\times 10^8$&$1.9\times 10^3$&$2.9\times 10^7$&$6.8\times 10^3$&$1.3\times 10^7$&$8.9\times 10^3$&$4.8\times 10^7$&$1.2\times 10^4$&$1.8\times 10^8$&$1.0\times 10^4$&$1.4\times 10^8$&$7.8\times 10^3$&$1.2\times 10^8$&$1.9\times 10^4$&$1.4\times 10^8$&---&---\\
    &Insolation &$9.7\times 10^3$&$1.2\times 10^8$&$1.8\times 10^3$&$1.3\times 10^8$&$5.8\times 10^3$&$1.0\times 10^8$&$5.8\times 10^3$&$5.8\times 10^7$&$1.0\times 10^4$&$1.1\times 10^8$&$8.4\times 10^3$&$7.9\times 10^7$&$5.9\times 10^3$&$7.8\times 10^7$&$1.7\times 10^4$&$1.9\times 10^8$&---&---\\
    &Inclination &0.77&600&0.010&67&0.037&310&$4.3\times 10^{-3}$&33&0.011&66&0.011&120&0.089&65&$9.7\times 10^{-4}$&27&---&---\\\hline
  \end{tabular}
}
\end{adjustwidth}
\end{table}

\begin{figure}[ht!]
  \centering
  \includegraphics[scale=0.6]{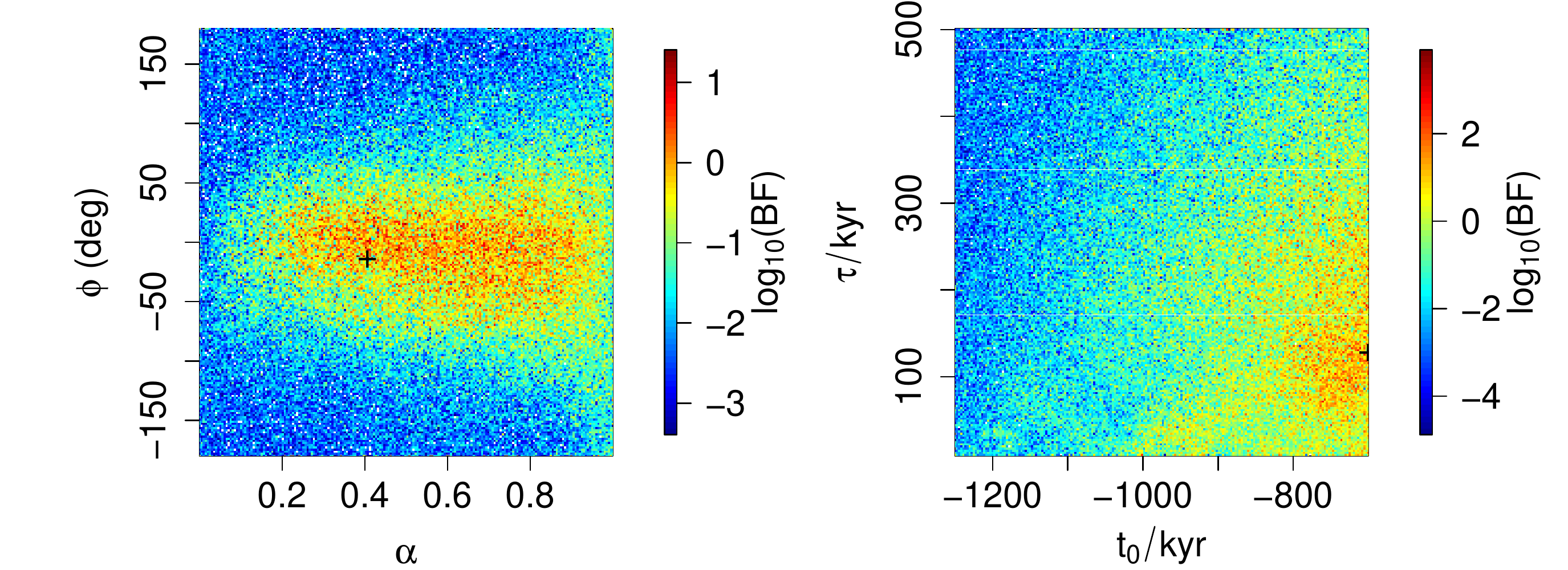}
  \caption{The $log_{10}$(BF) distributions for the PT model with $\gamma=1$ for the ML data set (left) and for the PT model with a sigmoid trend for the HAML data set (right). Left panel: the BF distribution over the contribution factor of precession ($\alpha$) and precession phase ($\phi$) over the last 1\,Myr. Right panel: the $log_{10}$(BF) distribution over the transition time ($t_0$) and transition time scale ($\tau$) over the past 2\,Myr. The point with the highest probability is denoted with a cross. For each panel, $log_{10}$(BF) is shown by the color bar, which is truncated slightly to highlight the regions with high values. Both samples have $10^6$ points.}
  \label{fig:likelihood_distribution}
\end{figure}

To specify the role of precession in pacing the major late-Pleistocene deglaciations, we show the BF distribution over the contribution factor of precession, $\alpha$, and the precession phase, $\phi$, in the PT model for the ML data set over the last 1\,Myr in the left panel of Fig. \ref{fig:likelihood_distribution}. It is evident that the main pace-maker under this model is the summer insolation in the Northern hemisphere which is dominated by precession with phase ranging from -50 to 50 degree. Although a fairly small contribution factor, $\alpha<0.1$, is strongly disfavored, $\alpha$ can be adjusted in a broad range to favor the data. This indicates that insolation is a multi-spatial pace-maker because $\alpha$ determines the combination of tilt and precession and thus the latitudinal insolation. That means, the main terminations over the last 1\,Myr are more likely to be paced by the insolation over the whole Northern summer and at multi-latitudes in the Northern Hemisphere than by the insolation at a specific spatial and temporal point. This is consistent with \citealt{huybers11}'s conclusion that ``the climate systems are thoroughly interconnected across temporal and spatial scales''.

Given that the PT model with a sigmoid trend is more favored than all models without a sigmoid trend, we confirm that the MPT is not gradual and can not be well modeled using a linear trend proposed by H07. We now find the optimized transition time scale $\tau$ and transition time $t_0$ based on the BF distribution over these two parameters for the PT model with a sigmoid trend. For the HAML data set over the past 2\,Myr and the PT model with a sigmoid trend, we marginalize the likelihood over all parameters except for $t_0$ and $\tau$, and show the BFs relative to the uniform model in logarithmic scale in color in the right panel of Fig. \ref{fig:likelihood_distribution}. We see that the region around $\tau=130$\,kyr and $t_0=-700$\,kyr have higher likelihoods than other regions. That means the MPT is a rather rapid transition which is consistent with the findings of \cite{honisch09,mudelsee97,tziperman03,martinez11} but seems to be inconsistent with the results of H07 and \cite{raymo04,liu04,medina05,blunier98}. This discrepancy may be superficial because we only model the time of terminations, which represents only one feature in $\delta^{18}$O variations. Thus we only conclude that the MPT was rapid for the transition from 41-kyr to 100-kyr cycles in terminations identified from $\delta^{18}$O records rather than for the transition of the average frequency, mean, standard deviation, time derivative and skewness of $\delta^{18}$O records (see H07 for details). In this context, we further conclude that the transition time scale is about 130\,kyr. We also find that the mid-point of the MPT is around 700\,kya based on our analysis of the HAML data set. This transition time is rather late compared with the mid-point of the MPT, $\sim$900\,kya, given by \cite{clark06}. Our value is based on the Bayesian inference for models which predict the time of terminations over the whole Pleistocene. In contrast, \cite{clark06} calculate the time of MPT using a frequentist approach, i.e. the time-frequency spectrogram which is obtained by dividing the Pleistocene into different time bins and calculate the power spectrogram for each. The above apparent or intrinsical discrepancies may all or partly be caused by different statistics. 

Finally, we change the sign of the contribution factor of forcing, $a$, to model possible anti-correlations between forcing models and the data over the late Pleistocene. We find that the previously disfavored models are still disfavored while the previously favored models have much lower evidences now. We conclude that our inference about the role of different orbital elements in pacing the deglaciations over the past 2\,Myr is robust to these changes of model priors. 

\section{Discussion and Conclusion}\label{sec:conclusion_climate}

With a Bayesian inference method, we confirm the dominant role of obliquity (i.e. tilt) in pacing the glacial terminations over the last 2\,Myr. This is consistent with the result of H07 that the bundle of obliquity cycles can explain the variation of the 100-kyr power in the climate over the course of the Pleistocene. However, unlike H07, the model with obliquity alone can model Pleistocene deglaciations, comprised of both minor and major deglaciations, better than the model combining obliquity with a trend. Thus, considering both minor and major terminations over the Pleistocene, obliquity is enough to explain at least the time of terminations before and after the MPT without re-parameterizing the model as done by H07 and \cite{raymo97,paillard98,ashkenazy04,paillard04,clark06}.

Performing model comparisons for models and data sets over different time scales, we observe that precession becomes important in pacing the $\sim$100-kyr glacial-interglacial cycles after the MPT. 
Based on the BF distribution for the precession-tilt (or PT) model, we confirm the conclusion drawn by \cite{huybers11} that the climate response to the precession-obliquity dominant insolation is interconnected over multiple spatial and temporal scales.

Through the comparison between models with a linear trend and models with a sigmoid trend, we find that the glacial terminations over the whole Pleistocene can be paced by a combination of precession, obliquity and a sigmoid trend. According to the BF distribution for the PT model with a sigmoid trend, the MPT has a time scale of $\sim$130\,kyr and a mid-point of around 700\,kya.
Thus the MPT seems to be caused by rapid internal changes in the climate system, and certain climate response modes may be switched on/off in this process \citep{paillard98,parrenin03,ashkenazy04,clark06}. 

In addition, the inclination forcing and geomagnetic forcing are very unlikely to cause climate changes over the last 2\,Myr. This at least weakens, if not excludes, the hypothesis that the Earth's orbital inclination relative to the invariant plane can influence the climate through Earth's accumulation of more interplanetary dust during the cross of the invariant plane \citep{muller97}. Our results also challenge the hypothesis that connects the geomagnetic paleointensity with climate changes over 100-kyr time scales \citep{channell09}. If the geomagnetic intensity does not cause climate change, the cosmic ray influx and the solar activity can only cause climate change through primordial variations rather than through the modulation from geomagnetic field.

Our conclusion based on model inference for different forcing models is robust to some changes of parameters, priors, time scales and data sets. The main uncertainty in our work comes from the identification of glacial terminations over the Pleistocene. But we have used different data sets of terminations to reduce this uncertainty. In future work, a more sophisticated Bayesian method (e.g. the method introduced by \citealt{bailer-jones12}) will be employed to compare more complex conceptual models for the full time series of climate proxies. Using this model inference approach, we may learn more about the mechanisms involved in the climate response to Milankovitch forcings.

\newpage

\chapter{The impact of the Gaia survey on researches of the Earth's extraterrestrial environment}\label{cha:gaia}
\section{Chapter summary}\label{sec:abstract_gaia}

Gaia will perform an astrometric and spectrophotometric survey of one billion stars in the Galaxy. This will benefit researches on the Earth's extraterrestrial environment by providing a large sample of Solar System minor bodies, and by accurately determining the motions of the Sun and nearby stars. I find that the Sun's orbit in the Galaxy can be integrated back to 100\,Myr with an uncertainty less than 6\%, leading to a reliable assessment of the extraterrestrial environment of the Earth. In addition, I calculate that Gaia will find almost all stellar encounters over the past 2\,Myr and about 90\% of encounters over the past 10\,Myr. This, together with simulations of the Oort cloud and analyses of current observed long period comets, will enable an unbiased assessment of whether the inner solar system is experiencing a comet shower. Gaia will determine the comet impact rate in the inner solar system with an uncertainty less than 50\% over the past 170 Myr. I also propose a Bayesian method to assess the link between LPCs and potential culprit stars. Gaia will improve this assessment mainly by determining accurate encountering times for stellar encounters. With astrometric and spectrophotometric observations of minor bodies in the Solar System, Gaia will advance our understanding of the origin of craters on terrestrial planets and satellites.

\section{Introduction}\label{sec:introduction_gaia}

The ESA Gaia satellite aims to observe one billion sources over the whole celestial sphere with visual brightness down to the 20th magnitude and provide astrometry, photometry and low resolution spectrophotometry for them. Designed with two three Mirror Anastigmat telescopes and equipped with an astrometric instrument (ASTRO), a blue/red photometer (BP/RP) and a Radial Velocity Spectrometer (RVS), Gaia was launched on 19 December 2013 and is mapping stars from the Sun-Earth Lagrange point L2. With astrometry measurements down to a precision of 10$\mu$as, Gaia will provide accurate positions, distances and proper motions, leading to a better understanding of the structure of the Galaxy, stellar structure and evolution, and will enable the discovery of many exoplanets and Near-Earth Objects (NEOs) \citep{perryman01,turon05,lindegren08,casertano08,bailer-jones08,tanga12}.

In this chapter, I will discuss how Gaia will improve my research on the influence of astronomical phenomena on the Earth. First, Gaia will provide a large sample of asteroids and comets, which will enable robust statistical studies of the characteristics of these objects. For example, the signal to noise ratio of the angular distribution of the perihelia of long period comets (LPCs) will be increased if future Gaia sample of LPCs is used \citep{horner02}. Second, Gaia will give an almost unbiased sample of stellar encounters which encounter the solar system in the past or future $\sim$10\,Myr \citep{feng14}. Calculated with robust Monte Carlo methods as done by \citep{bailer-jones15}, the times and distances of the closest approaches of these stellar encounters will enable us to identify potential comet showers and related terrestrial phenomena. Gaia will also detect bright stellar encounters over a longer time scale, possibly leading to the first confirmation of a link between terrestrial mass extinctions and stellar encounters (see chapter \ref{cha:comet} and \citealt{feng14}). Third, an accurate determination of the Galactic potential and the Sun's current phase space coordinates could be derived from future Gaia data, enabling a precise simulation of Sun's motion. An accurate simulation of the Sun's orbit will improve the investigation of the influence of the Sun's interstellar environment on the Earth's climate, biosphere and geology.

The following sections are structured as follows. In section \ref{sec:data_gaia}, I first describe the individual sources which Gaia will detect as well as how Gaia data can improve our knowledge of the composition and potential of the Galaxy. In section \ref{sec:research_gaia}, I investigate the benefit of Gaia to four topics about the influence of extraterrestrial pheonmena on the Earth. Finally, I conclude in section \ref{sec:conclusion_gaia}. 

\section{Big data from Gaia}\label{sec:data_gaia}

Gaia will provide astrometric, photometric and spectroscopic observations for 1 billion sources with each object observed $\sim$70 times. These objects include stars, asteroids, comets, brown dwarfs, supernovae and quasars. Here I mainly introduce the objects within or near the solar system. 

\subsection{Discrete sources}

\subsubsection{Stellar encounters}

Many previous studies have been performed to connect the catastrophic events, such as asteroid/comet impacts and mass extinctions, with stellar encounters of the solar system. Close and massive stellar encounters can penetrate the Oort cloud and produce comet showers on the Earth \citep{hoyle78,napier79,napier82,torbett89,shaviv03}. Although the Sun may also encounter large molecular clouds \citep{wickramasinghe08}, these clouds are too extended to exert strong perturbations on the Oort cloud over a $\sim$100\,Myr time scale \citep{thaddeus85}. Stars in the solar neighborhood are better candidates for possible culprits of comet showers. However, faint culprits, which encountered (or will encounter) the solar system over the past (or future) 10\,Myr, are not completely detected by past surveys such as Hipparcos \citep{rickman12}. Fortunately, according to my analysis, Gaia will provide a nearly unbiased sample of stars encountering the Sun within 5\,pc over the past 10\,Myr (see chapter \ref{cha:comet} and \citealt{feng14}).

To see the completeness of different types of stellar encounters in future Gaia catalogue, I quantify the strength of a stellar encounter using a proxy proportional to the impulse gained by the Sun from an encounter. This proxy is justified in our previous studies (see chapter \ref{cha:comet} and \citealt{feng14}), and is expressed here again as
\begin{equation}
  \gamma = \frac{M_{\rm enc}}{v_{\rm enc} r_{\rm enc}}
  \label{eqn:gamma2}
\end{equation}
where $M_{\rm enc}$ is the mass of a stellar encounter, $v_{\rm enc}$ is the velocity of an encounter with respect to the Sun, and $r_{\rm enc}$ is the closest distance from the Sun to the star's trajectory projected by $\boldsymbol{v}_{\rm enc}$. Because Oort cloud comets are isotropic and the total impulse gained by a comet is calculated in the heliocentric rest frame, $\gamma$, as a measure of the common impulse imposed on all comets, is appropriate to be used to predict the mean flux of LPCs induced by an encounter. However, more complex proxies can be obtained by applying different impulse approximations to study close and distant encounters \citep{fouchard11}. Considering the frequency and strength of encounters, the most probable culprits of LPCs are those encounters with $\gamma \in \lbrack 1\times 10^{-7}, 5\times 10^{-6}\rbrack{\rm M}_{\odot} ~{\rm km s}^{-1}~{\rm AU}^{-1}$. These encounters are defined as weak encounters \citep{feng14}. 

\begin{figure}[h!]
  \centering
  \includegraphics[scale=0.8]{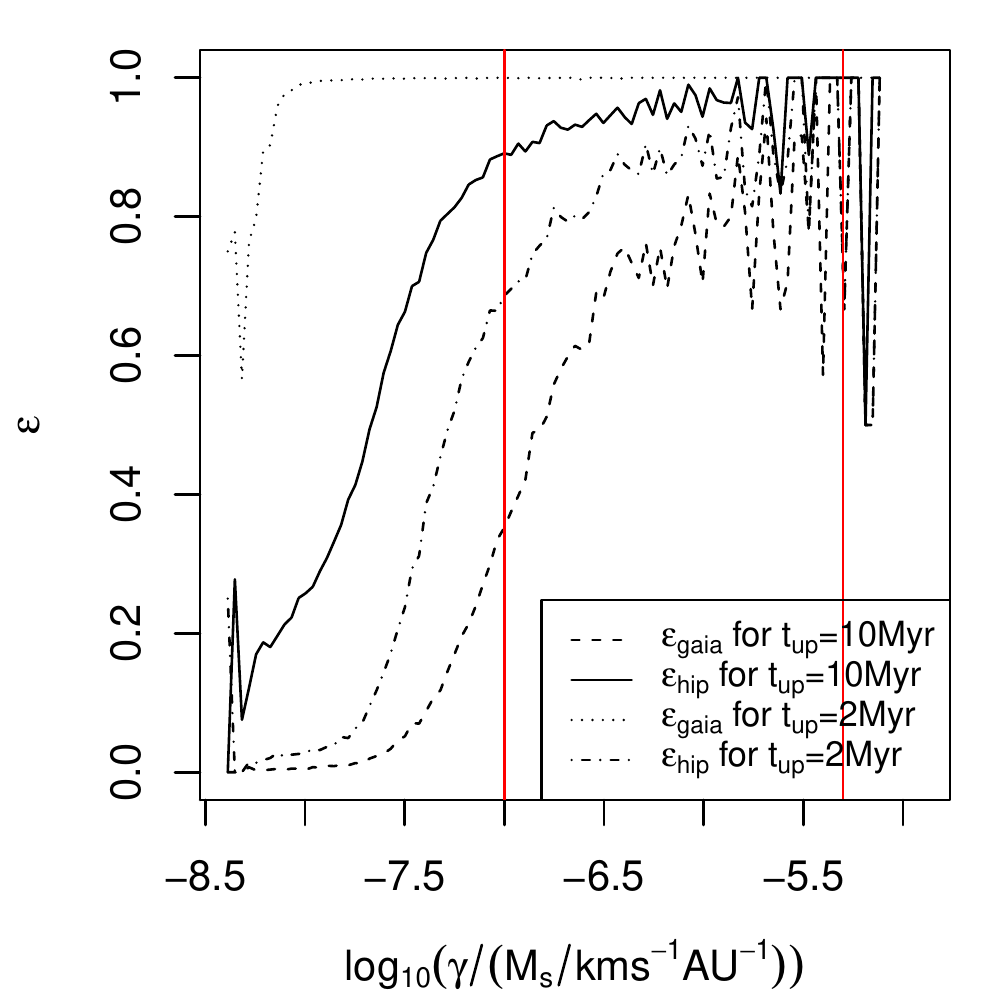}
  \caption{The completeness of the Hipparcos ($\epsilon_{\rm hip}$) and Gaia ($\epsilon_{\rm gaia}$) catalogues of stellar encounters over different time spans ($t_{\rm up}$). The two red lines are the $\gamma$ thresholds used to define weak encounters. The visual magnitude limits for Hipparcos and Gaia are set to be 12 and 20, respectively. I adopt the ranges of the magnitudes of the giants and white dwarfs in the Hipparcos catalogue to be $(-2,2)$ and $(10,15)$, respectively.}
  \label{fig:completeness}
\end{figure}

I generate 197,906 stellar encounters over 10\,Myr, according to the method introduced by \cite{rickman08} based on the incompleteness-corrected Hipparcos catalogue of stellar encounters provided in Table 8 of \cite{sanchez01}\footnote{The encountering frequencies of different stars in this catalog are calculated by modeling the luminosity, number density and velocity distribution of stars as a function of stellar types from quantities and data provided by \citep{mihalas81,allen85,gould97,jahreiss97}}. Then I calculate the current heliocentric distance of each encounter by assuming that the stellar velocity in the Heliocentric rest frame (HRF) ($v_{\rm enc}$) is constant from the time of closest approach ($t_{\rm enc}$) to the present. After calculating the current apparent magnitude from the heliocentric distance, mass, and absolute magnitude for each star, I truncate the apparent magnitude using the optimal magnitude limit of Hipparcos (visual magnitude of 12) and Gaia (visual magnitude of 20). Finally, I select those encounters that satisfy this limit, and calculate their population relative to the total population of encounters (defined as {\it completeness}, $\epsilon$) for different bins of $log_{10}{\gamma}$ ($\gamma$ is in unit of $M_{\odot} ~{\rm km}~{\rm s}^{-1}~{\rm AU}^{-1}$). I repeat the above processes for different time scales ($t_{\rm up}$), and show the results in Figure \ref{fig:completeness}.

I find that the Hipparcos catalogue has missed 46\% and 22\% of the weak encounters,  which have impacted the solar system over the past 10\,Myr and 2\,Myr, respectively. The missed encounters in the Hipparcos catalogue have high $\gamma$ but low magnitudes or masses. In other words, these missed encounters may have low masses and approach the solar system with small perihelia. The Gaia survey will find all those stars, which encounter the solar system within 2\,Myr of now, and find more than 90\% of encounters within 10\,Myr of now. This achievement is significant in terms of providing a real sample of recent encounters for Oort cloud simulations to answer the question of whether the solar system is experiencing a comet shower or not. 

\subsubsection{Asteroids, comets and objects in the outer solar system}

Gaia will observe about 400,000 asteroids, planetary satellites and comets, most of which are or will be discovered before the release of Gaia data \citep{mignard07}. Although the number of detected asteroids will be limited due both to window clipping and to apparent velocity of moving sources \citep{tanga12}, Gaia will directly measure sizes of about 1000 objects and masses of about 100 of them \citep{mignard07}. In addition, spin properties and overall shapes of about 10,000 objects can be derived from the disk-integrated photometry detected by several different epochs \citep{mignard07}. About 30\% of the NEO population and 25\% of the most hazardous population will be observed by Gaia \citep{bancelin12}. Designed mainly for doing astrometry, Gaia will improve the accuracy of orbital elements of these moving objects by a factor at least 2 orders of magnitude \citep{tanga12,eggl14}.

Although Gaia is not perfectly suited for comet discovery \citep{tozzi12}, Gaia will benefit comet research by improving the determination of cometary orbits, in particular when combined with ground-based observations \citep{rickman12}. These accurate orbital determination can be used to compute the non-gravitational forces imposed on comets and hence the mass of comets \citep{weiler10,tozzi12}. Most trans-Neptunian objects (TNOs) and Centaurs have a brightness below 20th magnitude and thus will not be detected by Gaia. However, Gaia will detect $~$50 Centaurs and TNOs on the whole sky, and will determine the sizes, albedos and even masses of some of these small bodies in the outer solar system.

Second, extrapolating from the experience of LINEAR, which has detected $\sim$20 comets per year and has a similar limiting magnitude with Gaia but only cover less than 1/10 of the sky \citep{horner02}, Gaia is expected to find about 1000 new LPCs. The improved dataset of LPCs will enhance the significance of any non-uniformity in the distribution of the LPCs' perihelia (or aphelia), leading to the confirmation or rejection of the hypothesis of Planet X in the solar system \citep{horner02}. 

\subsection{Galactic parameters inferred from Gaia}

Based on the physical parameters of more than 1 billion stars derived from Gaia observations, we can better understand the formation, evolution, structure and kinematics of different Galactic components, including bulge, bar(s), spiral arms, think and thick disks, and halo, and their gravitational potential \citep{perryman01,turon05}.

First, Gaia will observe more than 20 million bulge stars down to the main sequence photometrically, and observe some of them astrometrically despite the limitation from extinction and crowding in the Galactic bulge \citep{robin05, reyle08}. There is a bulge-star accessible window, where the extinction is not too high to make bulge stars too faint to be detected, and is not too low to make stars too crowded to be observed. In this window, Gaia will provide photometry (with a crowding limit of 600 000 stars deg$^{-2}$) of a sample of bulge stars and the RVS spectrometry (with the crowding limit of 40 000 stars deg$^{-2}$) of the clump giants in the sample \citep{reyle08}. This will allow 3-D structure studies of the interfaces between bulge, bar(s), spiral arms \citep{babusiaux10}, and the structure and mass of the inner Galaxy. The gravitational potential of the central region of the Galaxy can be well determined by the dynamical studies based on the radial velocities of bulge stars with RVS magnitude of $G_{\rm RVS}<16.2$.\footnote{$G_{\rm RVS}$ is the magnitude obtained by the Gaia RVS passband.} 

Second, the structure of the thin and thick disks will be well defined with Gaia's large sample of disk stars with 3D spatial and 2D/3D velocity measures \citep{vallenari05,bailer-jones08}. In addition, Gaia will measure the position and velocity of the spiral arms by observing 50,000 OB stars in the arms, and determine their distances to an accuracy about 10\% and their velocities to about 1 km/s \citep{bailer-jones08}. 

Third, combining with other surveys such as {\it Spitzer}, Gaia will determine the global potential of the dark matter halo \footnote{Yet the potential model of the substructure of the halo may not be strongly constrained by future Gaia data.}, with parameter uncertainties down to 2\%, by either providing accurate spatial and kinematic distributions of stellar streams (e.g. the Sagittarius debris system) or calibrating tracers (e.g. RR Lyrae stars) of these streams \citep{whelan13}.  

Finally, combining the proper motions and radial velocities derived from the Gaia astrometry and spectrometric measurements, precise simulations of the motions of the Sun and other stars in the Galaxy can be performed using the Galactic potential inferred from future Gaia data. 

\section{Extraterrestrial environment researches advanced by Gaia}\label{sec:research_gaia}

Gaia will not only increase our knowledge of the Milky Way but also improve our understanding of the influence of the extraterrestrial environment/phenomena on the Earth. The interstellar environment of the Earth can be precisely reconstructed with accurately simulated solar motion and Gaia's direct or indirect measurements of the time and location of different phenomena, such as supernovae (SNe) explosions, star formation regions and stellar/cloud/nebula encounters. As for the interplanetary environment of the Earth, Gaia will provide accurate kinematics for asteroids/comets and a better model of NEOs. This will bring forth an improved estimation of the terrestrial impact rate over the course of Earth's history (see section \ref{sec:impact_gaia}). 

\subsection{Improved reconstruction of the solar motion in the Galaxy}\label{sec:motion_gaia}

The Earth's climate, biosphere and geological activities can be influenced by various interstellar phenomena, such as cosmic rays, SNe explosions, gamma ray bursts and stellar encounters. For example, a large flux of cosmic rays from the Galaxy can kill organisms directly or damage their DNA/RNA. Cosmic rays may also influence the surface temperature on the Earth through enhancing the formation of condensation nuclei required for cloud formation \citep{thorsett95,scalo02,carslaw02,kirkby07,svensmark09,atri12}. The influence of these phenomena on the Earth can not be accurately assessed partly because of a lack of precise determination of the solar motion around the Galactic center. These assessments may be improved by the Gaia-improved Galaxy models and Sun's initial conditions.

This section aims to investigate how Gaia will improve the determination of the solar orbit, which is integrated from the Sun's current phase space coordinates. To do this, I integrate the Sun's orbit accounting for uncertainties in the Sun's initial conditions, and in the Galactic potential. As is done in \cite{feng14} and in chapter \ref{cha:comet}, I sample the Sun's initial conditions and parameters of Galaxy model. Considering that the gravitational force imposed on the Sun is mainly from the disc, I also randomly select the parameters of the disc potential, i.e. the total mass, $M_d$, the scale length, $a_d$, and the scale height, $b_d$, according to the Gaussian distributions with means given by \cite{feng14} and standard deviations equal to 20\% of corresponding parameters \citep{juric08,mcmillan11}. As a result, I generate 1000 sets of all the above parameters. For each parameter set, I simulate the solar orbit over the past 5\,Gyr by integrating the Newtonian equations using the {\tt lsoda} method implemented in the R package {\tt deSolve}, with a time step of 1\,Myr (see section \ref{sec:orbit_biodiversity} and \cite{feng13} for details). Then I calculate the deviation of each solar orbit from the reference orbit, which is simulated by adopting the mean values of parameters. The deviation of each solar motion in location and velocity relative to the reference solar motion are $|\boldsymbol{r}-\boldsymbol{r_0}|$ and $|\boldsymbol{v}-\boldsymbol{v_0}|$, respectively. $\boldsymbol{r}$ is the 3D position of the Sun relative to the Galactic center, and $\boldsymbol{v}$ is the 3D velocity of the Sun in the rest frame centered on the Galactic center. The subscript ``0'' represents the reference values of $\boldsymbol{r}$ or $\boldsymbol{v}$. 

\begin{figure}
  \centering
  \includegraphics[scale=0.7]{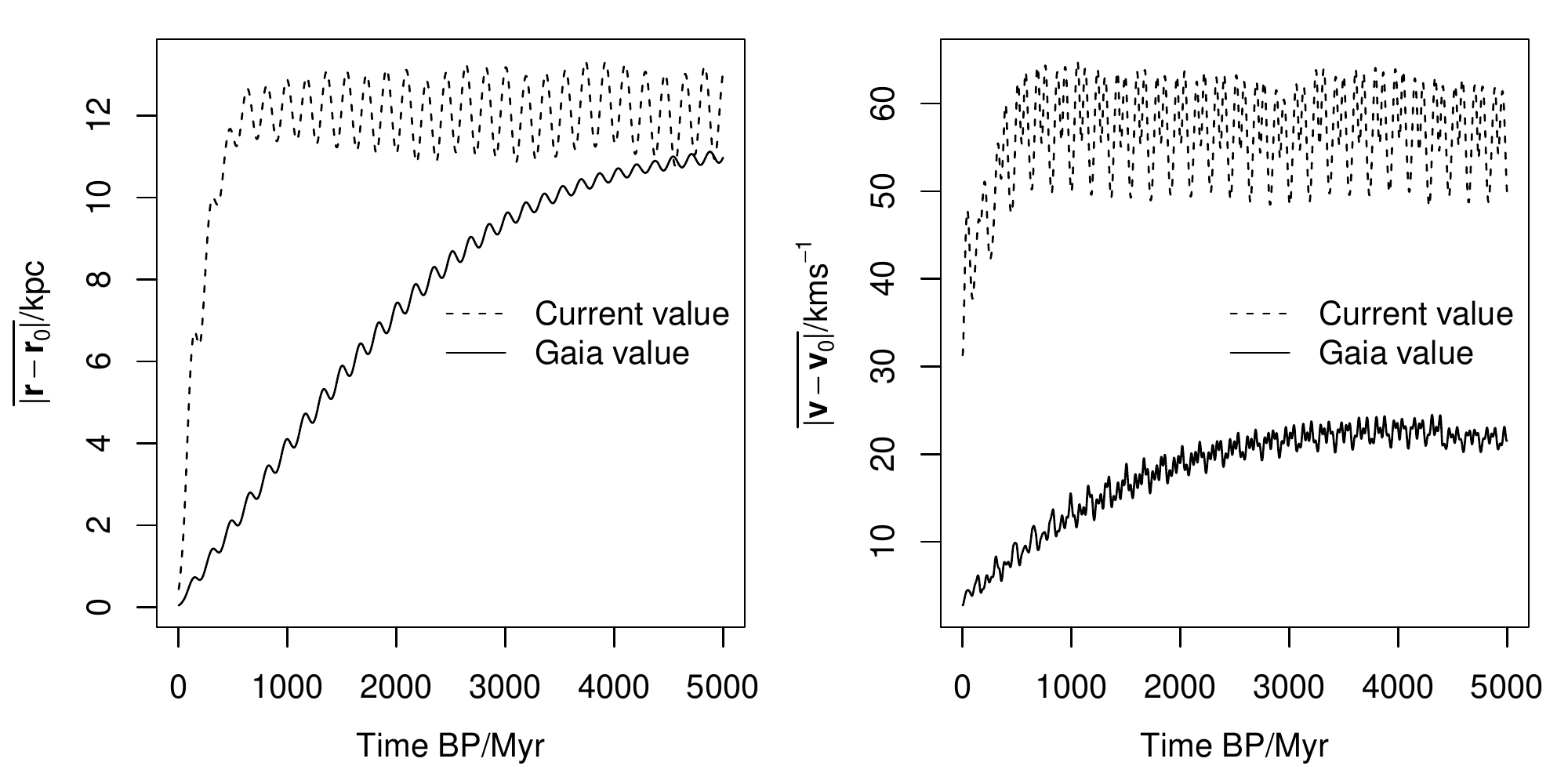}
  \caption{The mean deviation of reconstructed solar orbits in position (left) and velocity (right) relative to the reference solar orbit. The two deviation parameters, $\overline{|\boldsymbol{r}-\boldsymbol{r_0}|}$ and $\overline{|\boldsymbol{v}-\boldsymbol{v_0}|}$, are calculated based on simulations of 1000 solar orbits for current data and future Gaia data.}
  \label{fig:gaia_current_orbit_err}
\end{figure}

Then I repeat the above Monte Carlo procedure for the Gaia case by decreasing the uncertainties in initial conditions and model parameters by a factor of 10. This factor is a rough estimate, which account both for the ratio between the astrometry precisions of Hipparcos and Gaia, and for the limit of Gaia in measuring the current distance between the Sun and the Galactic center. Finally, I compare the deviation parameters based on current observations and those based on Gaia observations in Fig. \ref{fig:gaia_current_orbit_err}.

In Fig. \ref{fig:gaia_current_orbit_err}, we see that the deviation in the Sun's position increases over the past 500\,Myr and oscillates around 12\,kpc before that. That means the Sun's motion is unpredictable before 500\,Myr ago, given our current knowledge of the Galaxy. The Sun's velocity also deviates significantly from the reference value after 500\,Myr (about 2 orbital periods). In contrast, based on Gaia data, the solar orbit can be traced back to about 100\,Myr ago with position deviation less than 0.5\,kpc (about 6\% of current Sun's galactocentric distance) and to 500\,Myr ago with position uncertainties of around 2\,kpc (about 26\% of current Sun's galactocentric distance). These results assume an axisymmetric potential and an asymmetric and time-varying potential may cause the Sun's migration \citep{feng14}. Considering that Gaia may improve the accuracy of the axisymmetric and asymmetric component of the Galactic potential with a same factor, different potential models may not change the deviation parameters too much. Our results indicate that it is possible to study the extraterrestrial environment of the Earth based on an accurate reconstruction of the Sun's motion in the Galaxy over the past 100\,Myr (including the K-Pg extinction event). 

\subsection{The variation of comet impact rate in the inner solar system}\label{sec:impact_gaia}

Comet impacts on the Earth or other terrestrial planets and satellites in the inner solar system can be caused by Jupiter-family comets (JFCs), Halley-type comets (HTCs) or LPCs. Only the impact rate due to LPCs is modulated by the solar motion, which can modulate the tidal force imposed on the Oort cloud. The uncertainties in the variations of the comet impact rate over a long time scale (e.g. $>$100\,Myr) arise partly from numerical errors in simulations of the Oort cloud, and partly from the lack of knowledge of the tidal force from stellar encounters and the Galaxy. I will discuss Gaia's improvement on the accuracy of simulations of the LPC flux induced by the Galactic tide and stellar encounters as follows.

In section \ref{sec:motion_gaia}, I find that the uncertainties in the solar orbit over the past 100\,Myr are less than 6\%. As is demonstrated in \cite{feng14} and chapter \ref{cha:comet}, the vertical Galactic tide, which has uncertainties the same as the solar orbit, is nearly proportional to the comet flux induced by the tide alone. Thus Gaia will significantly reduce the uncertainty in the tide-induced comet flux.

The encounter-induced comet flux is highly dependent on the sample of stellar encounters. As is shown in Fig. \ref{fig:completeness}, Hipparcos may miss nearly 50\% weak encounters which encountered (or will encounter) the solar system over the past 10\,Myr (or future 10\,Myr) with an impact factor less than $4\times 10^5$\,AU. This is also the reason why I use a stochastic model of stellar encounters rather than a real sample to simulate the encounter-induced LPC flux in \cite{feng14}. However, Gaia will give us a much larger sample of stellar encounters, enabling the simulation of realistic encounter-induced LPC flux with a real sample of stellar encounters. I quantify the influence of Gaia's encounter sample on the calculation of comet impact rate from two aspects: the proxy-based LPC flux and the simulated LPC flux.

I use the encounter strength factor (see Eqn. \ref{eqn:gamma2}), $\gamma$, to predict the encounter-induced LPC flux. Thus the encounter-induced LPC flux over a certain period is proportional to the sum of $\gamma$ of all stars encountering the solar system over this period. To quantify our knowledge of the past LPC flux, I simulate 197,906 encounters according to \cite{rickman08}'s model and count the encounters which have brightness higher than Gaia's lower magnitude limit. Then I define the completeness of our knowledge of the past comet impact rate as
\begin{equation}
E=\sum\limits_i^{N_{\rm gaia}}\gamma_i/\sum\limits_j^N\gamma_j,  
\label{eqn:comp-LPC}
\end{equation}
where $N_{\rm gaia}$ is the number of encounters can be detected by Gaia in the encounter sample and $N$ is the total number of encounters in the sample.

\begin{figure}
  \centering
  \includegraphics[scale=0.7]{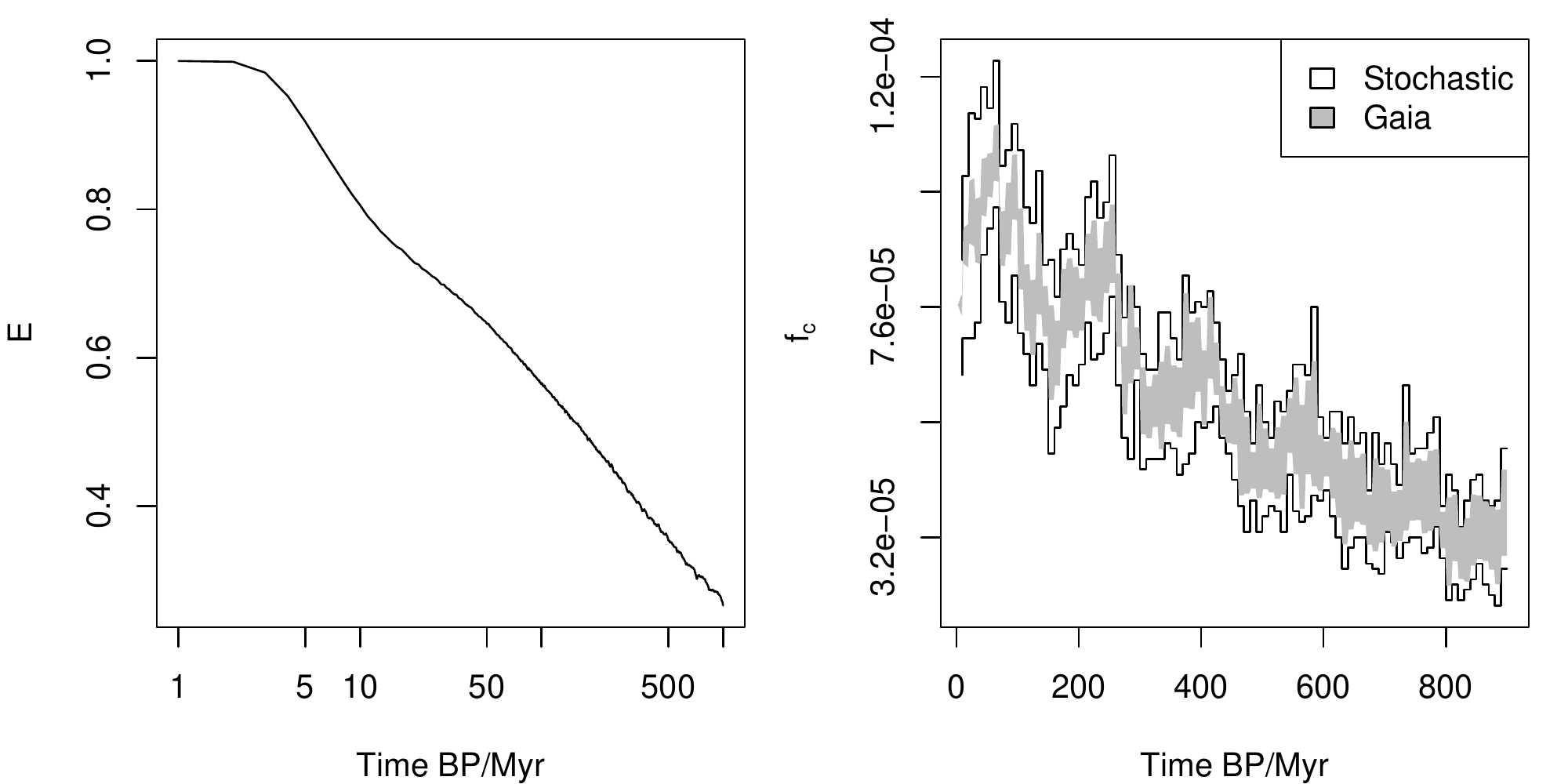}  
\caption{The completeness of our knowledge of the past LPC flux, E, (left panel) and the uncertainties in simulated LPC flux, $f_c$, (right panel). In the right panel, the two black lines represent the maximum and minimum LPC relative fluxes and the gap between them represents the uncertainty in our knowledge of the LPC flux. The grey region represents the uncertainty in the Gaia-updated knowledge of the LPC flux.}
\label{fig:comp-LPC}
\end{figure}

The left panel of Fig \ref{fig:comp-LPC} shows the completeness $E$ varying with time. It is evident that Gaia will give a nearly complete sample of encounters over the past 2\,Myr, and before 2\,Myr ago the completeness is inversely proportional to the logarithm of time before present. In other words, the completeness decrease rapidly over the past 170\,Myr and gradually decreases from 0.5 to 0 from 170\,Myr ago towards the past. Because the motions of stars are reversible in time, this analysis is also applicable to future stellar encounters of the solar system.

To see the improvement from Gaia more directly, I simulate the Oort Cloud under the perturbations from the Galactic tide and stellar encounters. I generate 10 samples (with 39,581 particles in each) of stellar encounters over the past 1\,Gyr from the stochastic model given by \cite{rickman08}, and simulate $10^5$ Oort cloud comets for each encounter sample according to the Oort cloud model given by \cite{dones04}. Then I adopt the axisymmetric Galactic potential expressed in Eqn. \ref{eqn:Phi_sym}, and use the AMUSE software to simulate the orbits of Oort cloud comets back towards 1\,Gyr ago. I count the comets entering the loss cone (i.e. perihelion less than 15\,AU) as LPCs\footnote{Although the observable zone, $q_c<5$\,AU, is always used as the criterion of defining observable LPCs, the loss cone is also a proper criterion to guarantee both a small sample noise in LPCs and an unbiased analysis of variations in the LPC flux \citep{feng14}. } and remove these particles in the following simulations. We choose the LPCs injected from 100\,Myr to 1\,Gyr for the following analyzation because the Oort cloud has a relaxation time of about 100\,Myr (see \citealt{feng14} and chapter \ref{cha:comet}). I divide the left time span into 10\,Myr time bins, count the LPCs over each time bin for each encounter sample and calculate the relative LPC flux $f_c$ by dividing the number of LPCs by the bin size and the total number of comets in the original Oort cloud sample. The range of $f_c$ for a given time bin is determined by the maximum and minimum $f_c$ for different encounter samples. The difference between these two relative LPC fluxes represents the uncertainty in the past LPC flux or cometary impact rate which is mainly caused by the incompleteness of the current sample of stellar encounters. This uncertainty is shown in the right panel of Fig. \ref{fig:comp-LPC} by the gap between the two $f_c$ curves.

Gaia will significantly reduce the uncertainty of our knowledge of past LPC flux by providing a real sample of encounters, although Gaia may miss half the stellar encounter that encountering the solar system more than 100\,Myr ago. The right panel of Figure \ref{fig:comp-LPC} shows the uncertainty in the LPC flux remaining in simulations based on the Gaia sample of encounters by multiplying the difference between maximum and minium $f_c$ with $1-E$. The mean LPC flux based on the Gaia sample is set to be equal to the mean LPC flux based on the 10 samples of stochastic encounters. In the right panel, we see that there is a semi-periodic variation (with a period around 200\,Myr) superposed on a trend in the LPC flux. The trend, which decreases toward the past, is caused by the depletion of the phase space of Oort cloud comets by the Galactic tide while the semi-periodic variation is caused by the change of the Galactic tide arising from the solar motion. In addition, the stochastic peaks in the LPC flux shown in Fig. \ref{fig:comp-LPC} are generated by stellar encounters, which can trigger comet showers with a duration around 2\,Myr. Gaia will provide a real sample of encounters to reduce the randomness of comet showers, and also improve the accuracy of reconstructed solar motion to reduce the uncertainties in the semi-periodic tide-induced LPC flux. In Fig. \ref{fig:comp-LPC}, we see that Gaia can decrease the uncertainties in the LPC flux significantly over the first 100\,Myr. However, even with the future Gaia data, we cannot obtain an accurate temporal distribution of the LPC flux over more than 200\,Myr, although Gaia may detect strong stellar encounters that triggered comet showers more than 200\,Myr ago. 

\subsection{The link between LPCs and stellar encounters}\label{sec:LPC_enc_gaia}

The LPCs are delivered from the Oort cloud through perturbations from the Galactic tide and stellar encounters of the solar system. Although the tide may play a main role in making Outer Oort Cloud (OOC) comets discernable at the present epoch \citep{heisler86,delsemme87,matese92,wiegert99,matese11}, the phase space of the OOC can be depleted by the tide and thus needs to be replenished by stellar impulses, which is so called ``synergy effect'' \citep{heisler87,rickman08}. This synergy effect makes a separate treatment of perturbations from the tide and stellar encounters inviable, and thus enable an investigation of the direct or indirect link between every observed LPC and the corresponding stellar encounters (or culprits) \citep{rickman12}.

\subsubsection{Bayesian method}\label{sec:LPC_statistics_gaia}

I apply a Bayesian method to infer the potential culprit of a LPC from the observables of the LPC and potential culprit stars. To do this, I treat each LPC as a data point and each encounter as a culprit candidate. The Bayesian method can be used to assign a list of weighted culprits for each LPC, with each weighting factor equal to the posterior of a culprit given a LPC. The posterior of model M can be calculated according to Bayes' theorem:
\begin{equation}
  P(M|D)=\frac{P(D|M)P(M)}{P(D)},
  \label{eqn:bayes_rule1}
\end{equation}
where $P(D|M)$ is the marginalized likelihood or evidence, $P(M)$ is the prior of model M, and $P(D)$ is a normalization factor. The reader is referred to section \ref{sec:bayes} and \cite{bailer-jones09} for an in-depth explanation of Bayesian statistics. 

In the case of finding culprits, model M contains observables of a stellar encounter (defined as culprit parameters, $\boldsymbol{\alpha}$), and parameters of models of the Galactic tide and the Oort cloud (defined as auxiliary parameters, $\boldsymbol{\beta}$). Parameter set $\boldsymbol{\alpha}$ is consisted of the stellar mass, $M_{\rm enc}$, the velocity of the stellar encounter relative to the Sun, $\boldsymbol{v_{\rm enc}}$, the closest heliocentric distance of the encounter, $\boldsymbol{r_{\rm enc}}$, and the time at the closest approach, $t_{\rm enc}$. The data, D, is the measured values of observables of a LPC, including the perihelion, $q_c$, the semi-major axis, $a_c$, the direction of the perihelion in terms of Galactic latitude, $b_p$, and longitude, $l_p$, the orbital inclination of the cometary orbit relative to the invariant plane, $i_c$, and the time at the perihelion, $t_p$. According to the above definitions, $D=\{q_c,a_c,b_p,l_p,i_c,t_p\}$ and $\boldsymbol{\alpha}=\{M_{\rm enc},\boldsymbol{v_{\rm enc}},\boldsymbol{r_{\rm enc}},t_{\rm enc}\}$. Then the evidence in Eqn. \ref{eqn:bayes_rule1} is expressed as
\begin{equation}
  P(D|M) = \int_{\boldsymbol{\theta}} P(D|\boldsymbol{\theta},M)P(\boldsymbol{\theta}|M) d\boldsymbol{\theta}
  \label{eqn:bayes_rule2}
\end{equation}
where $P(D|\boldsymbol{\theta},M)$ is the {\it likelihood} and $P(\boldsymbol{\theta}|M)$ is the prior distribution of culprit and auxiliary parameters, i.e.\ $\boldsymbol{\theta}=\{\boldsymbol{\alpha},\boldsymbol{\beta}\}$. 

Following \cite{bailer-jones11,bailer-jones11-err} and section \ref{sec:bayes}, I account for the measurement uncertainties of observables of LPCs and encounters by including additional model parameters, $\boldsymbol{\sigma}=\{\sigma_i\}$, $\boldsymbol{\delta}=\{\delta_j\}$, and $\boldsymbol{\mu}=\{\mu_j\}$, into the likelihood. First, I define the {\it measurement model} as the probability of obtaining a measured value of observable i, $D_i$, given both the true value, $q_i$, and the measurement uncertainty, $\sigma_i$, of the observable. The measurement model is a Gaussian distribution, which is
\begin{equation}
  P(D_i|\sigma_i,q_i) = \frac{1}{\sqrt{2\pi}\sigma_i} e^{-(D_i-q_i)^2/2\sigma_i^2}.
  \label{eqn:measurement_LPC}
\end{equation}
Although the true values, $\boldsymbol{q}$, of LPC observables may depend on each other, the measured value of an observable only depend on the corresponding measurement uncertainty and the true value. Considering these, the likelihood of model M is written as 
\begin{align}
  P(D|\boldsymbol{\sigma},\boldsymbol{\theta},M) &= \int_{\boldsymbol{q}} P(D|\boldsymbol{\sigma},\boldsymbol{q})P(\boldsymbol{q}|\boldsymbol{\theta},M)d \boldsymbol{q}\nonumber\\
  &=\int_{\boldsymbol{q}}\prod_i P(D_i|\sigma_i,q_i) P(\boldsymbol{q}|\boldsymbol{\theta},M)d \boldsymbol{q}. 
  \label{eqn:observable_like}
\end{align}

After defining the likelihood, I move on to define the prior distributions of model M, i.e.\ $P(\boldsymbol{\theta}|M)$. Here I only define the prior distributions of the culprit parameters, $P(\boldsymbol{\alpha}|M)$, based on the measured values, $\boldsymbol{\mu}$, and measurement uncertainties, $\boldsymbol{\sigma}$. Because these measurements give hyperparameters of prior distributions, I define $P(\boldsymbol{\alpha}|\boldsymbol{\mu},\boldsymbol{\sigma},M)$ as ``prior distributions'' rather than ``measurement model''. Then the prior distributions of parameters $\boldsymbol{\theta}$ become
\begin{equation}
  \begin{aligned}
    P(\boldsymbol{\theta}|\boldsymbol{\mu},\boldsymbol{\delta},M)&=P(\boldsymbol{\alpha},\boldsymbol{\beta}|\boldsymbol{\mu},\boldsymbol{\delta},M)\\
    &=P(\boldsymbol{\beta}|M)\prod_j P(\alpha_j|\mu_j,\delta_j)\\
    &=P(\boldsymbol{\beta}|M) \prod_j \frac{1}{\sqrt{2\pi}\delta_j} e^{-(\alpha_j-\mu_j)^2/2\delta_j^2},
  \end{aligned}
  \label{eqn:observable_evi}
\end{equation}
where $\delta_j$ and $\mu_j$ are hyperparameters for observable $\alpha_j$. In other words, the values of these hyperparameters are obtained from observations of stellar encounters. In the above equation, I assume that the auxiliary parameters, $\boldsymbol{\beta}=\{\beta_j\}$, and the culprit parameters, $\boldsymbol{\alpha}=\{\alpha_j\}$, are independent of each other. 

With the likelihood in Eqn. \ref{eqn:observable_like} and the prior distributions in eqn. \ref{eqn:observable_evi}, the evidence in Eqn. \ref{eqn:bayes_rule2} can be expressed as
\begin{align}
  P(D|\boldsymbol{\sigma},\boldsymbol{\mu},\boldsymbol{\delta},M) &= \int_{\boldsymbol{\theta}} P(D|\boldsymbol{\theta},\boldsymbol{\sigma},M)P(\boldsymbol{\theta}|\boldsymbol{\mu},\boldsymbol{\delta},M) d\boldsymbol{\theta} \nonumber\\
  &=\int_{\boldsymbol{\alpha}} \int_{\boldsymbol{\beta}} \int_{\boldsymbol{q}} [\prod_i P(D_i|\sigma_i,q_i)] [\prod_j P(\alpha_j|\mu_j,\delta_j,M)] P(\boldsymbol{q}|\boldsymbol{\alpha},\boldsymbol{\beta},M) P(\boldsymbol{\beta}|M) d\boldsymbol{\alpha}d\boldsymbol{\beta}d \boldsymbol{q}
\end{align}
The above evidence is calculated using a Monte Carlo method:
\begin{equation}
  P(D|\boldsymbol{\sigma},\boldsymbol{\delta},\boldsymbol{\mu},M) = \sum_{\boldsymbol{\alpha_k}\in \mathcal{N}(\boldsymbol{\mu},\boldsymbol{\delta})}\sum_{\boldsymbol{\beta_l}\in P(\boldsymbol{\beta}|M)} \sum_{\boldsymbol{q_m}\in \mathcal{N}(\boldsymbol{D},\boldsymbol{\sigma})} P(\boldsymbol{q_m}|\boldsymbol{\alpha_k},\boldsymbol{\beta_l},M)
  \label{eqn:observable_evidence}
\end{equation}
where $\boldsymbol{\alpha_k}\in \mathcal{N}(\boldsymbol{\mu},\boldsymbol{\delta})$ means that a parameter set, $\boldsymbol{\alpha_k}$, is drawn from Gaussian distributions with means and standard deviations equal to $\boldsymbol{\mu}$ and $\boldsymbol{\delta}$, respectively. Likewise, $\boldsymbol{\beta_l}\in P(\boldsymbol{\beta}|M)$ means that a parameter set, $\boldsymbol{\beta_l}$, is drawn from the corresponding prior distributions, $P(\boldsymbol{\beta}|M)$. Similarly, $\boldsymbol{q_m}\in \mathcal{N}(\boldsymbol{D},\boldsymbol{\sigma})$ means that a set of true values of LPC observables are drawn from Gaussian distributions with means and standard deviations equal to $\boldsymbol{D}$ and $\boldsymbol{\sigma}$.

Finally, the Bayes factor (BF) of culprit model $M_1$ relative to the reference model, $M_0$, is calculated as
\begin{equation}
  \rm BF_{\rm 10} = \frac{P(D|\boldsymbol{\sigma},\boldsymbol{\delta},\boldsymbol{\mu},M_1)}{P(D|\boldsymbol{\sigma},M_0)}.
  \label{eqn:LPC_BF}
\end{equation}
In the above equation, the reference model, $M_0$, predicts characteristics of a LPC delivered by the tide alone, and thus does not have any culprit parameter. Similar with the calculation of the evidence for the culprit model, the evidence for the reference model, $M_0$, is
\begin{equation}
  P(D|\boldsymbol{\sigma},M_0)=\sum_{\boldsymbol{\beta_l}\in P(\boldsymbol{\beta}|M_0)}\sum_{\boldsymbol{q_m}\in \mathcal{N}(\boldsymbol{D},\boldsymbol{\sigma})} P(\boldsymbol{q_m}|\boldsymbol{\beta_l},M_0).
  \label{eqn:reference_evidence}
\end{equation}

If the BF is larger than 10, I claim that a LPC is delivered by a combined perturbation from a culprit star and the Galactic tide rather than by the perturbation from the tide alone. I compare the evidences of these culprit models, and select out the most plausible culprit using the criterion of BF. In the case of finding one culprit for a few LPCs, we can average the evidences of culprits models for these LPCs, and compare these mean evidences for different culprit stars. In the following section, I will introduce culprit models, i.e. $P(\boldsymbol{q}|\boldsymbol{\theta},M)$.

\subsubsection{A culprit model }

The simulations of the Oort Cloud are very computationally expensive because they have to cover a large range of spacial and temporal scales. These simulations calculate gravitational forces over a spatial scale extending from the perihelion (about 5\,AU) to the aphelion (typically 10,000\,AU) of a LPC, and over a time scale extending from the duration of the encountering process of a stellar encounter (about 1\,kyr) to the evolution time scale of the Oort cloud (more than 100\,Myr). The limitation in computation power motivates me to find the link between LPCs and culprits using semi-analytical methods. To do this, I build a culprit model based on the results from \cite{matese02,dybczynski02,dybczynski02b,dybczynski05,dybczynski06,dybczynski02}.
This culprit model only aims to establish a framework for further investigations of the link between LPCs and potential culprits rather than to be complex enough to select out every LPC-culprit pairs. 

Because strong encounters are rare and super weak encounters, i.e. $\gamma <1\times 10^{-7}{\rm M}_{\odot} ~{\rm km}~{\rm s}^{-1}~{\rm AU}^{-1}$, cannot effectively perturbe the Oort Cloud, weak encounters are representative for building a culprit model. A culprit model is actually a series of probability distributions of observables of LPCs, $\boldsymbol{q}$, which depend on the observables of the culprit star, $\boldsymbol{\alpha}$, and other model parameters, $\boldsymbol{\beta}$. Because the observables of a LPC, such as the perihelion, $q_c$, are nearly independent of the characteristics of the culprit star, $\boldsymbol{\alpha}$, I only build a culprit model to predict the probability distributions of some LPC observables, $\boldsymbol{q}=\{t_p,a_c,l_c\}$, where $t_p$ is the time at perihelion, $a_c$ is the semi-major axis, $l_c$ is the Galactic longitude. Because these LPC observables are not very sensitive to the change of the parameters of the Galaxy model (see \citealt{feng14} and chapter \ref{cha:comet}), I assume that these observables are independent of the auxiliary parameters, $\boldsymbol{\beta}$ for a given Oort cloud model (e.g.\ the DLDW model introduced by \citealt{dones04}). The analytical functions of the distributions of these observables are described below. 

First, I use a log-normal distribution, $\ln\mathcal{N}(\mu,\sigma)$, to predict the time-varying flux of encounter-induced LPCs. Here I ignore the influence of the tide on the temporal distribution of LPC flux because the tide cannot significantly change the LPC flux over a short time scale, e.g.\ 2\,Myr. That is, if there is an abrupt change of LPC flux, it is more probable to be caused by a stellar encounter rather than the Galactic tide. Assuming that the time at perihelion, $t_p$, of a LPC only depends on the time at perihelion, $t_{\rm enc}$, of the culprit star, the probability of observing a LPC at time $t_p$ given a culprit star is
\begin{align}
  P(t_p|t_{\rm enc},M_{\rm enc}, r_{\rm enc}, v_{\rm enc},M)&=P(t_p|t_{\rm enc},\Theta_t,M)\nonumber\\
  &=\ln\mathcal{N}(t_p-t_{\rm enc}-t_0;\mu_t,\sigma_t)\nonumber\\
  &=\frac{1}{(t_p-t_{\rm enc}-t_0)\sqrt{2\pi}\sigma_t}e^{-[ln(t_p-t_{\rm enc}-t_0)-\mu_t]^2/(2\sigma_t^2)}
  \label{eqn:model_tp}
\end{align}
where some observables of an encounter are replaced by the substitutional parameters, $\Theta_t$, comprising of $t_0$, $\mu_t$ and $\sigma_t$. $t_0$ is the minimum time needed to deliver a LPC by an encounter, $\mu_t$ and $\sigma_t$ are the mean and standard deviation of a log-normal distribution, respectively.

To fit the parameters in the log-normal distribution, I use the temporal distributions of LPC fluxes given in \cite{dybczynski02} for three stellar encounters. \cite{dybczynski02} simulated the Oort cloud using the impulse approximation in a Monte Carlo simulation scheme. The three encounters all have one solar mass and an encountering velocity of 20\,km/s, and three different perihelia, 60 000\,AU, 30 000\,AU and 90 000\,AU (see Fig. 7,8,9 of \citealt{dybczynski02}). The Oort Cloud model for the first encounter is the DQT model introduced by \cite{duncan87} while the Oort cloud model for the other two encounters is the DLDW model based on the results in \cite{dones04}. However, as mentioned by \cite{dybczynski02}, the simulations based on different Oort cloud models change the amplitude of the LPC flux induced by an encounter rather than the shape of the temporal distribution of the LPC flux.    

I fit parameters $\Theta_t$, except for $t_{\rm enc}$, to the three examples, and give the optimized parameters in unit of Myr as follows,
\begin{align}
  t_0=1.2,~\mu_t=0.64,~\sigma_t=0.80\nonumber\\
  t_0=0.40,~\mu_t=0.52,~\sigma_t=1.2\\
  t_0=1.8,~\mu_t=1.1,~\sigma_t=0.45\nonumber~,
  \label{eqn:par_lognormal}
\end{align}
where $t_0$, $\mu_t$ and $\sigma_t$ are all in unit of Myr. Because I only have three data points and the encounter strength parameter, $\gamma$, is a reasonable proxy to predict the encounter-induced comet flux (see chapter \ref{cha:comet} and \citealt{feng14}), I will use linear functions of $\gamma$ to fit $t_0$, $\sigma_t$ and $\mu_t-\sigma_t^2$. Adopting the unit of $\gamma$ as $10^{-6}{\rm M}_\odot{\rm km s}^{-1}{\rm AU}^{-1}$, the fitted functions are
\begin{align}
  t_0&=-1.191\gamma+2.346\nonumber\\
  \sigma_t&=0.630\gamma+0.175\\
  \mu_t&=-1.522\gamma+1.551+\sigma^2\nonumber
  \label{eqn:fit_lognormal}
\end{align}

Replacing $t_0$, $\mu_t$ and $\sigma_t$ in Eqn. \ref{eqn:model_tp} with the above functions, I calculate the probability of observing an encounter-induced LPC at a certain time given the encounter strength parameter, $\gamma$, of an encounter.

Similarly, I also use a log-normal distribution to predict the distribution of semi-major axes, $a_c$, of LPCs induced by a culprit star based on the results of \cite{dybczynski02, matese02}. The distribution is
\begin{equation}
  \begin{array}{r@{}l}
  P(a_c|t_p,t_{\rm enc},M_{\rm enc}, r_{\rm enc}, v_{\rm enc},M)&{}\displaystyle =P(a_c|t_p,t_{\rm enc},\Theta_a,M)\\
  &{}\displaystyle =\ln\mathcal{N}(a_c-a_0;\mu_a,\sigma_a)  \\
  &{}\displaystyle =\frac{1}{(a_c-a_0)\sqrt{2\pi}\sigma_a}e^{-[ln(a_c-a_0)-\mu_a]^2/(2\sigma_a^2)},
  \end{array}
  \label{eqn:model_a}
\end{equation}
where $a_0$ is a start point where the probability is larger than 0, and $\Theta_a$ are substitutional parameters which determine the shape of a log-normal distribution. 
In the above equation, I have simplified the model by ignoring the dependence of $a_c$ on $t_p$, although there is a weak dependence as shown in Fig. 6 of \cite{matese02}.

Then I fit the above function to the distributions of LPCs' semi-major axes for the three stellar encounters presented in Figs. 10-11 of \cite{dybczynski02}. Using a method the same as that used for fitting the probability distribution of $t_p$, I obtain the following fitted functions:
\begin{equation}
\begin{array}{l}
  a_0 = -2.354\gamma+4.831\\
  \sigma_a = 0.6  \\
  \mu_a = -0.357+\sigma_a^2,
\end{array}
\label{eqn:fit_a}
\end{equation}
where $a_0$, $\sigma_a$ and $\mu_a$ are in unit of $10^4$\,AU. From Eqn. \ref{eqn:model_a} and \ref{eqn:fit_a}, we can calculate the probability of getting a value of the semi-major axis of a LPC, $a_c$, given a culprit. 

Finally, I analytically model the distribution of Galactic longitudes, $l_c$, of the perihelia of encounter-induced LPCs. Induced by a weak stellar encounter, the LPC flux can be enhanced along a broad (about $40^\circ$--$60^\circ$) track of aphelia directions extending about $150^\circ$ along the celestial sphere \citep{matese02}. This enhanced flux does not significantly change the overall shape of the latitudinal distribution of LPCs' perihelia \citep{matese02,dybczynski02b,feng14} because the latitudinal distribution is mainly shaped by the anisotropic tidal force from the Galactic tide. In addition, the synergy effect between the tide and encounters can not remove the signal of a weak shower in the longitude distribution \citep{dybczynski02b,rickman08,feng14}. Thus I use the Galactic longitude, $l_c$, rather than the Galactic latitude, $b_c$, of a LPC to infer the potential culprit.

The longitude distribution of LPCs' perihelia are time dependent and has a bar-shaped profile in the stellar reference frame \citep{dybczynski02b,dybczynski02}. Because the amplitude of the flux depends on the time at the LPC perihelion, $t_p$ and the time at the culprit perihelion, $t_{\rm enc}$, the probability of $l_c$ given a culprit is modeled using a scaled Gaussian distribution with a background,
\begin{equation}
  \begin{array}{r@{}l}
    P(l_c|t_{\rm p},t_{\rm enc},b_{\rm enc},l_{\rm enc},M_{\rm enc}, r_{\rm enc}, v_{\rm enc},M)&{}=P(l_c|t_p,t_{\rm enc},\Theta_l,M)\\
    &{}\displaystyle =\frac{A}{\sqrt{2\pi}\sigma}e^{-(l_c-\mu_l)^2/(2\sigma_l^2)}+B,
  \end{array}
  \label{eqn:model_lc}
\end{equation}
where $\Theta_l$ are substitutional parameters, including the amplitude of a Gaussian function $A$, the background $B$, the mean, $\mu_l$, and standard deviation, $\sigma_l$, of a Gaussian function. I fit the above model to the latitudinal distributions of LPCs' perihelia presented by \cite{matese02,dybczynski02}, and the fitted functions are
\begin{equation}
  \begin{array}{r@{}l}
  A&{}=400\gamma P(t_p|t_{\rm enc},M_{\rm enc}, r_{\rm enc}, v_{\rm enc},M)\\
  B&{}=1\\
  \mu_l&{}=l_{\rm enc}+180~{\rm if }~ l_{\rm enc}<180{\rm ,~otherwise }~ l_{\rm enc}-180,  \\
  \sigma_l&{}=20\gamma,
  \end{array}
\end{equation}
where $\gamma=M_{\rm enc}/(v_{\rm enc}r_{\rm enc})$ is in unit of $10^{-6}{\rm M}_{\odot} ~{\rm km}~{\rm s}^{-1}~{\rm AU}^{-1}$, $\mu_l$, $\sigma_l$ and $l_{\rm enc}$ are in unit of degree, and $P(t_p|t_{\rm enc},M_{\rm enc}, r_{\rm enc}, v_{\rm enc},M)$ can be calculated from Eqn. \ref{eqn:model_tp} and \ref{eqn:par_lognormal}.

Finally, from the models given in Eqn. \ref{eqn:model_tp}, \ref{eqn:model_a} and \ref{eqn:model_lc}, I derive the model of a culprit candidate, i.e.
\begin{eqnarray}
  P(\boldsymbol{q}|\boldsymbol{\theta},M)&=&P(t_p,a,l_c|t_{\rm enc},l_{\rm enc},M_{\rm enc},r_{\rm enc},v_{\rm enc},M)\nonumber\\
  &=&P(t_p|t_{\rm enc},M_{\rm enc},r_{\rm enc},v_{\rm enc},M)\nonumber\\
  &&\times P(a|M_{\rm enc},r_{\rm enc},v_{\rm enc},M)\nonumber\\
  &&\times P(l_c|t_p,t_{\rm enc},l_{\rm enc},M_{\rm enc},r_{\rm enc},v_{\rm enc},M).
\label{eqn:model_culprit}
\end{eqnarray}
However, this culprit model assumes the existence of a culprit for a LPC. To see whether there is a culprit star related to a LPC, it is necessary to build a reference model, which predicts the characteristics of a LPC without the effect of a culprit star. It is natural to use tide-induced LPCs to build such a model. I model the probability of $t_p$, $a_c$ and $l_c$ given a model of the Galactic tide and a model of the Oort cloud as 
\begin{equation}
  \begin{array}{r@{}l}
    P(t_p|M)&{}=1/10\\
    P(a_c|M)&{}=\mathcal{N}(1/a_c;\mu=1.5\times 10^{-5},\sigma=0.4\times 10^{-5})/a_c^2\\
    P(l_c|M)&{}=1/360,
  \end{array}
  \label{eqn:ref_model}
\end{equation}
where $P(a_c|M)$ is given according to the top-left panel of Fig. 5 in \cite{rickman08}, $P(t_p|M)$ is a uniform distribution over a typical time span of a weak comet shower, i.e. 10\,Myr, and $P(l_c|M)$ is uniform over the range of the Galactic longitude. 

By calculating the BF of the culprit model relative to the above reference model according to Eqn. \ref{eqn:LPC_BF}, I estimate the probability of a link between a LPC and a potential culprit using the criterion of BF \citep{kass95}. In the following section, I will use an example to specify how I find a culprit without and with the future Gaia data. 

\subsubsection{Finding culprits of LPCs}

Gaia will improve the assessment of the link between a stellar encounter and a LPC by providing a large (or even complete) sample of stellar encounters and accurate astrometry (or kinematics) for LPCs and encounters. The former will enable us to find culprit stars from a relatively complete sample of encounters. The latter, meanwhile, will decrease the measurement uncertainties of the observables of LPCs and encounters, and thus make the likelihood more sharp and the BF more capable to select out true culprits (see Eqn. \ref{eqn:observable_evidence}). 

To quantify the improvement from Gaia, I apply the Bayesian method to the LPCs injected by the tidal force from Algol. Despite a large uncertainty in its radial velocity, Algol is an appropriate example to show how Gaia will improve the assessment of the link between LPCs and culprits by decreasing the measurement errors of stellar encounters. I perform the following steps to recover the culprit star. (1) I adopt the mean values of the orbital elements of Algol based on current observations, and use the Galactic potential from \cite{feng14}, and simulate the motion of Algol relative to the Sun using the AMUSE software; (2) then I simulate Oort cloud comets under the gravitational perturbations from the Galactic tide and Algol in the AMUSE framework; (3) I count the comets injected into the loss cone as LPCs, and calculate the BF of the culprit model relative to the reference model for each LPC; (4) finally, I change the measurement uncertainties of the observables of Algol according to the precision of the Gaia astrometric observations, and repeat step (3) to see the improvement from Gaia. I further specify these steps as follows. 

First, for the Algol system, I adopt an initial heliocentric distance, $r_\star$, of 28.5\,pc, a radial velocity, $v_r$, of 4\,km/s and other orbital elements given in the Hypparcos catalog revised by \cite{leeuwen07}. Adopting the axisymmetric Galactic potential described in section \ref{sec:galaxy} and \cite{feng14}, and simulating the motion of Algol in a heliocentric reference frame, I find that Algol encountered the Sun with a closest distance of $\sim$2.7\,pc about 7.1\,Myr ago.  These values are a little different from those given by \cite{matese02, dybczynski06}, which may be caused by the uncertainties in initial conditions and the usage of a different Galactic potential.

\begin{figure}
  \centering
  \includegraphics[scale=0.5]{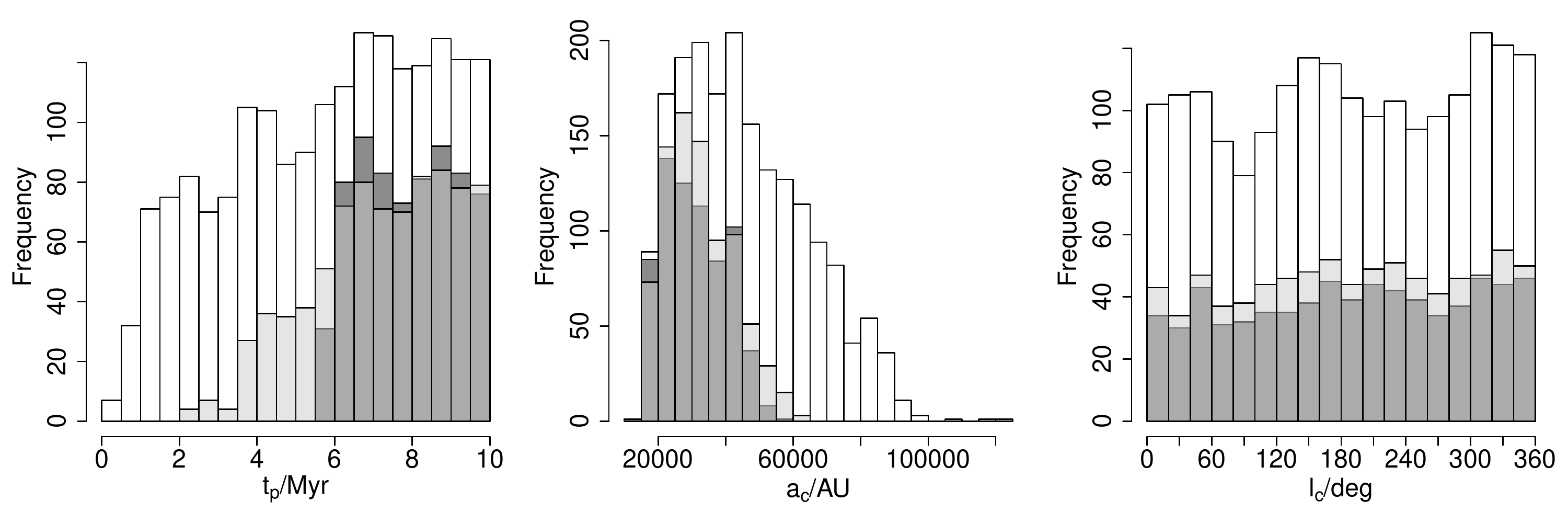}
  \caption{The distributions of the time at perihelion, $t_p$, the semi-major axis, $a_c$, and the Galactic longitude, $l_c$ of LPCs injected from the Oort cloud over the past 10\,Myr by the Galactic tide and Algol. The distributions for LPCs with BFs larger than 10 are shown in light gray and dark gray for current observations and Gaia observations, respectively. }
  \label{fig:algol_LPC}
\end{figure}

Second, in the AMUSE framework, I simulate the motions of 10 million Oort cloud comets generated from the DQT model backward to 10\,Myr ago under the perturbations from the tide and Algol with a time step of 0.01\,Myr. The comets with perihelia, $r_{\rm enc}$, less than 15\,AU (i.e. loss cone) are counted as LPCs, which is justified by the linear relationship between the flux of comets with $r_{\rm enc}<15$\,AU and that with $r_{\rm enc}<5$\,AU given in chapter \ref{cha:comet} and \cite{feng14}. The distributions of the time of the closest approach, the semi-major axes and the Galactic longitudes of LPCs are shown in Fig. \ref{fig:algol_LPC}.

Third, assuming that the distributions of observables of LPCs do not depend on the Galaxy model given a fixed Oort cloud model, I calculate the evidence for each LPC for the Algol according to Eqn. \ref{eqn:observable_evidence}. I sample the observables of Algol following Gaussian distributions with standard deviations given by \cite{sanchez01}, i.e. $\Delta r_{\rm enc}=0.93$\,pc, $\Delta v_{\rm enc}=0.7$\,km/s, $\Delta t_{\rm enc}=1.1$\,Myr. The error of $r_{\rm enc}$ can be approximately transformed into the error of $l_c$, i.e.
\begin{equation}
  \Delta l_c \simeq \Delta r_{\rm enc}/r_\star = 0.033~{\rm rad}.
  \label{eqn:delta_lc}
\end{equation}
I ignore the error of $M_{\rm enc}$ and uncertainties in parameters of LPCs because I use simulated LPCs, which do not have errors in their parameters.

\begin{figure}
  \centering
  \includegraphics[scale=0.8]{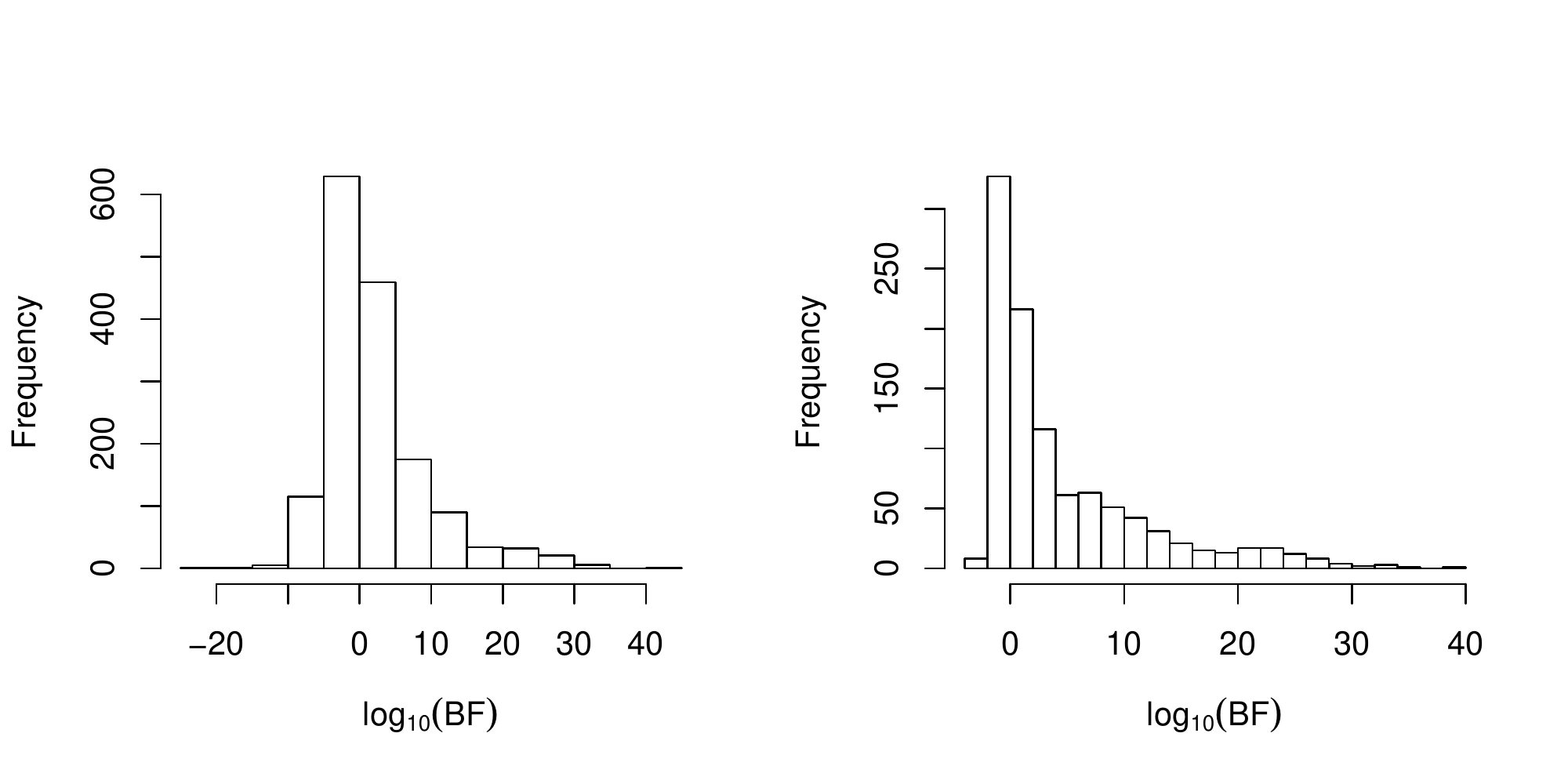}
  \vspace{-4ex}
  \caption{The distribution of BFs for all LPCs delivered by the tidal force from Algol based on current data (left) and future Gaia data (right). }
  \label{fig:BF_LPC}
\end{figure}

After sampling culprit parameters of the model, $\boldsymbol{\alpha}=\{r_{\rm enc}, v_{\rm enc}, t_{\rm enc}, M_{\rm enc}\}$, I calculate the distributions of LPC observables predicted by the culprit model with a set of culprit parameters, $P(\boldsymbol{q}|\boldsymbol{\alpha_k},M)$, according to Eqn. \ref{eqn:model_culprit}. Then the BF of the culprit model relative to the reference model for each LPC is calculated according to Eqn. \ref{eqn:LPC_BF}. I calculate the BFs for all LPCs, and show the distribution of BFs in Fig. \ref{fig:BF_LPC}. I also show the the distributions of parameters of the LPCs with BFs larger than 10 in Fig. \ref{fig:BF_LPC}.

Replacing the measurement errors of the parameters of Algol with errors reduced by Gaia, I repeat the above procedure of calculating evidences for LPCs. Assuming that current observations are more or less based on the astrometry measurements given by the Hipparcos survey, Gaia will improve the precision of the proper motions of nearby stars by a factor of at least 100 because its astrometry resolution is at least 100 times higher than the resolution of Hipparcos. Considering the fact that Gaia may not significantly improve the determination of the Galactic potential and the radial velocities of some stellar encounters, I propose that Gaia may improve the precision of culprit parameters with a factor of around 10. Although Gaia will not observe Algol, which has a brightness higher than the magnitude limit of Gaia, i.e. $5.7<G<20$, I assign Gaia errors to the culprit parameters of Algol to theoretically predict the improvement from Gaia. 

With new Gaia priors of culprit parameters, I calculate the BFs for LPCs and show the BF distribution in Fig. \ref{fig:BF_LPC}. I find that the BF distribution is more concentrate for the culprit model based on the Gaia data. Specifically, there are less LPCs with BFs lower than 1, indicating that some Algol-induced LPCs may be classified as tide-induced LPCs with current data of Algol. The distributions of observables, $t_p$, $a_c$ and $l_c$, of LPCs with BFs larger than 10 are shown in Fig. \ref{fig:algol_LPC}. I find that Gaia significantly improves the determination of the encountering time, $\Delta t_p=0.11$\,Myr, and thus reduce the probability of classifying LPCs which have perihelion time, $t_p$, before the encountering time, $t_{\rm enc}=7.1$\,Myr, as encounter-induced LPCs. However, distributions of other parameters are not sensitive to different astrometry resolutions.

In sum, we conclude that Gaia will mainly improve the assessment of the link between LPCs and stellar encounters by reducing the uncertainties in the determination of the encountering time of culprit stars. 

\subsection{Connecting craters with impactors}\label{sec:crater_imp_gaia}

Some craters on terrestrial planets and satellites result from cometary and asteroid impacts.
Gaia's precise astrometric measurement of comets and asteroids will enable precise calculation of the non-gravitational force imposed on the minor bodies in the solar system. In addition, the BP/RP spectrum of asteroids can be used to establish a new asteroid taxonomy, and thus disentangle between asteroids belonging to different dynamical families overlapping in the space of the orbital proper elements \citep{mignard07}. This will better constrain the disruption time of the sources of large asteroid families, and thus shape the debates on whether the Cretaceous-Tertiary (K/T) extinction event is caused by the impact of a member of the Baptistina asteroid family or a comet \citep{raup84,bottke07,moore13}. In addition, Gaia's accurate astrometric observations of NEOs will enable a more reliable calculation of the collision danger posed by these Earth-crossing objects.

There are three ways to find the link between craters and impactors: the link between the age of a crater and the planet-crossing (or satellite-crossing) time of an impactor, the link between the geochemical characteristics of a crater and the chemical composition of an impactor, the link between the location of a crater and the perihelion direction of a culprit star which could induce an anisotropic comet shower. The first method will benefit from the accurate planet-crossing time calculated from the orbital elements of minor bodies provided by Gaia. The impact of Gaia on the other two approaches is qualitatively assessed as follows. 

The physical parameters of an impactor can be inferred from the geochemical characteristics of the corresponding crater \citep{anders89,kyte98,mukhopadhyay01}. Gaia will perform multi-epoch spectrophotometry observations of the solar system's minor bodies, and provide a spatially unbiased sample of asteroids/comets. The surface properties and composition of some asteroids/comets in this sample will be derived by the low resolution spectrophotometry, i.e.\ BP/RP spectrum, provided by Gaia. In addition, Gaia will provide 3D velocities of many moving objects in the Solar system, leading to accurate distributions of mass, size and velocity for different types of asteroids and comets. With the Gaia-improved knowledge of chemical and physical parameters of different impactors, the geochemical properties of asteroids and comets can be better modeled to simulate impacts and cratering processes, and predict the geochemical features of impact craters. Comparing the characteristics of the simulated craters generated by different types of impactors with the geochemical evidences recorded in real craters, we may know better the origin of the impactor of a crater by analyzing the geochemical characteristics of the crater. 

Another method to assess the link between an impactor and a crater is to study the angular distribution of craters on the surface of terrestrial planets and satellites over a certain time span. As is concluded in chapter \ref{cha:comet} and \cite{feng14}, the solar apex motion and the Galactic tide will generate an anisotropic angular distribution of the LPCs' perihelia through imposing anisotropic tidal forces on the Oort cloud. This anisotropic LPC flux may cause some latitudinal variation of the impact rate of a planet, despite the long-term variation in inclination and obliquity of the orbit of the planet \citep{feuvre08,feuvre11}. Since the direction of the solar apex varies with time, the latitudinal distribution of the cometary impact rate will also vary with time. If the dating of craters are accurate enough (e.g. less than 10\,Myr uncertainties), we can infer the motion of the Sun relative to the local standard of rest and the temporal distribution of comet showers in the inner solar system. Yes these are all high order effects which would be difficult to convincingly detect even with future Gaia data.

Within the Bayesian framework introduced in section \ref{sec:LPC_enc_gaia}, it is possible to identify the impactors of craters using the above three methods simultaneously. With accurate astrometric and photometric observations of moving objects within and out of the solar system, Gaia will enable a reliable reconstruction of the bombardment history over at least the past 100\,Myr in the solar system and a comprehensive assessment of possible links between biological/climatic/geological changes on the Earth (or other terrestrial planets/satellites) and minor objects in the solar system and even stellar encounters. 

\section{Discussion and conclusion}\label{sec:conclusion_gaia}

In this work, I have quantified the impact of the Gaia survey on studies of the influence of extraterrestrial environment on the Earth. Based on simulations of the motions of stellar encounters from the catalog provided by \cite{sanchez01}, I find that Gaia can detect more than 90\% stellar encounters, which encountered or will encounter the solar system over the past or future 10\,Myr, compared with about 50\% detection efficiency of the Hipparcos survey. With the 6D (3D velocity plus 3D position) data provided by Gaia for each stellar encounter, it becomes feasible to assess the link between a LPC and the potential culprit star within a Bayesian framework. With accurate astrometric and spectrophotometric observations of 1 billion stars, Gaia will also constrain parameters of Galaxy models, and enable precise simulations of the Sun's motions with uncertainties less than 6\% in the Galaxy over the past 100\,Myr. However, the precision of the simulated solar motion can be influenced by the time-varying asymmetric Galactic potential, which may not be well constrained by Gaia data.

The Gaia-improved Galaxy model and integration of the solar orbit will shrink the uncertainties in the determination of the encountering time of stellar encounters, leading to stronger evidences for or against potential links between stellar encounters and LPCs. Assuming an isotropic Oort cloud model and an axisymmetric Galactic potential, I find that Gaia data will enable a reconstruction of the LPC flux (or LPC impact rate) in the inner solar system with an uncertainty less than 50\% over the past 170\,Myr. 

Furthermore, Gaia will provide astrometric and photometric observations of asteroids and comets, including NEOs, TNOs, LPCs, etc.., from which the kinematics, mass, albedo and even chemical composition can be derived. This will improve the assessment of the link between terrestrial craters and comet/asteroid impactors.
 Gaia may also determine the time-varying solar apex up to a high precision, and infer the impactors of terrestrial craters from the latitudinal distribution of craters. Considering these connections between the solar apex motion, stellar encounters, LPCs and craters, Gaia will advance our understanding of the bombard history, and even the evolution of the solar system.

However, even with the future Gaia sample of stellar encounters, we cannot reconstruct the cometary impact rate in the inner solar system with an uncertainty less than 50\% more than 200\,Myr ago. A possible solution of this problem is to combine simulations of the cometary impact rate with the cratering records on terrestrial planets and satellites. The craters, which are confirmed to be generated by comets, may be a proper proxy of cometary impact rate if they are dated accurately enough. 

There are also assumptions drawn in designing the Bayesian method for assessing the links between LPCs and encounters. One assumption is that the semi-major axis and the perihelion of a LPC are independent of each other. This assumption is justified by the fact that the distribution of LPCs' perihelia is rather flat from 0\,AU to 10\,AU, and that the correlation between perihelion and semi-major axis is not well modeled due to the uncertainties in Oort cloud models and limited computation power in performing Monte Carlo simulations of the Oort cloud. To apply the Bayesian method to find the culprit star for a LPC in the future Gaia sample of stellar encounters, it is necessary to provide a more comprehensive culprit model. However, scientists may not be concern about the culprit of one LPC but the culprit corresponding to many LPCs over a certain period. In this case, the evidence for a culprit model can be calculated by averaging the evidences for all LPCs. This method aims to investigate whether there is a comet shower caused by a stellar encounter at a given time.  

In sum, combined with accurate simulations of motions of stars, the Sun, and minor bodies in the solar system, Gaia will greatly promote the studies of the Earth's extraterrestrial environment within a Bayesian framework.

\newpage

\chapter{Conclusions}\label{cha:conclusion}
In this dissertation I have presented sets of time series models to predict the extinction rate of species, the temporal distribution of the impact cratering rate and ice sheet deglaciations. These models aim to model the relevance of the solar motion in the Galaxy and the variation of the geometry of the Earth's orbit in modulating the extraterrestrial mechanisms. I compare these orbital models with other time series models in a Bayesian framework, and draw conclusions according to their Bayes factors. This Bayesian method is capable to compare models with different complexities on the same footing by marginalizing the likelihood over parameters. 

The first set of models predict the terrestrial extinction rate using the stellar density local to the Sun as a proxy. The main conclusions and the insights provided by this work are as follows:
\begin{enumerate}[i]
  \item In the Monte Carlo simulations of the solar motion over the past 550\,Myr, most simulated solar motions perpendicular to the Galactic plane are not far from periodic, although a single period can not be accurately inferred due to the uncertainties in the coordinates of current solar phase space as well as the lack of an exact Galactic potential. In contrast, assuming a two-arm Galactic model, the crossing of the solar orbits with the spiral arms are not periodic at all, with many solar orbits crossing the spiral arms only once. Thus any attempt to connect the solar motion with geological phenomena such as mass extinctions has to consider the solar motion as quasi-periodic rather than strictly periodic. 
  \item Compare the orbital models with other time series models, I find that the extinction rate is consistent with being randomly distributed in time. In addition, the Sun's motion does not significantly influence the terrestrial extinction rate, and there is no notable periodicity in mass extinctions. However, this does not mean that there is no pattern in terrestrial extinction events. Rather, the uncertainties in reconstructed solar motions prevent us to draw a conclusion on the positive relevance of the solar motion based on current geological data, or there may be other untested models more suited for the time series of extinction events. 
  \item In this work, the time series models only predict the probability of the occurrence of an extinction event at a given time rather than the probability of the intensity of a mass extinction. In future work, models based on different extraterrestrial mechanisms would be developed to explain both the intensity and the time of the terrestrial extinctions.
\end{enumerate}

The second set of models predict the terrestrial cratering rate based on simulations of the Oort cloud. I use the vertical Galactic tide and the impulses gained by the Sun from stellar encounters as proxies to model the cometary impact rate (i.e.\ dynamical models). I also use a trend component to model the asteroid impact rate and the preservation bias in terrestrial craters. Comparing dynamical models, trend models with other models, I draw the following main conclusions.  
\begin{enumerate}[i]
   \item We find that the craters larger than 5\,km formed over the past 250\,Myr favor models with a trend component that predicts an impact rate increasing towards the present rate from about 200\,Myr ago. This is consistent with results of previous studies \citep{shoemaker98,gehrels94,mcewen97,bailer-jones11}, and can be caused by the disruption of a large asteroid into an asteroid family \cite{bottke07}.
   \item Since the pure trend model has a Bayes factor of the same magnitude as the combination of the trend model and dynamical models, we conclude that either the tide and encounter components are unnecessary in modelling the temporal distribution of craters, or the data cannot effectively discriminate between the models. 
   \item The gravitational force from Galactic bar and spiral arms can make the Sun migrate inward to its current location, leading to large deviations of the Sun's motion from a circular orbit. This will increase the variation of the local stellar density and hence the time-variation of the comet flux. It turns out that the non-axisymmetric components, particularly the spiral arms, can only change the comet impact rate significantly when the Sun is in co-rotation with the spiral arms. 
   \item The simulations of the Oort cloud show that the non-uniform latitude distribution of the perihelia of LPCs can be explained by the Galactic tide, which imposes anisotropic tidal forces on the Oort cloud. I find that the non-uniform longitude distribution can be properly modeled using stellar encounters with preferred directions induced by the solar apex motion. Thus, without invoking a Jupiter-mass solar companion as \cite{matese99,matese11} did, our model can reasonably explain the anisotropic perihelia of LPCs. My findings are consistent with the non-uniform angular distribution of the perihelia of stellar encounters found by \citep{sanchez01}, although the signal is rather weak due to the incompleteness of the Hipparcos catalog of encounters. 
   \item The anisotropic flux of LPCs may not cause a non-uniform longitude distribution of the comet impacts on terrestrial planets and satellites due to their rotations on their axes. However, some latitude variation may be expected, despite a long-term variation in inclination and obliquity of terrestrial planets \citep{feuvre08,werner10}. Considering the small contribution of the comet impact rate to the total impact rate in the inner Solar System, these higher order effects would be difficult to convincingly detect and relate to the solar orbit in the analysis of terrestrial impact craters.
\end{enumerate}

The third set of models, based on a simple climate response to the variations of the Earth's orbit, predict the times of the deglaciations over the past 2\,Myr. The main conclusions from this work are as follows:
\begin{enumerate}[i]
\item I find that obliquity plays a dominant role in pacing both minor and major deglaciations over the past 2\,Myr. In addition, my work demonstrates that it is unnecessary to re-parameterize the obliquity-based model, such as adding a linear trend component as done by \citep{huybers07}, to explain the mid-Pleistocene transition (MPT). I also find that precession becomes important in triggering major deglaciations over the past 1\,Myr after the MPT. This may result from internal changes in the climate system such as the erosion of a continental regolith \citep{clark06}. 
\item I also find that the MPT occurred around 700\,kyr ago with a duration of about 130\,kyr. The mid-time of the MPT inferred using our Bayesian inference method is rather late compared with that found by \cite{clark06} using the method of time-varying power spectrum. This discrepancy may be caused either by different inference methods or by different data sets. In contrast to the hypothesis of gradual MPT \citep{huybers07}, I conclude that the MPT is rather rapid (with a duration slightly more than one glacial-interglacial cycle), and thus the possible internal changes in the climate system are rapid as well. 
\item The geomagnetic field and the Earth's orbital inclination are not likely to cause the Pleistocene deglaciations. This challenges the hypotheses proposed by \cite{muller97} that the variation of the Earth's orbital inclination can influence the climate through the accumulation of interplanetary dust in the Earth's atmosphere. 
\end{enumerate}

Finally, I have quantified the impact of the Gaia survey on researches on the influence of extraterrestrial phenomena on the Earth. The main conclusions are as follows:
\begin{enumerate}[i]
\item  Future Gaia data will allow us to reconstruct the solar orbit with an uncertainty less than 6\% within 100\,Myr of now, under the assumption that models of the asymmetric components in the Galactic potential can be improved by Gaia with the same magnitude as models of axisymmetric components. A Gaia-improved reconstruction of the solar motion may enable convincing studies of the connections between extinction events and extraterrestrial phenomena over the past 100\,Myr. 
\item  Gaia will detect more than 90\% stellar encounters that encountered (will encounter) the solar system over the past (future) 10\,Myr, compared with about 50\% detection efficiency of Hipparcos survey. With an almost complete list of recent stellar encounters provided by Gaia, it will become possible to test my hypothesis that the non-uniform longitude distribution of LPCs' perihelia is caused by the solar apex motion \citep{feng14}. I have also proposed a Bayesian method to assess the link between LPCs and stellar encounters. Gaia will improve this assessment by providing a large sample of stellar encounters, and by determining the encountering times of stars more precisely. 
\item Gaia will detect the masses, sizes and non-gravitational effects of hundreds of minor bodies in the solar system, leading to a new asteroid taxonomy and an accurate reconstruction of the evolution of asteroid dynamical families. This may aid to find the origin of terrestrial extinctions such as the K-Pg extinction event. 
\end{enumerate}

\newpage

%

 \bibliographystyle{mn2e}
\bibliography{astro}
\newpage

\chapter*{\centerline{Acknowledgements}}
\addcontentsline{toc}{chapter}{Acknowledgements}
I would like to thank my wife for supporting me in my pursuit of a PhD, and also for giving birth to two children who sometimes make trouble but always make fun. I am privileged to share her love and companionship. I also thank my parents for bringing me up, and for pushing me toward learning and education while growing up. Without their love and support, I wouldn't have the opportunity to pursue my career as a professional scientist. \\

In addition, I'd like to thank my supervisor, Coryn Bailer-Jones, for his patient guidance as I equipped myself with professional skills. He has given me enough freedom to choose topics, and provided valuable career advice on seeking jobs. I also appreciate his patience through many paper revisions when I didn't seem to be making much improvement. Although I was not very interested in my PhD project at the beginning of my PhD studies, I became more and more excited about interdisciplinary research when I learned about the organic connections between the Earth and the extraterrestrial environment and the impact those links have on terrestrial life. The invaluable experience of touching on different topics in different fields eventually enabled me to complete work of which I am proud of and that will hopefully provide new insights to other researchers.\\

I'd also like to thank the other advisors in my thesis committee, Glenn van de Ven and Anthony Brown, for generously spending time writing reference letters, and for providing valuable suggestions during my PhD studies. I also thank the rest of the faculty and students at MPIA for friendly interactions and for making the institute fun to work at.\\

I would like to thank the IMPRS, the European Union Seventh Framework Programme (FP7/2007-2013), and the Gaia Research for European Astronomy Training (GREAT-ITN) network for supporting my study in Heidelberg, and providing me with opportunities to attend numerous conferences, workshops and schools around the world. I am also grateful to many researchers in the GREAT-ITN network for their help and advice, and for building a sense of belonging among those of us who was just setting out on a career in science. \\

Finally, and most importantly, I thank God. After all, he created the world and give me gifts to explore it.

\end{document}